\RequirePackage[T1]{fontenc}
\documentclass[12pt]{article}

\usepackage[height=8.85in,width=6.25in]{geometry}

\usepackage[utf8]{inputenc}
\usepackage{amsmath}
\usepackage{amssymb}
\usepackage{mathtools}
\numberwithin{equation}{section}
\usepackage{slashed}
\usepackage{braket}
\usepackage[svgnames]{xcolor}
\usepackage[colorlinks,citecolor=DarkGreen,linkcolor=FireBrick,urlcolor=FireBrick,linktocpage,breaklinks=true]{hyperref}
\usepackage{cite}
\usepackage[pdftex]{graphicx}
\usepackage{tikz}
\usepackage{tikz-cd}
\usepackage{times}
\usepackage{courier}
\usepackage{bm}
\usepackage{subfig}

\usepackage{xcolor}
\usepackage{mdframed}

\usetikzlibrary{decorations.pathreplacing,decorations.markings}

\usepackage{mathrsfs} 
\DeclareMathAlphabet{\mymathbb}{U}{BOONDOX-ds}{m}{n}

\usetikzlibrary{arrows.meta}
\newcommand{\midarrow}{\tikz[baseline] \draw[-{Stealth[scale=1.5]}] (0,0) -- +(.1,0);}

\definecolor{dgreen}{rgb}{0, 0.55, 0}
\definecolor{dorange}{rgb}{0.8, 0.33, 0.0}
\definecolor{dorange}{rgb}{1.0, 0.5, 0.31}
\definecolor{dgrey}{rgb}{0.3,0.3,0.3}
\definecolor{grey}{rgb}{0.9,0.9,0.9}

\tikzset{
  on each segment/.style={
    decorate,
    decoration={
      show path construction,
      moveto code={},
      lineto code={
        \path [#1]
        (\tikzinputsegmentfirst) -- (\tikzinputsegmentlast);
      },
      curveto code={
        \path [#1] (\tikzinputsegmentfirst)
        .. controls
        (\tikzinputsegmentsupporta) and (\tikzinputsegmentsupportb)
        ..
        (\tikzinputsegmentlast);
      },
      closepath code={
        \path [#1]
        (\tikzinputsegmentfirst) -- (\tikzinputsegmentlast);
      },
    },
  },
  mid arrow/.style={postaction={decorate,decoration={
        markings,
        mark=at position .5 with {\arrow[#1]{stealth}}
      }}},
}

\tikzset{line/.style={line width=0.25mm},
curve/.style={line,smooth,tension=1},
->-/.style={decoration={
  markings,
  mark=at position #1 with {\arrow[>=stealth]{>}}},postaction={decorate}},
-<-/.style={decoration={
  markings,
  mark=at position #1 with {\arrow[>=stealth]{<}}},postaction={decorate}},
}

\renewenvironment{figure}[1][]{
  \begin{originalfigure}[#1]
    \begin{mdframed}[linecolor=black!0,backgroundcolor=black!1]
}{
    \end{mdframed}
  \end{originalfigure}
}

\newcommand{\ie}{\begin{equation}\begin{aligned}}
\newcommand{\fe}{\end{aligned}\end{equation}}

\definecolor{dgreen}{rgb}{0, 0.55, 0}
\definecolor{dorange}{rgb}{0.8, 0.33, 0.0}
\definecolor{dorange}{rgb}{1.0, 0.5, 0.31}

\newcommand\oSigma{\overline{\Sigma}}

\newcommand{\oQ}{\overline{Q}}

\usetikzlibrary{shapes.geometric}
\usetikzlibrary{shapes.multipart}
\usetikzlibrary{shapes.multipart}
\usetikzlibrary{decorations.pathmorphing}
\tikzset{snake it/.style={decorate, decoration=snake}}

\def\ZZ{\mathbb{Z}}

\def\p{\partial}
\newcommand{\bea}{\begin{eqnarray}}
\newcommand{\eea}{\end{eqnarray}}
\def\no{\nonumber}
\def\m{\mu}

\def\a{\alpha}
\def\b{\beta}
\def\g{\gamma}

\def\cL{\mathcal{L}}
\def\cR{\mathcal{R}}
\def\cX{\mathcal{X}}

\def\bv{{\mathbf{v}}}
\def\bw{{\mathbf{w}}}
\def\bu{{\mathbf{u}}}

\def\bm{{\mathbf{m}}}

\def\eps{\epsilon}
\def\half{{1\over 2}}
\def\cN{\mathcal{N}}
\def\cM{\mathcal{M}}
\def\cC{\mathcal{C}}
\def\cA{\mathcal{A}}

\def\cV{\mathcal{V}}
\def\ST{\mathsf{ST}}
\def\sfC{\mathsf{C}}
\def\sfT{\mathsf{T}}

\def\A{\mathsf{A}}
\def\B{\mathsf{B}}
\def\C{\mathsf{C}}
\def\D{\mathsf{D}}

\usepackage{dsfont}

\begin{document}
\begin{titlepage}

\begin{flushright}
KYUSHU-HET-335 \\
OU-HET-1286
\end{flushright}

\vskip 3cm

\begin{center}

{\Large \bfseries The SymTFT for $N$-ality defects}

{\large \bfseries Part I}

\vskip 2cm
Justin Kaidi$^{1,2}$,
Xiaoyi Shi$^3$,
Soichiro Shimamori$^4$, Zhengdi Sun$^5$
\vskip 2cm

\begin{tabular}{ll}
1 &Institute for Advanced Study, Kyushu University, Fukuoka 812-8581, Japan \\
2 &Department of Physics, Kyushu University, Fukuoka 819-0395, Japan\\
3 & Department of Physics, University of Washington, Seattle, WA, 98195, USA\\
4 & Department of Physics, The University of Osaka, Toyonaka 560-0043, Japan \\
5 & Mani L. Bhaumik Institute for Theoretical Physics,
UCLA, CA 90095, USA \\
\end{tabular}

\vskip 1cm

\end{center}

\noindent\textbf{Abstract:} In order to obtain the SymTFT for a theory with an $N$-ality extension of a discrete, Abelian group $G$, 
one begins by considering a bulk $G$-gauge theory, and then gauges an appropriate $\ZZ_N$ symmetry. This procedure involves three choices: 
the choice of a suitable bulk $\ZZ_N$ symmetry, of a fractionalization class, and of a discrete torsion. The first choice is, 
somewhat surprisingly, the most involved, and in this paper we discuss it in detail. 
In particular, we show that the choice of bulk $\ZZ_N$ symmetry determines all boundary $F$-symbols with a single incoming $N$-ality defect,
and that any theory with an $N$-ality symmetry is invariant under a certain twisted gauging given in terms of these $F$-symbols. 
These $F$-symbols can furthermore be input into the pentagon identities to obtain the other $F$-symbols, up to freedoms related 
to the choices appearing in the second and third steps of bulk gauging. 
Although many of our results hold for general $N$, we restrict ourselves in some places to the case of $N=p$ prime. In particular, for generic triality defects, we acquire explicit $F$-symbols which are reminiscent of those in Tambara-Yamagami fusion categories.

\end{titlepage}

\setcounter{tocdepth}{2}

\newpage

\tableofcontents

\section{Introduction}
In recent years, the notion of symmetry has been generalized in a number of ways, including to so-called ``non-invertible symmetries.'' 
Non-invertible symmetries were originally studied in two dimensions, for which there is a rather extensive literature, see e.g. \cite{verlinde1988fusion,Petkova:2000ip,Fuchs:2002cm,Bhardwaj:2017xup,Chang:2018iay,Lin:2022dhv,Komargodski:2020mxz,Tachikawa:2017gyf, Frohlich:2004ef, Frohlich:2006ch, Frohlich:2009gb, Carqueville:2012dk, Brunner:2013xna, Huang:2021zvu, Thorngren:2019iar, Thorngren:2021yso, Lootens:2021tet, Huang:2021nvb, Inamura:2022lun}, though recent progress has extended these results to higher dimensions  \cite{Kaidi:2021xfk,Choi:2021kmx, Koide:2021zxj,Choi:2022zal,Hayashi:2022fkw,Arias-Tamargo:2022nlf,Roumpedakis:2022aik,Bhardwaj:2022yxj,Kaidi:2022uux,Choi:2022jqy,Cordova:2022ieu,Antinucci:2022eat,Bashmakov:2022jtl,Damia:2022rxw,Damia:2022bcd,Choi:2022rfe,Lu:2022ver,Bhardwaj:2022lsg,Lin:2022xod,Bartsch:2022mpm,Apruzzi:2022rei,GarciaEtxebarria:2022vzq, Benini:2022hzx, Wang:2021vki, Chen:2021xuc, DelZotto:2022ras, Heckman:2022muc,Kaidi:2022cpf,Bashmakov:2022uek,Bonetti:2024etn}.

As with standard group-like symmetries, an important problem for physicists is to understand the representation theory of non-invertible symmetries \cite{Bartsch:2022mpm,Bartsch:2022ytj,Bartsch:2023wvv}. One of the most important tools in the study of the representation theory of non-invertible symmetries is the so-called \textit{Symmetry TFT} (SymTFT), which has been developed in a variety of directions in a vast series of works  including but not limited to \cite{Freed:2012bs,Freed:2018cec,Kong:2020cie,Gaiotto:2020iye,Burbano:2021loy, Apruzzi:2021nmk,Apruzzi:2022dlm, Chatterjee:2022kxb, Chatterjee:2022tyg, Moradi:2022lqp, Freed:2022qnc,Kaidi:2023maf,DelZotto:2024tae,Putrov:2025xmw,DelZotto:2025yoy,Schafer-Nameki:2025fiy,Qi:2025tal,Jia:2025bui}. 

Amongst non-invertible symmetries, the simplest are those of Tambara-Yamagami type \cite{tambara1998tensor}, which contain only a single non-invertible element known as a ``duality defect''. The SymTFT for duality defects was discussed in great detail in \cite{Kaidi:2022cpf}, in both 2d/3d and 4d/5d; additional results can be found in \cite{Bhardwaj:2023idu,Antinucci:2022vyk,Bhardwaj:2024xcx}.\footnote{Previous results from the mathematics literature include \cite{izumi2001structure,gelaki2009centers}, and from the physics literature include \cite{Barkeshli:2014cna,2015arXiv150306812T}. } In addition, for some special cases where the non-invertible symmetry can be obtained from discrete gauging of certain invertible symmetries, SymTFTs are simply finite gauge theories and some explicit examples have been studied in \cite{Lu:2024lzf,Bhardwaj:2025jtf,Bhardwaj:2025piv}. Beyond these simplest examples, the list of explicit SymTFTs appearing in the Physics literature grows more sparse, though some examples include the SymTFT for  $G$-ality defects discussed in \cite{Lawrie:2023tdz,Apruzzi:2024cty}, as well as that for Haagerup symmetry discussed in \cite{Bottini:2025hri,Albert:2025umy}. The current series of papers aims to analyze in full detail the SymTFT for what is perhaps the most natural generalization of duality defects.

In particular, the focus of this paper is on the SymTFT for $N$-ality defects (occasionally restricting to $N=p$ being a prime number, though most of our results hold for more general $N$).\footnote{Earlier studies of the structure of $N$-ality, and more generally $G$-ality defects can be found in \cite{Lu:2024ytl,Ando:2024hun,Lu:2024lzf,Maeda:2025rxc}. }
The fusion category we are interested in has $|G|$ elements  $g \in G$ with fusion rules given by the group multiplication for the Abelian group $G$, as well as $N-1$ non-invertible elements $Q_I$ for $I=1, \dots, N-1$ and $I \simeq I + N$, with fusion rules given by,
\bea
Q_I \times Q_J = \left\{ \begin{matrix}  \a_{IJ}\, Q_{I+J}~, &&J \neq N-I  \\
\sum_{g \in G} g&& J = N-I~
\end{matrix} \right.  ~, \hspace{0.5 in} g \times Q_I = Q_I \times g = Q_I~.
\eea 

From these fusion rules, we note that $\overline{Q_I} = Q_{N-I}$. 
The coefficients $\a_{IJ}$ in the fusion rules can be fixed by comparing the quantum dimension $\langle Q \rangle = \langle \overline Q \rangle$ computed from the first and second equations, giving the result $\a_{IJ} = \sqrt{|G|}$. Since the coefficients in the fusion rules should be integers, this tells us that the $N$-ality fusion ring for $N>2$ is only consistent if $|G|$ is a perfect square. 
For the rest of this paper, we will consider the case of $G = \ZZ_M^{n}$, with $n \in2 \mathbb{N}$ when $N>2$ and $M$ is not a perfect square.

\begin{figure}[!tbp]
	\centering
	\begin{tikzpicture}[scale=0.8]
	
		\shade[line width=2pt, top color=blue!30, bottom color=blue!5] 
	(0,0) to [out=90, in=-90]  (0,3)
	to [out=0,in=180] (6,3)
	to [out = -90, in =90] (6,0)
	to [out=180, in =0]  (0,0);
	
	\draw[very thick, violet] (-7,0) -- (-7,1.5);
	\draw[very thick, blue] (-7,1.5) -- (-7,3);
	\node[above] at (-7,3) {$Z_{\cX}[A]$};
	\node[below] at (-7,0) {$Z_{\cX/\ZZ_N}[A]$};
	\node at (-7,1.5) [circle,fill,dgreen, inner sep=1.5pt]{};
	\node[left,dgreen] at (-7,1.5) {$Q$};

	\draw[thick, snake it, <->] (-1.7,1.5) -- (-5, 1.5);
	\node[above] at (-3.3,1.6) {Expand\,/\,Shrink};
	
	\draw[thick] (0,0) -- (0,3);
	\draw[thick] (6,0) -- (6,3);
	\draw[ultra thick,orange] (3,0) -- (3,1.5);
	\node[above] at (3,1.7) {$\frac{2\pi}{M} \widehat{\mathbf{a}} \cdot \cup \delta \mathbf{a}$};
	\node at (3,1.5) [circle,fill,dgreen, inner sep=1.5pt]{};
	\node[below] at (0,0) {$\langle \mathbf{A}| $};
	\node[below] at (6,0) {$|\cX\rangle $}; 
	
	
	\end{tikzpicture}
	
	\caption{A $(1+1)$d QFT $\cX$ with $\ZZ_M^n$ symmetry and another $(1+1)$d QFT $\cX/\ZZ_M^n$  are separated by a topological interface $Q$. This setup can be expanded into a $(2+1)$d slab, where the $(2+1)$d $\ZZ_N$ SymTFT has an insertion of a twist defect parallel to the Dirichlet boundary. The particular choice of gauging (discrete torsion etc.) is determined by the choice of bulk condensation defect ending on the twist defect.  }
	\label{fig:twistdefshrink}
\end{figure}
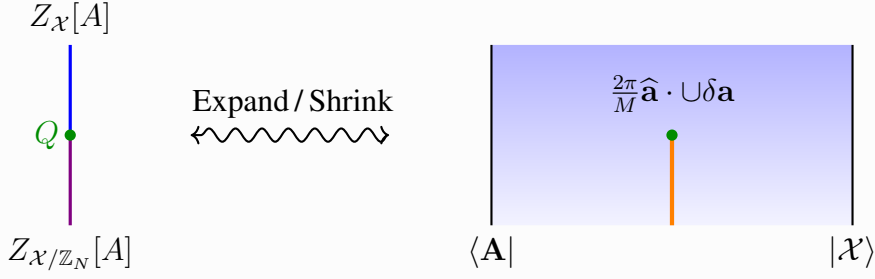

Amongst other things, in this work we will be interested in how different $N$-ality categories in 2d arise from the bulk 3d SymTFT, and in particular 
how boundary $F$-symbols can be computed via bulk data. In order to obtain an $N$-ality defect $Q$ on the boundary, one begins by considering an $N$-ality \textit{interface} separating a theory $\cX$ from the gauged theory  $\cX/\ZZ_M^n$. Of course, there are multiple different choices of gauging, differing by discrete torsion, the coupling of dynamical gauge fields to dual background gauge fields, and stacking with SPT phases for the dual background gauge fields. From the bulk perspective, the setup of interest is the insertion of a twist defect in the bulk, with the attached codimension-1 defect running parallel to the topological boundary; see Figure \ref{fig:twistdefshrink}. As we will see, this codimension-1 defect is a condensation defect generating a certain bulk $\ZZ_N$ symmetry, with the different choices of bulk $\ZZ_N$ symmetry corresponding to different gaugings on the boundary. In order to turn $Q$ from an interface into a defect, we must require that the theories  $\cX$ and  $\cX/\ZZ_M^n$ are equivalent, and from the bulk perspective this means that we must gauge the bulk $\ZZ_N$ symmetry, so that the corresponding condensation defect becomes transparent, thereby making the twist defect into a genuine line operator in the bulk.

To summarize then,  in order to obtain an $N$-ality defect on the boundary, one must consider the gauging of an appropriate $\ZZ_N$ symmetry in the bulk. 
This gauging proceeds in three steps,
\begin{enumerate}
\item First, one must choose a $\ZZ_N$ symmetry (up to equivalence transformation). As will be discussed below, the set of symmetries of the bulk $\ZZ_M^n$ gauge theory is $O(n,n;M)$, and there may be many choice of $\ZZ_N \subset O(n,n;M)$. Note that this choice already determines all boundary $F$-symbols involving a single incoming $N$-ality defect, and these in turn will be shown to determine the particular gauging of $\ZZ_M^n$ under which the boundary theory is invariant.

\item Next, we must choose a fractionalization class given by the twisted cohomology group $H^2_T (\ZZ_N, \ZZ_M^{2n})$, where $T$ describes the action of $\ZZ_N$ on $\ZZ_M^{2n}$.

\item Finally, we must choose a discrete torsion in $H^3(\ZZ_N, U(1))$ with which to perform the gauging.
\end{enumerate}

In fact, the first step turns out to be the hardest part. 
Once the first step is completed, the data of the remaining steps can be strongly constrained and easily organized. 
As such, we will mainly focus on the first step, and perform the second and third steps only in a concrete example: namely, the case of triality with $N=3$. 

Because this paper is rather technical, we will begin with a brief summary of all of our results, hopefully serving as a guide through the main text. 
Let us reemphasize that although most of our results hold for generic $N$, in certain places we do restrict to $N=p$ being prime. 
The steps in which it is necessary to assume this will be made explicit. 

\subsection{$N$-ality and invariance under twisted gauging}

We begin in Section \ref{sec:invarianceundergauging} by showing that any 2d theory with an $N$-ality defect $Q$ is invariant under an appropriate twisted gauging. The particular discrete torsions and couplings 
between dynamic and background gauge fields specifying this gauging are determined entirely in terms of the $F$-symbols involving a single incoming and outgoing $Q$. At the level of the $(g,h)$ twisted-twined torus partition function $Z_{\mathcal{X}}[g,h]$ defined in \eqref{eq:ghpf}, we find concretely the following result,
\bea
\label{eq:introgauging}
    Z_{\mathcal{X}}[g,h] = {1 \over d_Q^2}\frac{F^{Q}_{Q,h,g}}{F^{Q}_{Q,g,h}} \sum_{\widetilde{g},\widetilde{h} \in G} \frac{F^{Q}_{\widetilde{h},Q,g^{-1}}}{F^{Q}_{\widetilde{g},Q,h^{-1}}} \frac{F^{Q}_{\widetilde{h},\widetilde{g},Q}}{F^Q_{\widetilde{g},\widetilde{h},Q}} Z_{\mathcal{X}}[\widetilde{g},\widetilde{h}] ~. 
\eea
The coefficients above may be split up into three pieces. First, the ratio ${F^Q_{\widetilde h, \widetilde g, Q}/ F^Q_{\widetilde g, \widetilde h, Q}}$ captures the discrete torsion for the gauging. Second, the ratio ${F^Q_{\widetilde h, Q, g^{-1}} / F^Q_{\widetilde g, Q, h^{-1}}}$ specifies how the background field of the dual symmetry couples to the dynamical gauge field. Finally, the ratio ${F^Q_{Q,h,g} /  F^Q_{Q,g,h}}$ specifies the SPT stacked after gauging.  

The proof of this formula proceeds in a similar way to the usual proof that a Kramers-Wannier-type duality defect leads to invariance under gauging---namely, a bubble of the defect is nucleated on the torus and deformed to wrap the sum of all cycles, upon which the fusion rules can be used to reduce the configuration to an appropriate mesh of invertible lines. Because in the $N$-ality case the orientation of the $Q$ line is important, the rigorous computation is quite a bit more involved than for the  duality case.

\subsection{$F$-symbols with a single incoming $Q$ from the bulk}
\label{sec:introFsymbols}

As it turns out, the $F$-symbols involving a single incoming $Q$ are completely determined by the choice of bulk symmetry $A \subset O(n,n;M)$ whose twist defect gives the boundary $N$-ality defect. There are two ways in which this can be shown.
\begin{enumerate}
\item First, we may work out the global action of the $A$ condensation defect $D_A$ on the topological boundary, which should implement the twisted gauging in (\ref{eq:introgauging}). In particular, if we parameterize the $2n \times 2n$ dimensional matrix $A$ in terms of $n\times n$ dimensional matrices via 
\bea
A =\left(\begin{matrix}\A & \C \\ \B & \D \end{matrix}  \right) ~,
\eea
and define the following quantities 
\bea
    \chi := (\B^{-1})^T ~, \hspace{0.3 in} \varphi := \A\B^{-1} ~, \hspace{0.3 in} \widetilde{\varphi} := \B^{-1} \D ~,
\eea
then we may work out the following action of $D_A$ on the topological boundary labelled by background field $\mathbf{A}$, 
\bea
    \langle \mathbf{A}| D_{A} = \sum_{\mathbf{a}} \langle \mathbf{a}| \exp\left(\frac{\pi i}{M}\int \mathbf{a}\cdot \widetilde{\varphi}^T \cup \mathbf{a} -\frac{2\pi i}{M}\int \mathbf{a} \cdot \chi \cup \mathbf{A} + \frac{\pi i}{M}\int \mathbf{A}\cdot \varphi^T \cup \mathbf{A}\right) ~.\hspace{0.4 in}
\eea
Taking the inner product with the dynamical boundary $|\mathcal{X}\rangle$ to obtain the partition function and comparing to (\ref{eq:introgauging}), we then read off the $F$-symbols directly in terms of the matrix $A$, namely
\bea
\label{intro:bdyF1}
F_{Q, g,h}^Q = \omega^{\sum_{i<j} g_i \varphi_{ij} h_j }~, \hspace{0.5 in} F_{g, Q, h}^Q = \omega^{-g^T \chi h}~, \hspace{0.5 in} F_{g,h,Q}^Q = \omega^{\sum_{i<j} g_i\widetilde \varphi_{ij} h_j }~.
\eea
This method is described in detail in Section \ref{sec:globalactionA}. 

\item Alternatively, we may first work out the $F$-symbols of twist defects directly in the bulk, and then take the boundary limit. In particular, say that we have a series of bulk twist defects $\Sigma^{(i)}$, with $i$ an arbitrary label for the moment. In general, the boundary $N$-ality defect $Q$ can be expanded in terms of these $\Sigma^{(i)}$, but one must be careful to note that the hom-spaces $\mathrm{Hom}(Q, \Sigma^{(i)}|_\p)$ may have order larger than one.  This makes the boundary limit somewhat involved, but the computation is nevertheless possible, and is performed in Section \ref{sec:bulktoboundaryFsymb}, with some useful results about trivalent junctions relegated to Appendix \ref{app:trivalentvertices}. 
\end{enumerate}

Of the two methods described above, the second one is significantly more involved, and in particular relies on a number of results about the module structure of twist defects in the bulk SymTFT, which we now summarize. 

\subsection{Condensation defects, twist defects, and bimodules}

For any element $A \in O(n,n;M)$, we first work out the explicit form of the condensation defect $D_A$ corresponding to it. As will be shown in Section \ref{sec:conddefect}, the concrete expression for the operator $D_A$ wrapping a submanifold $M_2$ is given by 
\bea
\label{eq:introconddef}
 D_A(M_2):= {1\over |H^0(M_2, \ZZ_M)|^{r_A}}\sum_{\widetilde \g_1, \dots,\widetilde\g_{r_A} \in H_1(M_2, \ZZ_M)} e^{{2 \pi i \over M} \half \sum_{i,j=1}^{r_A} \widetilde Q^A_{ij} \langle \widetilde\g_i, \widetilde\g_j \rangle} \prod_{i=1}^{r_A} L_{R_i}(\widetilde\gamma_i)~,
\eea
where $r_A$ is the rank of the matrix $1-A$, the vectors $R_i$ for $i = 1, \dots, r_A$ are a basis for the image of $1-A$, and $\widetilde Q_{ij}^A$ is a matrix defined in and around (\ref{eq:widehatQexpression}).\footnote{Note that for non-square-free $M$, it is possible that the vectors $R_i$ do not generate $\mathbb{Z}_M^{r_A}$, but rather a proper subset of it. For simplicity, we will restrict to the case where the $R_i$ do generate $\mathbb{Z}_M^{r_A}$.} The fusion rules for these condensation defects are computed in Section \ref{sec:conddefect} and Appendices \ref{app:DAfusions} and \ref{app:matrixnotation}, and are 
\bea
D_A(M_2) \times {D_A}^\dagger (M_2) &=& \chi[M_2, \ZZ_M]^{-r_A}~,
\no\\
L_{(\mathbf{e}, \mathbf{m})}(\gamma)\times D_A(M_2)   &=& D_A(M_2) \times  L_{A(\mathbf{e}, \mathbf{m})}(\gamma)~,
\no\\
D_{A^I}(M_2) \times D_{A^J}(M_2) &=& \chi[M_2; \ZZ_M]^{-r_A} D_{A^{I+J}}(M_2)~,
\eea
where $\chi[M_2; \ZZ_M]$ is the Euler character of the manifold $M_2$ with coefficients in $\ZZ_M$. Note that the third fusion rule holds only for $N=p$---for more generic $N$, the fusion of condensation defects $D_A$ and $D_B$ is more complicated, as discussed around (\ref{eq:complicatedfusions}). 

Of course, these condensation defects can be obtained by condensing (i.e. higher gauging) an appropriate algebra object on the codimension-1 surface $M_2$. In particular, the algebra is given by the following sum,
\bea
\cA_A := \bigoplus_{\bu \in \mathrm{im}(1-A)} L_\bu~. 
\eea
To fully specify the algebra object though, we must also specify a multiplication (and comultiplication) operation $\mu_A(\bu , \widetilde \bu)$, subject to various consistency conditions discussed in Section \ref{sec:algebraobjs}. In fact, this data is directly related to the coefficients appearing in the definition of the condensation defects in (\ref{eq:introconddef}), and in Section \ref{sec:algebraobjs} we derive the following result, 
\bea
\label{eq:intromudef}
\mu_{A^x}(\bu, \widetilde \bu) = \omega^{\sum_{i< j} u_i \left\{ \left[ (R^{-1})^T \widetilde Q^{A^x} R^{-1}\right]_{ij} - \sum_{m<n} (R^T \Omega^T R)_{mn} (R^{-1}_{mi} R^{-1}_{nj} -R^{-1}_{mj} R^{-1}_{ni})  \right\} {\widetilde u}_j}
\eea
where $R$ is a $2n \times 2n$ matrix whose first $r_A$ columns are the $R_i$, and whose final $2n - r_A$ columns are all zero.

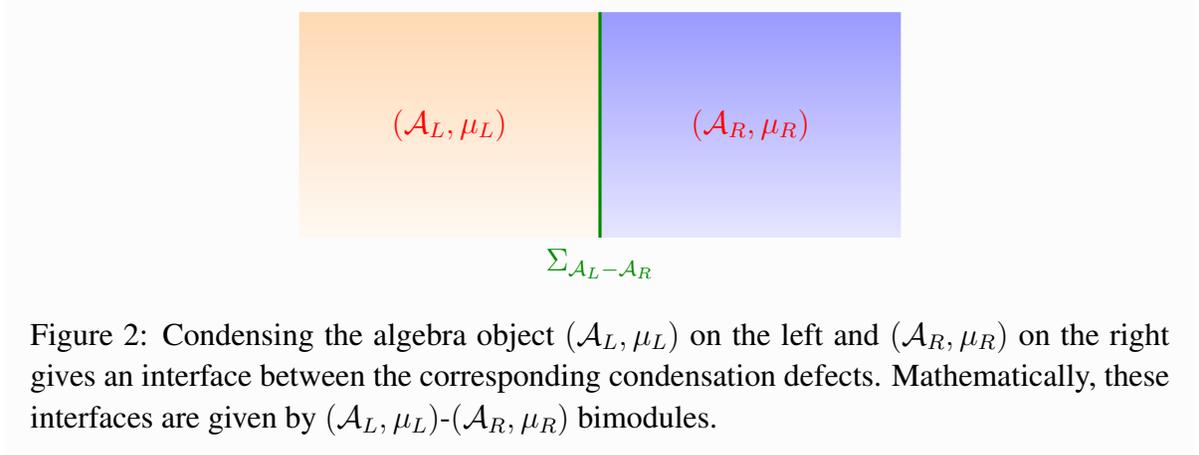
\begin{figure}[t]
	\centering
	\begin{tikzpicture}

	\shade[line width=2pt, top color=orange!30, bottom color=orange!5] 
	(-4,0) to [out=90, in=-90]  (-4,3)
	to [out=0,in=180] (0,3)
	to [out = -90, in =90] (0,0)
	to [out=180, in =0]  (-4,0);
	
	\shade[line width=2pt, top color=blue!40, bottom color=blue!10] 
	(0,0) to [out=90, in=-90]  (0,3)
	to [out=0,in=180] (4,3)
	to [out = -90, in =90] (4,0)
	to [out=180, in =0]  (0,0);

	\draw[dgreen, very thick] (0,0) -- (0,3);

	\node[red] at (-2,1.5) {$(\cA_L, \mu_L)$};
	\node[red] at (2,1.5) {$(\cA_R, \mu_R)$};
	\node[dgreen, below] at (0,0) {$\Sigma_{\cA_L - \cA_R}$};

	\end{tikzpicture}
	
	\caption{Condensing the algebra object $(\cA_L, \mu_L)$ on the left and $(\cA_R, \mu_R)$ on the right gives an interface between the corresponding condensation defects. Mathematically, these interfaces are given by $(\cA_L, \mu_L)$-$(\cA_R, \mu_R)$ bimodules. }
	\label{fig:basicsetup}
\end{figure}

Having discussed condensation defects, we next discuss the interfaces between different condensation defects, including their boundaries (i.e. twist defects). An interface between condensation defects obtained by condensing $\cA_L$ on the left and $\cA_R$ on the right---see Figure \ref{fig:basicsetup}---is described mathematically by an $(\cA_L, \cA_R)$-bimodule. In the case of relevance to us, which are $(\cA_{A^{x-I}}, \cA_{A^x})$-bimodules for arbitrary $x,I \in \ZZ_{N}$, we denote the bimodules by $\Sigma_{A^{x-I}-A^x}^{[\bw], \rho}$, where $[\bw] \in \mathrm{coker}(1-A)$ and $\rho$ denotes a non-trivial projective representation of $\mathrm{im}(1-A)$, satisfying, 
\bea
\rho(\bu) \rho(\widetilde \bu) = {\mu_{A^x}(\bu, \widetilde \bu) \over \mu_{A^{x-I}}( \bu, \widetilde \bu)} \rho(\bu + \widetilde \bu)~. 
\eea
The projective phase is given explicitly in terms of $A$ using (\ref{eq:intromudef}). Intuitively, this structure arises since interfaces can absorb lines in $\mathrm{im}(1-A)$, and hence interfaces should be labelled by cosets $[\bw] \in \mathrm{coker}(1-A) = \ZZ_{M}^{2n}/\mathrm{im}(1-A)$, such that 
\bea
\Sigma^{[\bw], \rho}_{A^{x-I}-A^x} = \bigoplus_{\bw \in [\bw]} n_\bw L_\bw~
\eea
for some coefficients $n_\bw = \mathrm{dim}\, \mathrm{Hom}(\Sigma^{[\bw], \rho}, L_\bw)$. These coefficients can be non-trivial, and the matrix $\rho$ captures data about them. 
Multiplication of two bimodules is given by the \textit{balanced tensor product}, as discussed in Section \ref{sec:twistdefects2}.

After having developed all of the above machinery, one is finally able to compute the bulk $F$-symbols. As computed in Section \ref{sec:bulktoboundaryFsymb}, the $F$-symbols involving a single incoming twist defect are found to be as in (\ref{eq:bulkFsymbols}). As a special case, the simplest twist defect $\Sigma^{[0]}$ (with trivial $\rho$) has the following $F$-symbols, 
\bea
    F^{\Sigma^{[0]}}_{\Sigma^{[0]}, \bu, \widetilde{\bu}} = \frac{R^{\bu,\Sigma^{[0]}}R^{\widetilde{\bu},\Sigma^{[0]}}}{R^{\bu + \widetilde{\bu},\Sigma^{[0]}}} \frac{\mu_A(\widetilde{\bu},\bu)}{R^{\widetilde{\bu},\bu}} ~, \quad F^{\Sigma^{[0]}}_{\bu,\Sigma^{[0]}, \widetilde{\bu}} = \frac{1}{R^{\widetilde{\bu},\bu}} \frac{\mu_A(\widetilde{\bu},\bu)}{\mu_A(\bu,\widetilde{\bu})} ~, \quad F^{\Sigma^{[0]}}_{\bu,\widetilde{\bu},\Sigma^{[0]}} = \mu_A(\bu,\widetilde{\bu})^{-1} ~,\no\\
\eea
where $\bu, \widetilde{\bu} \in \mathrm{im}(1-A)$ (lest the outgoing twist defect cease to be $\Sigma^{[0]}$) and $R^{\bv, \Sigma}$ is the braiding between the bulk line $L_{\bv}$ and the twist defect $\Sigma$, defined in (\ref{eq:RvSigmadef}).  Taking the boundary limit and making a particular choice of gauge then allows one to match these expressions with those in (\ref{intro:bdyF1}), though this matching is highly non-trivial and we content ourselves with checking it in just a few concrete examples. 

\subsection{Computing the remaining $F$-symbols}

Finally, having computed the $F$-symbols with a single incoming $Q$ in two distinct ways, we then show how the remaining $F$-symbols can be computed by explicitly solving the pentagon identities. Here, for concreteness, we restrict to the example of triality, i.e. $N=3$, for which things are worked out in Section \ref{sec:pentagons}, with the final result given in (\ref{eq:fullFsymbols1}). We want to highlight that $F_{g,Q,h}^Q$, viewed as a non-degenerate bicharacter of $G$, plays a similar role as the symmetric non-degenerate bicharacter in the TY fusion category, except that it now satisfies a more exotic condition \eqref{eq:Upsilondef}, instead of just being symmetric.  This bicharacter is the only input data needed to specify the choice of the bulk $\mathbb{Z}_3$ symmetry. Note that in general there can be multiple inequivalent solutions to the pentagon identity for given inputs (\ref{intro:bdyF1}), and these different choices correspond to different choices of symmetry fractionalization class and bulk discrete torsion, mentioned above. The result that we give is one solution, corresponding to trivial fractionalization class. Extensions to non-trivial fractionalization class will be discussed in future work \cite{upcoming}. 

\subsection{The SymTFT for $N$-ality defects}
Up until this point, the entire discussion involved the analysis of a bulk $\ZZ_M^n$ gauge theory, which is the SymTFT for a boundary $\ZZ_M^n$ symmetry.
As mentioned above, in order to get the SymTFT for the full $N$-ality symmetry, we must now gauge one of the $\ZZ_N$ symmetries discussed above. 
In Section \ref{sec:SymTFT}, we finally discuss this gauging for the special case where $N = p$ is a prime number, which gives the following spectrum of line operators,
\begin{itemize}
\item Line operators $\widehat L^k_{[\![\mathbf{v}_d]\!]}$ with $k = 0, \dots, d-1$, where $\mathbf{v}_d \in \cV_{d,M}(A)$ for $d =1,p$ and
the spaces $\cV_{d,M}(A)$ are defined in (\ref{eq:cVdef}). 
These lines have quantum dimension ${p\over d}$. 

\item Line operators $\widehat \Sigma^{[\mathbf{w}], k}$ with $k = 0, \dots, p-1$, where $\mathbf{v} \in \mathrm{coker}_M(1-A^i)$. 
These lines have quantum dimension $M^{r_{A^i}\over 2}$. 
\end{itemize}
The fusion of these line operators is also analyzed in several examples. 
\newline
\newline
Finally, let us note that the current paper is the first in a series of papers. Upcoming work will include a more detailed discussion of the SymTFT, including the complete set of fusion rules for general cases and the braiding of bulk lines (thereby giving the full representation category of the tube algebra). It will also incorporate non-trivial symmetry fractionalization, and will present some physical applications \cite{upcoming}.

\subsection*{Conventions:} 
Throughout this paper, we use the following conventions for the labels on bulk lines,
\begin{itemize}
\item $\bv, \widetilde \bv, \dots \in \ZZ_M^{2n}$
\item $\bu, \widetilde\bu, \dots \in \mathrm{im}(1-A)$
\item $[\bw], [\widetilde\bw], \dots \in \mathrm{coker}(1-A)$
\item $\bw, \widetilde\bw, \dots  \in \ZZ_M^{2n}$ such that $\bw \in [\bw], \widetilde\bw \in [\widetilde\bw], \dots$
\item $\mathbf{e}, \widetilde{\mathbf{e}},\dots \in \ZZ_M^{2n}$ are electric labels with $\mathbf{e}= (e_1, \dots, e_n; 0,\dots, 0)$
 \item $\mathbf{m}, \widetilde{\mathbf{m}}, \dots \in \ZZ_M^{2n}$ are magnetic labels with $\mathbf{m}= (0, \dots, 0; m_1, \dots, m_n)$
\end{itemize}
For example, a generic bulk line will be written as $L_\bv$, while the lines making up the algebra object $\cA_A$ are written as $L_\bu$.

\section{$N$-ality defects and invariance under twisted gauging}

\label{sec:invarianceundergauging}
We begin our discussions by showing that the existence of an $N$-ality extension of a discrete symmetry means that the theory is invariant under a certain twisted gauging of that symmetry, as discussed around (\ref{eq:introgauging}). In order to do so, let us first establish some of our conventions.

Given a fusion algebra $a \times b = \sum_{c} N_{ab}^c \,c$, we assign to each element $a$ a topological line, and define trivalent junctions of topological lines as follows,
\bea
 \begin{tikzpicture}[baseline={([yshift=-1ex]current bounding box.center)},vertex/.style={anchor=base,
    circle,fill=black!25,minimum size=18pt,inner sep=2pt},scale=0.4]
   \draw[->-=0.5,thick] (-2,-2) -- (0,0);
   \draw[->-=0.5,thick] (2,-2) -- (0,0);
   \draw[->-=0.7,thick] (0,0)--(0,2) ;
     \node[below] at (-2,-2.2) {$a$};
     \node[below] at (2,-2) {$b$};
     \node[above] at (0,2) {$c$};
     \node[right] at (0,0) {\scriptsize$\mu$};
    \end{tikzpicture} \in\, \mathrm{Hom}(a\times b, c)~, \hspace{0.4 in}
     \begin{tikzpicture}[baseline={([yshift=-1ex]current bounding box.center)},vertex/.style={anchor=base,
    circle,fill=black!25,minimum size=18pt,inner sep=2pt},scale=0.4]
   \draw[->-=0.6,thick] (0,-2) -- (0,0);
   \draw[->-=0.6,thick]  (0,0) -- (2,2);
   \draw[->-=0.6,thick] (0,0)--(-2,2) ;
     \node[above] at (2,2) {$b$};
     \node[above] at (-2,2) {$a$};
     \node[below] at (0,-2) {$c$};
       \node[right] at (0,-0.2) {\scriptsize$\overline \mu$};
    \end{tikzpicture}
    \in\, \mathrm{Hom}(c,a\times b)~,
\eea
which take values in the complex vector spaces $\mathrm{Hom}(a\times b, c)$ and $\mathrm{Hom}(c, a\times b)$ of dimension $N_{ab}^c$.  
Basis vectors of the hom-spaces are chosen such that they satisfy the following completeness and orthogonality relations,
\bea
\label{eq:basisconventions}
\begin{tikzpicture}[baseline={([yshift=-1ex]current bounding box.center)},vertex/.style={anchor=base,
    circle,fill=black!25,minimum size=18pt,inner sep=2pt},scale=0.4]
   \draw[->-=0.5,thick] (0,0) -- (0,4);
   \draw[->-=0.5,thick] (2,0) -- (2,4);
     \node[below] at (0,-0.2) {$a$};
     \node[below] at (2,0) {$b$};
    \end{tikzpicture} 
    =\sum_{c} \sum_{\mu=1}^{N^c_{ab}} \sqrt{d_c \over d_a d_b} \,\,\,
    \begin{tikzpicture}[baseline={([yshift=-1ex]current bounding box.center)},vertex/.style={anchor=base,
    circle,fill=black!25,minimum size=18pt,inner sep=2pt},scale=0.4]
   \draw[->-=0.7,thick] (0,0) -- (1,1);
   \draw[->-=0.7,thick] (2,0) -- (1,1);
    \draw[->-=0.6,thick] (1,1) -- (1,3);
     \draw[->-=0.7,thick] (1,3) -- (0,4);
       \draw[->-=0.7,thick] (1,3) -- (2,4);
     \node[below] at (0,-0.2) {$a$};
     \node[below] at (2,0) {$b$};
      \node[above] at (0,4) {$a$};
     \node[above] at (2,4) {$b$};
      \node[right] at (1,2) {$c$};
      
      \node[left] at (1,1.1) {\scriptsize$\mu$};
       \node[left] at (1,2.9) {\scriptsize$\overline\mu$};
    \end{tikzpicture} ~, 
    \hspace{0.5 in}
\begin{tikzpicture}[baseline={([yshift=-2ex]current bounding box.center)},vertex/.style={anchor=base,
    circle,fill=black!25,minimum size=18pt,inner sep=2pt},scale=0.4]
  
    \draw[thick] (0,0) to [out = 180, in = 180,distance = 1.2 cm]  node[rotate=90]{\midarrow} (0,2);
     \draw[thick] (0,0) to [out = 0, in = 0,distance=1.2 cm]  node[rotate=90]{\midarrow}(0,2);
   \draw[thick, ->-=0.5] (0,-1.5) -- (0,0); 
   \draw[thick, ->-=0.9] (0,2) -- (0,3.5);

    \node[left] at (-0.9,1) {\footnotesize $a$};
        \node[right] at (0.9,1) {\footnotesize $ b$};
    \node[below] at (0,-1.5) {\footnotesize $c$};
    \node[above] at (0,3.5) {\footnotesize $d$};
       
       \node[left] at (0,2.2) {\scriptsize$\mu$};
       \node[left] at (0,-0.3) {\scriptsize$\overline\nu$};
      
\end{tikzpicture}
\quad=\delta_{cd} \delta_{\mu \nu} \sqrt{{d_a d_b \over d_c}} \,\, 
\begin{tikzpicture}[baseline={([yshift=-1ex]current bounding box.center)},vertex/.style={anchor=base,
    circle,fill=black!25,minimum size=18pt,inner sep=2pt},scale=0.4]
  
   \draw[thick, ->-=0.5] (0,-1.5) -- (0,3.5); 
    \node[below] at (0,-1.5) {\footnotesize $c$};
    
\end{tikzpicture}~,
\eea
where $d_a$ is the quantum dimension of the line $a$. Our conventions for the $F$- and $G$-symbols are as follows, 
\bea
   \begin{tikzpicture}[baseline={([yshift=-1ex]current bounding box.center)},vertex/.style={anchor=base,
    circle,fill=black!25,minimum size=18pt,inner sep=2pt},scale=0.4]
   \draw[->-=0.2,->-=0.6,->-=0.95,thick] (0,0) -- (4,4);
   \draw[->-=0.7,thick] (2.6,0) -- (1.3,1.3);
    \draw[->-=0.6,thick] (6,0) -- (3,3);
          \node[below] at (0,-0.2) {$a$};
     \node[below] at (3,0) {$b$};
      \node[below] at (6,-0.2) {$c$};
     \node[above] at (4,4) {$d$};
      \node[right] at (2,2) {$e$};
      \node[left] at (1.3,1.3) {\scriptsize$\mu$};
       \node[left] at (3,3) {\scriptsize$\nu$};
    \end{tikzpicture} 
    = 
    \sum_{f} \sum_{\sigma = 1}^{N_{bc}^f} \sum_{\rho = 1}^{N_{af}^d} (F_{a,b,c}^d)_{(e\mu\nu)(f \rho\sigma)}\,\,\,
       \begin{tikzpicture}[baseline={([yshift=-1ex]current bounding box.center)},vertex/.style={anchor=base,
    circle,fill=black!25,minimum size=18pt,inner sep=2pt},scale=0.4]
   \draw[->-=0.45,->-=0.95,thick] (0,0) -- (4,4);
   \draw[->-=0.6,thick] (3,0) -- (4.5,1.5);
    \draw[->-=0.3,->-=0.8,thick] (6,0) -- (3,3);
          \node[below] at (0,-0.2) {$a$};
     \node[below] at (3,0) {$b$};
      \node[below] at (6,-0.2) {$c$};
     \node[above] at (4,4) {$d$};
      \node[right] at (2.5,1.8) {$f$};
      \node[left] at (3,3) {\scriptsize $\rho$};
      \node[right] at (4.4,1.5) {\scriptsize $\sigma$};
    \end{tikzpicture} ~,
    \\
       \begin{tikzpicture}[baseline={([yshift=-1ex]current bounding box.center)},vertex/.style={anchor=base,
    circle,fill=black!25,minimum size=18pt,inner sep=2pt},scale=0.5]
   \draw[->-=0.2,->-=0.6,->-=0.95,thick] (4,0) -- (0,4);
   \draw[->-=0.7,thick] (2.5,1.5) -- (5,4);
   \draw[->-=0.7,thick] (1,3) -- (2,4);
   \node[below] at (4,0) {$d$};
    \node[above] at (0,4) {$a$};
    \node[above] at (2,4) {$b$};
    \node[above] at (5,4) {$c$};
        \node[left] at (2.5,1.3) {\scriptsize $\overline\nu$};
      \node[left] at (1,2.8) {\scriptsize $\overline\mu$};
       \node[right] at (1.7,2.6) {$e$};
           \end{tikzpicture} 
    = 
    \sum_{f} \sum_{\sigma = 1}^{N_{bc}^f} \sum_{\rho = 1}^{N_{af}^d} (G^{a,b,c}_d)_{(e\overline\mu\overline\nu)(f \overline\rho\overline\sigma)}\,\,\,
       \begin{tikzpicture}[baseline={([yshift=-1ex]current bounding box.center)},vertex/.style={anchor=base,
    circle,fill=black!25,minimum size=18pt,inner sep=2pt},scale=0.5]
   \draw[->-=0.2,->-=0.6,->-=0.95,thick] (4,0) -- (0,4);
   \draw[->-=0.8,->-=0.2,thick] (2.5,1.5) -- (5,4);
   \draw[->-=0.7,,thick] (3.5,2.5) -- (2,4);
   \node[below] at (4,0) {$d$};
    \node[above] at (0,4) {$a$};
    \node[above] at (2,4) {$b$};
    \node[above] at (5,4) {$c$};
        \node[left] at (2.5,1.3) {\scriptsize $\overline\rho$};
      \node[right] at (3.5,2.5) {\scriptsize $\overline\sigma$};
       \node[right] at (2,2.4) {$f$};
           \end{tikzpicture} ~,
\eea
which are related by 
\bea
(G^{a,b,c}_d)_{(e\overline\mu\overline\nu)(f \overline\rho\overline\sigma)} = (F_{a,b,c}^d)_{(f \rho\sigma)(e\mu\nu)}^{-1}~.
\eea

As follows from (\ref{eq:basisconventions}) and $d_a = d_{ \overline a}$, we have 
\bea
\begin{tikzpicture}[baseline={([yshift=-0.5ex]current bounding box.center)},vertex/.style={anchor=base,
    circle,fill=black!25,minimum size=18pt,inner sep=2pt},scale=0.4]
  
    \draw[thick] (0,0) to [out = 180, in = 180,distance = 1.25 cm] node[rotate=90]{\midarrow} (0,2);
     \draw[thick] (0,0) to [out = 0, in = 0,distance=1.25 cm] node[rotate=90]{\midarrow} (0,2);
   \draw[thick, dashed] (0,-1) -- (0,0); 
   \draw[thick, dashed] (0,2) -- (0,3); 

  \filldraw  (0,0) circle (2pt);
  \filldraw  (0,2) circle (2pt);
  
    \node[left] at (-1,1.1) {\footnotesize $ \overline a$};
        \node[right] at (1,1) {\footnotesize $ a$};
       
\end{tikzpicture} = \,\,\, d_a ~,
\eea
and using this together with an $F$-move shows that
\bea
\label{eq:FSisotopydef}
\begin{tikzpicture}[scale=0.8,baseline=20]
    \begin{scope}[ thick, every node/.style={sloped,allow upside down}]
         \draw[ postaction={decorate}, decoration={markings,
      mark=at position 0.4 with {\arrow[scale=0.9]{Stealth}}, mark=at position 0.99 with {\arrow[scale=0.9]{Stealth[reversed]}} }] (-1,-0.5+0.5) .. controls (-1,1.8+0.5) and (0,1.4+0.5) .. (0,0.5+0.5);
      \draw[ postaction={decorate}, decoration={markings,
      mark=at position 0.7 with {\arrow[scale=0.9]{Stealth}} }] (0,0.5+0.5) .. controls (0,-0.3+0.5) and (1,-0.8+0.5) .. (1,1.6+0.5);
      \filldraw  (-0.45,1.21+0.5) circle (1pt);
      \filldraw  (0.45,-0.15+0.5) circle (1pt);
        \node[ below] at (-1,-0.4+0.5) {$a$};
        \node[ right] at (-0.1,0.9+0.5) {$\overline{a}$};
        \node[right] at (1,1.5+0.5) {$a$};
        
       \draw[dashed] (0.46,0.2)   -- (0.46,0.2-0.5)  ;
       \draw[dashed] (-0.46,1.8)   -- (-0.46,1.8+0.5)  ;
    \end{scope}
\end{tikzpicture}
\hspace{0.1 in}= \quad d_a \, (F_{a, \overline a, a}^a)_{11} \hspace{0.1 in} 
\begin{tikzpicture}[scale=0.8,baseline=20]
\begin{scope}[ thick, every node/.style={sloped,allow upside down}]
   \draw(0,-0.7+0.5) -- node{\midarrow} (0,1.5+0.5);
   \node[right] at (0,0) {$a$};
\end{scope}
\end{tikzpicture}~.
\eea
The quantity $\kappa_a := d_a (F_{a, \overline a, a}^a)_{11}$ is referred to as the Frobenius-Schur indicator of $a$, 
and is gauge invariant as long as $a$ is self-dual. When $a$ is not self-dual, we may always work in a gauge in which $\kappa_a=1$.

 Having established our conventions, let us now return to the question of invariance under twisted gauging. Our starting point is the following twisted partition function,
\begin{equation}\label{eq:ghpf}
    Z_{\mathcal{X}}[g,h] = \,\,\, \begin{tikzpicture}[baseline={([yshift=+.5ex]current bounding box.center)},vertex/.style={anchor=base,
    circle,fill=black!25,minimum size=18pt,inner sep=2pt},scale=0.5]
    \filldraw[grey] (-2,-2) rectangle ++(4,4);
    \draw[thick, dgrey] (-2,-2) -- (-2,+2);
    \draw[thick, dgrey] (-2,-2) -- (+2,-2);
    \draw[thick, dgrey] (+2,+2) -- (+2,-2);
    \draw[thick, dgrey] (+2,+2) -- (-2,+2);
    \draw[thick, black, -stealth] (0,-2) -- (0.354,-1.354);
    \draw[thick, black] (0,-2) -- (0.707,-0.707);
    \draw[thick, black, -stealth] (2,0) -- (1.354,-0.354);
    \draw[thick, black] (2,0) -- (0.707,-0.707);
    \draw[thick, black, -stealth] (-0.707,0.707) -- (-0.354,1.354);
    \draw[thick, black] (0,2) -- (-0.707,0.707);
    \draw[thick, black, -stealth] (-0.707,0.707) -- (-1.354,0.354);
    \draw[thick, black] (-2,0) -- (-0.707,0.707);
    \draw[thick, black, -stealth] (0.707,-0.707) -- (0,0);
    \draw[thick, black] (0.707,-0.707) -- (-0.707,0.707);

    \node[black, below] at (0,-2) {\scriptsize $g$};
    \node[black, right] at (2,0) {\scriptsize $h$};
    \node[black, above] at (0.2,0) {\scriptsize $gh$};
\end{tikzpicture} ~.
\end{equation}
We wish to show that this is equivalent to the $G$-gauged partition function, for appropriate discrete torsion and stacking with SPT phases.
To do so, we begin by nucleating a $Q,\overline{Q}$-bubble in the middle of the diagram and flipping the orientation of the $\overline{Q}$-line,
\begin{equation}
\label{eq:firststep}
     Z_{\mathcal{X}}[g,h] =\,\, \frac{1}{d_{Q}} \,\,\, \begin{tikzpicture}[baseline={([yshift=+.5ex]current bounding box.center)},vertex/.style={anchor=base,
    circle,fill=black!25,minimum size=18pt,inner sep=2pt},scale=0.6]
    \filldraw[grey] (-2,-2) rectangle ++(4,4);
    \draw[thick, dgrey] (-2,-2) -- (-2,+2);
    \draw[thick, dgrey] (-2,-2) -- (+2,-2);
    \draw[thick, dgrey] (+2,+2) -- (+2,-2);
    \draw[thick, dgrey] (+2,+2) -- (-2,+2);
    \draw[thick, black, ->-=0.5] (0,-2) -- (0.707,-0.707);
    \draw[thick, black, ->-=0.5] (2,0) -- (0.707,-0.707);
    \draw[thick, black, ->-=0.5] (-0.707,0.707) -- (0,2);
    \draw[thick, black, -<-=0.5] (-2,0) -- (-0.707,0.707);
    \draw[thick, black, ->-=0.7] (0.707,-0.707) -- (0.293,-0.293);
    \draw[thick, black, -<-=0.7] (-0.707,0.707) -- (-0.293,0.293);
    \draw[thick, red, ->-=0.55] (0.293,-0.293) arc(-45:135:0.414);
    \draw[thick, red, ->-=0.55] (0.293,-0.293) arc(-45:-225:0.414);
    \node[black, below] at (0,-2) {\footnotesize $g$};
    \node[black, right] at (2,0) {\footnotesize $h$};
    \node[red] at (0.58,0.58) {\footnotesize $Q$};
    \node[red] at (-0.62,-0.62) {\footnotesize $\overline{Q}$};
    
\end{tikzpicture} = \,\,\, \frac{\kappa_Q}{d_Q} \,\,\, \begin{tikzpicture}[baseline={([yshift=+.5ex]current bounding box.center)},vertex/.style={anchor=base,
    circle,fill=black!25,minimum size=18pt,inner sep=2pt},scale=0.6]
    \filldraw[grey] (-2,-2) rectangle ++(4,4);
    \draw[thick, dgrey] (-2,-2) -- (-2,+2);
    \draw[thick, dgrey] (-2,-2) -- (+2,-2);
    \draw[thick, dgrey] (+2,+2) -- (+2,-2);
    \draw[thick, dgrey] (+2,+2) -- (-2,+2);
    \draw[thick, black, ->-=0.5] (0,-2) -- (0.707,-0.707);
    \draw[thick, black, ->-=0.5] (2,0) -- (0.707,-0.707);
    \draw[thick, black, ->-=0.5] (-0.707,0.707) -- (0,2);
    \draw[thick, black, -<-=0.5] (-2,0) -- (-0.707,0.707);
    \draw[thick, black, ->-=0.7] (0.707,-0.707) -- (0.293,-0.293);
    \draw[thick, black, -<-=0.7] (-0.707,0.707) -- (-0.293,0.293);
    \draw[thick, red, ->-=0.55] (0.293,-0.293) arc(-45:135:0.414);
    \draw[thick, red, -<-=0.55] (0.293,-0.293) arc(-45:-225:0.414);
    \node[black, below] at (0,-2) {\footnotesize $g$};
    \node[black, right] at (2,0) {\footnotesize $h$};
    \node[red] at (0.58,0.58) {\footnotesize $Q$};
    \node[red] at (-0.58,-0.58) {\footnotesize $Q$};
    
\end{tikzpicture}~,
\end{equation}
where $\kappa_Q$ is the aforementioned Frobenius-Schur indicator of $Q$ (which is only potentially non-trivial in the case of duality). More concretely, this factor arises from the following manipulations. First, we note that we have 
\bea
\label{eq:FSdefinition}
 \begin{tikzpicture}[baseline={([yshift=+.5ex]current bounding box.center)},vertex/.style={anchor=base,
    circle,fill=black!25,minimum size=18pt,inner sep=2pt},scale=0.4]
  
    \draw[thick, red, ->-=0.125, -<-=0.5, ->-=0.925] (0,0)--(0,4);
    \draw[red, fill=red] (0,1) circle (.5ex);
    \draw[red, fill=red] (0,3) circle (.5ex);

    \node[red, right] at (0,2) {\footnotesize $\overline Q$};
    \node[red,above] at (0,4) {\footnotesize $Q$};
    \node[red,below] at (0,0) {\footnotesize $Q$};
    
\end{tikzpicture}
\quad=\quad \kappa_Q \,\,\,
 \begin{tikzpicture}[baseline={([yshift=+.5ex]current bounding box.center)},vertex/.style={anchor=base,
    circle,fill=black!25,minimum size=18pt,inner sep=2pt},scale=0.4]
  
    \draw[thick, red, ->-=0.5] (0,0)--(0,4);
    \node[red,below] at (0,0) {\footnotesize $Q$};
    
\end{tikzpicture}~
\eea
by definition of the Frobenius-Schur indicator; this is just shorthand for (\ref{eq:FSisotopydef}). We then have that 
\bea
\begin{tikzpicture}[baseline={([yshift=+.5ex]current bounding box.center)},vertex/.style={anchor=base,
    circle,fill=black!25,minimum size=18pt,inner sep=2pt},scale=0.5]
  
    \draw[thick, red, ->-=0.55] (0,0) to [out = 180, in = 180,distance = 1.25 cm] (0,2);
     \draw[thick, red, ->-=0.55] (0,0) to [out = 0, in = 0,distance=1.25 cm] (0,2);
   \draw[thick, ->-=0.5] (0,-1.5) -- (0,0); 
   \draw[thick, ->-=0.5] (0,2) -- (0,3.5);

    \node[red, left] at (-0.9,1) {\footnotesize $\overline Q$};
        \node[red, right] at (0.9,1) {\footnotesize $ Q$};
    \node[below] at (0,-1.5) {\footnotesize $gh$};
    \node[above] at (0,3.5) {\footnotesize $gh$};
    
\end{tikzpicture}
\quad=\quad \kappa_Q \,\,\,
\begin{tikzpicture}[baseline={([yshift=+.5ex]current bounding box.center)},vertex/.style={anchor=base,
    circle,fill=black!25,minimum size=18pt,inner sep=2pt},scale=0.5]
  
    \draw[thick, red, ->-=0.25,-<-=0.55,->-=0.85] (0,0) to [out = 180, in = 180,distance = 1.25 cm] (0,2);
     \draw[thick, red, ->-=0.55] (0,0) to [out = 0, in = 0,distance=1.25 cm] (0,2);
   \draw[thick, ->-=0.5] (0,-1.5) -- (0,0); 
   \draw[thick, ->-=0.5] (0,2) -- (0,3.5); 

\draw[thick, dashed,dgreen] (-0.87,1.4) to [out = -30 , in = 210] (0.87,1.4);
\draw[thick, dashed,dgreen] (-0.87,0.6) to [out = 30 , in = 150] (0.87,0.6);

 \draw[red, fill=red] (-0.87,0.6) circle (.5ex);
  \draw[red, fill=red] (-0.87,1.4)  circle (.5ex);
  
    \node[red, left] at (-0.9,1) {\footnotesize $ Q$};
     \node[red, left] at (-0.4,2.2) {\footnotesize $\overline Q$};
      \node[red, left] at (-0.4,-0.2) {\footnotesize $\overline Q$};
        \node[red, right] at (0.9,1) {\footnotesize $ Q$};
    \node[below] at (0,-1.5) {\footnotesize $gh$};
    \node[above] at (0,3.5) {\footnotesize $gh$};
\end{tikzpicture}
\quad=\quad \kappa_Q \sum_{x,y} (F^Q_{Q,\overline Q, Q})^{-1}_{y,1} (F^Q_{Q,\overline Q, Q})_{1,x}\,\,\,
\begin{tikzpicture}[baseline={([yshift=+.5ex]current bounding box.center)},vertex/.style={anchor=base,
    circle,fill=black!25,minimum size=18pt,inner sep=2pt},scale=0.4]
  
    \draw[thick, red, ->-=0.55] (0,0) to [out = 180, in = 180,distance = 0.625 cm] (0,1);
     \draw[thick, red, ->-=0.55] (0,0) to [out = 0, in = 0,distance=0.625 cm] (0,1);
     
       \draw[thick, red,-<-=0.55] (0,2) to [out = 180, in = 180,distance = 0.625 cm] (0,3);
     \draw[thick, red, ->-=0.55] (0,2) to [out = 0, in = 0,distance=0.625 cm] (0,3);
     
            \draw[thick, red, ->-=0.55] (0,4) to [out = 180, in = 180,distance = 0.625 cm] (0,5);
     \draw[thick, red, ->-=0.55] (0,4) to [out = 0, in = 0,distance=0.625 cm] (0,5);

   \draw[thick, ->-=0.55] (0,-1.5) -- (0,0); 
   \draw[thick, ->-=0.55] (0,1) -- (0,2); 
   \draw[thick, ->-=0.55] (0,3) -- (0,4); 
      \draw[thick, ->-=0.55] (0,5) -- (0,6.5); 
      
        \node[red, left] at (-0.4,4.5) {\footnotesize $ \overline Q$};
        \node[red, right] at (0.4,4.5) {\footnotesize $ Q$};

   \node[red, left] at (-0.4,2.5) {\footnotesize $ Q$};
        \node[red, right] at (0.4,2.5) {\footnotesize $ Q$};
 
      \node[red, left] at (-0.4,0.5) {\footnotesize $\overline Q$};
        \node[red, right] at (0.4,0.5) {\footnotesize $ Q$};
        
          \node[right] at (0.2,1.5) {\footnotesize $x$};
           \node[right] at (0.2,3.5) {\footnotesize $y$};
          
    \node[below] at (0,-1.5) {\footnotesize $gh$};
    \node[above] at (0,6.5) {\footnotesize $gh$};
\end{tikzpicture}
\eea
where the green lines in the middle diagram are identity lines, and we have performed $F$-moves around them to put the diagram in the form on the right. 
Shrinking the upper and lower bubbles then gives 
\bea
\begin{tikzpicture}[baseline={([yshift=+.5ex]current bounding box.center)},vertex/.style={anchor=base,
    circle,fill=black!25,minimum size=18pt,inner sep=2pt},scale=0.5]
  
    \draw[thick, red, ->-=0.55] (0,0) to [out = 180, in = 180,distance = 1.25 cm] (0,2);
     \draw[thick, red, ->-=0.55] (0,0) to [out = 0, in = 0,distance=1.25 cm] (0,2);
   \draw[thick, ->-=0.5] (0,-1.5) -- (0,0); 
   \draw[thick, ->-=0.5] (0,2) -- (0,3.5);

    \node[red, left] at (-0.9,1) {\footnotesize $\overline Q$};
        \node[red, right] at (0.9,1) {\footnotesize $ Q$};
    \node[below] at (0,-1.5) {\footnotesize $gh$};
    \node[above] at (0,3.5) {\footnotesize $gh$};
    
\end{tikzpicture}
 = \kappa_Q d_Q^2 (F^Q_{Q, \overline Q, Q})_{1,gh}(F^Q_{Q, \overline Q, Q})_{gh,1}^{-1} \,\,\,
 \begin{tikzpicture}[baseline={([yshift=+.5ex]current bounding box.center)},vertex/.style={anchor=base,
    circle,fill=black!25,minimum size=18pt,inner sep=2pt},scale=0.5]
  
    \draw[thick, red, -<-=0.55] (0,0) to [out = 180, in = 180,distance = 1.25 cm] (0,2);
     \draw[thick, red, ->-=0.55] (0,0) to [out = 0, in = 0,distance=1.25 cm] (0,2);
   \draw[thick, ->-=0.5] (0,-1.5) -- (0,0); 
   \draw[thick, ->-=0.5] (0,2) -- (0,3.5);

    \node[red, left] at (-0.9,1) {\footnotesize $ Q$};
        \node[red, right] at (0.9,1) {\footnotesize $ Q$};
    \node[below] at (0,-1.5) {\footnotesize $gh$};
    \node[above] at (0,3.5) {\footnotesize $gh$};
    
\end{tikzpicture}
=  \kappa_Q \,\,\,
 \begin{tikzpicture}[baseline={([yshift=+.5ex]current bounding box.center)},vertex/.style={anchor=base,
    circle,fill=black!25,minimum size=18pt,inner sep=2pt},scale=0.5]
  
    \draw[thick, red, -<-=0.55] (0,0) to [out = 180, in = 180,distance = 1.25 cm] (0,2);
     \draw[thick, red, ->-=0.55] (0,0) to [out = 0, in = 0,distance=1.25 cm] (0,2);
   \draw[thick, ->-=0.5] (0,-1.5) -- (0,0); 
   \draw[thick, ->-=0.5] (0,2) -- (0,3.5);

    \node[red, left] at (-0.9,1) {\footnotesize $ Q$};
        \node[red, right] at (0.9,1) {\footnotesize $ Q$};
    \node[below] at (0,-1.5) {\footnotesize $gh$};
    \node[above] at (0,3.5) {\footnotesize $gh$};
    
\end{tikzpicture}~,
\eea
as claimed. Note that in the second equality, we have used that $d_Q^2 (F^Q_{Q, \overline Q ,Q})_{1,gh}(F^Q_{Q ,\overline Q, Q})_{gh,1}^{-1} = 1$. This can be shown in a variety of ways, for example by using the pentagon identities in (\ref{eq:usefulpent1}) and (\ref{eq:usefulpent2}) to rewrite it as $d_Q^2 (F^Q_{Q, \overline Q, Q})_{1,1}(F^Q_{Q, \overline Q, Q})_{1,1}^{-1} = 1$, and then assuming that we are working with a unitary gauge such that $(F^Q_{Q ,\overline Q, Q})_{g,h}$ is a unitary matrix and hence $d_Q (F^Q_{Q,\overline Q, Q})_{1,1}$ is a phase.

With the diagram in the form of (\ref{eq:firststep}), we may now perform a pair of $F$-moves (or rather an $F$- and $G$-move) to obtain
\begin{equation}
     Z_{\mathcal{X}}[g,h] = \frac{\kappa_Q}{d_{Q}} (F_{Q,g,h}^Q)_{gh,Q}^{-1} (F^Q_{Q,h,g})_{Q,hg}\quad \begin{tikzpicture}[baseline={([yshift=+.5ex]current bounding box.center)},vertex/.style={anchor=base,
    circle,fill=black!25,minimum size=18pt,inner sep=2pt},scale=0.6]
    \filldraw[grey] (-2,-2) rectangle ++(4,4);
    \draw[thick, dgrey] (-2,-2) -- (-2,+2);
    \draw[thick, dgrey] (-2,-2) -- (+2,-2);
    \draw[thick, dgrey] (+2,+2) -- (+2,-2);
    \draw[thick, dgrey] (+2,+2) -- (-2,+2);
    \draw[thick, black, ->-=0.5] (0,-2) -- (0,-0.5);
    \draw[thick, black, ->-=0.5] (0,0.5) -- (0,2);
    \draw[thick, black, ->-=0.5] (-0.5,0) -- (-2,0) ;
    \draw[thick, black, ->-=0.5] (2,0) -- (0.5,0) ;
    
 \draw[thick, red, ->-=0.6] (0,0.5) to [out = 180, in = 90] (-0.5, 0);
      \draw[thick, red, -<-=0.6] (0,0.5) to [out = 0, in = 90] (0.5, 0);
        \draw[thick, red, -<-=0.6] (0,-0.5) to [out = 180, in = -90] (-0.5, 0);
         \draw[thick, red, -<-=0.6 ] (0.5, 0) to [out = -90, in = 0] (0,-0.5);
         
    \node[black, below] at (0,-2) {\footnotesize $g$};
    \node[black, right] at (2,0) {\footnotesize $h$};
    \node[red] at (0.65,0.65) {\footnotesize $Q$};
    \node[red] at (-0.7,-0.7) {\footnotesize $Q$};
    
\end{tikzpicture}~.
\end{equation}

We next rewrite the diagram using a topological manipulation, as well as (\ref{eq:FSdefinition}) and a pair of $F$-moves to once again flip the orientation of one segment of the $Q$ line, 
\begin{equation}
\begin{tikzpicture}[baseline={([yshift=+.5ex]current bounding box.center)},vertex/.style={anchor=base,
    circle,fill=black!25,minimum size=18pt,inner sep=2pt},scale=0.6]
    \filldraw[grey] (-2,-2) rectangle ++(4,4);
    \draw[thick, dgrey] (-2,-2) -- (-2,+2);
    \draw[thick, dgrey] (-2,-2) -- (+2,-2);
    \draw[thick, dgrey] (+2,+2) -- (+2,-2);
    \draw[thick, dgrey] (+2,+2) -- (-2,+2);
    \draw[thick, black, ->-=0.5] (0,-2) -- (0,-0.5);
    \draw[thick, black, ->-=0.5] (0,0.5) -- (0,2);
    \draw[thick, black, ->-=0.5] (-0.5,0) -- (-2,0) ;
    \draw[thick, black, ->-=0.5] (2,0) -- (0.5,0) ;
  
 \draw[thick, red, ->-=0.6] (0,0.5) to [out = 180, in = 90] (-0.5, 0);
      \draw[thick, red, -<-=0.6] (0,0.5) to [out = 0, in = 90] (0.5, 0);
        \draw[thick, red, -<-=0.6] (0,-0.5) to [out = 180, in = -90] (-0.5, 0);
         \draw[thick, red, -<-=0.6] (0.5, 0) to [out = -90, in = 0] (0,-0.5);

    \node[black, below] at (0,-2) {\footnotesize $g$};
    \node[black, right] at (2,0) {\footnotesize $h$};
    \node[red] at (0.65,0.65) {\footnotesize $Q$};
    \node[red] at (-0.7,-0.7) {\footnotesize $Q$};
    
\end{tikzpicture}=  \begin{tikzpicture}[baseline={([yshift=+.5ex]current bounding box.center)},vertex/.style={anchor=base,
    circle,fill=black!25,minimum size=18pt,inner sep=2pt},scale=0.6]
    \filldraw[grey] (-2,-2) rectangle ++(4,4);
    \draw[thick, dgrey] (-2,-2) -- (-2,+2);
    \draw[thick, dgrey] (-2,-2) -- (+2,-2);
    \draw[thick, dgrey] (+2,+2) -- (+2,-2);
    \draw[thick, dgrey] (+2,+2) -- (-2,+2);
    
     \draw[thick, black, ->-=0.6] (0.72,-1.25)  [out = 150, in = 30] to (-0.72,-1.25); 
      \draw[thick, black, ->-=0.6] (-1.25,-0.72)   [out = 60, in = 300] to (-1.25,0.72); 
     
   \draw[thick, red, ->-=0.2, ->-=0.7] (-2,0.5)  [out = 0, in = -90] to (-0.5,2); 
      \draw[thick, red, ->-=0.5] (0.5,2) to [out = -90, in = 180]  (2,0.5);
      \draw[thick, red, ->-=0.4,->-=0.9] (2,-0.5)  [out = 180, in = 90] to (0.5,-2); 
    \draw[thick, red, ->-=0.23, ->-=0.55, ->-=0.9] (-0.5,-2)  [out = 90, in = 0] to (-2,-0.5);

    \node[black, above] at (0,-1) {\footnotesize $h$};
    \node[black, right] at (-1,0) {\footnotesize $g$};
   
    \node[red] at (-0.7,-0.7) {\footnotesize $Q$};
     \node[red,below] at (-0.5,-2) {\footnotesize $Q$};
      \node[red,below] at (0.5,-2) {\footnotesize $Q$};
       \node[red,above] at (0.5,2) {\footnotesize $Q$};
       \node[red,above] at (-0.5,2) {\footnotesize $Q$};
       \node[red,left] at (-2,0.5) {\footnotesize $Q$};
       \node[red,left] at (-2,-0.5) {\footnotesize $Q$};
       \node[red,right] at (2,-0.5) {\footnotesize $Q$};
    
\end{tikzpicture}
= \kappa_Q (F^g_{\overline Q, Q, g})^{-1} (F^h_{\overline Q, Q, h})
\begin{tikzpicture}[baseline={([yshift=+.5ex]current bounding box.center)},vertex/.style={anchor=base,
    circle,fill=black!25,minimum size=18pt,inner sep=2pt},scale=0.6]
    \filldraw[grey] (-2,-2) rectangle ++(4,4);
    \draw[thick, dgrey] (-2,-2) -- (-2,+2);
    \draw[thick, dgrey] (-2,-2) -- (+2,-2);
    \draw[thick, dgrey] (+2,+2) -- (+2,-2);
    \draw[thick, dgrey] (+2,+2) -- (-2,+2);
    
     \draw[thick, black, ->-=0.6] (0.72,-1.25)  [out = 150, in = 30] to (-0.72,-1.25); 
      \draw[thick, black, ->-=0.6] (-1.25,-0.72)   [out = 60, in = 300] to (-1.25,0.72); 
     
   \draw[thick, red, -<-=0.2, ->-=0.7] (-2,0.5)  [out = 0, in = -90] to (-0.5,2); 
      \draw[thick, red, -<-=0.5] (0.5,2) to [out = -90, in = 180]  (2,0.5);
      \draw[thick, red, ->-=0.4,-<-=0.9] (2,-0.5)  [out = 180, in = 90] to (0.5,-2); 
    \draw[thick, red, ->-=0.23, ->-=0.55, ->-=0.9] (-0.5,-2)  [out = 90, in = 0] to (-2,-0.5);

    \node[black, above] at (0,-1) {\footnotesize $h$};
    \node[black, right] at (-1,0) {\footnotesize $g$};
   
    \node[red] at (-0.7,-0.7) {\footnotesize $Q$};
     \node[red,below] at (-0.5,-2) {\footnotesize $Q$};
      \node[red,below] at (0.5,-2) {\footnotesize $\overline Q$};
       \node[red,above] at (0.5,2) {\footnotesize $\overline Q$};
       \node[red,above] at (-0.5,2) {\footnotesize $Q$};
       \node[red,left] at (-2,0.5) {\footnotesize $\overline Q$};
       \node[red,left] at (-2,-0.5) {\footnotesize $Q$};
       \node[red,right] at (2,-0.5) {\footnotesize $Q$};
   
\end{tikzpicture} ~.
\end{equation}
A further sequence of F- and $G$-moves allows one to simplify this configuration as follows,
\bea
\begin{tikzpicture}[baseline={([yshift=+.5ex]current bounding box.center)},vertex/.style={anchor=base,
    circle,fill=black!25,minimum size=18pt,inner sep=2pt},scale=0.6]
    \filldraw[grey] (-2,-2) rectangle ++(4,4);
    \draw[thick, dgrey] (-2,-2) -- (-2,+2);
    \draw[thick, dgrey] (-2,-2) -- (+2,-2);
    \draw[thick, dgrey] (+2,+2) -- (+2,-2);
    \draw[thick, dgrey] (+2,+2) -- (-2,+2);
    
     \draw[thick, black, ->-=0.6] (0.72,-1.25)  [out = 150, in = 30] to (-0.72,-1.25); 
      \draw[thick, black, ->-=0.6] (-1.25,-0.72)   [out = 60, in = 300] to (-1.25,0.72); 
     
   \draw[thick, red, -<-=0.2, ->-=0.7] (-2,0.5)  [out = 0, in = -90] to (-0.5,2); 
      \draw[thick, red, -<-=0.5] (0.5,2) to [out = -90, in = 180]  (2,0.5);
      \draw[thick, red, ->-=0.4,-<-=0.9] (2,-0.5)  [out = 180, in = 90] to (0.5,-2); 
    \draw[thick, red, ->-=0.23, ->-=0.55, ->-=0.9] (-0.5,-2)  [out = 90, in = 0] to (-2,-0.5);

    \node[black, above] at (0,-1) {\footnotesize $h$};
    \node[black, right] at (-1,0) {\footnotesize $g$};
   
    \node[red] at (-0.7,-0.7) {\footnotesize $Q$};
     \node[red,below] at (-0.5,-2) {\footnotesize $Q$};
      \node[red,below] at (0.5,-2) {\footnotesize $\overline Q$};
       \node[red,above] at (0.5,2) {\footnotesize $\overline Q$};
       \node[red,above] at (-0.5,2) {\footnotesize $Q$};
       \node[red,left] at (-2,0.5) {\footnotesize $\overline Q$};
       \node[red,left] at (-2,-0.5) {\footnotesize $Q$};
       \node[red,right] at (2,-0.5) {\footnotesize $Q$};
   
\end{tikzpicture} 
&= &\sum_{\widetilde g} (F^Q_{Q ,\overline Q, Q})_{\widetilde g g}\quad
\begin{tikzpicture}[baseline={([yshift=+.5ex]current bounding box.center)},vertex/.style={anchor=base,
    circle,fill=black!25,minimum size=18pt,inner sep=2pt},scale=0.6]
    \filldraw[grey] (-2,-2) rectangle ++(4,4);
    \draw[thick, dgrey] (-2,-2) -- (-2,+2);
    \draw[thick, dgrey] (-2,-2) -- (+2,-2);
    \draw[thick, dgrey] (+2,+2) -- (+2,-2);
    \draw[thick, dgrey] (+2,+2) -- (-2,+2);
    
     \draw[thick, black, ->-=0.6] (0.5,-1) -- (-0.5,-1); 
       \draw[thick, black, ->-=0.6] (-0.5,0.5) -- (-2,0.5); 
       \draw[thick, black, ->-=0.6] (2,0.5) -- (0.5,0.5); 
     
   \draw[thick, red, ->-=0.15, ->-=0.5, ->-=0.85] (-0.5,-2) -- (-0.5, 2); 
      \draw[thick, red, ->-=0.15, -<-=0.5, ->-=0.85] (0.5,-2) -- (0.5, 2);    
        
    \node[black, below] at (0,-1) {\footnotesize $h$};
    \node[black, below] at (-1.25,0.5) {\footnotesize $\widetilde g$};
      \node[black, below] at (1.25,0.5) {\footnotesize $\widetilde g$};
   
    \node[red,below] at (-0.5,-2) {\footnotesize $Q$};
     \node[red,left] at (-0.5,-0.5) {\footnotesize $Q$};
      \node[red,above] at (-0.5,2) {\footnotesize $ Q$};
       \node[red,below] at (0.5,-1.9) {\footnotesize $\overline Q$};
       \node[red,right] at (0.5,-0.5) {\footnotesize $Q$};
       \node[red,above] at (0.5,2) {\footnotesize $\overline Q$};

\end{tikzpicture} 
\\
&=& \sum_{\widetilde g, \widetilde h} (F^Q_{Q ,\overline Q, Q})_{\widetilde g g} \left(F^Q_{Q,\overline Q, Q}\right)^{-1}_{h \widetilde h} \quad
\begin{tikzpicture}[baseline={([yshift=+.5ex]current bounding box.center)},vertex/.style={anchor=base,
    circle,fill=black!25,minimum size=18pt,inner sep=2pt},scale=0.6]
    \filldraw[grey] (-2,-2) rectangle ++(4,4);
    \draw[thick, dgrey] (-2,-2) -- (-2,+2);
    \draw[thick, dgrey] (-2,-2) -- (+2,-2);
    \draw[thick, dgrey] (+2,+2) -- (+2,-2);
    \draw[thick, dgrey] (+2,+2) -- (-2,+2);
    \draw[thick, black, ->-=0.5] (0,-2) -- (0,-0.5);
    \draw[thick, black, ->-=0.5] (0,0.5) -- (0,2);
    \draw[thick, black, ->-=0.5] (-0.5,0) -- (-2,0) ;
    \draw[thick, black, ->-=0.5] (2,0) -- (0.5,0) ;
    \draw[thick, red, -<-=0.6] (0,0.5) to [out = 180, in = 90] (-0.5, 0);
      \draw[thick, red, -<-=0.6] (0,0.5) to [out = 0, in = 90] (0.5, 0);
        \draw[thick, red, ->-=0.6] (0,-0.5) to [out = 180, in = -90] (-0.5, 0);
         \draw[thick, red, ->-=0.6] (0.5, 0) to [out = -90, in = 0] (0,-0.5);
    \node[black, below] at (0,-2) {\footnotesize $\widetilde h$};
    \node[black, right] at (2,0) {\footnotesize $\widetilde g$};
    \node[red] at (0.65,0.65) {\footnotesize $\oQ$};
    \node[red] at (-0.7,-0.7) {\footnotesize $Q$};
\end{tikzpicture}
\no
\eea
\bea
\hspace{1.5 in}&=& \sum_{\widetilde g, \widetilde h} (F^Q_{Q, \overline Q, Q})_{\widetilde g g} \left(F^Q_{Q,\overline Q, Q}\right)^{-1}_{h \widetilde h} F_{\widetilde h, Q, \overline Q}^{\widetilde h \widetilde g} (F^{\widetilde g \widetilde h}_{\widetilde g, Q, \overline Q})^{-1}\quad
 \begin{tikzpicture}[baseline={([yshift=+.5ex]current bounding box.center)},vertex/.style={anchor=base,
    circle,fill=black!25,minimum size=18pt,inner sep=2pt},scale=0.6]
    \filldraw[grey] (-2,-2) rectangle ++(4,4);
    \draw[thick, dgrey] (-2,-2) -- (-2,+2);
    \draw[thick, dgrey] (-2,-2) -- (+2,-2);
    \draw[thick, dgrey] (+2,+2) -- (+2,-2);
    \draw[thick, dgrey] (+2,+2) -- (-2,+2);
    \draw[thick, black, ->-=0.5] (0,-2) -- (0.707,-0.707);
    \draw[thick, black, ->-=0.5] (2,0) -- (0.707,-0.707);
    \draw[thick, black, ->-=0.5] (-0.707,0.707) -- (0,2);
    \draw[thick, black, -<-=0.5] (-2,0) -- (-0.707,0.707);
    \draw[thick, black, ->-=0.7] (0.707,-0.707) -- (0.293,-0.293);
    \draw[thick, black, -<-=0.7] (-0.707,0.707) -- (-0.293,0.293);
    \draw[thick, red, ->-=0.55] (0.293,-0.293) arc(-45:135:0.414);
    \draw[thick, red, ->-=0.55] (0.293,-0.293) arc(-45:-225:0.414);
    \node[black, below] at (0,-2) {\footnotesize $\widetilde h$};
    \node[black, right] at (2,0) {\footnotesize $\widetilde g$};
    \node[red] at (0.65,0.65) {\footnotesize $\overline Q$};
    \node[red] at (-0.7,-0.7) {\footnotesize ${Q}$};
\end{tikzpicture}
\no\\
&=& \sum_{\widetilde g, \widetilde h} d_Q (F^Q_{Q, \overline Q, Q})_{\widetilde g g} \left(F^Q_{Q,\overline Q, Q}\right)^{-1}_{h \widetilde h} F_{\widetilde h, Q, \overline Q}^{\widetilde h \widetilde g} (F^{\widetilde g \widetilde h}_{\widetilde g, Q, \overline Q})^{-1} \,\,\, 
\begin{tikzpicture}[baseline={([yshift=+.5ex]current bounding box.center)},vertex/.style={anchor=base,
    circle,fill=black!25,minimum size=18pt,inner sep=2pt},scale=0.6]
    \filldraw[grey] (-2,-2) rectangle ++(4,4);
    \draw[thick, dgrey] (-2,-2) -- (-2,+2);
    \draw[thick, dgrey] (-2,-2) -- (+2,-2);
    \draw[thick, dgrey] (+2,+2) -- (+2,-2);
    \draw[thick, dgrey] (+2,+2) -- (-2,+2);
    \draw[thick, black, -stealth] (0,-2) -- (0.354,-1.354);
    \draw[thick, black] (0,-2) -- (0.707,-0.707);
    \draw[thick, black, -stealth] (2,0) -- (1.354,-0.354);
    \draw[thick, black] (2,0) -- (0.707,-0.707);
    \draw[thick, black, -stealth] (-0.707,0.707) -- (-0.354,1.354);
    \draw[thick, black] (0,2) -- (-0.707,0.707);
    \draw[thick, black, -stealth] (-0.707,0.707) -- (-1.354,0.354);
    \draw[thick, black] (-2,0) -- (-0.707,0.707);
    \draw[thick, black, -stealth] (0.707,-0.707) -- (0,0);
    \draw[thick, black] (0.707,-0.707) -- (-0.707,0.707);

    \node[black, below] at (0,-2) {\scriptsize $\widetilde h$};
    \node[black, right] at (2,0) {\scriptsize $\widetilde g$};
    \node[black, above] at (0.2,0) {\scriptsize $\widetilde g \widetilde h$};
\end{tikzpicture} ~,
\no
\eea
where in the third equality we have used 
\bea
\label{eq:intermediatethingbubble}
\begin{tikzpicture}[baseline={([yshift=+.5ex]current bounding box.center)},vertex/.style={anchor=base,
    circle,fill=black!25,minimum size=18pt,inner sep=2pt},scale=0.5]

    \draw[red, thick, ->- = 0.5] (-2,-2) --(-1,-1);; 
     \draw[red, thick, ->- = 0.5] (2,-2) --(1,-1);
     
      \draw[ thick,->-=0.5] (-1,-1)-- (-1,1); 
  \draw[ thick,->-=0.5]  (1,-1)  -- (1,1); 
 \draw[red, thick, ->-=0.5] (-1,-1) --(1,-1); 
 
 \node[above] at (-1,1) {$\widetilde g$};
 \node[above] at (1,1) {$\widetilde h$};
  \node[below,red] at (-2,-2) {$Q$};
    \node[below,red] at (0,-1) {$Q$};
 \node[below,red] at (2,-1.8) {$\overline Q$};
\end{tikzpicture}~
\hspace{0.3 in} = \hspace{0.3 in} 
\begin{tikzpicture}[baseline={([yshift=+.5ex]current bounding box.center)},vertex/.style={anchor=base,
    circle,fill=black!25,minimum size=18pt,inner sep=2pt},scale=0.5]

    \draw[red, thick, ->- = 0.5] (-2,-2) --(-1,-1); 
     \draw[red, thick, ->- = 0.5] (2,-2) --(1,-1);
     
 \draw[red, thick, ->-=0.5] (-1,-1) --(1,-1); 
 
 \draw[thick, ->-=0.7 ]  (-1,-1) to[out = 90, in = 180] (0,-0.5); 
  \draw[thick, ->-=0.7 ]  (1,-1) to[out = 90, in = 0] (0,-0.5); 
    \draw[thick, ->-=0.7 ]  (0,-0.5)-- (0,0.5); 
      \draw[thick, ->-=0.7 ]  (0,0.5) -- (-1,1); 
     \draw[thick, ->-=0.7 ]  (0,0.5) -- (1,1);

 \node[above] at (-1,1) {$\widetilde g$};
 \node[above] at (1,1) {$\widetilde h$};
  \node[left] at (-0.9,-0.7) {\footnotesize$\widetilde g $};
 \node[right] at (0.9,-0.7) {\footnotesize$\widetilde h $};
  \node[left] at (0,0.1) {\footnotesize{$\widetilde g \widetilde h$}};
  \node[below,red] at (-2,-2) {$Q$};
    \node[below,red] at (0,-1) {$Q$};
 \node[below,red] at (2,-1.8) {$\overline Q$};
 
\end{tikzpicture}
\hspace{0.3 in} = \,\,\,
(F_{\widetilde g, Q, \overline Q}^{\widetilde g \widetilde h})^{-1}\hspace{0.1 in} 
\begin{tikzpicture}[baseline={([yshift=+.5ex]current bounding box.center)},vertex/.style={anchor=base,
    circle,fill=black!25,minimum size=18pt,inner sep=2pt},scale=0.5]

    \draw[red, thick, ->- = 0.5] (-2,-2) --(0,-1);
     \draw[red, thick, ->- = 0.5] (2,-2) --(0,-1);
     
 \draw[ thick, ->- =0.7] (0,-1) -- (0,0); 
 
      \draw[thick, ->-=0.7 ]  (0,0) -- (-1,1); 
     \draw[thick, ->-=0.7 ]  (0,0) -- (1,1);

 \node[above] at (-1,1) {$\widetilde g$};
 \node[above] at (1,1) {$\widetilde h$};

  \node[left] at (0,-0.4) {$\widetilde g \widetilde h$};
  
    \node[below,red] at (-2,-2) {$Q$};
 \node[below,red] at (2,-1.8) {$\overline Q$};
 
\end{tikzpicture}~,
\eea
in the upper left corner and a similar set of moves in the lower right corner, and in last step we have eliminated the bubble of $Q$-$\overline Q$ to produce a factor of the quantum dimension.
Assembling the ingredients, we find that 
\bea
\begin{tikzpicture}[baseline={([yshift=+.5ex]current bounding box.center)},vertex/.style={anchor=base,
    circle,fill=black!25,minimum size=18pt,inner sep=2pt},scale=0.5]
    \filldraw[grey] (-2,-2) rectangle ++(4,4);
    \draw[thick, dgrey] (-2,-2) -- (-2,+2);
    \draw[thick, dgrey] (-2,-2) -- (+2,-2);
    \draw[thick, dgrey] (+2,+2) -- (+2,-2);
    \draw[thick, dgrey] (+2,+2) -- (-2,+2);
    \draw[thick, black, -stealth] (0,-2) -- (0.354,-1.354);
    \draw[thick, black] (0,-2) -- (0.707,-0.707);
    \draw[thick, black, -stealth] (2,0) -- (1.354,-0.354);
    \draw[thick, black] (2,0) -- (0.707,-0.707);
    \draw[thick, black, -stealth] (-0.707,0.707) -- (-0.354,1.354);
    \draw[thick, black] (0,2) -- (-0.707,0.707);
    \draw[thick, black, -stealth] (-0.707,0.707) -- (-1.354,0.354);
    \draw[thick, black] (-2,0) -- (-0.707,0.707);
    \draw[thick, black, -stealth] (0.707,-0.707) -- (0,0);
    \draw[thick, black] (0.707,-0.707) -- (-0.707,0.707);

    \node[black, below] at (0,-2) {\scriptsize $g$};
    \node[black, right] at (2,0) {\scriptsize $h$};
    \node[black, above] at (0.2,0) {\scriptsize $gh$};
\end{tikzpicture}  
= \,\,\,
 {F^Q_{Q,h,g} F^h_{\overline Q, Q, h} \over F^Q_{Q,g,h}F^g_{\overline Q, Q, g}}  \sum_{\widetilde g, \widetilde h} (F^Q_{Q, \overline Q, Q})_{\widetilde g g} \left(F^Q_{Q,\overline Q, Q}\right)^{-1}_{h \widetilde h} {F_{\widetilde h, Q ,\overline Q}^{\widetilde h \widetilde g} \over F^{\widetilde g \widetilde h}_{\widetilde g, Q, \overline Q}} \,\,\,
\begin{tikzpicture}[baseline={([yshift=+.5ex]current bounding box.center)},vertex/.style={anchor=base,
    circle,fill=black!25,minimum size=18pt,inner sep=2pt},scale=0.5]
    \filldraw[grey] (-2,-2) rectangle ++(4,4);
    \draw[thick, dgrey] (-2,-2) -- (-2,+2);
    \draw[thick, dgrey] (-2,-2) -- (+2,-2);
    \draw[thick, dgrey] (+2,+2) -- (+2,-2);
    \draw[thick, dgrey] (+2,+2) -- (-2,+2);
    \draw[thick, black, -stealth] (0,-2) -- (0.354,-1.354);
    \draw[thick, black] (0,-2) -- (0.707,-0.707);
    \draw[thick, black, -stealth] (2,0) -- (1.354,-0.354);
    \draw[thick, black] (2,0) -- (0.707,-0.707);
    \draw[thick, black, -stealth] (-0.707,0.707) -- (-0.354,1.354);
    \draw[thick, black] (0,2) -- (-0.707,0.707);
    \draw[thick, black, -stealth] (-0.707,0.707) -- (-1.354,0.354);
    \draw[thick, black] (-2,0) -- (-0.707,0.707);
    \draw[thick, black, -stealth] (0.707,-0.707) -- (0,0);
    \draw[thick, black] (0.707,-0.707) -- (-0.707,0.707);

    \node[black, below] at (0,-2) {\scriptsize $\widetilde h$};
    \node[black, right] at (2,0) {\scriptsize $\widetilde g$};
    \node[black, above] at (0.2,0) {\scriptsize $\widetilde g \widetilde h$};
\end{tikzpicture} ~.
\eea

In order to simplify this result, we now use the pentagon identity to eliminate the terms of the form $(F^Q_{Q ,\overline Q, Q})_{\bullet\bullet}$. In particular, the pentagon identity gives rise to the following equations, 
\bea
(F^Q_{Q, \overline Q, Q})_{\widetilde g, g} = (F^Q_{Q, \overline Q, Q})_{\widetilde g, g y} {F_{\widetilde g, Q, y}^Q \over F^Q_{Q, g, y} F^{gy}_{\overline Q, Q, y}}~, \hspace{0.2 in}(F^Q_{Q, \overline Q, Q})_{\widetilde g g} = (F^Q_{Q, \overline Q, Q})_{x^{-1} \widetilde g, g} {F_{x, x^{-1} \widetilde g, Q}^Q F^{\widetilde g}_{x, Q, \overline Q}  \over F^{Q}_{x, Q, g}}~
\hspace{0.2 in}\eea
for any $x$, $y$.
For example, the sequence of moves giving rise to the first equation is as follows, where the orientation of all lines is pointing upwards,
\bea
\begin{tikzpicture}[baseline=0,scale = 0.5, baseline=-10,rotate=180]
\draw [very thick] (0,0) to (0,-2);
\draw [very thick] (0,0) to (2.5,2.5);
\draw [very thick] (0,0) to (-2.5,2.5);
\draw [very thick] (-0.75,0.75) to (1.75-0.75,1.75+0.75);
\draw [very thick] (-1.5,1.5) to (1-1.5,1+1.5);
\node[above] at (0,-2) {$Q$};
\node[below] at (-2.5,2.5) {$y$};
\node[below] at (1.75-0.75,1.75+0.65) {$\overline Q$};
\node[below] at (1-1.5,1+1.5) {$Q$};
\node[below] at (2.5,2.5) {$Q$};
\node[right] at (-0.1,0) {$gy$};
\node[right] at (-0.77,0.76) {$Q$};

\begin{scope}[xshift=5in, yshift = 2in]
\draw [very thick] (0,0) to (0,-2);
\draw [very thick] (0,0) to (2.5,2.5);
\draw [very thick] (0,0) to (-2.5,2.5);
\draw [very thick] (1.5,1.5) to (-1+1.5,1+1.5);
\draw [very thick] (-1.5,1.5) to (1-1.5,1+1.5);
\node[above] at (0,-2) {$Q$};
\node[below] at (-2.5,2.5) {$y$};
\node[below] at (1.75-0.75,1.75+0.65) {$\overline Q$};
\node[below] at (1-1.5,1+1.5) {$Q$};
\node[below] at (2.5,2.5) {$Q$};
\node[right] at (-0.5,0.3) {$Q$};
\node[left] at (0.5,0.3) {$\widetilde g$};
\end{scope}

\begin{scope}[xshift=10in, yshift = 0in]
\draw [very thick] (0,0) to (0,-2);
\draw [very thick] (0,0) to (2.5,2.5);
\draw [very thick] (0,0) to (-2.5,2.5);
\draw [very thick] (0.75,0.75) to (-1.75+0.75,1.75+0.75);
\draw [very thick] (1.5,1.5) to (-1+1.5,1+1.5);
\node[above] at (0,-2) {$Q$};
\node[below] at (-2.5,2.5) {$y$};
\node[below] at (1.75-0.75,1.75+0.65) {$\overline Q$};
\node[below] at (1-1.5,1+1.5) {$Q$};
\node[below] at (2.5,2.5) {$Q$};
\node[left] at (0.2,0) {$Q$};
\node[left] at (0.85,0.78) {$\widetilde g$};
\end{scope}

\begin{scope}[xshift=7in, yshift = -1in]
\draw [very thick] (0,0) to (0,-2);
\draw [very thick] (0,0) to (2.5,2.5);
\draw [very thick] (0,0) to (-2.5,2.5);
\draw [very thick] (0.75,0.75) to (-1.75+0.75,1.75+0.75);
\draw [very thick] (-0.125,1.625) to (0.875-0.125,0.875+1.625);
\node[above] at (0,-2) {$Q$};
\node[below] at (-2.5,2.5) {$y$};
\node[below] at (1.75-0.75,1.75+0.65) {$\overline Q$};
\node[below] at (1-1.5,1+1.5) {$Q$};
\node[below] at (2.5,2.5) {$Q$};
\node[left] at (0.2,0) {$Q$};
\node[right] at (0.5,0.9) {$g$};
\end{scope}

\begin{scope}[xshift=3in, yshift = -1in]
\draw [very thick] (0,0) to (0,-2);
\draw [very thick] (0,0) to (2.5,2.5);
\draw [very thick] (0,0) to (-2.5,2.5);
\draw [very thick] (-0.75,0.75) to (1.75-0.75,1.75+0.75);
\draw [very thick] (0.125,1.625) to (-0.875+0.125,0.875+1.625);
\node[above] at (0,-2) {$Q$};
\node[below] at (-2.5,2.5) {$y$};
\node[below] at (1.75-0.75,1.75+0.56) {$\overline Q$};
\node[below] at (1-1.5,1+1.5) {$Q$};
\node[below] at (2.5,2.5) {$Q$};
\node[right] at (-0.2,0) {$gy$};
\node[left] at (-0.6,0.9) {$g$};
\end{scope}

\draw[thick,stealth-] (2,-2) -- (5,-3);
\draw[thick,stealth-] (11,-3) -- (14.5,-3);
\draw[thick,stealth-] (20,-3) -- (23,-2);
\draw[thick,stealth-] (2,4) -- (8.5,6);
\draw[thick,stealth-] (16,6) -- (6.5+16,4);

\node[right] at (4.5, -3.5) {$F^{gy}_{\overline Q, Q, y}$};
\node[above] at (12.75, -3) {$F^Q_{Q, g, y}$};
\node[right] at (23.5, -3.5) {$(F^Q_{Q,\overline Q, Q})_{\widetilde g, g}$};
\node[above] at (4.5, 7) {$(F^Q_{Q,\overline Q, Q})_{\widetilde g, gy}$};
\node[above] at (20, 7) {$F^Q_{\widetilde g, Q, y}$};

\end{tikzpicture} ~.
\eea

The above two constraints together imply that
\bea
\label{eq:usefulpent1}
(F^Q_{Q, \overline Q, Q})_{\widetilde gg} = (F^Q_{Q ,\overline Q ,Q})_{11} \, {F^{\widetilde g}_{\widetilde g, Q, \overline Q} F^Q_{\widetilde g, Q, g^{-1}} \over F^Q_{Q, g, g^{-1}} F^1_{\overline Q, Q,g^{-1}}}~,
\eea
and similarly one finds that 
\bea
\label{eq:usefulpent2}
(F^Q_{Q, \overline Q ,Q})_{h \widetilde h}^{-1} = (F^Q_{Q, \overline Q ,Q})^{-1}_{11} \,  { F^Q_{Q, h, h^{-1}} F^1_{\overline Q, Q ,h^{-1}} \over F^{\widetilde h}_{\widetilde h, Q ,\overline Q} F^Q_{\widetilde h, Q, h^{-1}} }~.
\eea
Together with the fact that $(F^Q_{Q, \overline Q, Q})_{11} = (F^Q_{Q, \overline Q, Q})_{11}^{-1} = {\kappa_Q \over d_Q}$, this allows us to remove all terms of the form $(F^Q_{Q, \overline Q, Q})_{\bullet \bullet}$, giving 
\bea
\begin{tikzpicture}[baseline={([yshift=+.5ex]current bounding box.center)},vertex/.style={anchor=base,
    circle,fill=black!25,minimum size=18pt,inner sep=2pt},scale=0.5]
    \filldraw[grey] (-2,-2) rectangle ++(4,4);
    \draw[thick, dgrey] (-2,-2) -- (-2,+2);
    \draw[thick, dgrey] (-2,-2) -- (+2,-2);
    \draw[thick, dgrey] (+2,+2) -- (+2,-2);
    \draw[thick, dgrey] (+2,+2) -- (-2,+2);
    \draw[thick, black, -stealth] (0,-2) -- (0.354,-1.354);
    \draw[thick, black] (0,-2) -- (0.707,-0.707);
    \draw[thick, black, -stealth] (2,0) -- (1.354,-0.354);
    \draw[thick, black] (2,0) -- (0.707,-0.707);
    \draw[thick, black, -stealth] (-0.707,0.707) -- (-0.354,1.354);
    \draw[thick, black] (0,2) -- (-0.707,0.707);
    \draw[thick, black, -stealth] (-0.707,0.707) -- (-1.354,0.354);
    \draw[thick, black] (-2,0) -- (-0.707,0.707);
    \draw[thick, black, -stealth] (0.707,-0.707) -- (0,0);
    \draw[thick, black] (0.707,-0.707) -- (-0.707,0.707);

    \node[black, below] at (0,-2) {\scriptsize $g$};
    \node[black, right] at (2,0) {\scriptsize $h$};
    \node[black, above] at (0.2,0) {\scriptsize $gh$};
\end{tikzpicture}  
= \,\,\,
{1\over d_Q^2}{F^Q_{Qhg} \over  F^Q_{Q,g,h}} { F^h_{\overline Q, Q, h} F^1_{\overline Q, Q, h} \over F^g_{\overline Q, Q, g} F^1_{\overline Q, Q, g^{-1}}}  \sum_{\widetilde g, \widetilde h} {F^Q_{\widetilde g, Q, g^{-1}} F^Q_{Q,h, h^{-1}} \over F^Q_{\widetilde h, Q, h^{-1}} F^Q_{Q,g,g^{-1}}} {F^{\widetilde g}_{\widetilde g, Q, \overline Q} F_{\widetilde h,Q ,\overline Q}^{\widetilde h \widetilde g} \over F^{\widetilde g \widetilde h}_{\widetilde g, Q ,\overline Q} F^{\widetilde h}_{\widetilde h, Q ,\overline Q}} \,\,\,
\begin{tikzpicture}[baseline={([yshift=+.5ex]current bounding box.center)},vertex/.style={anchor=base,
    circle,fill=black!25,minimum size=18pt,inner sep=2pt},scale=0.5]
    \filldraw[grey] (-2,-2) rectangle ++(4,4);
    \draw[thick, dgrey] (-2,-2) -- (-2,+2);
    \draw[thick, dgrey] (-2,-2) -- (+2,-2);
    \draw[thick, dgrey] (+2,+2) -- (+2,-2);
    \draw[thick, dgrey] (+2,+2) -- (-2,+2);
    \draw[thick, black, -stealth] (0,-2) -- (0.354,-1.354);
    \draw[thick, black] (0,-2) -- (0.707,-0.707);
    \draw[thick, black, -stealth] (2,0) -- (1.354,-0.354);
    \draw[thick, black] (2,0) -- (0.707,-0.707);
    \draw[thick, black, -stealth] (-0.707,0.707) -- (-0.354,1.354);
    \draw[thick, black] (0,2) -- (-0.707,0.707);
    \draw[thick, black, -stealth] (-0.707,0.707) -- (-1.354,0.354);
    \draw[thick, black] (-2,0) -- (-0.707,0.707);
    \draw[thick, black, -stealth] (0.707,-0.707) -- (0,0);
    \draw[thick, black] (0.707,-0.707) -- (-0.707,0.707);

    \node[black, below] at (0,-2) {\scriptsize $\widetilde h$};
    \node[black, right] at (2,0) {\scriptsize $\widetilde g$};
    \node[black, above] at (0.2,0) {\scriptsize $\widetilde g \widetilde h$};
\end{tikzpicture} ~.
\no\\
\eea
Finally, we again simplify using the pentagon identity, which in particular tells us that 
\bea
F^z_{\overline Q, Q, y} F^{z y^{-1}}_{\overline Q, Q, x} F^Q_{Q,x,y} = F^z_{\overline Q, Q, xy}~, \hspace{0.3in} F^Q_{x,y,Q} F^z_{x,Q,\overline Q}F^{x^{-1} z}_{y, Q,\overline Q} = F^z_{xy, Q, \overline Q}~.
\eea
The result after simplification and change of summation variables is as follows,
\bea
\begin{tikzpicture}[baseline={([yshift=+.5ex]current bounding box.center)},vertex/.style={anchor=base,
    circle,fill=black!25,minimum size=18pt,inner sep=2pt},scale=0.5]
    \filldraw[grey] (-2,-2) rectangle ++(4,4);
    \draw[thick, dgrey] (-2,-2) -- (-2,+2);
    \draw[thick, dgrey] (-2,-2) -- (+2,-2);
    \draw[thick, dgrey] (+2,+2) -- (+2,-2);
    \draw[thick, dgrey] (+2,+2) -- (-2,+2);
    \draw[thick, black, -stealth] (0,-2) -- (0.354,-1.354);
    \draw[thick, black] (0,-2) -- (0.707,-0.707);
    \draw[thick, black, -stealth] (2,0) -- (1.354,-0.354);
    \draw[thick, black] (2,0) -- (0.707,-0.707);
    \draw[thick, black, -stealth] (-0.707,0.707) -- (-0.354,1.354);
    \draw[thick, black] (0,2) -- (-0.707,0.707);
    \draw[thick, black, -stealth] (-0.707,0.707) -- (-1.354,0.354);
    \draw[thick, black] (-2,0) -- (-0.707,0.707);
    \draw[thick, black, -stealth] (0.707,-0.707) -- (0,0);
    \draw[thick, black] (0.707,-0.707) -- (-0.707,0.707);

    \node[black, below] at (0,-2) {\scriptsize $g$};
    \node[black, right] at (2,0) {\scriptsize $h$};
    \node[black, above] at (0.2,0) {\scriptsize $gh$};
\end{tikzpicture}  
= \,\,\,
{1\over d_Q^2}{F^Q_{Q,h,g} \over  F^Q_{Q,g,h}}  \sum_{\widetilde g, \widetilde h} {F^Q_{\widetilde h, Q, g^{-1}} \over F^Q_{\widetilde g, Q, h^{-1}}} {F^Q_{ \widetilde h,\widetilde g, Q}\over F^Q_{\widetilde g, \widetilde h, Q}}  \,\,\,
\begin{tikzpicture}[baseline={([yshift=+.5ex]current bounding box.center)},vertex/.style={anchor=base,
    circle,fill=black!25,minimum size=18pt,inner sep=2pt},scale=0.5]
    \filldraw[grey] (-2,-2) rectangle ++(4,4);
    \draw[thick, dgrey] (-2,-2) -- (-2,+2);
    \draw[thick, dgrey] (-2,-2) -- (+2,-2);
    \draw[thick, dgrey] (+2,+2) -- (+2,-2);
    \draw[thick, dgrey] (+2,+2) -- (-2,+2);
    \draw[thick, black, -stealth] (0,-2) -- (0.354,-1.354);
    \draw[thick, black] (0,-2) -- (0.707,-0.707);
    \draw[thick, black, -stealth] (2,0) -- (1.354,-0.354);
    \draw[thick, black] (2,0) -- (0.707,-0.707);
    \draw[thick, black, -stealth] (-0.707,0.707) -- (-0.354,1.354);
    \draw[thick, black] (0,2) -- (-0.707,0.707);
    \draw[thick, black, -stealth] (-0.707,0.707) -- (-1.354,0.354);
    \draw[thick, black] (-2,0) -- (-0.707,0.707);
    \draw[thick, black, -stealth] (0.707,-0.707) -- (0,0);
    \draw[thick, black] (0.707,-0.707) -- (-0.707,0.707);

    \node[black, below] at (0,-2) {\scriptsize $\widetilde g$};
    \node[black, right] at (2,0) {\scriptsize $\widetilde h$};
    \node[black, above] at (0.2,0) {\scriptsize $\widetilde g \widetilde h$};
\end{tikzpicture} ~.
\eea

In summary then, we have seen that the presence of an $N$-ality defect implies an invariance under a certain twisted gauging,  
\bea
\label{eq:twistedgauging}
    Z_{\mathcal{X}}[g,h] = {1 \over d_Q^2}\frac{F^{Q}_{Q,h,g}}{F^{Q}_{Q,g,h}} \sum_{\widetilde{g},\widetilde{h}} \frac{F^{Q}_{\widetilde{h},Q,g^{-1}}}{F^{Q}_{\widetilde{g},Q,h^{-1}}} {F^Q_{ \widetilde h,\widetilde g, Q}\over F^Q_{\widetilde g ,\widetilde h, Q}} Z_{\mathcal{X}}[\widetilde{g},\widetilde{h}] ~,
\eea
matching with (\ref{eq:introgauging}). 
As claimed in the introduction, this twisted gauging is determined in terms of $F$-symbols involving a single incoming $Q$, namely $F_{Q,g,h}^Q$, $F_{g,Q,h}^Q$, and $F_{g,h,Q}^Q$. Each of these pieces has a simple physical interpretation. The term ${F^Q_{\widetilde h, \widetilde g, Q}/ F^Q_{\widetilde g ,\widetilde h ,Q}}$ captures the SPT stacked before gauging, namely the discrete torsion. The term ${F^Q_{\widetilde h ,Q ,g^{-1}} / F^Q_{\widetilde g, Q, h^{-1}}}$ gives the identification of the dual symmetry, i.e. it dictates how the background field of the dual symmetry couples to the dynamical gauge field. Finally, ${F^Q_{Q,h,g} /  F^Q_{Q,g,h}}$ specifies the SPT stacked after gauging. Our next goal will be to show that these $F$-symbols follow directly from the choice of bulk $\ZZ_N$ symmetry. 

\section{ $F$-symbols and the choice of bulk symmetries}
\label{sec:globalactionA}
In this section, we show that the $F$-symbols involving a single incoming $Q$ follow from the choice of the $\ZZ_N$ symmetry $A$ that is gauged in the bulk in order to obtain the SymTFT. 
In other words, we must choose the symmetry $A$ whose condensation defects' twist defects become the boundary $N$-ality defect $Q$. 
In particular, we need not actually gauge $A$ in order to obtain these $F$-symbols---it suffices for us to discuss the pre-gauged bulk $\ZZ_M^{ n}$ gauge theory, which we now review. 

\subsection{$\ZZ_M^{ n}$ gauge theory}
Our starting point is a (1+1)d theory with anomaly-free $\ZZ_M^{ n}$ global symmetry. In this case, the (2+1)d SymTFT is known to be $\ZZ_M^{ n}$ gauge theory. We take the action to be 
\bea
S = {2 \pi \over M} \sum_{i=1}^{n} \int_{X_2} \widehat a_i \cup \delta a_i ~
\eea
in discrete cochain notation. This theory has $M^{2n}$ topological line operators, which can be written as
\bea
L_{(\mathbf{e}, \mathbf{m}) }(\gamma) = \mathrm{exp}\left({2 \pi i \over M}\oint_\gamma \mathbf{e} \cdot \mathbf{ a}\right)\mathrm{exp}\left({2 \pi i \over M} \oint_\g \mathbf{m}\cdot \mathbf{\widehat a}\right)~,\hspace{0.5 in} (\mathbf{e}; \mathbf{m})\in \ZZ_M^{2n}~,
\eea
where $(\mathbf{e}; \mathbf{m}) = (e_1, \dots, e_n; m_1, \dots, m_n)$ and $(\mathbf{a}; \mathbf{\widehat a}) = (  a_1, \dots, a_n; \widehat a_1,\dots, \widehat a_n)$ are electric and magnetic charges/gauge fields.
In fact, it will often be more useful for us to work in terms of the following basis of operators, 
\bea
\label{eq:simplerbasis}
 L_i(\gamma) := e^{{2\pi i \over M} \oint_\gamma  a_i}~, \hspace{0.5 in}  \widehat L_i(\gamma) := e^{{2\pi i \over M} \oint_\gamma \widehat a_i}~,\hspace{0.5 in} i = 1, \dots, n
\eea
which obey the following commutation relations, 
\bea
 L_i(\g) \times \widehat L_j(\g') = \mathrm{exp} \left( - {2 \pi i \over M} \delta_{ij} \langle \g, \g' \rangle \right) \widehat  L_j(\g') \times  L_i(\g)~. 
\eea
From these commutation relations, together with the relationship,
\bea
\label{eq:Ldecomposition}
L_{(\mathbf{e}, \mathbf{m}) }(\gamma)  = \prod_{i=1}^n L_i^{e_i}(\gamma) \prod_{i=1}^n \widehat L_i^{m_i}(\gamma) ~,
\eea
 one straightforwardly derives the commutation relations for $L_{(\mathbf{e}, \mathbf{m}) }(\gamma)$, 
    \bea
      \label{eq:Lcomm}
     L_{(\mathbf{e}, \mathbf{m})}(\gamma) \times L_{(\mathbf{e'}, \mathbf{m'})}(\gamma') = e^{-{2 \pi i \over M}\langle \g, \g'\rangle \sum_{i=1}^n (e_i m_i' + m_i e_i')}L_{(\mathbf{e'}, \mathbf{m'})}(\gamma')\times L_{(\mathbf{e}, \mathbf{m})}(\gamma)~,
    \eea
as well as the so-called quantum torus algebra,
    \bea
    \label{eq:Lquantumtorus}
    L_{(\mathbf{e}, \mathbf{m})}(\gamma) \times L_{(\mathbf{e}, \mathbf{m})}(\gamma') = e^{- {2 \pi i \over M} \,\mathbf{e} \cdot \mathbf{m}\, \langle \g, \g' \rangle}  L_{(\mathbf{e}, \mathbf{m})}(\gamma+\gamma') ~.
    \eea

 For simplicity, we will often denote a generic charge vector as $\bv = (\mathbf{e}; \mathbf{m}) \in \ZZ_M^{2n}$. Throughout this paper, we will work in a gauge in which the $F$- and $R$-symbols of the lines $L_\bv$ are as given in Figure \ref{fig:FR-move}, explicitly 
    \bea
    F_{\bv_1, \bv_2, \bv_3}^{\bv_1+\bv_2+\bv_3} = 1~, \hspace{0.5 in} R^{\bv, \widetilde \bv} = \omega^{- \bv^T \Omega \widetilde \bv}~,
    \eea
 where we have defined the $2n \times 2n$ matrix
    \bea
    \Omega := \left(\begin{matrix} 0 & \mathds{1}_{n \times n} \\ 0 & 0 \end{matrix}  \right)~.
    \eea

 \begin{figure}[tbp]
\begin{center}
{\begin{tikzpicture}[baseline=0,square/.style={regular polygon,regular polygon sides=4},scale=0.6]
\draw [blue, thick, decoration={markings, mark=at position 0.5 with {\arrow{stealth}}}, postaction={decorate}] (1,1) -- (2,2);
\draw [blue, thick, decoration={markings, mark=at position 0.5 with {\arrow{stealth}}}, postaction={decorate}] (0,0) -- (1,1);
\draw [blue, thick, decoration={markings, mark=at position 0.5 with {\arrow{stealth}}}, postaction={decorate}] (-1,-1) -- (0,0);
\draw [blue, thick, decoration={markings, mark=at position 0.5 with {\arrow{stealth}}}, postaction={decorate}] (1,-1) -- (0,0);
\draw [blue, thick, decoration={markings, mark=at position 0.5 with {\arrow{stealth}}}, postaction={decorate}] (3,-1) -- (1,1);

\filldraw[blue] (0,0) circle (2pt);
\filldraw[blue] (1,1) circle (2pt);

\node[below] at (-1.2,-1) {\footnotesize \color{blue}$L_{\mathbf{v}_1}$};
\node[below] at (1.1,-1) {\footnotesize \color{blue}$L_{\mathbf{v}_2}$};
\node[below] at (3.6,-1) {\footnotesize \color{blue}$L_{\mathbf{v}_3}$};
\node[left] at (0.4,0.8) {\footnotesize \color{blue}$L_{\mathbf{v}_1+\mathbf{v}_2}$};
\end{tikzpicture}}
= \hspace{0.1 in}
{\begin{tikzpicture}[baseline=0,square/.style={regular polygon,regular polygon sides=4},scale=0.6]
\draw [blue, thick, decoration={markings, mark=at position 0.5 with {\arrow{stealth}}}, postaction={decorate}] (1,1) -- (2,2);
\draw [blue, thick] (0,0) -- (1,1);
\draw [blue, thick, ->] (-1,-1) -- (0,0);
\draw [blue, thick, decoration={markings, mark=at position 0.5 with {\arrow{stealth}}}, postaction={decorate}] (1,-1) -- (2,0);
\draw [blue, thick, decoration={markings, mark=at position 0.5 with {\arrow{stealth}}}, postaction={decorate}] (3,-1) -- (2,0);
\draw [blue, thick, decoration={markings, mark=at position 0.5 with {\arrow{stealth}}}, postaction={decorate}] (2,0) -- (1,1);

\filldraw[blue] (2,0) circle (2pt);
\filldraw[blue] (1,1) circle (2pt);

\node[below] at (-1.2,-1) {\footnotesize \color{blue}$L_{\mathbf{v}_1}$};
\node[below] at (1.1,-1) {\footnotesize \color{blue}$L_{\mathbf{v}_2}$};
\node[below] at (3.6,-1) {\footnotesize \color{blue}$L_{\mathbf{v}_3}$};
\node[right] at (1.3,0.8) {\footnotesize \color{blue}$L_{\mathbf{v}_2+\mathbf{v}_3}$};
\end{tikzpicture}}
\\\vspace{0.2 in}
{\begin{tikzpicture}[baseline=-5,rotate=180]
\draw [blue,thick,decoration={markings, mark=at position 0.5 with {\arrow{stealth}}}, postaction={decorate}] (0,-.5)  to (0,0);
\draw [blue,thick] (0,0) arc [radius=.3, start angle=240, end angle=120];
\draw [blue,thick] (0,0) arc [radius=.3, start angle=-60, end angle=42];

\draw [blue,thick, decoration={markings, mark=at position 0.8 with {\arrow{stealth}}}, postaction={decorate}] (-0.1,0.42) -- (0.4,0.7) ;

\draw [white, line width = 0.15cm] (0.1,0.42) -- (-0.4,0.7) ;

\draw [blue,thick, decoration={markings, mark=at position 0.8 with {\arrow{stealth}}}, postaction={decorate}] (0.1,0.42) -- (-0.4,0.7) ;

\filldraw[blue] (0,0) circle (1.2pt);

\node[below] at (0.6,0.7) {\footnotesize \color{blue}$L_{\mathbf{v}}$};
\node[below] at (-0.7,0.7) {\footnotesize \color{blue}$L_{\mathbf{\widetilde v}}$};
\end{tikzpicture}}
\hspace{0 in}= \,\,\,\,$\omega^{\mathbf{v}^T \Omega^T \mathbf{\widetilde v}}$\hspace{-0.2 in}
{\begin{tikzpicture}[baseline=5,rotate=180]
\draw [blue,thick,decoration={markings, mark=at position 0.5 with {\arrow{stealth}}}, postaction={decorate}] (0,-.8) to (0,0);
\draw [blue,thick, decoration={markings, mark=at position 0.7 with {\arrow{stealth}}}, postaction={decorate}] (0,0) to (-.6,.5);
\draw [blue,thick, decoration={markings, mark=at position 0.7 with {\arrow{stealth}}}, postaction={decorate}] (0,0)--(.6,.5) ;
\filldraw[blue] (0,0) circle (1.2pt);

\node[below] at (0.8,0.4) {\footnotesize \color{blue}$L_{\mathbf{v}}$};
\node[below] at (-0.9,0.4) {\footnotesize \color{blue}$L_{\mathbf{\widetilde v}}$};
\end{tikzpicture}}
\hspace{0.2 in}
{\begin{tikzpicture}[baseline=-5,rotate=180]
\draw [blue,thick,decoration={markings, mark=at position 0.5 with {\arrow{stealth}}}, postaction={decorate}] (0,-.5)  to (0,0);
\draw [blue,thick] (0,0) arc [radius=.3, start angle=240, end angle=120];
\draw [blue,thick, decoration={markings, mark=at position 0.7 with {\arrow{stealth}}}, postaction={decorate}] (0,0.52) to +(30:.5) ;
\draw [blue,thick] (0,0) arc [radius=.3, start angle=-60, end angle=42];
\draw [blue,thick, decoration={markings, mark=at position 0.6 with {\arrow{stealth}}}, postaction={decorate}] (0,0.52) ++(150:.1) to +(150:.4) ;

\filldraw[blue] (0,0) circle (1.2pt);

\node[below] at (0.6,0.7) {\footnotesize \color{blue}$L_{\mathbf{v}}$};
\node[below] at (-0.7,0.7) {\footnotesize \color{blue}$L_{\mathbf{\widetilde v}}$};
\end{tikzpicture}}
\hspace{0 in}= \,\,\,\,$\omega^{-\mathbf{v}^T \Omega \mathbf{\widetilde v}}$\hspace{-0.2 in}
{\begin{tikzpicture}[baseline=5,rotate=180]
\draw [blue,thick,decoration={markings, mark=at position 0.5 with {\arrow{stealth}}}, postaction={decorate}] (0,-.8) to (0,0);
\draw [blue,thick, decoration={markings, mark=at position 0.7 with {\arrow{stealth}}}, postaction={decorate}] (0,0) to (-.6,.5);
\draw [blue,thick, decoration={markings, mark=at position 0.7 with {\arrow{stealth}}}, postaction={decorate}] (0,0)--(.6,.5) ;
\filldraw[blue] (0,0) circle (1.2pt);

\node[below] at (0.8,0.4) {\footnotesize \color{blue}$L_{\mathbf{v}}$};
\node[below] at (-0.9,0.4) {\footnotesize \color{blue}$L_{\mathbf{\widetilde v}}$};
\end{tikzpicture}}
\caption{F- and R-moves of the line operators $L_{\mathbf{\bv}}$.}\label{fig:FR-move}
\end{center}
\end{figure}

\subsection{Symmetries of the bulk}
\label{sec:symmofbulk}

 In order to go from $\ZZ_M^{n}$ gauge theory to the SymTFT for $N$-ality defects, one must gauge an appropriate bulk action, as discussed in the introduction.  
The set of allowed bulk symmetries is given by all actions on the lines $L_\mathbf{v}$ that preserve the Dirac pairing. In three-dimensions, the Dirac pairing is given by the symmetric matrix 
 \bea
 \label{eq:matrixI}
 \mathfrak{I} := \Omega + \Omega^T = \left( \begin{matrix} 0 & \mathds{1}_{n \times n} \\ \mathds{1}_{n \times n} & 0 \end{matrix} \right) ~, 
 \eea
 and the set of matrices which leave this invariant (modulo $M$) is given by\footnote{ The finite orthogonal group defined in this way is sometimes also denoted by $O^+(n,n;M)$. Note that for $M\neq 2$, it is equivalent to 
 \bea
 O(n,n; M) = \left\{ A\in GL(2n; M) \,| \,\, A^T \mathrm{diag}(\mathds{1}_n, - \mathds{1}_n) A = \mathrm{diag}(\mathds{1}_n, - \mathds{1}_n)\, \right\} ~.
 \eea
As explicit examples, we have 
\bea
O(1,1; 3 ) = \ZZ_2 \times \ZZ_2~, \hspace{0.3 in} O(1,1; 5) = D_8~, \hspace{0.3 in}O(1,1; 7) = S_3\times \ZZ_2~, \hspace{0.3 in}O(1,1; 9) = D_{16}~.
\eea
 } 
 \bea
 O(n,n; M) := \left\{ A\in GL(2n; M) \,| \, A^T \mathfrak{I} A = \mathfrak{I}\, \right\} ~.
 \eea
Hence in order to get the SymTFT for $N$-ality defects, we would like to gauge an order $N$ element of $ O(n,n; M) $. For each conjugacy class of order $N$ elements, we may obtain the SymTFT for a different $N$-ality category. Physically, the elements of $ O(n,n; M) $ can be generated by three types of operations\cite{Etingof:2009yvg,nikshych2014categorical,fuchs2015brauer,Moradi:2022lqp}:
  \begin{enumerate}
  
\item \textbf{Automorphisms of the boundary $\mathbb{Z}_M^n$ symmetry lines:} these correspond to elements in $\mathrm{Aut}(\ZZ_M^{n})$. In this case, the symmetry $A$ is of the form
    \begin{equation} \label{eq:bsC}
        \sfC_\alpha = \begin{pmatrix} \alpha & 0 \\ 0 & (\alpha^{-1})^T\end{pmatrix} ~,
    \end{equation}
    where $\alpha$ is an invertible $n\times n$-matrix. We denote the symmetry as $\sfC_\alpha$ because it is a generalization of charge conjugation, in the sense of being a generic automorphism. It is straightforward to check that this preserves the Dirac pairing,
    \begin{equation}
        \begin{pmatrix} \alpha^T & 0 \\ 0 & \alpha^{-1} \end{pmatrix} \begin{pmatrix} 0 & \mathds{1} \\ \mathds{1} & 0   \end{pmatrix} \begin{pmatrix} \alpha & 0 \\ 0 & (\alpha^{-1})^T \end{pmatrix} =\begin{pmatrix} 0 & \mathds{1} \\ \mathds{1} & 0   \end{pmatrix}~.
    \end{equation}

\item \textbf{Stacking with SPT phases:} these correspond to elements in\footnote{More generally, for $G= \prod_{i=1}^n \ZZ_{M_i}$, we have 
\bea
H^2(G, U(1)) = \prod_{i<j}Z_{\mathrm{gcd}(M_i, M_j )}~,
\eea
}
\bea
H^2(\ZZ_M^{n}, U(1)) = \ZZ_M^{\half n(n-1)}~.
\eea
In this case, the symmetry $A$ is of the form
    \begin{equation}\label{eq:bsT}
        \sfT_\varphi = \begin{pmatrix} \mathds{1} & \varphi \\ 0 & \mathds{1} \end{pmatrix} ~,
    \end{equation}
    where $\varphi$ is a $n\times n$ anti-symmetric matrix. We denote this symmetry as $\sfT_\varphi$, as is standard notation for stacking with an SPT phase \cite{Witten:2003ya,Gaiotto:2020iye}. It is straightforward to check that the anti-symmetry of $\varphi$ guarantees that the Dirac pairing is preserved. 

\item \textbf{Electromagnetic dualities:} there is one such duality for each factor of $\ZZ_M$. In this case $A$ is an off-diagonal block matrix of the form
    \begin{equation}
        \begin{pmatrix} 0 & \chi \\ 
        \beta & 0\end{pmatrix} ~.
    \end{equation}
    It is straightforward to show that for this matrix to be a symmetry, that is, 
    \begin{equation}
        \begin{pmatrix} 0 & \beta^T \\ 
        \chi^T & 0\end{pmatrix} \begin{pmatrix}  0 & \mathds{1} \\ \mathds{1} & 0 \end{pmatrix} \begin{pmatrix} 0 & \chi \\ 
        \beta & 0\end{pmatrix} = \begin{pmatrix}  0 & \mathds{1} \\ \mathds{1} & 0 \end{pmatrix} ~,
    \end{equation}
    we must have $\beta^T \chi = \mathbf{1}$. Thus, we denote these matrices as 
    \begin{equation}
        \mathsf{S}_\chi = \begin{pmatrix} 0 & \chi \\ (\chi^{-1})^{T} & 0\end{pmatrix} ~.
    \end{equation}
  As will be seen below, these implement electromagnetic dualities, for which $\mathsf{S}$ is standard notation \cite{Witten:2003ya,Gaiotto:2020iye}. 

\end{enumerate}

Having understood these ingredients, we now show that any symmetry\footnote{The $n \times n$ matrix $\mathsf{C}$ appearing here is unrelated to the autmorphism $\C_\alpha$ appearing above.}
\begin{equation}
\label{eq:decompofA}
    A = \begin{pmatrix} \A & \C \\ \B & \D \end{pmatrix} ~,
\end{equation}
with $\B$ assumed to be invertible in order to ensure maximal gauging, can be decomposed as
\begin{equation}
\label{eq:Adecomposition}
    A = \sfT_{\varphi} \mathsf{S}_\chi \sfT_{\widetilde{\varphi}} ~
\end{equation}
for appropriate $\varphi$, $\widetilde \varphi$, and $\chi$. 

To see this, it suffices to prove that there always exist $\varphi$ and $\widetilde{\varphi}$ such that
\begin{equation}
    \sfT_{-\varphi} A \sfT_{-\widetilde{\varphi}} = \mathsf{S}_\chi ~.
\end{equation}
Notice that
\begin{equation}
    \sfT_{-\varphi} A \sfT_{-\widetilde{\varphi}} = \begin{pmatrix} \A - \varphi \B & \A - \varphi \B - \C \widetilde{\varphi} + \varphi \D \widetilde{\varphi} \\ \B & -\B \widetilde{\varphi} + \D  \end{pmatrix} ~,
\end{equation}
and since $\B$ is invertible, we can always choose $\varphi = \A \B^{-1}$ and $\widetilde{\varphi} = \B^{-1} \D$, which then gives a matrix of the correct form with $\chi = (\B^{-1})^T$. It then only remains to show that both $\varphi$ and $\widetilde{\varphi}$ are anti-symmetric. This follows from the fact that $A$ is a symmetry, namely
\begin{equation}
    \begin{pmatrix} 0 & \mathds{1} \\ \mathds{1} & 0 \end{pmatrix} = \begin{pmatrix} \A^T & \B^T \\ \C^T & \D^T \end{pmatrix} \begin{pmatrix} 0 & \mathds{1} \\ \mathds{1} & 0 \end{pmatrix} \begin{pmatrix} \A & \C \\ \B & \D \end{pmatrix} = \begin{pmatrix} \B^T \A + \A^T \B & \B^T \C + \A^T \D \\ \D^T \A + \C^T \B & \D^T \C + \C^T \D \end{pmatrix} 
\end{equation}
which implies that
\begin{equation}
    \A\B^{-1} = -(\B^{-1})^T \A^T ~, \quad \B^T \C + \A^T \D = \mathds{1} ~, \quad \D^T \C= - \C^T \D ~.
\end{equation}
It is straightforward to see that $\varphi = \A \B^{-1}$ is anti-symmetric from the first equation; to see that $\widetilde{\varphi}$ is anti-symmetric, we notice that using the second equation, we can express it as a sum of two anti-symmetric matrices,
\begin{equation}
    \widetilde{\varphi} = \B^{-1} \D = \C^T \D + \D^T \A \B^{-1} \D ~.
\end{equation}
To summarize then, we may writes $A$ given in (\ref{eq:decompofA}) in the factorized form (\ref{eq:Adecomposition}), where $\varphi$, $\widetilde \varphi$, and $\chi$ are defined as follows,
\begin{equation}
\label{eq:chiphidefs}
   \varphi = \A\B^{-1} ~, \hspace{0.2 in} \widetilde{\varphi} = \B^{-1} \D ~, \hspace{0.2 in} \chi = (\B^{-1})^T.
\end{equation}
Physically, we have split the action of $A$ on the boundary into three pieces: a stacking with an SPT, a gauging, and another stacking.

\subsection{Action of $A$ on boundaries}

We now compute the action of $A$ on the topological boundary.  Recall that we assign the physical state $|\mathcal{X}\rangle$ to the right boundary, given by
\begin{equation}
    |\mathcal{X}\rangle = \sum_{\mathbf{a} \in H^1(\mathbb{Z}_M^n,U(1))} Z_{\mathcal{X}}[\mathbf{a}] \,| \mathbf{a}\rangle ~,
\end{equation}
and boundary states representing Dirichlet or Neumann boundary conditions to the left boundary,
\begin{equation}
    \langle D(\mathbf{A}) | = \langle \mathbf{A} | ~, \hspace{0.5 in} \langle N(\mathbf{A}) | = \sum_{\mathbf{a}} e^{i\int  \mathbf{a} \cup \mathbf{A}} \langle \mathbf{a}| ~.
\end{equation}
We will focus on the Dirichlet boundary condition here. 
The simple anyons act on the Dirichlet boundary state via
\begin{equation}
\begin{aligned}
    L_{(\mathbf{e},\mathbf{m})}(\gamma)|\mathbf{A}\rangle &= \exp\left(\frac{2\pi i}{M}\oint_\gamma \mathbf{e}\cdot \mathbf{A} \right) |\mathbf{A} - \mathbf{m}\,\omega_\gamma\rangle ~, 
\end{aligned}
\end{equation}
where $\omega_\gamma$ denotes the Poincar{\'e} dual of $\gamma$, defined as
\begin{equation}
    \exp\left(\frac{2\pi i}{M}\oint_{\gamma'}\omega_\gamma \right) = \exp\left(\frac{2\pi i}{M} \langle \gamma',\gamma\rangle \right) ~, \qquad \forall\,\, \gamma' ~.
\end{equation}
It is straightforward to check that the above gives the correct algebra of anyons, that is,
\begin{equation}
    L_{(\mathbf{e},\mathbf{m})}(\gamma) L_{(\mathbf{e}',\mathbf{m}')}(\gamma') |\mathbf{A}\rangle = \exp\left(-\frac{2\pi i}{M}(\mathbf{e}\cdot \mathbf{m}' + \mathbf{e}'\cdot \mathbf{m}) \langle \gamma,\gamma'\rangle\right) L_{(\mathbf{e}',\mathbf{m}')}(\gamma') L_{(\mathbf{e},\mathbf{m})}(\gamma) |\mathbf{A}\rangle ~.
\end{equation}

Next, we compute the symmetry action on the boundary states. We separately compute three types of actions,

\begin{enumerate}
    \item \textbf{$\mathrm{Aut}(\mathbb{Z}_M^n)$:}  Denote the condensation defect corresponding to $\sfC_\alpha$ by $D_{\sfC_\alpha}$.   From the fact that $\sfC_\alpha$ simply permutes the pure electric lines (and pure magnetic lines) among themselves, it follows that
    \begin{equation}
        D_{\sfC_\alpha} |\mathbf{A}\rangle = |\alpha^T \cdot \mathbf{A}\rangle = |\mathbf{A}_i \alpha_{ij}\rangle ~.
    \end{equation}
    Likewise, the action of $D_{\sfC_\alpha}$  on a bra can be computed as
    \begin{equation}
        \langle \mathbf{A}| D_{\sfC_{\alpha}} = \sum_{\mathbf{B}}\langle \mathbf{A}|D_{\sfC_\alpha}|\mathbf{B}\rangle \langle \mathbf{B}| = \langle (\alpha^{-1})^T \cdot \mathbf{A}| ~.
    \end{equation}
    
    As a consistency check, note that the operator equation 
    \begin{equation}
        L_{(\mathbf{e},\mathbf{m})}(\gamma) \cdot D_{\sfC_\alpha} = D_{\sfC_\alpha} \cdot L_{(\alpha \cdot \mathbf{e}, (\alpha^{-1})^T\cdot \mathbf{m})}   
    \end{equation}
    holds for every state $|\mathbf{A}\rangle$,
    \begin{equation}
    \begin{aligned}
        L_{(\mathbf{e},\mathbf{m})}(\gamma) \cdot D_{\sfC_\alpha} | \mathbf{A}\rangle &= \exp\left(\frac{2\pi i}{M}\oint_{\gamma} \mathbf{e}\cdot \alpha^T \cdot \mathbf{A}\right) |\alpha^T (\mathbf{A} - (\alpha^T)^{-1}\cdot \mathbf{m}\,\omega_\gamma)\rangle ~, \\
        D_{\sfC_\alpha} \cdot L_{(\alpha\cdot \mathbf{e},(\alpha^{-1})^T \cdot \mathbf{m})}(\gamma) |\mathbf{A}\rangle &= \exp\left(\frac{2\pi i}{M}\oint_{\gamma} \mathbf{e}\cdot \alpha^T \cdot \mathbf{A}\right) |\alpha^T (\mathbf{A} - (\alpha^T)^{-1}\cdot \mathbf{m}\,\omega_\gamma)\rangle ~.
    \end{aligned}
    \end{equation}
    \item \textbf{$\mathrm{H}^2(\mathbb{Z}_M^n,U(1))$:} Denote the condensation defect corresponding to $\sfT_\varphi$ by $D_{\sfT_\varphi}$. Since $\sfT_\varphi$ does not change the background fields, but only stacks a non-trivial SPT given by $\varphi$, it must act on $|\mathbf{A}\rangle$ by a phase, and in particular
    \begin{equation}
        D_{\sfT_\varphi} | \mathbf{A}\rangle = \exp \left(\frac{\pi i}{M} \int \mathbf{A}\cdot \varphi^T \cup \mathbf{A}\right) |\mathbf{A}\rangle = \exp \left(\frac{\pi i}{M} \int \mathbf{A}_i (\varphi^T)_{ij} \cup \mathbf{A}_j\right) |\mathbf{A}\rangle  ~,
    \end{equation}
    where since $\varphi$ and $\cup$ are anti-symmetric, each term is actually evaluated twice, and hence the phase with coefficient $\frac{\pi i}{M}$ is well-defined. 

    We can check explicitly that this action is consistent with the algebra
    \begin{equation}
        L_{(\mathbf{e},\mathbf{m})}(\gamma) \cdot D_{\sfT_\varphi} = D_{\sfT_\varphi} \cdot L_{(\mathbf{e}+\varphi\cdot\mathbf{m},\mathbf{m})}(\gamma) ~,
    \end{equation}
    by using the fact that
    \begin{equation}
    \begin{aligned}
         & \exp\left(\frac{\pi i}{M} \int (\mathbf{A} - \mathbf{m}\,\omega_\gamma)\cdot \varphi^T \cup (\mathbf{A} - \mathbf{m}\,\omega_\gamma)\right) \\
        =& \exp\left(\frac{\pi i}{M} \int \mathbf{A}\cdot \varphi^T \cup \mathbf{A} - \omega_\gamma\cup \mathbf{m}\cdot \varphi^T  \cdot \mathbf{A} - \mathbf{A}\cdot \varphi^T\cdot \mathbf{m}\cup \omega_\gamma + \mathbf{m}\cdot \varphi^T\cdot \mathbf{m}\,\omega_\gamma\cup \omega_\gamma\right) \\
        =& \exp\left(\frac{\pi i}{M} \int \mathbf{A}\cdot \varphi^T \cup \mathbf{A}\right) \exp\left(-\frac{\pi i}{M}\oint_\gamma (\mathbf{m}\cdot\varphi^T \cdot \mathbf{A} -  \mathbf{A}\cdot \varphi^T \cdot \mathbf{m})\right) \\
        =& \exp\left(\frac{\pi i}{M} \int \mathbf{A}\cdot \varphi^T \cup \mathbf{A}\right) \exp\left(-\frac{2 \pi i}{M}\oint_\gamma \mathbf{m}\cdot\varphi^T \cdot \mathbf{A} \right) ~,
    \end{aligned}
    \end{equation}
    where we have used the anti-symmetry of $\varphi$ and the cup product.

    It is not hard to see that 
    \begin{equation}
        \langle \mathbf{A}| D_{\sfT_\varphi} = \langle \mathbf{A}| \exp\left(\frac{\pi i}{M} \int \mathbf{A} \cdot \varphi^T \cup \mathbf{A}\right) ~.
    \end{equation}

    \item \textbf{Electromagnetic dualities:} Finally, denote the condensation defect corresponding to $\mathsf{S}_\chi$ by $D_{\mathsf{S}_\chi}$.     %
   The action on the states $|\mathbf{A}\rangle$ is given by,
    \begin{equation}
        D_{\mathsf{S}_\chi} |\mathbf{A}\rangle = \sum_{\mathbf{a}} \exp\left(\frac{2\pi i}{M}\int \mathbf{a}\cup \chi^T \mathbf{A}\right)|\mathbf{a}\rangle 
    \end{equation}
   which reproduces the correct algebra relations
    \begin{equation}
        L_{(\mathbf{e},\mathbf{m})}(\gamma) \cdot D_{\mathsf{S}_\chi}|\mathbf{A}\rangle = D_{\mathsf{S}_\chi}\cdot L_{(\chi\cdot \mathbf{m},(\chi^{-1})^T\cdot \mathbf{e})}(\gamma) |\mathbf{A}\rangle ~.
    \end{equation}
    Similarly, we have that $D_{\mathsf{S}_\chi}$ acts on the bra $\langle \mathbf{A}|$ as
    \begin{equation}
        \langle \mathbf{A}| D_{\mathsf{S}_\chi} = \sum_{\mathbf{a}} \langle \mathbf{a}| \exp\left(-\frac{2\pi i}{M}\int \mathbf{a} \cdot \chi \cup \mathbf{A}\right) ~.
    \end{equation}
\end{enumerate}

The above results, together with the decomposition of $A$ in (\ref{eq:Adecomposition}), allow us to straightforwardly compute the action of a generic $\ZZ_N$ element of $O(n,n;M)$ on the topological boundary, giving
\begin{equation}
\begin{aligned}
    \langle \mathbf{A}| D_{A} &= \langle \mathbf{A}| D_{\sfT_{\varphi}} D_{\mathsf{S}_{\chi}} D_{\sfT_{\widetilde{\varphi}}} \\
    &= \langle \mathbf{A}| \exp\left(\frac{\pi i}{M}\int \mathbf{A}\cdot \varphi^T\cup \mathbf{A}\right) D_{\mathsf{S}_{\chi}} D_{\sfT_{\widetilde{\varphi}}} \\
    &= \sum_{\mathbf{a}} \langle \mathbf{a}| \exp\left(\frac{\pi i}{M}\int \mathbf{a}\cdot \widetilde{\varphi}^T \cup \mathbf{a} -\frac{2\pi i}{M}\int \mathbf{a} \cdot \chi \cup \mathbf{A} + \frac{\pi i}{M}\int \mathbf{A}\cdot \varphi^T \cup \mathbf{A}\right) ~,
\end{aligned}
\end{equation}
or, at the level of the partition function,
\begin{equation}\label{eq:act_on_part}
     Z_{\mathcal{X}}[\mathbf{a}]= \langle \mathbf{A}|D_A|\mathcal{X}\rangle = \sum_{\mathbf{a}} Z_{\mathcal{X}}[\mathbf{a}] \exp\left(\frac{\pi i}{M}\int \mathbf{a}\cdot \widetilde{\varphi}^T \cup \mathbf{a} -\frac{2\pi i}{M}\int \mathbf{a} \cdot \chi \cup \mathbf{A} + \frac{\pi i}{M}\int \mathbf{A}\cdot \varphi^T \cup \mathbf{A}\right) ~.
\end{equation}
We may now fix the $F$-symbols involving a single incoming $Q$ in terms of the choice of bulk symmetry $A$ by comparing to the result in (\ref{eq:twistedgauging}). 
In particular, let us take standard conventions for the $\mathfrak{A}$ and $\mathfrak{B}$ cycles on the torus, illustrated below, 
\begin{equation}
   \begin{tikzpicture}[baseline={([yshift=+.5ex]current bounding box.center)},vertex/.style={anchor=base,
    circle,fill=black!25,minimum size=18pt,inner sep=2pt},scale=0.5]
    \filldraw[grey] (-2,-2) rectangle ++(4,4);
    \draw[thick, dgrey] (-2,-2) -- (-2,+2);
    \draw[thick, dgrey] (-2,-2) -- (+2,-2);
    \draw[thick, dgrey] (+2,+2) -- (+2,-2);
    \draw[thick, dgrey] (+2,+2) -- (-2,+2);
    \draw[very thick, blue, ->-=0.3 ] (0,-2) -- (0,2);
       \draw[very thick, red, ->-=0.3 ] (-2,0) -- (2,0);
   
    \node[blue, below] at (0,-2) { $\mathfrak{B}$};
    \node[red, right] at (2,0) { $\mathfrak{A}$};
\end{tikzpicture} ~,
\end{equation}
with intersection pairing  $\langle \mathfrak{A}, \mathfrak{B} \rangle = - \langle \mathfrak{B}, \mathfrak{A} \rangle = 1$. In this case, we have that 
\bea
\int_{\mathfrak{A} } A_i = g_i~, \hspace{0.4 in} \int_{\mathfrak{B} } A_i = h_i~, \hspace{0.4in}\int_{\mathfrak{A} } a_i = \widetilde g_i~, \hspace{0.4 in} \int_{\mathfrak{B} } a_i =\widetilde h_i~, 
\eea
or in other words 
\bea
A_i = - h_i\, \omega_\mathfrak{A} + g_i\, \omega_{\mathfrak{B}}~, \hspace{0.4 in}a_i = - \widetilde h_i\, \omega_\mathfrak{A} + \widetilde g_i\, \omega_{\mathfrak{B}}~. 
\eea
It is then straightforward to rewrite (\ref{eq:act_on_part}) in terms of $g_i, h_i, \widetilde g_i, \widetilde h_i$ and compare to  (\ref{eq:twistedgauging}), giving
\bea\label{eq:boundaryFmethod2}
{F_{Q gh}^Q \over  F_{Q hg}^Q} = \omega^{g^T \varphi h }~, \hspace{0.5 in} F_{g Q h }^Q = \omega^{-g^T \chi h}~, \hspace{0.5 in} {F_{ghQ}^Q\over  F_{hgQ}^Q}= \omega^{g^T \widetilde \varphi h }~,
\eea
with $\varphi$, $\widetilde \varphi$, and $\chi$ defined in (\ref{eq:chiphidefs}). In particular, we make the following choice of gauge for the $F$-symbols,
\bea
\label{eq:boundaryFmethod1}
F_{Q gh}^Q = \omega^{\sum_{i<j} g_i \varphi_{ij} h_j }~, \hspace{0.5 in} F_{g Q h }^Q = \omega^{-g^T \chi h}~, \hspace{0.5 in} F_{ghQ}^Q = \omega^{\sum_{i<j} g_i\widetilde \varphi_{ij} h_j }~,
\eea
Thus the choice of $A$ determines all boundary $F$-symbols with only a single incoming $Q$, as claimed in Section \ref{sec:introFsymbols}.

\section{Condensation defects, interfaces, and bimodules}

Our next goal is to provide an alternative derivation of the boundary $F$-symbols, by first computing the bulk $F$-symbols and then taking the boundary limit. 
This derivation is significantly more involved though, and requires a more thorough analysis of the bulk theory (and in particular the bulk twist defects). To prepare for this computation, in the current section we discuss various technical aspects about the bulk theory, including a description of condensation defects and the interfaces between them. We return to the connection with boundary $F$-symbols in Section \ref{sec:bulktoboundaryFsymb}.

\subsection{Condensation defects} 
\label{sec:conddefect}
 
 Our first goal will be to give an explicit form for the condensation defect $D_A(M_2)$ for an arbitrary element $A \in O(n,n; M) $. The defining property of the condensation defect $D_A(M_2)$ is that, when the line $L_{(\mathbf{e}, \mathbf{m})}(\gamma)$ passes through it on the right, it is modified to $L_{A(\mathbf{e}, \mathbf{m})}(\gamma)$. By the folding trick, this means that $L_{(1-A)(\mathbf{e}, \mathbf{m})}(\gamma)$ can be absorbed by the defect. This is true for every choice of $(\mathbf{e}, \mathbf{m})$, and in particular holds when $(\mathbf{e}, \mathbf{m})$ is taken to be any element of the canonical basis of $\mathbb{R}^{2n}$, e.g. $(1,0,\dots, 0)$, $(0,1,0\dots, 0)$ and so on. Denoting the $i$-th column of $(1-A)$ by $(1-A)_i$, we conclude that $D_A(M_2)$ can be written as a condensate of $L_{(1-A)_i}(\gamma)$ for $i=1,\dots, 2n$. 
 
 A subtlety arises in the case that $(1-A)_i$ and $(1-A)_j$ for $i\neq j$ are equivalent, or more generally whenever one of the $(1-A)_i$ is a linear combination of the others, since in this case including all of $L_{(1-A)_i}(\gamma)$ for $i=1,\dots, 2n$ would lead to a redundancy. First let us assume that the mod $M$ kernel is trivial, i.e. $\mathrm{ker}_M(1-A)=0$, so that such redundancies do not occur; this is equivalent to demanding that $\mathrm{gcd}(\mathrm{det}(1-A), M) = 1$.  
 With this assumption, the general form of the condensation defect is as follows, 
 \bea
 \label{eq:defofDA}
 D_A(M_2):= {1\over |H^0(M_2, \ZZ_M)|^{2n}}\sum_{\g_1, \dots, \g_{2n} \in H_1(M_2, \ZZ_M)} e^{{2 \pi i \over M} \half \sum_{i,j=1}^{2n} Q^A_{ij} \langle \g_i, \g_j \rangle} \prod_{i=1}^{2n} L_{(1-A)_i}(\gamma_i)~,
 \eea
 where we take the matrix of coefficients $Q^A$ to be anti-symmetric.
This matrix can be fixed by requiring that the desired fusion rules are satisfied, 
\bea
\label{eq:DAfusions}
D_A(M_2) \times {D_A}^\dagger (M_2) &=& \chi[M_2, \ZZ_M]^{-2n}~,
\no\\
L_{(\mathbf{e}, \mathbf{m})}(\gamma)\times D_A(M_2)   &=& D_A(M_2) \times  L_{A(\mathbf{e}, \mathbf{m})}(\gamma)~,
\eea
where $\chi[M_2; \ZZ_M]$ is the Euler character of the manifold $M_2$ with coefficients in $\ZZ_M$.

In particular, denoting the matrix elements of $1-A$ by
\bea
\label{eq:abdef}
1- A = \left(\begin{matrix}  \a_1^1 & \dots & \a_1^{2n} \\ \vdots & & \vdots \\ \a_n^1 &\dots & \a_n^{2n} \\ \b^1_1 & \dots & \b_1^{2n} \\ \vdots & & \vdots \\ \b_n^1 & \dots & \b_n^{2n}\end{matrix}  \right) ~, 
\eea
the fusion rules fix the coefficients $Q^A$ to be as follows, 
\bea
\label{eq:QAdefab}
Q^A_{ij} = \left\{\begin{matrix}  \beta_j^i && j \leq n\\
\a_{j-n}^i&&j > n
\end{matrix}  \right. ~, \hspace{0.5 in} i < j~, 
\eea
and the analogous quantities with an overall minus sign for $i >j$, such that $Q^A_{ij}$ is anti-symmetric. The derivation of these coefficients is rather lengthy and is relegated to Appendices \ref{app:DAfusions} and \ref{app:matrixnotation}. 

Next, when the mod $M$ kernel of $(1-A)$ is \textit{non-trivial},\footnote{\label{ft:SizeofKer} 
Generically, the order of $\mathrm{ker}_M(1-A)$ can be described as follows. Given the prime factorization of the modulus $M=\prod_i p_i^{k_i}$, the ring isomorphism $\mathbb{Z}_M\cong \prod_i\mathbb{Z}_{p_i^{k_i}}$ induces a $\mathbb{Z}_M$ module isomorphism $\mathbb{Z}_M^{2n}\cong \prod_i\mathbb{Z}_{p_i^{k_i}}^{2n}$. A vector $\mathbf{v}\in\mathbb{Z}_M^{2n}$ then corresponds to a unique tuple of vectors $(\mathbf{v}_1,\mathbf{v}_2,\dots)$ where $\mathbf{v}_i\in\mathbb{Z}_{p_i^{k_i}}^{2n}$. Under this isomorphism, the single kernel condition $(1-A)\mathbf{v}=0\,\,(\mathrm{mod}\,M)$ is equivalent to a system of simultaneous linear equations $(1-A)\mathbf{v}_i=0\,\,(\mathrm{mod}\,p_i^{k_i})$.
In general, the structure and cardinality of this module are determined by the Smith normal form (SNF) of the matrix $1-A$ (see Theorem 15.9 in \cite{Brown1992-ds} and use the fact that $\mathbb{Z}_N$ is a principal ideal ring for any $N$),
\bea
P(1-A)Q=D=\mathrm{diag(}s_1,s_2,\dots,s_{2n})~,
\eea
where $P,Q$ are invertible matrices with determinant $\pm 1$ mod $N$. Notice that $s_i$ can be $0$ and those entries always appear at the end, namely, there exists an $m$ such that $s_{i} = 0 \mod N$ for $i > m$ and $s_i \neq 0 \mod N$ for $i \leq m$. The non-zero diagonal entries $s_i$ are called the elementary divisors of $1-A$ and satisfy the divisibility chain condition $s_1|s_2|\dots|s_{m}$. 

In SNF, the system of equations can be rewritten as
\ie
P^{-1}D\mathbf{y}=\mathbf{0} \,\,\, \mathrm{mod}\,\,p_i^{k_i}~,
\fe
where $\mathbf{y}=Q^{-1}\mathbf{v}_i$. 
Since $P,Q$ are invertible mod $p_i^{k_i}$, the solutions $\mathbf{v}_i$ are in one-to-one corrrespondence with $\mathbf{y}$, whose components decouple and each satisfy a much simpler condition,
\ie
s_j y_{j}=0~ \,\,\, \mathrm{mod}\,\,p_i^{k_i}~, \hspace{0.3 in}j=1,\dots,2n~.
\fe
When $s_j \neq 0$, the number of solutions for the component $y_j$ is $\mathrm{gcd}(s_j,p_i^{k_i})$, and when $s_j = 0$ the number is $p_i^{k_i}$. Written in terms of the $p$-adic valuation, the number of solutions is
\ie
\label{eq:component}
    p_i^{\sum_{j=1}^{2n}\mathrm{min}(\nu_{p_i}(s_j),k_i)}~,
\fe
where the $p$-adic valuation is defined as
\bea
\nu_p(n) = \begin{cases} \max\{k \in \mathbb{N}_0 : p^k \mid n\} & \text{if } n \neq 0 \\ ~\infty & \text{if } n = 0 \end{cases}~. 
\eea
The total number of solutions is then the product of the number for each decoupled equation. Furthermore, the total number of invariant vectors modulo $M$ is the product of the numbers for each prime factor, as guaranteed by the Chinese Remainder Theorem, from which we obtain 
\ie
\label{eqn:KernelSize}
|\mathrm{ker}_M(1-A)|=\prod_i p_i^{\sum_{j=1}^{2n}\mathrm{min}(\nu_{p_i}(s_j),k_i)} ~.
\fe
For entries $s_j = 0 \mod M$, we have $\mathrm{min}(\nu_{p_i}(0),k_i)=k_i$, and hence $\prod_i p_i^{\mathrm{min}(\nu_{p_i}(0),k_i)}=M$. On the other hand, for entries coprime with $M$ (therefore with every $p_i$), we have $\nu_{p_i}(s_j)=0$ so that $\prod_i p_i^{\mathrm{min}(\nu_{p_i}(s_j),k_i)}=1$. When only these two types of entries exist, the dimensionality is like that over a field, and we obtain $|\mathrm{ker}_M(1-A)|=M^{n_A}$, where $n_A=2n-r_A$ is the rank of the matrix $1-A$. As mentioned in the introduction, we will restrict ourselves to such cases for the rest of the paper.} then as mentioned above some of the $L_{(1-A)_i}$ are products of the others, and including all of them in the sum leads to redundancies. In this case, we should instead include lines labelled by a minimal basis for the image of $1-A$. In particular, let us say that the image is spanned by the linearly independent vectors $\eps_i$ for $i=1, \dots, r_A$, where $r_A:= \mathrm{rank}(1-A)$. For such a matrix, there exists a column operation $E$ such that $R:= (1-A)E$ has $2n-r_A$ columns set to zero, with the remaining non-zero columns reorganized to be on the left side of the matrix; as with all column operations, $E$ is invertible. Clearly $\eps_i = R_i$ for $i=1, \dots, r_A$ span the image, where $R_i$ denotes the $i$-th column of $R$. We may now write $D_A(M_2)$ as 
\bea
 \label{eq:generaldefofDA}
 D_A(M_2):= {1\over |H^0(M_2, \ZZ_M)|^{r_A}}\sum_{\widetilde \g_1, \dots,\widetilde\g_{r_A} \in H_1(M_2, \ZZ_M)} e^{{2 \pi i \over M} \half \sum_{i,j=1}^{r_A} \widetilde Q^A_{ij} \langle \widetilde\g_i, \widetilde\g_j \rangle} \prod_{i=1}^{r_A} L_{R_i}(\widetilde\gamma_i)~,
\eea
where we again assume that the matrix $\widetilde Q^A_{ij}$ is anti-symmetric. The matrix $\widetilde Q^A_{ij}$ is obtained from $ Q^A_{ij}$ as follows. Begin by noting that we may rewrite the above as 
\bea
 D_A(M_2)= {|H_1(M_2, \ZZ_M)|^{r_A-2n}\over |H^0(M_2, \ZZ_M)|^{r_A}}\sum_{\widetilde \g_1, \dots,\widetilde\g_{2n} \in H_1(M_2, \ZZ_M)} e^{{2 \pi i \over M}\half  \sum_{i,j=1}^{2n} \widehat Q^A_{ij} \langle \widetilde\g_i, \widetilde\g_j \rangle} \prod_{i=1}^{2n} L_{R_i}(\widetilde\gamma_i)~,
 \no\\
\eea
where $\widehat Q^A_{ij}$ is an anti-symmetric block-diagonal $2n \times 2n$ matrix with the upper left block being $\widetilde Q^A_{ij} $ and all other elements being zero. That we may rewrite things as such follows from the fact that $R_i= 0$ for $i > r_A$, and hence the corresponding $\widetilde\gamma_i$ for $i > r_A$ can be summed over to give a factor of $|H_1(M_2, \ZZ_M)|^{2n-r_A}$. Focusing on the line operators in the above sum, we have that 
\bea
 \prod_{i=1}^{2n} L_{R_i}(\widetilde\gamma_i) &=& \prod_{i=1}^{2n} L_{((1-A)E)_i}(\widetilde\gamma_i)
   =e^{- {2\pi i \over M} \sum_{i<j}^{2n} \left( \sum_{k < \ell}^{2n} (\a^k \cdot \b^\ell + \a^\ell \cdot \b^k)E_{\ell i} E_{k j} \right) \langle \widetilde \g_i, \widetilde \g_j\rangle } 
   \no\\
   &\vphantom{.}& \times\, e^{-{2\pi i \over M} \sum_{i<j}^{2n} \left(\sum_{k=1}^{2n} \a^k \cdot \b^k E_{ki} E_{kj} \right) \langle \widetilde \g_i, \widetilde \g_j \rangle }\prod_{j=1}^{2n} L_{(1-A)_j} \left(\sum_{i=1}^{2n} E_{ji}\widetilde \gamma_i \right) ~,
\eea
where the overall factor was obtained by repeated use of the decomposition (\ref{eq:Ldecomposition}) and the commutation relations in (\ref{eq:Lcomm}). 
Now since $E$ is invertible, we may perform a change of variables to $\gamma:= E \widetilde \gamma$, which gives 
\bea\label{eq:condens_non_full}
D_A(M_2)&=& {|H_1(M_2,\ZZ_M)|^{r_A-2n}\over |H^0(M_2, \ZZ_M)|^{r_A}}\sum_{ \g_1, \dots,\g_{2n} \in H_1(M_2, \ZZ_M)} e^{{2 \pi i \over M} \half \sum_{i,j=1}^{2n}  ({E^{-1}}^T \widehat Q^A E^{-1})_{ij} \langle \g_i, \g_j \rangle} 
\no\\
&\vphantom{.}& \times 
e^{- {2\pi i \over M}\sum_{i,j=1}^{2n} \sum_{p<q}^{2n}  \sum_{k < \ell}^{2n} (\a^k \cdot \b^\ell + \a^\ell \cdot \b^k)E_{\ell p} E_{k q} E_{pi}^{-1}E_{qj}^{-1} \langle  \g_i,  \g_j\rangle } 
   \no\\
   &\vphantom{.}& \times\, e^{-{2\pi i \over M} \sum_{i,j=1}^{2n}\sum_{p<q}^{2n} \sum_{k=1}^{2n} \a^k \cdot \b^k E_{kp} E_{kq}E_{pi}^{-1}E_{qj}^{-1} \langle  \g_i,  \g_j \rangle }\prod_{j=1}^{2n} L_{(1-A)_j}(\gamma_j)~.\no\\
\eea
This is now a defect of the same form as that in (\ref{eq:defofDA}), with $Q^A$ replaced by a more complicated exponential factor. 
Demanding that we have 
\bea
\label{eq:DAfusions1}
D_A(M_2) \times {D_A}^\dagger (M_2) &=& \chi[M_2, \ZZ_M]^{-r_A}~,
\no\\
L_{(\mathbf{e}, \mathbf{m})}(\gamma)\times D_A(M_2)   &=& D_A(M_2) \times  L_{A(\mathbf{e}, \mathbf{m})}(\gamma)~
\eea
and making use of the discussion around (\ref{eq:subtletyinnormalization}), 
the same computations as in the appendix fix this more complicated exponential factor to be equal to $Q^A$ given in (\ref{eq:QAdefab}), from which we conclude that
\bea
\label{eq:widehatQexpression}
\widehat Q^A_{ij} = (E^T Q^A E)_{ij} + \sum_{k < \ell}^{2n} (\a^k \cdot \b^\ell + \a^\ell \cdot \b^k)E_{k j} E_{\ell i} + \sum_{k=1}^{2n} \a^k \cdot \b^k E_{kj} E_{ki} ~,\quad i <j
\eea
and the analogous quantity with an overall minus sign for $i >j$, such that $\widehat Q^A_{ij} $ is anti-symmetric. 
The matrix $\widetilde  Q^A_{ij}$ is then given by the upper left $r_A \times r_A$ block of this matrix.\footnote{It is not obvious from this expression that all entries besides the upper left $r_A \times r_A$ block of $\widehat Q^A_{ij}$ are zero mod $M$, but this can be verified using (\ref{eq:abconstraints}).}

Note that changing the column operation from $E$ to $E'$ by including extra permutations of the columns changes the above exponential factor, but also changes the matrix $R$, and hence also the order of the line operators in the defect $D_A$. Reorganizing the line operators to put them back in the order corresponding to $E$ produces exponential factors which precisely compensate those obtained by switching from $E$ to $E'$, and hence both $E$ and $E'$ give equivalent defects.



   \subsubsection{Explicit examples}
   \label{sec:explicitexamples1}
   
Since the expressions above are somewhat technical, let us give some explicit examples.
   
   \paragraph{$N=2$ and $n=1$: } Let us first consider the case of $n=1$ and 
   \bea
   A = \mathsf{S} = \left(\begin{matrix} 0 & 1 \\ 1 & 0  \end{matrix}  \right) \hspace{0.3 in} \Rightarrow \hspace{0.3 in}  1- A =  \left(\begin{matrix} 1 & -1 \\ -1 & 1  \end{matrix}  \right) ~.
   \eea
   Clearly the matrix $1-A$ is rank 1 regardless of $M$, and hence the corresponding condensation defect $D_A$ will be a condensate of only one type of line operator. 
    In order to write down the explicit defect, we first introduce the reduced matrix $R$ related to the matrix $1-A$ by a column operation $E$ given as follows, 
   \bea
   R = \left( \begin{matrix} 1 & 0 \\ -1 & 0 \end{matrix} \right) ~, \hspace{0.5 in} E  = \left(\begin{matrix} 1 & 1 \\ 0 & 1 \end{matrix}  \right)~. 
   \eea
   From $1-A$ we compute the matrix 
   \bea
   Q^{\mathsf{S}} = \left( \begin{matrix} 0 & 1 \\ -1 & 0 \end{matrix} \right) ~, 
   \eea
   and combining this with $E$ using (\ref{eq:widehatQexpression}) we find that $\widehat Q^{\mathsf{S}} = \mymathbb{0}_{2 \times 2}$, which means that $\widetilde Q^{\mathsf{S}} = 0$. The condensation defect is thus given by 
   \bea
   D_{\mathsf{S}}(M_2) = {1\over |H^0 (M_2, \ZZ_M)|} \sum_{\g \in H_1(M_2, \ZZ_M)} L_{(1,-1)}(\gamma)~, 
   \eea
   which matches the expression given in \cite{Kaidi:2022cpf}.

\paragraph{$N=3$ and $n=2$: }
  We next consider the example of $n=2$, namely $\ZZ_M \times \ZZ_M$ gauge theory, and study elements of $O(2,2;M)$ of order three. 
   These are relevant to the construction of the SymTFT for triality defects. 
   Two such elements are as follows \cite{Lu:2024lzf}, 
   \bea
   \label{eq:STdef}
   \mathsf{ST}_1 = \left( \begin{matrix}-1 & 0 & 0 & -1 \\ 0 & -1 & 1 & 0 \\ 0 & -1 & 0 & 0 \\ 1 & 0 & 0 & 0  \end{matrix} \right) ~, \hspace{0.5 in} \mathsf{ST}_2  = \left(\begin{matrix}0 & 1 & 1 & 0 \\ - 1 & 1 & 1 & 1 \\ 1 & - 1& 0 & 0 \\ 0 & 1 & 0 & 0  \end{matrix} \right) ~. 
   \eea
Starting with $A =  \mathsf{ST}_1$, in this case the rank depends on the field $\ZZ_M$ over which the matrix is defined. Indeed, since $\mathrm{ker}_M(1-A) = 0$ if and only if
   \bea
   \mathrm{gcd} \left( \mathrm{det}(1-A), M\right) = 1~,
   \eea
and since $\mathrm{det}(1- \mathsf{ST}_1) = 9$, we conclude that the matrix $ \mathsf{ST}_1$ is full rank if and only if $\mathrm{gcd}(M, 3) = 1$. In this case, computing $Q^A$ from (\ref{eq:QAdefab}) gives 
\bea
Q^{\mathsf{ST}_1} =  \left(\begin{matrix} 0 & -1 & 2 & 0 \\ 1 & 0 & 0 & 2 \\ -2 & 0 & 0 & -1 \\ 0 & -2 & 1& 0 \end{matrix} \right)~,
\eea
and the expression for $D_{\mathsf{ST}_1}$ is then obtained from (\ref{eq:defofDA}), giving
   \bea
   D_{\mathsf{ST}_1}(M_2) &=& {1\over |H^0(M_2, \ZZ_M)|^{4}}\sum_{\gamma_1, \dots, \gamma_4} e^{{2\pi i \over M}\left[-\langle \g_1, \g_2 \rangle + 2 \langle \g_1, \g_3 \rangle + 2 \langle \g_2, \g_4 \rangle -\langle \g_3, \g_4 \rangle \right]}
   \\
  &\vphantom{.}&\hspace{0.5 in} 
   \times\, L_{(2,0,0,-1)}(\g_1) L_{(0,2,1,0)}(\g_2) L_{(0,-1,1,0)}(\g_3) L_{(1,0,0,1)}(\g_4)~.
   \no
      \eea

      On the other hand, when $\mathrm{gcd}(M, 3) \neq 1$, then $1- \mathsf{ST}_1$ is not full rank, and hence some of the lines being summed over on the right-hand side are redundant. For example, for $M=3$ we have that 
      \bea
      L_{(2,0,0,-1)}(\g) = L_{(1,0,0,1)}(\g)^{-1}~, \hspace{0.5 in}  L_{(0,2,1,0)}(\g) = L_{(0,-1,1,0)}(\g) ~
      \eea
and hence in this case the matrix is rank 2. In this case we may reduce $(1-A)$ to $R$ via the matrix $E$ given below,
\bea
R = \left(\begin{matrix} 2 & 0 & 0 & 0 \\ 0 & 2 & 0 & 0 \\ 0 & 1 & 0 & 0 \\ -1& 0 & 0 & 0 \end{matrix} \right) ~, \hspace{0.5 in}E = \left(\begin{matrix} 1 & 0 & 0 & 1 \\ 0 & 1 & 1 & 0 \\ 0 & 0 & -1 & 0 \\ 0 & 0 & 0 & 1\end{matrix} \right) ~,
\eea
from which we find that 
\bea
\widehat Q^{\mathsf{ST}_1} = \left(\begin{matrix} 0 & -1 & 0 & 0 \\ 1 & 0 & 0 & 0 \\ 0 & 0 & 0 & 0 \\ 0 & 0 & 0 & 0 \end{matrix} \right) ~,
\eea
whence $\widetilde Q^{\mathsf{ST}_1}$ is obtained by restricting to the upper left $2 \times 2 $ block. The expression in (\ref{eq:generaldefofDA}) then gives us the following form for the defect,
      \bea
       D_{\mathsf{ST}_1}(M_2) &=& {1\over |H^0(M_2, \ZZ_3)|^{2}}\sum_{\gamma_1, \g_2 \in H_1(M_2, \ZZ_3)} e^{-{2\pi i \over 3}\langle \g_1, \g_2 \rangle} L_{(2,0,0,-1)}(\g_1) L_{(0,2,1,0)}(\g_2)~.\qquad
      \eea

Moving on to the case of $\mathsf{ST}_2$, in this case for any $M$ we have
\bea
L_{(1- \mathsf{ST}_2)_3}(\gamma) = L^{-1}_{(1- \mathsf{ST}_2)_1}(\gamma)~, \hspace{0.5 in} L_{(1- \mathsf{ST}_2)_4}(\gamma) = L^{-1}_{(1- \mathsf{ST}_2)_1}(\gamma)L^{-1}_{(1- \mathsf{ST}_2)_2}(\gamma)~,
\eea
with $(1- \mathsf{ST}_2)_i$ representing the $i$-th column of $1- \mathsf{ST}_2$, and hence the matrix is of rank 2; in other words, the defect can be written in terms of the lines $L_{(1- \mathsf{ST}_2)_1}$ and $L_{(1- \mathsf{ST}_2)_2}$ alone. 
In this case we may reduce $(1-A)$ to $R$ via the matrix $E$ given below,
\bea
R = \left(\begin{matrix} 1 & -1 & 0 & 0 \\ 1 & 0 & 0 & 0 \\ -1 & 1 & 0 & 0 \\ 0& -1 & 0 & 0 \end{matrix} \right) ~, \hspace{0.5 in}E = \left(\begin{matrix} 1 & 0 & 1 & 1 \\ 0 & 1 & 0 & 1 \\ 0 & 0 & 1 & 0 \\ 0 & 0 & 0 & 1\end{matrix} \right) ~,
\eea
from which we find that $ \widehat Q^{\mathsf{ST}_2} = \mymathbb{0}_{4\times 4}$, and hence the final form of the defect is as follows,
\bea
D_{\mathsf{ST}_2}(M_2) = {1 \over |H^0(M_2, \ZZ_M)|^2} \sum_{\g_1, \g_2 \in H_1(M_2, \ZZ_M)}  L_{(1,1,-1,0)}(\gamma_1)L_{(-1,0,1,-1)}(\gamma_2)~.
\eea

\subsubsection{Restricting to $N = p$}\label{subsubsec:restrict_primary}

While the defects $D_A(M_2)$ defined above satisfy the fusion rules in (\ref{eq:DAfusions}), 
it is in general \textit{not} the case that the fusion of $D_A(M_2)$ and $D_B(M_2)$ is equal to $D_{BA}(M_2)$. 
This can be seen as follows. Because $D_A(M_2)$ is a condensate of elements of $\mathrm{im}_M(1-A)$ and $D_B(M_2)$ is a condensate
of elements of $\mathrm{im}_M(1-B)$, it follows that $D_A(M_2) \times D_B(M_2)$ must be able to absorb $L_\mathbf{v}$ for any $\mathbf{v} \in \mathrm{im}_M(1-A) \cup \mathrm{im}_M(1-B)$. 
On the other hand, the defect $D_{BA}(M_2)$ is a condensate of elements of $\mathrm{im}_M(1-BA)$, and hence can in general only absorb lines $L_\mathbf{v}$ for $\mathbf{v} \in \mathrm{im}_M(1-BA)$. As such, the fusion rules would instead take the form 
\bea
\label{eq:complicatedfusions}
D_A(M_2) \times D_B(M_2) = \cC_{A,B}(M_2)  D_{BA}(M_2)~, 
\eea
where $\cC_{A,B}(M_2)$ is an appropriate condensate of elements in $\mathrm{im}_M(1-A) \cup \mathrm{im}_M(1-B) - \mathrm{im}_M(1-BA)$.\footnote{A related discussion can be found in \cite{Antinucci:2022cdi}.}

In the current paper we are interested in $\ZZ_{N}$ subgroups of $O(n,n;M)$, whose elements can be written as $A^I$ for $I=1,\dots, N$, 
where $A$ is a primitive generator of $\ZZ_{N}$. A significant simplification is then attained when the order of $A$ is taken to be prime, i.e. $N = p$. In this case 
it is easy to see that $\mathrm{ker}(1-A^I)$ is the same for any $I=1,\dots, N-1$. Indeed, a vector $\mathbf{v} \in \mathrm{ker}(1-A^I)$ by definition satisfies 
$A^I \mathbf{v} = \mathbf{v}$. Because $\mathrm{gcd}(I,p)=1$, by Bezout's identity there exist integers $x$ and $y$ such that $x I + y p = 1$, and hence we 
have that $A \mathbf{v} = (A^I)^x \mathbf{v} = \mathbf{v} $. Thus for arbitrary $I$, every $\mathbf{v}$ in $\mathrm{ker}(1-A^I)$ is also in $ \mathrm{ker}(1-A)$, 
and the converse holds as well since $A$ generates $\ZZ_{p}$. 

By the first isomorphism theorem, the equality of $\mathrm{ker}_M(1-A^I)$ for every $I$ implies that the spaces $\mathrm{im}_M(1-A^I)$ for every $I$ are also identical. 
In particular, this means that the space $\mathrm{im}_M(1-A^I) \cup \mathrm{im}_M(1-A^J)$  is equivalent to the space $\mathrm{im}_M(1-A^{I+J})$, 
and hence the fusion rules
\bea
D_{A^{I}}(M_2)\times D_{A^{J}}(M_2)=\chi[M_{2}, \ZZ_{M}]^{-n}D_{A^{I+J}}(M_2)
\eea
are consistent (the overall factor is computed in Appendix \ref{app:matrixnotation}). 
Throughout this paper, we will occasionally assume that $N=p$ so that such a simplification occurs. 
We will make it explicit at which steps such an assumption is necessary.

\subsection{Twist defects: a first pass}
\label{sec:twistdefects1}
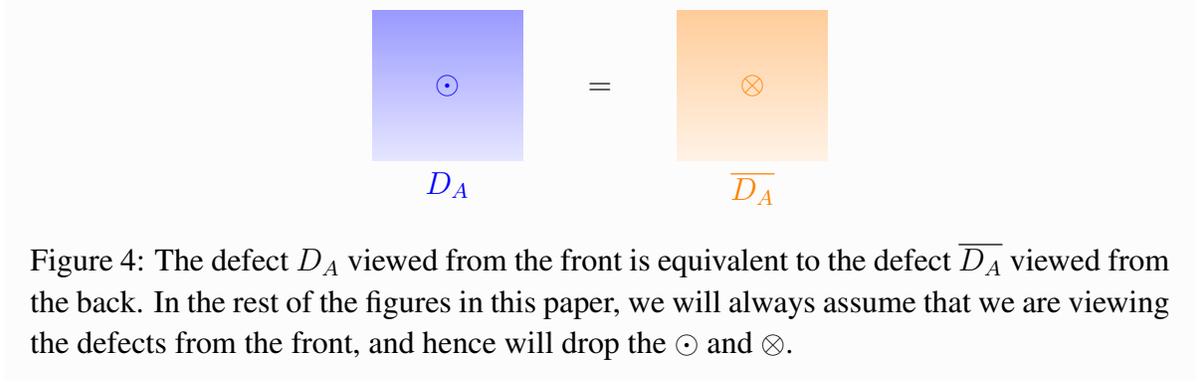
\begin{figure}[t]
    \centering
{\begin{tikzpicture}[baseline=25]

  \shade[top color=blue!40, bottom color=blue!10]  (0,0)--(0,2)--(2,2)--(2,0)--(0,0);
 
\begin{scope}[decoration={
    markings,
    mark=at position 0.5 with {\arrow{stealth}}}]

 \end{scope}

\node[blue] at (1,1) {$\odot$};

\node[below,blue] at (1,0) {$D_{{A}}$};

\end{tikzpicture}}
\hspace{0.3 in}$=$\hspace{0.3 in}
{\begin{tikzpicture}[baseline=25]

  \shade[top color=orange!40, bottom color=orange!10]  (0,0)--(0,2)--(2,2)--(2,0)--(0,0);
 
\begin{scope}[decoration={
    markings,
    mark=at position 0.5 with {\arrow{stealth}}}]

 \end{scope}

\node[orange] at (1,1) {$\otimes$};

\node[below,orange] at (1,0) {$\overline{D_{{A}}}$};

\end{tikzpicture}}
    \caption{The defect $D_{{A}}$ viewed from the front is equivalent to the defect $\overline{D_{{A}}}$ viewed from the back. In the rest of the figures in this paper, we will always assume that we are viewing the defects from the front, and hence will drop the $\odot$ and $\otimes$. }
    \label{fig:DSTorientation}
\end{figure}

We next discuss the topological boundaries for the defect $D_{A}$, also known as twist defects. Twist defects are defined by placing the symmetry operator on a manifold $M_2$ with boundary $M_1$, with Dirichlet boundary conditions imposed on $M_1$. Since the twist defects are boundaries of order-$N$ defects, which have an orientation for $N>2$, it is important that we specify the orientation used to define the boundary. First note that the defect $D_{A}$, when viewed from the front, is equivalent to the defect $\overline{D_{A}}:=\chi[M_2,  \ZZ_M]^{r_A} D_A^\dagger$ when viewed from the back, c.f. Figure \ref{fig:DSTorientation}. From now on we will always assume that we are viewing our defects $D_{A}$ from the front, where by ``front'' we mean the side such that, upon passing through that face, line operators experience the action of $A$ (and not  $A^{-1}$).

\begin{figure}[t]
    \centering
{\begin{tikzpicture}[baseline=25]
    
  \shade[top color=blue!40, bottom color=blue!10]  (0,0)--(0,2)--(2,2)--(2,0)--(0,0);
 
\begin{scope}[decoration={
    markings,
    mark=at position 0.5 with {\arrow{stealth}}}]
    
 \draw[very thick,dgreen,postaction={decorate}] (0,0)--(0,2);

 \end{scope}

\node[below,dgreen] at (0,0) {$\Sigma_{A}^{[\mathbf{0}]}$};

\node[below,blue] at (2,0) {$D_{{A}}$};

\end{tikzpicture}}
\hspace{0.3 in}$\Leftrightarrow$\hspace{0.3 in}
{\begin{tikzpicture}[baseline=25]

  \shade[top color=orange!40, bottom color=orange!10]  (0,0)--(0,2)--(2,2)--(2,0)--(0,0);
 
\begin{scope}[decoration={
    markings,
    mark=at position 0.5 with {\arrow{stealth}}}]
    
 \draw[very thick,dgreen,postaction={decorate}] (2,0)--(2,2);

 \end{scope}

\node[below,dgreen] at (2,0) {$\Sigma_{A}^{[\mathbf{0}]}$};

\node[below,orange] at (0,0) {$\overline{D_{{A}}}$};

\end{tikzpicture}}
    \caption{Our choice for the orientation of the twist defect. As mentioned before, each surface is viewed from the front, i.e. there is an implicit $\odot$. The orientation reversal twist defects $\overline \Sigma_{A}^{[\mathbf{0}]}$ can be obtained by flipping these pictures upside down.}
    \label{fig:orientationconvention}
\end{figure}
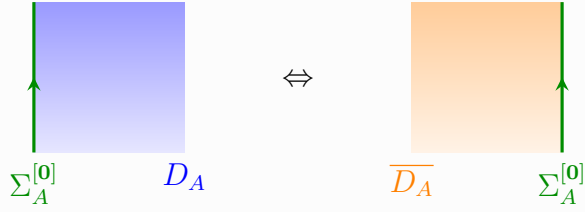

We now consider the minimal boundary $\Sigma_{A}^{[\mathbf{0}]}$ of the order-$N$ symmetry defect. We work with the convention given in Figure \ref{fig:orientationconvention}, where the twist defect $\Sigma_{A}^{[\mathbf{0}]}$ is defined to be the left-boundary of $D_{{A}}$, or the right-boundary of $\overline{D}_{{A}}$.

With this in mind, we may now write the explicit form of the minimal twist defect, namely 
\ie
\Sigma_{A}^{[\mathbf{0}]}(M_1,M_2)&:=&{1\over |H^0(M_2, M_1,\ZZ_M)|^{r_A}}\sum_{\g_1, \dots, \g_{r_A} \in H_1(M_2, \ZZ_M)} e^{{2 \pi i \over M} \half \sum_{i,j=1}^{r_A} \widetilde Q^A_{ij} \langle \g_i, \g_j \rangle} \prod_{i=1}^{r_A} L_{R_i}(\gamma_i)~,
\fe
where $\widetilde Q^A$ is given in terms of the matrix elements of $A$ as in (\ref{eq:QAdefab}) and (\ref{eq:widehatQexpression}). Note that 
this expression is nearly identical to the expression for the defects $D_{{A}}$ given in (\ref{eq:generaldefofDA}), with the exception of the overall prefactor, which is modified due to the Dirichlet boundary conditions along $M_1$ to the relative cohomology group $H^0(M_2, M_1,\ZZ_M)$.\footnote{A more detailed discussion of this factor in the case of duality can be found in \cite{Kaidi:2022cpf}. }

New twist defects can be constructed by fusing these minimal ones with additional line operators. Note that the lines $L_{R_i}$ can be absorbed into $M_1$ due to the Dirichlet boundary conditions. On the other hand, for every element $[\mathbf{w}] \in \mathrm{coker}_M(1-A):=\mathbb{Z}_M^{2n} / \mathrm{im}(1-A)$, we may define a new twist defect
\ie
\label{eq:generaltwistdefectdef}
\Sigma_{A}^{[\bw]}(M_1, M_2)&:=&L_{\bw}(M_1)\times\Sigma_{A}^{[\mathbf{0}]}(M_1, M_2)~
\fe
where $\bw \in [\bw]$ is an arbitrary element of the coset. The total number of such defects is given by the number of elements of $\mathrm{coker}_M(1-A)$, and since $1-A$ is a square matrix, this is equivalent to 
the number of elements in $\mathrm{ker}_M(1-A)$. 
In a similar way, we may define twist defects $\Sigma_{A^I}^{[\mathbf{w}]}$ for any $I$ up to $I=|A|-1$. 

The local fusions for these defects are given by\bea
\label{eq:twistdefectfusions}
\Sigma_{A^I}^{[\bw]} \otimes \Sigma_{A^J}^{[\bw']} = \left\{\begin{matrix}L_{\mathbf{w} +  \mathbf{w}'} \sum_{\g_1, \dots, \g_{r_A}} \prod_{i=1}^{r_A} L_{R_i} && I+J = 0 \,\,\mathrm{mod}\,\,|A| 
\\
|M|^{r_A\over 2}\,\, \Sigma_{A^{I+J}}^{[\mathbf{w} +\mathbf{w}']} && \mathrm{otherwise}\end{matrix} \right. 
\eea
where the factor appearing in the second line is required for consistency of the quantum dimension. The global fusions are similar to these, with appropriate factors of the Euler characteristic added in. 

\subsection{Algebra objects}
\label{sec:algebraobjs}

In addition to the twist defects defined above, which are boundaries of $D_A$ (i.e. interfaces between $D_A$ and the identity surface) we could also consider 
interfaces between two non-trivial surfaces. To treat all such interfaces on equal footing, a more general treatment will be necessary. 
In order to set up the more general problem, we first begin with a brief review of algebra objects. In the current case, we are interested in algebras of the form\bea
    \mathcal{A}_{A^x}:=\bigoplus_{\mathbf{u}\,\in\, \text{im}_M(1-A^x)}L_{\mathbf{u}}\  
\eea
for some $x \in \ZZ_N$.
Clearly the condensation defect $D_{A^x}(M_2)$ is a condensate of $\cA_{A^x}$.\footnote{Note that the braiding of $\cA$ with itself is non-trivial, which means that it cannot be gauged in the full three-dimensional bulk. This is fine for our purposes, since we will only be considering the higher-gauging of $\cA$ on a two-dimensional surface, for which the braiding does not pose an obstruction.} 

Because $\cA_{A^x}$ is 
composed of $L_\bu$ for $\bu \in \mathrm{im}(1-A^x)$, there exists non-zero (and in fact one-dimensional) hom-spaces $\mathrm{Hom}(L_\bu, \cA_{A^x})$ 
and $\mathrm{Hom}(\cA_{A^x},L_\bu)$. We write the basis vectors for these spaces graphically as follows, 
\begin{equation}\label{eq:ALvbasis}
{\begin{tikzpicture}[baseline=0]
\draw[blue, thick, decoration={markings, mark=at position 0.5 with {\arrow{stealth}}}, postaction={decorate}] (0,-0.5) -- (0,0);
\draw[red, thick, decoration={markings, mark=at position 0.7 with {\arrow{stealth}}}, postaction={decorate}] (0,0) -- (0,0.5);
\filldraw[blue] (0,0) circle (1pt);

\node[blue, below] at (0,-0.5) {\footnotesize $L_\bu$};
\node[red, above] at (0,0.5) {\footnotesize $\cA_{A^x}$};
\end{tikzpicture}} \in \mathrm{Hom}(L_\bu, \cA_{A^x})~, \hspace{0.5 in} {\begin{tikzpicture}[baseline=0]
\draw[red, thick, decoration={markings, mark=at position 0.5 with {\arrow{stealth}}}, postaction={decorate}] (0,-0.5) -- (0,0);
\draw[blue, thick,, decoration={markings, mark=at position 0.7 with {\arrow{stealth}}}, postaction={decorate}] (0,0) -- (0,0.5);
\filldraw[blue] (0,0) circle (1pt);

\node[red, below] at (0,-0.5) {\footnotesize $\cA_{A^x}$};
\node[blue, above] at (0,0.5) {\footnotesize $L_\bu$};
\end{tikzpicture}} \in \mathrm{Hom}(\cA_{A^x}, L_\bu)~.
\end{equation}
We may choose a complete orthonormal basis such that the following conditions are satisfied, 
\begin{equation}\label{eq:orthonormalA}
{\begin{tikzpicture}[baseline=0,square/.style={regular polygon,regular polygon sides=4}]
\draw[red, thick, decoration={markings, mark=at position 0.3 with {\arrow{stealth}}}, postaction={decorate}] (0,0) -- (0,0.5);
\draw[blue, thick, decoration={markings, mark=at position 0.7 with {\arrow{stealth}}}, postaction={decorate}] (0,0.5) -- (0,1);
\filldraw[blue] (0,0.5) circle (1pt);
\draw[blue, thick,  decoration={markings, mark=at position 0.5 with {\arrow{stealth}}}, postaction={decorate}] (0,-1) -- (0,-0.5);
\draw[red,thick] (0,-0.5) -- (0,0);
\filldraw[blue] (0,-0.5) circle (1pt);

\node[red, left] at (0,0) {\footnotesize $\cA_{A^x}$};
\node[blue, above] at (0,1) {\footnotesize $L_\bu$};
\node[blue, below] at (0,-1) {\footnotesize $L_{\bu'}$};
\end{tikzpicture}} \,\,=\,\, \delta^{\bu,\bu'}{\begin{tikzpicture}[baseline=0]
\draw[blue, thick, decoration={markings, mark=at position 0.6 with {\arrow{stealth}}}, postaction={decorate}] (0,-1) -- (0,1);

\node[blue, above] at (0,1) {\footnotesize $L_\bu$};
\node[blue, below] at (0,-1) {\footnotesize $L_\bu$};
\end{tikzpicture}}~, \hspace{0.7 in} \sum_{\bu} {\begin{tikzpicture}[baseline=0,square/.style={regular polygon,regular polygon sides=4}]
\draw[blue, thick] (0,0) -- (0,0.5);
\draw[red, thick, decoration={markings, mark=at position 0.7 with {\arrow{stealth}}}, postaction={decorate}] (0,0.5) -- (0,1);
\filldraw[blue] (0,0.5) circle (1pt);
\draw[red, thick, decoration={markings, mark=at position 0.5 with {\arrow{stealth}}}, postaction={decorate}] (0,-1) -- (0,-0.5);
\draw[blue, thick, decoration={markings, mark=at position 1 with {\arrow{stealth}}}, postaction={decorate}] (0,-0.5) -- (0,0);
\filldraw[blue] (0,-0.5) circle (1pt);

\node[blue, right] at (0,0) {\footnotesize $L_\bu$};
\node[red, above] at (0,1) {\footnotesize $\cA_{A^x}$};
\node[red, below] at (0,-1) {\footnotesize $\cA_{A^x}$};
\end{tikzpicture}} \,\,=\,\, {\begin{tikzpicture}[baseline=0,square/.style={regular polygon,regular polygon sides=4}] 
\draw[red, thick, decoration={markings, mark=at position 0.6 with {\arrow{stealth}}}, postaction={decorate}] (0,-1) -- (0,1);

\node[red, above] at (0,1) {\footnotesize $\cA_{A^x}$};
\node[red, below] at (0,-1) {\footnotesize $\cA_{A^x}$};
\end{tikzpicture}}~.
\end{equation}
We are always free to rescale these junction vectors by a phase, 
\bea
\label{eq:ALvgaugeredundancy}
{\begin{tikzpicture}[baseline=0]
\draw[blue, thick, decoration={markings, mark=at position 0.5 with {\arrow{stealth}}}, postaction={decorate}] (0,-0.5) -- (0,0);
\draw[red, thick, decoration={markings, mark=at position 0.7 with {\arrow{stealth}}}, postaction={decorate}] (0,0) -- (0,0.5);
\filldraw[blue] (0,0) circle (1pt);

\node[blue, below] at (0,-0.5) {\footnotesize $L_\bu$};
\node[red, above] at (0,0.5) {\footnotesize $\cA_{A^x}$};
\end{tikzpicture}} \rightarrow \chi(\bu) {\begin{tikzpicture}[baseline=0]
\draw[blue, thick, decoration={markings, mark=at position 0.5 with {\arrow{stealth}}}, postaction={decorate}] (0,-0.5) -- (0,0);
\draw[red, thick, decoration={markings, mark=at position 0.7 with {\arrow{stealth}}}, postaction={decorate}] (0,0) -- (0,0.5);
\filldraw[blue] (0,0) circle (1pt);

\node[blue, below] at (0,-0.5) {\footnotesize $L_\bu$};
\node[red, above] at (0,0.5) {\footnotesize $\cA_{A^x}$};
\end{tikzpicture}}~, \hspace{0.5 in} {\begin{tikzpicture}[baseline=0]
\draw[red, thick, decoration={markings, mark=at position 0.5 with {\arrow{stealth}}}, postaction={decorate}] (0,-0.5) -- (0,0);
\draw[blue, thick,, decoration={markings, mark=at position 0.7 with {\arrow{stealth}}}, postaction={decorate}] (0,0) -- (0,0.5);
\filldraw[blue] (0,0) circle (1pt);

\node[red, below] at (0,-0.5) {\footnotesize $\cA_{A^x}$};
\node[blue, above] at (0,0.5) {\footnotesize $L_\bu$};
\end{tikzpicture}} \rightarrow \chi(\bu)^{-1} {\begin{tikzpicture}[baseline=0]
\draw[red, thick, decoration={markings, mark=at position 0.5 with {\arrow{stealth}}}, postaction={decorate}] (0,-0.5) -- (0,0);
\draw[blue, thick,, decoration={markings, mark=at position 0.7 with {\arrow{stealth}}}, postaction={decorate}] (0,0) -- (0,0.5);
\filldraw[blue] (0,0) circle (1pt);

\node[red, below] at (0,-0.5) {\footnotesize $\cA_{A^x}$};
\node[blue, above] at (0,0.5) {\footnotesize $L_\bu$};
\end{tikzpicture}}
\eea
which preserve the orthonormality conditions. This is referred to as a ``gauge redundancy.''

In the case of $N=p$ being prime, we see that the lines $\mathcal{A}_{A^x}$ for any $x$ are the same since all of the image spaces are the same, and hence we denote them all by $\mathcal{A}_{A}$ in this case.  However, one should be careful to note that specification of the algebra object involves not just specifying the non-simple line $\mathcal{A}_{A}$ itself, but also a multiplication $\mu_{A^x}$ for the $\cA_A$, illustrated in Figure \ref{fig:three_algebra_objects}, as well as its dual $\mu^\vee_{A^x}$. Note that these phases are not necessarily the same for every $A^x$---indeed, we will compute $\mu_{A^x}$ explicitly in a moment, and see that they do depend on $x$. Thus the algebra object is really specified by a pair $(\cA_{A^x}, \mu_{A^x})$, and depends on $x$ even when $N=p$ is prime.

 \begin{figure}[tbp]
\begin{center}
{\begin{tikzpicture}[baseline=0,scale=0.9]
\draw [red, thick, decoration={markings, mark=at position 0.5 with {\arrow{stealth}}}, postaction={decorate}] (-1,-1) -- (0,0);
\draw [red, thick, decoration={markings, mark=at position 0.5 with {\arrow{stealth}}}, postaction={decorate}] (1,-1) -- (0,0);
\draw [red, thick, decoration={markings, mark=at position 0.6 with {\arrow{stealth}}}, postaction={decorate}] (0,0) -- (0,1);

\filldraw[red] (0,0) circle (1.5pt);
\node[red,left] at (-0.1,0) {$\mu_{A^x}$};

\node[below] at (-1.2,-1) {\footnotesize \color{red}$\cA_{A}$};
\node[below] at (1.1,-1) {\footnotesize \color{red}$\cA_{A}$};
\node[above] at (0,1) {\footnotesize \color{red}$\cA_{A}$};
\end{tikzpicture}}
$\displaystyle = \frac{1}{M^{r_A/2}}\sum_{\mathbf{u}, \mathbf{u'} \in \mathrm{im}_M(1-A)} \,\,\mu_{A^x}(\mathbf{u}, \mathbf{u'})$\hspace{-0.1 in}
{\begin{tikzpicture}[baseline=0,scale=0.9]

\draw [blue, thick, decoration={markings, mark=at position 0.5 with {\arrow{stealth}}}, postaction={decorate}] (0.7,-0.7) -- (0,0);
\draw [red, thick, decoration={markings, mark=at position 0.5 with {\arrow{stealth}}}, postaction={decorate}] (1,-1) -- (0.7,-0.7);

 \draw [red, thick, decoration={markings, mark=at position 0.5 with {\arrow{stealth}}}, postaction={decorate}] (-1,-1) -- (-0.7,-0.7);
\draw [blue, thick, decoration={markings, mark=at position 0.5 with {\arrow{stealth}}}, postaction={decorate}] (-0.7,-0.7) -- (0,0);
\draw [blue, thick, decoration={markings, mark=at position 0.5 with {\arrow{stealth}}}, postaction={decorate}] (0,0) -- (0,0.6);
\draw [red, thick, decoration={markings, mark=at position 0.8 with {\arrow{stealth}}}, postaction={decorate}] (0,0.6) -- (0,1);

\node[below] at (-1.2,-1) {\footnotesize \color{red}$\cA_A$};
\node[below] at (1.1,-1) {\footnotesize \color{red}$\cA_A$};
\node[above] at (0,1) {\footnotesize \color{red}$\cA_A$};

\node[below] at (-0.8,0.1) {\footnotesize \color{blue}$L_{\mathbf{u}}$};
\node[below] at (0.8,0.1) {\footnotesize \color{blue}$L_{\mathbf{u'}}$};
\node[right] at (-0.1,0.4) {\footnotesize \color{blue}$L_{\mathbf{u}+\mathbf{u'}}$};

\filldraw[blue] (-0.7,-0.7) circle (1pt);
\filldraw[blue] (0.7,-0.7) circle (1pt);
\filldraw[blue] (0,0.6) circle (1pt);

\end{tikzpicture}}
\\\vspace{0.2 in}
{\begin{tikzpicture}[baseline=0,scale=0.9]
\draw [red, thick, -<-=0.5] (-1,1) -- (0,0);
\draw [red, thick, -<-=0.5] (1,1) -- (0,0);
\draw [red, thick, -<-=0.6] (0,0) -- (0,-1);

\filldraw[red] (0,0) circle (1.5pt);
\node[red,left] at (-0.1,-0.2) {$\mu_{A^x}^\vee$};

\node[above] at (-1.2,1) {\footnotesize \color{red}$\cA_{A}$};
\node[above] at (1.1,1) {\footnotesize \color{red}$\cA_{A}$};
\node[below] at (0,-1) {\footnotesize \color{red}$\cA_{A}$};
\end{tikzpicture}}
$\displaystyle  = \frac{1}{M^{r_A/2}} \sum_{\mathbf{u}, \mathbf{u'} \in \mathrm{im}_M(1-A)} \,\,\mu_{A^x}^\vee(\mathbf{u}, \mathbf{u'})$\hspace{-0.1 in}
{\begin{tikzpicture}[baseline=0,scale=0.9]
 \draw [red, thick, , -<-=0.5] (-1,1) -- (-0.7,0.7);
 \draw [blue, thick, , -<-=0.5] (-0.7,0.7) -- (0,0);
\draw [blue, thick, , -<-=0.5] (0.7,0.7) -- (0,0);
\draw [red, thick, , -<-=0.5] (1,1) -- (0.7,0.7);

\draw [blue, thick, , -<-=0.5] (0,0) -- (0,-0.6);
\draw [red, thick, , -<-=0.7] (0,-0.6) -- (0,-1);

\filldraw[blue] (-0.7,0.7) circle (1pt);
\filldraw[blue] (0.7,0.7) circle (1pt);
\filldraw[blue] (0,-0.6) circle (1pt);

\node[left] at (-0.3,0.2) {\footnotesize \color{blue}$L_{\mathbf{u}}$};
\node[right] at (0.3,0.2) {\footnotesize \color{blue}$L_{\mathbf{u'}}$};
\node[right] at (0,-0.4) {\footnotesize \color{blue}$L_{\mathbf{u}+\mathbf{u'}}$};

\node[above] at (-1.2,1) {\footnotesize \color{red}$\cA_A$};
\node[above] at (1.1,1) {\footnotesize \color{red}$\cA_A$};
\node[below] at (0,-1) {\footnotesize \color{red}$\cA_A$};

\end{tikzpicture}}
\caption{Multiplication and comultiplication for the algebra object. We factor out an overall $M^{-r_A/2}$ so that the coefficients $\mu_{A^x}(\bu,\bu')$ and $\mu_{A^x}^{\vee}(\bu,\bu')$ are phases.}\label{fig:three_algebra_objects}
\end{center}
\end{figure}

 \begin{figure}[!tbp]
\begin{center}

{\begin{tikzpicture}[baseline=0,scale=0.6]
\draw [red, thick, decoration={markings, mark=at position 0.5 with {\arrow{stealth}}}, postaction={decorate}] (1,1) -- (2,2);
\draw [red, thick, decoration={markings, mark=at position 0.5 with {\arrow{stealth}}}, postaction={decorate}] (0,0) -- (1,1);
\draw [red, thick, decoration={markings, mark=at position 0.5 with {\arrow{stealth}}}, postaction={decorate}] (-1,-1) -- (0,0);
\draw [red, thick, decoration={markings, mark=at position 0.5 with {\arrow{stealth}}}, postaction={decorate}] (1,-1) -- (0,0);
\draw [red, thick, decoration={markings, mark=at position 0.5 with {\arrow{stealth}}}, postaction={decorate}] (3,-1) -- (1,1);

\filldraw[red] (0,0) circle (2pt);
\node[red,left] at (0,0) {$\mu_{A}$};
\filldraw[red] (1,1) circle (2pt);
\node[red,left] at (1,1) {$\mu_{A}$};

\node[below] at (-1.2,-1) {\footnotesize \color{red}$\cA_A$};
\node[below] at (1.1,-1) {\footnotesize \color{red}$\cA_A$};
\node[below] at (3.6,-1) {\footnotesize \color{red}$\cA_A$};
\end{tikzpicture}}
\hspace{0.05 in}= \hspace{0.1 in}
{\begin{tikzpicture}[baseline=0,square/.style={regular polygon,regular polygon sides=4},scale=0.6]
\draw [red, thick, decoration={markings, mark=at position 0.5 with {\arrow{stealth}}}, postaction={decorate}] (1,1) -- (2,2);
\draw [red, thick] (0,0) -- (1,1);
\draw [red, thick, ->-=0.5] (-1,-1) -- (0,0);
\draw [red, thick, decoration={markings, mark=at position 0.5 with {\arrow{stealth}}}, postaction={decorate}] (1,-1) -- (2,0);
\draw [red, thick, decoration={markings, mark=at position 0.5 with {\arrow{stealth}}}, postaction={decorate}] (3,-1) -- (2,0);
\draw [red, thick, decoration={markings, mark=at position 0.5 with {\arrow{stealth}}}, postaction={decorate}] (2,0) -- (1,1);

\filldraw[red] (2,0) circle (2pt);
\node[red,left] at (1,1) {$\mu_A$};
\filldraw[red] (1,1) circle (2pt);
\node[red,right] at (2,0) {$\mu_A$};

\node[below] at (-1.2,-1) {\footnotesize \color{red}$\cA_{A}$};
\node[below] at (1.1,-1) {\footnotesize \color{red}$\cA_{A}$};
\node[below] at (3.6,-1) {\footnotesize \color{red}$\cA_{A}$};
\end{tikzpicture}}
\hspace{0.7 in}
{\begin{tikzpicture}[baseline=0,scale=0.5]
\draw [red, thick, decoration={markings, mark=at position 0.5 with {\arrow{stealth}}}, postaction={decorate}] (0,-2) -- (0,-0.6);
\draw [red, thick, decoration={markings, mark=at position 0.5 with {\arrow{stealth}}}, postaction={decorate}] (0,0.6) -- (0,2);
\draw [red,thick,decoration={markings, mark=at position 0.5 with {\arrow{stealth}}}, postaction={decorate}] (0,-0.6) arc [radius=.6, start angle=-90, end angle=90];
\draw [red,thick, -<-=0.5] (0,0.6) arc [radius=.6, start angle=90, end angle=270];
\filldraw[red] (0,0.6)  circle (2pt);
\node[red,left] at (-0.1,0.8)  {$\mu_A$};
\filldraw[red] (0,-0.6) circle (2pt);
\node[red,right] at (0,-0.8)  {$\mu_A^\vee$};

\node[above] at (0,2) {\footnotesize \color{red}$\cA_A$};
\node[below] at (0,-2) {\footnotesize \color{red}$\cA_A$};
\end{tikzpicture}}
\hspace{0.1 in}= \hspace{0.2 in}
{\begin{tikzpicture}[baseline=0,scale=0.5]
\draw [red, thick, decoration={markings, mark=at position 0.5 with {\arrow{stealth}}}, postaction={decorate}] (0,-2) -- (0,2);
\node[above] at (0,2) {\footnotesize \color{red}$\cA_A$};
\node[below] at (0,-2) {\footnotesize \color{red}$\cA_A$};
\end{tikzpicture}}
\caption{The junction $\mu_A$ and its dual $\mu_A^\vee$ are subject to the consistency conditions shown above. These are the associativity and separability conditions, respectively. There is also a Frobenius condition which we do not draw, as well as the requirement of the existence of a unit.}\label{fig:Aconsistency}
\end{center}
\end{figure}

Let us now give more detail on the multiplications $\mu_{A^x}(\mathbf{u}, \mathbf{u'})$. In general, these junctions are subject to the consistency conditions illustrated in Figure \ref{fig:Aconsistency}, from which we conclude that 
\bea
\label{eq:muAconstraint1}
\mu_{A^x}^\vee(\mathbf{u}, \mathbf{u'}) = \mu_{A^x}^{-1}(\mathbf{u}, \mathbf{u'})~, 
\eea
as well as that
\bea
\label{eq:muAconstraint2}
\mu_{A^x}(\mathbf{u}, \mathbf{u'})  \mu_{A^x}(\mathbf{u} + \mathbf{u'}, \mathbf{u''}) =    \mu_{A^x}(\mathbf{u} , \mathbf{u'}+\mathbf{u''})\mu_{A^x}(\mathbf{u'}, \mathbf{u''}) ~.
\eea
The multiplications are also subject to a gauge redundancy following from junction redefinitions in (\ref{eq:ALvgaugeredundancy}), which shift
\bea
\label{eq:mugaugetransf}
\mu_{A^x}(\bu, \bu') \rightarrow \mu_{A^x}(\bu, \bu') { \chi(\bu + \bu') \over \chi(\bu) \chi(\bu')}~. 
\eea

We may now fix the precise form of $\mu_{A^x}(\bu, \bu')$ in terms of the phase appearing in the definition of the condensation defect $D_{A^x}(M_2)$ in (\ref{eq:generaldefofDA}) as follows. 
Say that the surface $M_2$ on which $D_{A^x}(M_2)$ lives is a genus-$g$ Riemann surfaces with cycles  $\mathfrak{A}_i$, $\mathfrak{B}_i$ for $i=1,\dots, g$ satisfying $\langle \mathfrak{A}_i, \mathfrak{B}_j \rangle = \delta_{ij}$. Denoting 
\bea
\widetilde \gamma_i = \sum_{j=1}^g a_{ij} \mathfrak{A}_j + \sum_{j=1}^g b_{ij} \mathfrak{B}_k~,
\eea
we have that 
\bea
\prod_{i=1}^{r_A} L_{R_i} (\widetilde \g_i) &=& \omega^{\sum_{i=1}^{r_A} R_i^T \Omega R_i \sum_{j=1}^g a_{ij} b_{ij}} \omega^{\sum_{i<j}^{r_A} R_i^T \mathcal{I} R_j \sum_{k=1}^g b_{ik} a_{jk}}
\no\\
&\vphantom{.}& \hspace{0.2 in}\times \prod_{j=1}^g \left( \prod_{i=1}^{r_A} L_{R_i}^{a_{ij}}(\mathfrak{A}_j)\right) \left( \prod_{i=1}^{r_A} L_{R_i}^{b_{ij}}(\mathfrak{B}_j)\right) ~,
\eea
or pictorially 
\bea
\prod_{i=1}^{r_A} L_{R_i} (\widetilde \g_i) &=& \omega^{\sum_{i=1}^{r_A} R_i^T \Omega R_i \sum_{j=1}^g a_{ij} b_{ij}} \omega^{\sum_{i<j}^{r_A} R_i^T \mathcal{I} R_j \sum_{k=1}^g b_{ik} a_{jk}}
\no\\&\vphantom{.}&\hspace{0.5 in}\times\,\,\,{\begin{tikzpicture}[baseline=30]
\draw [thick] (0,0)--(0,2.5)--(2.5,2.5) -- (2.5,0) -- (0,0);
\draw [blue, thick,decoration={markings, mark=at position 0.8 with {\arrow{stealth}}}, postaction={decorate}]  (0,1)--(2.5,1);
\draw [blue,thick] (1,0)--(1,0.9);
\node[blue,right] at (1,2) {\footnotesize{$\prod_i L_{R_i}^{b_{i1}}$}};
\draw [blue, thick,decoration={markings, mark=at position 0.5 with {\arrow{stealth}}}, postaction={decorate}]  (1,1.1)--(1,2.5);
\node[blue,below] at (1.9,1) {\footnotesize{$\prod_i L_{R_i}^{a_{i1}}$}};
\end{tikzpicture}}
\,\,\,\#\,\, \dots \,\,\#\,\,\, 
{\begin{tikzpicture}[baseline=30]
\draw [thick] (0,0)--(0,2.5)--(2.5,2.5) -- (2.5,0) -- (0,0);
\draw [blue,thick,decoration={markings, mark=at position 0.8 with {\arrow{stealth}}}, postaction={decorate}]  (0,1)--(2.5,1);
\draw [blue,thick] (1,0)--(1,0.9);
\node[blue,right] at (1,2) {\footnotesize{$\prod_i L_{R_i}^{b_{ig}}$}};
\draw [blue,thick,decoration={markings, mark=at position 0.5 with {\arrow{stealth}}}, postaction={decorate}]  (1,1.1)--(1,2.5);
\node[blue,below] at (1.9,1) {\footnotesize{$\prod_i L_{R_i}^{a_{ig}}$}};
\end{tikzpicture}}~,
\eea
where we have drawn $M_2$ as a connected sum of tori. We may now use the braiding of simple lines to rewrite this as 
\bea
\prod_{i=1}^{r_A} L_{R_i} (\widetilde \g_i) &=& \omega^{\sum_{i=1}^{r_A} R_i^T \Omega R_i \sum_{j=1}^g a_{ij} b_{ij}} \omega^{\sum_{i<j}^{r_A} R_i^T \mathcal{I} R_j \sum_{k=1}^g b_{ik} a_{jk}}
\\&\vphantom{.}&\hspace{0.0 in}\times\,\omega^{- \sum_{i,j,k=1}^{r_A} R_j^T \Omega R_i a_{jk}b_{ik}}\,\,\,{\begin{tikzpicture}[baseline=30]
\draw [thick] (0,0)--(0,2.5)--(2.5,2.5) -- (2.5,0) -- (0,0);
\draw [blue,thick,decoration={markings, mark=at position 0.5 with {\arrow{stealth}}}, postaction={decorate}]  (0,1)--(1,1);
\draw [blue,thick,decoration={markings, mark=at position 0.5 with {\arrow{stealth}}}, postaction={decorate}] (1,0)--(1,1);
\draw [blue,thick,decoration={markings, mark=at position 0.5 with {\arrow{stealth}}}, postaction={decorate}] (1,1)--(1.5,1.5);
\draw [blue,thick,decoration={markings, mark=at position 0.5 with {\arrow{stealth}}}, postaction={decorate}] (1.5,1.5)--(2.5,1.5);
\draw [blue,thick,decoration={markings, mark=at position 0.5 with {\arrow{stealth}}}, postaction={decorate}] (1.5,1.5)--(1.5,2.5);
\node[blue,right] at (1,0.5) {\footnotesize{$\prod_i L_{R_i}^{b_{i1}}$}};
\node[blue,above] at (0.6,1)  {\footnotesize{$\prod_i L_{R_i}^{a_{i1}}$}};
\end{tikzpicture}}
\,\,\,\#\,\, \dots \,\,\#\,\,\, 
{\begin{tikzpicture}[baseline=30]
\draw [thick] (0,0)--(0,2.5)--(2.5,2.5) -- (2.5,0) -- (0,0);
\draw [blue,thick,decoration={markings, mark=at position 0.5 with {\arrow{stealth}}}, postaction={decorate}]  (0,1)--(1,1);
\draw [blue,thick,decoration={markings, mark=at position 0.5 with {\arrow{stealth}}}, postaction={decorate}] (1,0)--(1,1);
\draw [blue,thick,decoration={markings, mark=at position 0.5 with {\arrow{stealth}}}, postaction={decorate}] (1,1)--(1.5,1.5);
\draw [blue,thick,decoration={markings, mark=at position 0.5 with {\arrow{stealth}}}, postaction={decorate}] (1.5,1.5)--(2.5,1.5);
\draw [blue,thick,decoration={markings, mark=at position 0.5 with {\arrow{stealth}}}, postaction={decorate}] (1.5,1.5)--(1.5,2.5);
\node[blue,right] at (1,0.5) {\footnotesize{$\prod_i L_{R_i}^{b_{ig}}$}};
\node[blue,above] at (0.6,1)  {\footnotesize{$\prod_i L_{R_i}^{a_{ig}}$}};
\end{tikzpicture}}~.
\no
\eea
We may thus write 
\bea
D_{A^x}(M_2) &=& {1 \over |H^0(M_2, \ZZ_M)|^{r_A}} \sum_{a_{ij}, b_{ij} \in \ZZ_M} \omega^{\half \sum_{i,j=1}^{r_A} \widetilde Q^{A^x}_{ij} \sum_{k=1}^g (a_{ik} b_{jk} - b_{ik} a_{jk}) + \sum_{j < i}^{r_A} R_j^T \Omega R_i \sum_{k=1}^g (b_{jk} a_{ik} -a_{jk} b_{ik})}
\no\\&\vphantom{.}&\hspace{0.5 in}\times
{\begin{tikzpicture}[baseline=30]
\draw [thick] (0,0)--(0,2.5)--(2.5,2.5) -- (2.5,0) -- (0,0);
\draw [blue,thick,decoration={markings, mark=at position 0.5 with {\arrow{stealth}}}, postaction={decorate}]  (0,1)--(1,1);
\draw [blue,thick,decoration={markings, mark=at position 0.5 with {\arrow{stealth}}}, postaction={decorate}] (1,0)--(1,1);
\draw [blue,thick,decoration={markings, mark=at position 0.5 with {\arrow{stealth}}}, postaction={decorate}] (1,1)--(1.5,1.5);
\draw [blue,thick,decoration={markings, mark=at position 0.5 with {\arrow{stealth}}}, postaction={decorate}] (1.5,1.5)--(2.5,1.5);
\draw [blue,thick,decoration={markings, mark=at position 0.5 with {\arrow{stealth}}}, postaction={decorate}] (1.5,1.5)--(1.5,2.5);
\node[blue,right] at (1,0.5) {\footnotesize{$\prod_i L_{R_i}^{b_{i1}}$}};
\node[blue,above] at (0.6,1)  {\footnotesize{$\prod_i L_{R_i}^{a_{i1}}$}};
\end{tikzpicture}}
\,\,\,\#\,\, \dots \,\,\#\,\,\, 
{\begin{tikzpicture}[baseline=30]
\draw [thick] (0,0)--(0,2.5)--(2.5,2.5) -- (2.5,0) -- (0,0);
\draw [blue,thick,decoration={markings, mark=at position 0.5 with {\arrow{stealth}}}, postaction={decorate}]  (0,1)--(1,1);
\draw [blue,thick,decoration={markings, mark=at position 0.5 with {\arrow{stealth}}}, postaction={decorate}] (1,0)--(1,1);
\draw [blue,thick,decoration={markings, mark=at position 0.5 with {\arrow{stealth}}}, postaction={decorate}] (1,1)--(1.5,1.5);
\draw [blue,thick,decoration={markings, mark=at position 0.5 with {\arrow{stealth}}}, postaction={decorate}] (1.5,1.5)--(2.5,1.5);
\draw [blue,thick,decoration={markings, mark=at position 0.5 with {\arrow{stealth}}}, postaction={decorate}] (1.5,1.5)--(1.5,2.5);
\node[blue,right] at (1,0.5) {\footnotesize{$\prod_i L_{R_i}^{b_{ig}}$}};
\node[blue,above] at (0.6,1)  {\footnotesize{$\prod_i L_{R_i}^{a_{ig}}$}};
\end{tikzpicture}}~.
\eea
On the other hand, we know that by the definition of the condensation defect we must have 
\bea
D_{A^x}(M_2) &=& {M^{r_A \cdot g} \over |H^0(M_2, \ZZ_M)|^{r_A}}\,\,\, {\begin{tikzpicture}[baseline=30]
\draw [thick] (0,0)--(0,2.5)--(2.5,2.5) -- (2.5,0) -- (0,0);
\draw [thick,red,decoration={markings, mark=at position 0.5 with {\arrow{stealth}}}, postaction={decorate}]  (0,1)--(1,1);
\draw [thick,red,decoration={markings, mark=at position 0.5 with {\arrow{stealth}}}, postaction={decorate}] (1,0)--(1,1);
\draw [thick,red,decoration={markings, mark=at position 0.5 with {\arrow{stealth}}}, postaction={decorate}] (1,1)--(1.5,1.5);
\draw [thick,red,decoration={markings, mark=at position 0.5 with {\arrow{stealth}}}, postaction={decorate}] (1.5,1.5)--(2.5,1.5);
\draw [thick,red,decoration={markings, mark=at position 0.5 with {\arrow{stealth}}}, postaction={decorate}] (1.5,1.5)--(1.5,2.5);
\node[right,red] at (1,0.8) {\footnotesize{$\mu$}};
\filldraw[red] (1,1) circle (1.5pt);
\node[left,red] at (1.6,1.6) {\footnotesize{$\mu^\vee$}};
\filldraw[red] (1.5,1.5) circle (1.5pt);
\end{tikzpicture}}
\,\,\,\#\,\, \dots \,\,\#\,\,\, 
{\begin{tikzpicture}[baseline=30]
\draw [thick] (0,0)--(0,2.5)--(2.5,2.5) -- (2.5,0) -- (0,0);
\draw [thick,red,decoration={markings, mark=at position 0.5 with {\arrow{stealth}}}, postaction={decorate}]  (0,1)--(1,1);
\draw [thick,red,decoration={markings, mark=at position 0.5 with {\arrow{stealth}}}, postaction={decorate}] (1,0)--(1,1);
\draw [thick,red,decoration={markings, mark=at position 0.5 with {\arrow{stealth}}}, postaction={decorate}] (1,1)--(1.5,1.5);
\draw [thick,red,decoration={markings, mark=at position 0.5 with {\arrow{stealth}}}, postaction={decorate}] (1.5,1.5)--(2.5,1.5);
\draw [thick,red,decoration={markings, mark=at position 0.5 with {\arrow{stealth}}}, postaction={decorate}] (1.5,1.5)--(1.5,2.5);
\node[right,red] at (1,0.8) {\footnotesize{$\mu$}};
\filldraw[red] (1,1) circle (1.5pt);
\node[left,red] at (1.6,1.6) {\footnotesize{$\mu^\vee$}};
\filldraw[red] (1.5,1.5) circle (1.5pt);
\end{tikzpicture}}~
\eea
where all red lines represent a copy of the algebra object $\cA_A$. Comparing the two now allows us to fix the junctions $\mu_{A^x}$ and $\mu^\vee_{A^x}$. 
In particular, we see that 
\bea
{\begin{tikzpicture}[baseline=30]
\draw [thick] (0,0)--(0,2.5)--(2.5,2.5) -- (2.5,0) -- (0,0);
\draw [thick,red,decoration={markings, mark=at position 0.5 with {\arrow{stealth}}}, postaction={decorate}]  (0,1)--(1,1);
\draw [thick,red,decoration={markings, mark=at position 0.5 with {\arrow{stealth}}}, postaction={decorate}] (1,0)--(1,1);
\draw [thick,red,decoration={markings, mark=at position 0.5 with {\arrow{stealth}}}, postaction={decorate}] (1,1)--(1.5,1.5);
\draw [thick,red,decoration={markings, mark=at position 0.5 with {\arrow{stealth}}}, postaction={decorate}] (1.5,1.5)--(2.5,1.5);
\draw [thick,red,decoration={markings, mark=at position 0.5 with {\arrow{stealth}}}, postaction={decorate}] (1.5,1.5)--(1.5,2.5);
\node[right,red] at (1,0.8) {\footnotesize{$\mu$}};
\filldraw[red] (1,1) circle (1.5pt);
\node[left,red] at (1.6,1.6) {\footnotesize{$\mu^\vee$}};
\filldraw[red] (1.5,1.5) circle (1.5pt);
\end{tikzpicture}}~ &=& \frac{1}{M^{r_A \cdot g}}\sum_{\bu, \widetilde\bu \in \mathrm{im}(1-A)} \omega^{\half \bu^T (R^{-1})^T \widetilde Q^{A^x} R^{-1} \widetilde\bu - \half \widetilde\bu^T (R^{-1})^T \widetilde Q^{A^x} R^{-1} \bu}
\\
&\vphantom{.}&\hspace{0.5 in}\times\,
\omega^{- \sum_{i<j}^{r_A} R_i^T \Omega R_j \left((R^{-1} \bu)_i (R^{-1} \widetilde\bu)_j - (R^{-1} \widetilde\bu)_i (R^{-1} \bu)_j \right) }
\,\,{\begin{tikzpicture}[baseline=30]
\draw [thick] (0,0)--(0,2.5)--(2.5,2.5) -- (2.5,0) -- (0,0);
\draw [blue,thick,decoration={markings, mark=at position 0.5 with {\arrow{stealth}}}, postaction={decorate}]  (0,1)--(1,1);
\draw [blue,thick,decoration={markings, mark=at position 0.5 with {\arrow{stealth}}}, postaction={decorate}] (1,0)--(1,1);
\draw [blue,thick,decoration={markings, mark=at position 0.5 with {\arrow{stealth}}}, postaction={decorate}] (1,1)--(1.5,1.5);
\draw [blue,thick,decoration={markings, mark=at position 0.5 with {\arrow{stealth}}}, postaction={decorate}] (1.5,1.5)--(2.5,1.5);
\draw [blue,thick,decoration={markings, mark=at position 0.5 with {\arrow{stealth}}}, postaction={decorate}] (1.5,1.5)--(1.5,2.5);
\node[blue,right] at (1,0.5) {\footnotesize{$L_{\widetilde\bu}$}};
\node[blue,above] at (0.6,1)  {\footnotesize{$L_{\bu}$}};
\end{tikzpicture}}~,\,\,\no
\eea
where we have dropped the $k$ index and set $\bu = R \mathbf{a}$ and $\widetilde\bu = R \mathbf{b}$. Note that $R$ is not actually an invertible matrix, 
but it does admit a left pseudo-inverse, which is what we have denoted by $R^{-1}$ above. Finally, making use of (\ref{eq:muAconstraint1}) and 
(\ref{eq:muAconstraint2}) we can fix $\m_{A^x}(\bu, \widetilde\bu)$ up to a term of the form $ \omega^{\half \bu^T S \widetilde\bu}$, with $S$ a symmetric matrix, 
which can be set to zero via a gauge transformation (\ref{eq:mugaugetransf}). The final result is
\bea
\label{eq:mufinalresult}
\mu_{A^x}(\bu, \widetilde \bu) = \omega^{\sum_{i< j} u_i \left\{ \left[ (R^{-1})^T \widetilde Q^{A^x} R^{-1}\right]_{ij} - \sum_{m<n} (R^T \Omega R)_{mn} (R^{-1}_{mi} R^{-1}_{nj} -R^{-1}_{mj} R^{-1}_{ni})  \right\} {\widetilde u}_j}~.
\eea
Clearly, there is an explicit dependence on $x$ through the coefficient matrix $\widetilde Q^{A^x}$.

\subsection{Twist defects: revisited}
\label{sec:twistdefects2}

We now return to the study of interfaces between general condensation defects, say $D_{A^{x-I}}$ and $D_{A^{x}}$, focusing on the case of $N=p$ being prime.  We denote such an interface by $\Sigma^{[\bw], \rho}_{A^{x-I}-A^x}$, with the notation to be explained shortly (we will often drop the subscript when it is clear from context). As in the case of usual twist defects introduced above, 
interfaces can absorb lines in $\mathrm{im}_M(1-A)$, and thus should be labelled by cosets $[\mathbf{w}] \in \mathrm{coker}_M(1-A)$, such that 
\bea
\Sigma^{[\bw], \rho}_{A^{x-I}-A^x} = \bigoplus_{\bw \in [\bw]} n_\bw L_\bw~
\eea
for some coefficients $n_\bw = \mathrm{dim}\, \mathrm{Hom}(\Sigma^{[\bw], \rho}, L_\bw)$. In fact, in order to be able to absorb arbitrary elements in $\mathrm{im}_M(1-A)$,  these coefficients should also be independent of the particular representative $\bw \in [\bw]$, and hence we have a simpler expression,
\bea
\Sigma^{[\bw], \rho}_{A^{x-I}-A^x} = n_{[\bw]} \bigoplus_{\bw \in [\bw]}  L_\bw~.
\eea
As opposed to the algebra objects, here it is not necessarily the case that the coefficient $n_{[\bw]}$ is equal to one,
which means that the hom-spaces can be higher-dimensional. Graphically, we can represent the
bases for these spaces as 
\begin{equation}\label{eq:SigmaLwbasis}
{\begin{tikzpicture}[baseline=0,square/.style={regular polygon,regular polygon sides=4}]
\draw[blue, thick, decoration={markings, mark=at position 0.5 with {\arrow{stealth}}}, postaction={decorate}] (0,-0.5) -- (0,0);
\draw[dgreen, thick, decoration={markings, mark=at position 0.7 with {\arrow{stealth}}}, postaction={decorate}] (0,0) -- (0,0.5);
\filldraw[blue] (0,0) circle (1pt);

\node[blue, below] at (0,-0.5) {\footnotesize $L_\bw$};
\node[dgreen, above] at (0,0.5) {\footnotesize $\Sigma^{[\bw],\rho}$};
\node[blue, left] at (0,0) {\footnotesize $i$};
\end{tikzpicture}} \in \mathrm{Hom}(L_\bw, \Sigma^{[\bw],\rho})~, \hspace{0.5 in} {\begin{tikzpicture}[baseline=0,square/.style={regular polygon,regular polygon sides=4}]
\draw[dgreen, thick, decoration={markings, mark=at position 0.5 with {\arrow{stealth}}}, postaction={decorate}] (0,-0.5) -- (0,0);
\draw[blue, thick,, decoration={markings, mark=at position 0.7 with {\arrow{stealth}}}, postaction={decorate}] (0,0) -- (0,0.5);
\filldraw[blue] (0,0) circle (1pt);

\node[dgreen, below] at (0,-0.5) {\footnotesize $\Sigma^{[\bw],\rho}$};
\node[blue, above] at (0,0.5) {\footnotesize $L_\bw$};
\node[blue, left] at (0,0) {\footnotesize $j$};
\end{tikzpicture}} \in \mathrm{Hom}(\Sigma^{[\bw],\rho}, L_\bw)~,
\end{equation}
where $i=1, \dots, n_{[\bw]}$ and likewise for $j$. 
We may choose a complete orthonormal basis, such that the following conditions are satisfied, 
\begin{equation}\label{eq:orthonormal}
{\begin{tikzpicture}[baseline=0,square/.style={regular polygon,regular polygon sides=4}]
\draw[dgreen, thick, decoration={markings, mark=at position 0.3 with {\arrow{stealth}}}, postaction={decorate}] (0,0) -- (0,0.5);
\draw[blue, thick, decoration={markings, mark=at position 0.7 with {\arrow{stealth}}}, postaction={decorate}] (0,0.5) -- (0,1);
\filldraw[blue] (0,0.5) circle (1pt);
\draw[blue, thick,  decoration={markings, mark=at position 0.5 with {\arrow{stealth}}}, postaction={decorate}] (0,-1) -- (0,-0.5);
\draw[dgreen,thick] (0,-0.5) -- (0,0);
\filldraw[blue] (0,-0.5) circle (1pt);

\node[dgreen, left] at (0,0) {\footnotesize $\Sigma^{[\bw],\rho}$};
\node[blue, left] at (0,0.5) {\footnotesize $j$};
\node[blue, left] at (0,-0.5) {\footnotesize $i$};
\node[blue, above] at (0,1) {\footnotesize $L_\bw$};
\node[blue, below] at (0,-1) {\footnotesize $L_{\bw'}$};
\end{tikzpicture}} \,\,=\,\, \delta^{ij} \delta^{\bw,\bw'}{\begin{tikzpicture}[baseline=0]
\draw[blue, thick, decoration={markings, mark=at position 0.6 with {\arrow{stealth}}}, postaction={decorate}] (0,-1) -- (0,1);

\node[blue, above] at (0,1) {\footnotesize $L_\bw$};
\node[blue, below] at (0,-1) {\footnotesize $L_\bw$};
\end{tikzpicture}}~, \hspace{0.7 in} \sum_{\bw,i} {\begin{tikzpicture}[baseline=0,square/.style={regular polygon,regular polygon sides=4}]
\draw[blue, thick] (0,0) -- (0,0.5);
\draw[dgreen, thick, decoration={markings, mark=at position 0.7 with {\arrow{stealth}}}, postaction={decorate}] (0,0.5) -- (0,1);
\filldraw[blue] (0,0.5) circle (1pt);
\draw[dgreen, thick, decoration={markings, mark=at position 0.5 with {\arrow{stealth}}}, postaction={decorate}] (0,-1) -- (0,-0.5);
\draw[blue, thick, decoration={markings, mark=at position 1 with {\arrow{stealth}}}, postaction={decorate}] (0,-0.5) -- (0,0);
\filldraw[blue] (0,-0.5) circle (1pt);

\node[blue, right] at (0,0) {\footnotesize $L_\bw$};
\node[dgreen, above] at (0,1) {\footnotesize $\Sigma^{[\bw],\rho}$};
\node[dgreen, below] at (0,-1) {\footnotesize $\Sigma^{[\bw],\rho}$};
\node[blue, left] at (0,0.5) {\footnotesize $i$};
\node[blue, left] at (0,-0.5) {\footnotesize $i$};
\end{tikzpicture}} \,\,=\,\, {\begin{tikzpicture}[baseline=0,square/.style={regular polygon,regular polygon sides=4}] 
\draw[dgreen, thick, decoration={markings, mark=at position 0.6 with {\arrow{stealth}}}, postaction={decorate}] (0,-1) -- (0,1);

\node[dgreen, above] at (0,1) {\footnotesize $\Sigma^{[\bw],\rho}$};
\node[dgreen, below] at (0,-1) {\footnotesize $\Sigma^{[\bw],\rho}$};
\end{tikzpicture}}~.
\end{equation}
Of course, the choice of basis is not unique, and one can in particular change bases as
 \bea
 \label{eq:Sigmagaugetransf}
 {\begin{tikzpicture}[baseline=0,square/.style={regular polygon,regular polygon sides=4}]
\draw[blue, thick] (0,-0.5) -- (0,0);
\draw[dgreen, thick] (0,0) -- (0,0.5);
\filldraw[blue] (0,0) circle (1pt);

\node[blue, below] at (0,-0.5) {\footnotesize $L_\bw$};
\node[dgreen, above] at (0,0.5) {\footnotesize $\Sigma^{[\bw],\rho}$};
\node[blue, left] at (0,0) {\footnotesize $i$};
\end{tikzpicture}} \longrightarrow 
 \hspace{0 in} \sum_{k} S_{ki}(\bw) {\begin{tikzpicture}[baseline=0,square/.style={regular polygon,regular polygon sides=4}]
\draw[blue, thick] (0,-0.5) -- (0,0);
\draw[dgreen, thick] (0,0) -- (0,0.5);
\filldraw[blue] (0,0) circle (1pt);

\node[blue, below] at (0,-0.5) {\footnotesize $L_\bw$};
\node[dgreen, above] at (0,0.5) {\footnotesize $\Sigma^{[\bw],\rho}$};
\node[blue, left] at (0,0) {\footnotesize $k$};
\end{tikzpicture}} ~, 
\hspace{0.5 in}{\begin{tikzpicture}[baseline=0,square/.style={regular polygon,regular polygon sides=4}]
\draw[dgreen, thick] (0,-0.5) -- (0,0);
\draw[blue, thick] (0,0) -- (0,0.5);
\filldraw[blue] (0,0) circle (1pt);

\node[dgreen, below] at (0,-0.5) {\footnotesize $\Sigma^{[\bw],\rho}$};
\node[blue, above] at (0,0.5) {\footnotesize $L_\bw$};
\node[blue, left] at (0,0) {\footnotesize $j$};
\end{tikzpicture}}
\rightarrow 
\sum_{\ell} S^\vee_{j\ell}(\bw){\begin{tikzpicture}[baseline=0,square/.style={regular polygon,regular polygon sides=4}]
\draw[dgreen, thick] (0,-0.5) -- (0,0);
\draw[blue, thick] (0,0) -- (0,0.5);
\filldraw[blue] (0,0) circle (1pt);

\node[dgreen, below] at (0,-0.5) {\footnotesize $\Sigma^{[\bw],\rho}$};
\node[blue, above] at (0,0.5) {\footnotesize $L_\bw$};
\node[blue, left] at (0,0) {\footnotesize $\ell$};
\end{tikzpicture}}~,
 \eea
which preserves the conditions (\ref{eq:orthonormal}) as long as $S^\vee_{j\ell}(\bw) = S^{-1}_{j\ell}(\bw)$.

\begin{figure}[tbp]
\begin{center}
{\begin{tikzpicture}[baseline=0,square/.style={regular polygon,regular polygon sides=4},scale=1.2]

\shade[top color=orange!40, bottom color=orange!10]  (0,-1) -- (-2,-1) -- (-2,2) -- (0,2)-- (0,-1);
\shade[top color=blue!40, bottom color=blue!10]  (0,-1) -- (2,-1) -- (2,2) -- (0,2)-- (0,-1);
\draw[dgreen,thick,  decoration={markings, mark=at position 0.5 with {\arrow{stealth}}}, postaction={decorate}](0,-1) -- (0,0); 
\draw[dgreen,thick,  decoration={markings, mark=at position 0.5 with {\arrow{stealth}}}, postaction={decorate}](0,0) -- (0,1); 
\draw[dgreen,thick,  decoration={markings, mark=at position 0.5 with {\arrow{stealth}}}, postaction={decorate}](0,1) -- (0,2); 
\draw [red, thick,  -<-=0.5] (0,0) to[out=180,in=90 ]  (-1,-1);
\draw [red, thick, -<-=0.5] (0,1) to[out=180,in=90 ]  (-2,-1);
\node[dgreen, above] at (0,1.9) {\footnotesize{$\Sigma^{[\bw], \rho}$}};
\node[dgreen, below] at (0,-1) {\footnotesize{$\Sigma^{[\bw], \rho}$}};
\node[red, below] at (-0.9,-1) {\footnotesize{$\cA_{A}$}};
\node[red, below] at (-1.9,-1) {\footnotesize{$\cA_{A}$}};

\filldraw[red] (0,0) circle (1.5pt);
\node[red, right] at (0,0) {\footnotesize{$\mu^{A^{x-I}-A^x}_L$}};
\filldraw[red] (0,1) circle (1.5pt);
\node[red, right] at (0,1) {\footnotesize{$\mu^{A^{x-I}-A^x}_L$}};
   
 \node[orange,right] at (-2,1.7) {\footnotesize{$D_{A^{x-I}}$}};
  \node[blue,left] at (2,1.7) {\footnotesize{$D_{A^{x}}$}};
\end{tikzpicture}} 
\hspace{0.2 in}$=$\hspace{0.2 in}
{\begin{tikzpicture}[baseline=0,square/.style={regular polygon,regular polygon sides=4},scale=1.2]

\shade[top color=orange!40, bottom color=orange!10]  (0,-1) -- (-2,-1) -- (-2,2) -- (0,2)-- (0,-1);
\shade[top color=blue!40, bottom color=blue!10]  (0,-1) -- (2,-1) -- (2,2) -- (0,2)-- (0,-1);
\draw[dgreen,thick,  decoration={markings, mark=at position 0.5 with {\arrow{stealth}}}, postaction={decorate}](0,-1) -- (0,0.5); 
\draw[dgreen,thick,  decoration={markings, mark=at position 0.5 with {\arrow{stealth}}}, postaction={decorate}](0,0.5) -- (0,2); 
\draw [red, thick,  -<-=0.5] (0,0.5) to[out=180,in=90 ]  (-1.1,-0.2);
\draw [red, thick, distance = 1 cm, ->-=0.2, -<-=0.85] (-1.6,-1) to[out=90,in=90 ]  (-0.6,-1);
\node[dgreen, above] at (0,1.9) {\footnotesize{$\Sigma^{[\bw], \rho}$}};
\node[dgreen, below] at (0,-1) {\footnotesize{$\Sigma^{[\bw], \rho}$}};
\node[red, below] at (-0.7,-1) {\footnotesize{$\cA_{A}$}};
\node[red, below] at (-1.7,-1) {\footnotesize{$\cA_{A}$}};

\filldraw[red] (0,0.5) circle (1.5pt);
\node[red, right] at (0,0.5) {\footnotesize{$\mu_L^{A^{x-I}-A^x}$}};
\filldraw[red]  (-1.1,-0.25) circle (1.5pt);
\node[red, right] at (-1.1,-0.1) {\footnotesize{$\mu_{A^{x-I}}$}};

 \node[orange,right] at (-2,1.7) {\footnotesize{$D_{A^{x-I}}$}};
  \node[blue,left] at (2,1.7) {\footnotesize{$D_{A^{x}}$}};
\end{tikzpicture}} \caption{Consistency condition to which the left junction $\mu_{L}$ is subject. This is the associativity axiom for left $(\cA_{A},\mu_{A^{x-I}})$ modules.}\label{fig:consistency_cond1_left}
\end{center}
\end{figure}

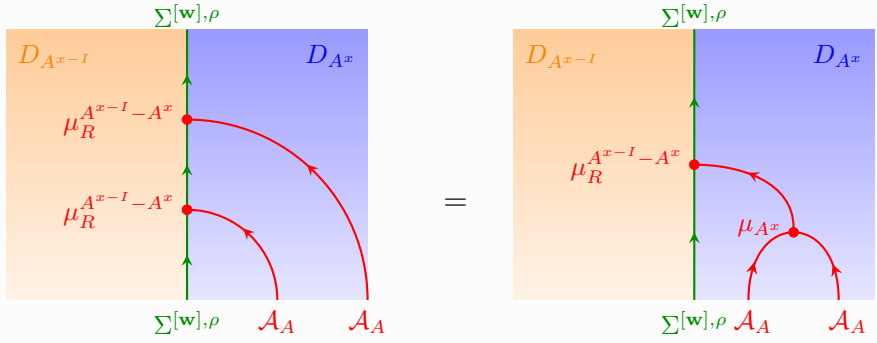
\begin{figure}[tbp]
\begin{center}
{\begin{tikzpicture}[baseline=0,square/.style={regular polygon,regular polygon sides=4},scale=1.2]

\shade[top color=orange!40, bottom color=orange!10]  (0,-1) -- (-2,-1) -- (-2,2) -- (0,2)-- (0,-1);
\shade[top color=blue!40, bottom color=blue!10]  (0,-1) -- (2,-1) -- (2,2) -- (0,2)-- (0,-1);
\draw[dgreen,thick,  decoration={markings, mark=at position 0.5 with {\arrow{stealth}}}, postaction={decorate}](0,-1) -- (0,0); 
\draw[dgreen,thick,  decoration={markings, mark=at position 0.5 with {\arrow{stealth}}}, postaction={decorate}](0,0) -- (0,1); 
\draw[dgreen,thick,  decoration={markings, mark=at position 0.5 with {\arrow{stealth}}}, postaction={decorate}](0,1) -- (0,2); 
\draw [red, thick, -<-=0.5] (0,0) to[out=0,in=90 ]  (1,-1);
\draw [red, thick, -<-=0.5] (0,1) to[out=0,in=90 ]  (2,-1);
\node[dgreen, above] at (0,1.9) {\footnotesize{$\Sigma^{[\bw], \rho}$}};
\node[dgreen, below] at (0,-1) {\footnotesize{$\Sigma^{[\bw], \rho}$}};
\node[red, below] at (1,-1) {\footnotesize{$\cA_{A}$}};
\node[red, below] at (2,-1) {\footnotesize{$\cA_{A}$}};

\filldraw[red] (0,0) circle (1.5pt);
\node[red, left] at (0,0) {\footnotesize{$\mu_R^{A^{x-I}-A^x}$}};
\filldraw[red] (0,1) circle (1.5pt);
\node[red, left] at (0,1) {\footnotesize{$\mu_R^{A^{x-I}-A^x}$}};
   
 \node[orange,right] at (-2,1.7) {\footnotesize{$D_{A^{x-I}}$}};
  \node[blue,left] at (2,1.7) {\footnotesize{$D_{A^{x}}$}};
\end{tikzpicture}} 
\hspace{0.2 in}$=$\hspace{0.2 in}
{\begin{tikzpicture}[baseline=0,square/.style={regular polygon,regular polygon sides=4},scale=1.2]

 \shade[top color=orange!40, bottom color=orange!10]  (0,-1) -- (-2,-1) -- (-2,2) -- (0,2)-- (0,-1);
  \shade[top color=blue!40, bottom color=blue!10]  (0,-1) -- (2,-1) -- (2,2) -- (0,2)-- (0,-1);
\draw[dgreen,thick,  decoration={markings, mark=at position 0.5 with {\arrow{stealth}}}, postaction={decorate}](0,-1) -- (0,0.5); 
\draw[dgreen,thick,  decoration={markings, mark=at position 0.5 with {\arrow{stealth}}}, postaction={decorate}](0,0.5) -- (0,2); 
\draw [red, thick, -<-=0.5] (0,0.5) to[out=0,in=90 ]  (1.1,-0.2);
\draw [red, thick, distance = 1 cm, -<-=0.85, ->-=0.2] (1.6,-1) to[out=90,in=90 ]  (0.6,-1);
\node[dgreen, above] at (0,1.9) {\footnotesize{$\Sigma^{[\bw], \rho}$}};
\node[dgreen, below] at (0,-1) {\footnotesize{$\Sigma^{[\bw], \rho}$}};
\node[red, below] at (0.65,-1) {\footnotesize{$\cA_{A}$}};
\node[red, below] at (1.7,-1) {\footnotesize{$\cA_{A}$}};

\filldraw[red] (0,0.5) circle (1.5pt);
\node[red, left] at (0,0.5) {\footnotesize{$\mu_R^{A^{x-I}-A^x}$}};
\filldraw[red]  (1.1,-0.25) circle (1.5pt);
\node[red, left] at (1.1,-0.2) {\footnotesize{$\mu_{A^x}$}};

 \node[orange,right] at (-2,1.7) {\footnotesize{$D_{A^{x-I}}$}};
  \node[blue,left] at (2,1.7) {\footnotesize{$D_{A^{x}}$}};
\end{tikzpicture}} 
\caption{Consistency condition to which the right junction $\mu_{R}$ is subject. This is the associativity axiom for right $(\cA_{A},\mu_{A^{x}})$ modules.}\label{fig:consistency_cond1_right}
\end{center}
\end{figure}

\begin{figure}[tbp]
\begin{center}
{\begin{tikzpicture}[baseline=0,square/.style={regular polygon,regular polygon sides=4},scale=1.2]

 \shade[top color=orange!40, bottom color=orange!10]  (0,-1) -- (-2,-1) -- (-2,2) -- (0,2)-- (0,-1);
  \shade[top color=blue!40, bottom color=blue!10]  (0,-1) -- (2,-1) -- (2,2) -- (0,2)-- (0,-1);
\draw[dgreen,thick,  decoration={markings, mark=at position 0.5 with {\arrow{stealth}}}, postaction={decorate}](0,-1) -- (0,0); 
\draw[dgreen,thick,  decoration={markings, mark=at position 0.5 with {\arrow{stealth}}}, postaction={decorate}](0,0) -- (0,1); 
\draw[dgreen,thick,  decoration={markings, mark=at position 0.5 with {\arrow{stealth}}}, postaction={decorate}](0,1) -- (0,2); 
\draw [red, thick, -<-=0.5] (0,0) to[out=180,in=90 ]  (-1,-1);
\draw [red, thick, -<-=0.5] (0,1) to[out=0,in=90 ]  (2,-1);
\node[dgreen, above] at (0,1.9) {\footnotesize{$\Sigma^{[\bw], \rho}$}};
\node[dgreen, below] at (0,-1) {\footnotesize{$\Sigma^{[\bw], \rho}$}};
\node[red, below] at (-1.15,-1) {\footnotesize{$\cA_{A}$}};
\node[red, below] at (1.9,-1) {\footnotesize{$\cA_{A}$}};

\filldraw[red] (0,0) circle (1.5pt);
\node[red, right] at (0,0) {\footnotesize{$\mu_L^{A^{x-I}-A^x}$}};
\filldraw[red] (0,1) circle (1.5pt);
\node[red, left] at (0,1) {\footnotesize{$\mu_R^{A^{x-I}-A^x}$}};
   
 \node[orange,right] at (-2,1.7) {\footnotesize{$D_{A^{x-I}}$}};
  \node[blue,left] at (2,1.7) {\footnotesize{$D_{A^{x}}$}};
\end{tikzpicture}} 
\hspace{0.2 in}$=$\hspace{0.2 in}
{\begin{tikzpicture}[baseline=0,square/.style={regular polygon,regular polygon sides=4},scale=1.2]

 \shade[top color=orange!40, bottom color=orange!10]  (0,-1) -- (-2,-1) -- (-2,2) -- (0,2)-- (0,-1);
  \shade[top color=blue!40, bottom color=blue!10]  (0,-1) -- (2,-1) -- (2,2) -- (0,2)-- (0,-1);
\draw[dgreen,thick,  decoration={markings, mark=at position 0.5 with {\arrow{stealth}}}, postaction={decorate}](0,-1) -- (0,0); 
\draw[dgreen,thick,  decoration={markings, mark=at position 0.5 with {\arrow{stealth}}}, postaction={decorate}](0,0) -- (0,1); 
\draw[dgreen,thick,  decoration={markings, mark=at position 0.5 with {\arrow{stealth}}}, postaction={decorate}](0,1) -- (0,2); 
\draw [red, thick, -<-=0.5] (0,0) to[out=0,in=90 ]  (1,-1);
\draw [red, thick, -<-=0.5] (0,1) to[out=180,in=90 ]  (-2,-1);
\node[dgreen, above] at (0,1.9) {\footnotesize{$\Sigma^{[\bw], \rho}$}};
\node[dgreen, below] at (0,-1) {\footnotesize{$\Sigma^{[\bw], \rho}$}};
\node[red, below] at (1,-1) {\footnotesize{$\cA_{A}$}};
\node[red, below] at (-2,-1) {\footnotesize{$\cA_{A}$}};

\filldraw[red] (0,0) circle (1.5pt);
\node[red, left] at (0,0) {\footnotesize{$\mu_R^{A^{x-I}-A^x}$}};
\filldraw[red] (0,1) circle (1.5pt);
\node[red, right] at (0,1) {\footnotesize{$\mu_L^{A^{x-I}-A^x}$}};
   
 \node[orange,right] at (-2,1.7) {\footnotesize{$D_{A^{x-I}}$}};
  \node[blue,left] at (2,1.7) {\footnotesize{$D_{A^{x}}$}};
\end{tikzpicture}} 
\caption{Consistency condition to which both left and right junctions are subject. This makes the interface into an $(\cA_A, \mu_{A^{x-I}}) - (\cA_A, \mu_{A^{x}})$-bimodule.}\label{fig:consistency_cond1_left_right}
\end{center}
\end{figure}
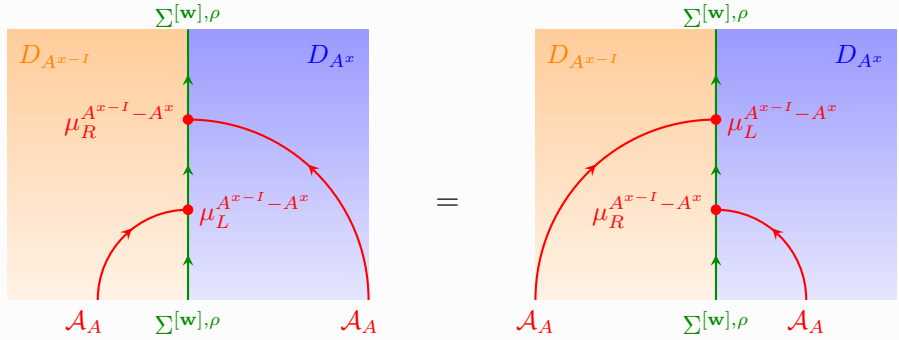

Just as the specification of the algebra object involved not only the line $\cA_{A^x}$, but also the multiplication $\mu_{A^x}$, in the current case the interfaces also depend on more data than just the number $n_{[\bw]}$. Mathematically, interfaces between $D_{A^{x-I}}$ and $D_{A^x}$ must be $(\cA_A, \mu_{A^{x-I}}) - (\cA_A, \mu_{A^{x}})$ bimodules, which means that there exist junctions $\mu_L^{A^{x-I}-A^x}$ and $\mu_R^{A^{x-I} - A^x}$ such that the conditions in Figures \ref{fig:consistency_cond1_left}, \ref{fig:consistency_cond1_right}, and \ref{fig:consistency_cond1_left_right} are satisfied (the superscripts will often be dropped when they are clear from context). The dual junctions are constrained in terms of these as in Figure \ref{fig:consistencymuvee}. Specification of the interface also involves specification of these junctions, but like in the case of the multiplication $\mu_{A^x}$, the junctions are subject to gauge redundancies which may be characterized as follows.

\begin{figure}[tbp]
\begin{center}
{\begin{tikzpicture}[baseline=0,square/.style={regular polygon,regular polygon sides=4},scale=0.9]

 \shade[top color=orange!40, bottom color=orange!10]  (0,-1) -- (-2,-1) -- (-2,2) -- (0,2)-- (0,-1);
  \shade[top color=blue!40, bottom color=blue!10]  (0,-1) -- (2,-1) -- (2,2) -- (0,2)-- (0,-1);
\draw[dgreen,thick,  decoration={markings, mark=at position 0.5 with {\arrow{stealth}}}, postaction={decorate}](0,-1) -- (0,0); 
\draw[dgreen,thick,  decoration={markings, mark=at position 0.5 with {\arrow{stealth}}}, postaction={decorate}](0,0) -- (0,1); 
\draw[dgreen,thick,  decoration={markings, mark=at position 0.5 with {\arrow{stealth}}}, postaction={decorate}](0,1) -- (0,2); 
\draw [gray!30, line width=3pt, distance = 0.3 in] (0,0) to[out=180,in=180 ]  (0,1);
\draw [red, thick,  decoration={markings, mark=at position 0.5 with {\arrow{stealth}}}, postaction={decorate}, distance = 0.3 in] (0,0) to[out=180,in=180 ]  (0,1);
\node[dgreen, above] at (0,1.9) {\footnotesize{$\Sigma^{[\mathbf{w}],\rho}$}};
\node[dgreen, below] at (0,-1) {\footnotesize{$\Sigma^{[\mathbf{w}],\rho}$}};
\node[red, left] at (-0.6,0.5) {\footnotesize{$\cA_{A}$}};

\filldraw[red] (0,0) circle (1.5pt);
\node[red, right] at (0,0) {\footnotesize{$\mu_L^{\vee\,A^{x-I}-A^x}$}};
\filldraw[red] (0,1) circle (1.5pt);
\node[red, right] at (0,1) {\footnotesize{$\mu_L^{A^{x-I}-A^x}$}};
   
 \node[orange,right] at (-2,1.7) {\footnotesize{$D_{A^{x-I}}$}};
  \node[blue,left] at (2,1.7) {\footnotesize{$D_{A^{x}}$}};
\end{tikzpicture}} 
\hspace{0.2 in}$=$\hspace{0.2 in}
{\begin{tikzpicture}[baseline=0,square/.style={regular polygon,regular polygon sides=4},scale=0.9]

 \shade[top color=orange!40, bottom color=orange!10]  (0,-1) -- (-2,-1) -- (-2,2) -- (0,2)-- (0,-1);
  \shade[top color=blue!40, bottom color=blue!10]  (0,-1) -- (2,-1) -- (2,2) -- (0,2)-- (0,-1);
\draw[dgreen,thick,  decoration={markings, mark=at position 0.5 with {\arrow{stealth}}}, postaction={decorate}](0,-1) -- (0,2); 
\node[dgreen, above] at (0,1.9) {\footnotesize{$\Sigma^{[\mathbf{w}],\rho}$}};
\node[dgreen, below] at (0,-1) {\footnotesize{$\Sigma^{[\mathbf{w}],\rho}$}};

 \node[orange,right] at (-2,1.7) {\footnotesize{$D_{A^{x-I}}$}};
  \node[blue,left] at (2,1.7) {\footnotesize{$D_{A^{x}}$}};
\end{tikzpicture}} 
\\
{\begin{tikzpicture}[baseline=0,square/.style={regular polygon,regular polygon sides=4},scale=0.9]

 \shade[top color=orange!40, bottom color=orange!10]  (0,-1) -- (-2,-1) -- (-2,2) -- (0,2)-- (0,-1);
  \shade[top color=blue!40, bottom color=blue!10]  (0,-1) -- (2,-1) -- (2,2) -- (0,2)-- (0,-1);
\draw[dgreen,thick,  decoration={markings, mark=at position 0.5 with {\arrow{stealth}}}, postaction={decorate}](0,-1) -- (0,0); 
\draw[dgreen,thick,  decoration={markings, mark=at position 0.5 with {\arrow{stealth}}}, postaction={decorate}](0,0) -- (0,1); 
\draw[dgreen,thick,  decoration={markings, mark=at position 0.5 with {\arrow{stealth}}}, postaction={decorate}](0,1) -- (0,2); 
\draw [gray!30, line width=3pt, distance = 0.3 in] (0,0) to[out=0,in=0 ]  (0,1);
\draw [red, thick,  decoration={markings, mark=at position 0.5 with {\arrow{stealth}}}, postaction={decorate}, distance = 0.3 in] (0,0) to[out=0,in=0 ]  (0,1);
\node[dgreen, above] at (0,1.9) {\footnotesize{$\Sigma^{[\mathbf{w}],\rho}$}};
\node[dgreen, below] at (0,-1) {\footnotesize{$\Sigma^{[\mathbf{w}],\rho}$}};
\node[red, right] at (0.6,0.5) {\footnotesize{$\cA_{A}$}};

\filldraw[red] (0,0) circle (1.5pt);
\node[red, left] at (0,0) {\footnotesize{$\mu_R^{\vee\,A^{x-I}-A^x}$}};
\filldraw[red] (0,1) circle (1.5pt);
\node[red, left] at (0,1) {\footnotesize{$\mu_R^{A^{x-I}-A^x}$}};
   
 \node[orange,right] at (-2,1.7) {\footnotesize{$D_{A^{x-I}}$}};
  \node[blue,left] at (2,1.7) {\footnotesize{$D_{A^{x}}$}};
\end{tikzpicture}} 
\hspace{0.2 in}$=$\hspace{0.2 in}
{\begin{tikzpicture}[baseline=0,square/.style={regular polygon,regular polygon sides=4},scale=0.9]

 \shade[top color=orange!40, bottom color=orange!10]  (0,-1) -- (-2,-1) -- (-2,2) -- (0,2)-- (0,-1);
  \shade[top color=blue!40, bottom color=blue!10]  (0,-1) -- (2,-1) -- (2,2) -- (0,2)-- (0,-1);
\draw[dgreen,thick,  decoration={markings, mark=at position 0.5 with {\arrow{stealth}}}, postaction={decorate}](0,-1) -- (0,2); 
\node[dgreen, above] at (0,1.9) {\footnotesize{$\Sigma^{[\mathbf{w}],\rho}$}};
\node[dgreen, below] at (0,-1) {\footnotesize{$\Sigma^{[\mathbf{w}],\rho}$}};
   
 \node[orange,right] at (-2,1.7) {\footnotesize{$D_{A^{x-I}}$}};
  \node[blue,left] at (2,1.7) {\footnotesize{$D_{A^{x}}$}};
\end{tikzpicture}} 
\caption{Consistency conditions determining $\mu^\vee_{L,R}$.}\label{fig:consistencymuvee}
\end{center}
\end{figure}
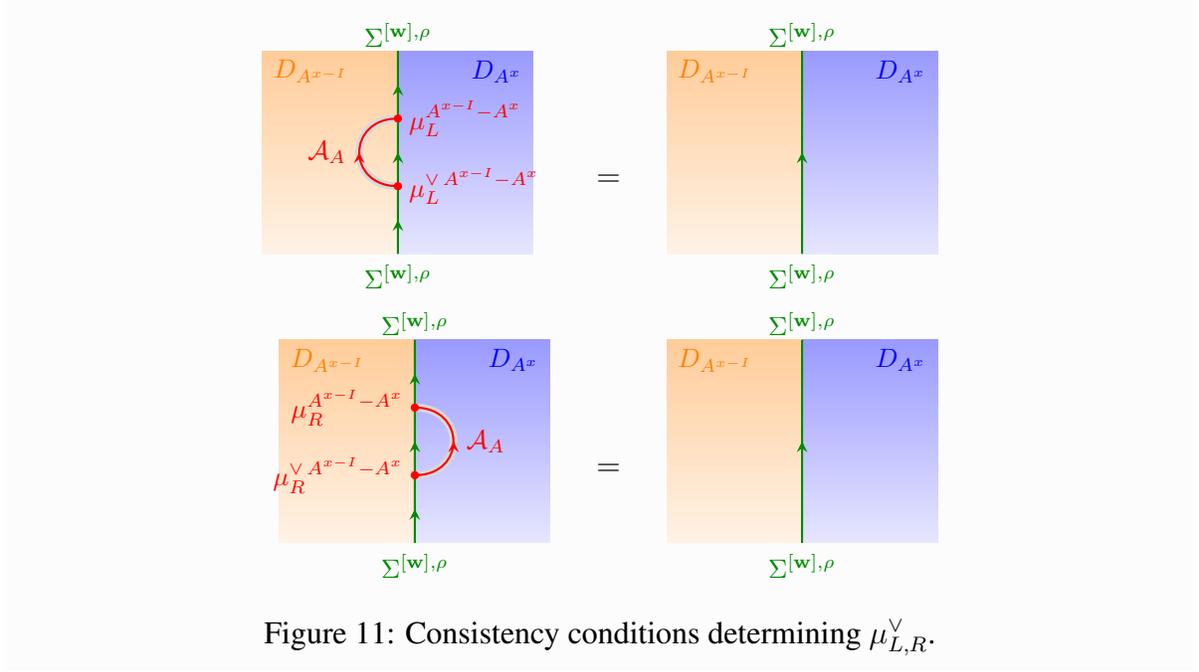

\begin{figure}[tbp]
\begin{center}
{\begin{tikzpicture}[baseline=0,scale=1.4]

 \shade[top color=orange!40, bottom color=orange!10]  (0,-1) -- (-1.5,-1) -- (-1.5,1) -- (0,1)-- (0,-1);
  \shade[top color=blue!40, bottom color=blue!10]  (0,-1) -- (1.5,-1) -- (1.5,1) -- (0,1)-- (0,-1);
\draw[dgreen,thick,  decoration={markings, mark=at position 0.5 with {\arrow{stealth}}}, postaction={decorate}](0,-1) -- (0,0); 
\draw[dgreen,thick,  decoration={markings, mark=at position 0.5 with {\arrow{stealth}}}, postaction={decorate}](0,0) -- (0,1); 
\draw [red, thick, -<-=0.5] (0,0) to[out=180,in=90 ]  (-1,-1);
\node[dgreen, above] at (0,0.9) {\footnotesize{$\Sigma^{[\mathbf{w}],\rho}$}};
\node[dgreen, below] at (0,-1) {\footnotesize{$\Sigma^{[\mathbf{w}],\rho}$}};
\node[red, below] at (-1.15,-1) {\footnotesize{$\cA_{A}$}};

\filldraw[red] (0,0) circle (1.5pt);
\node[red, right] at (0,0) {\footnotesize{$\mu^{A^{x-I}-A^x}_L$}};
   
 \node[orange,right] at (-1.5,0.7) {\footnotesize{$D_{A^{x-I}}$}};
  \node[blue,left] at (1.5,0.7) {\footnotesize{$D_{A^{x}}$}};
\end{tikzpicture}} 
\hspace{0.1 in}$\displaystyle =\,\, \frac{1}{M^{r_A/2}}\sum_{\substack{\bw \in [\bw]\\\mathbf{u}\in \mathrm{im}(1-A)}}\sum_{i,j}\,[\cL^{A^{x-I} - A^x}_{ [\bw]}(\bw, \bu)]_{ij} $\hspace{-0.1 in}
{\begin{tikzpicture}[baseline=0,scale=1.2]

\draw[dgreen,thick,  decoration={markings, mark=at position 0.5 with {\arrow{stealth}}}, postaction={decorate}](0,-1) -- (0,-0.6); 
\draw[blue,thick,  decoration={markings, mark=at position 0.5 with {\arrow{stealth}}}, postaction={decorate}](0,-0.6) -- (0,0); 
\draw[blue,thick,  decoration={markings, mark=at position 0.5 with {\arrow{stealth}}}, postaction={decorate}](0,0) -- (0,0.6); 
\draw[dgreen,thick,  decoration={markings, mark=at position 0.7 with {\arrow{stealth}}}, postaction={decorate}](0,0.6) -- (0,1); 
\draw [blue, thick, -<-=0.3] (0,0) to[out=180,in=90 ]  (-1,-1);
\draw [red, thick, -<-=0.5] (-0.825,-0.433) to[out=239,in=90 ]  (-1,-1);
\node[blue, right] at (0,0.4) {\footnotesize{$L_{\mathbf{w} + \mathbf{u}}$}};
\node[blue, right] at (0,-0.4) {\footnotesize{$L_{\mathbf{w}}$}};
\node[blue, left] at (-0.4,0.1) {\footnotesize{$L_{\mathbf{u}}$}};

\node[dgreen, above] at (0,0.9) {\footnotesize{$\Sigma^{[\mathbf{w}],\rho}$}};
\node[red, below] at (-1.15,-1) {\footnotesize{$\cA_A$}};
\node[dgreen, below] at (0,-1) {\footnotesize{$\Sigma^{[\mathbf{w}],\rho}$}};

\node[blue, left] at (0,-0.6) {\footnotesize{$i$}};
\node[blue, left] at (0,0.6) {\footnotesize{$j$}};

\filldraw[blue] (0,0) circle (1pt);
\filldraw[blue] (0,-0.6) circle (1pt);
\filldraw[blue] (0,0.6) circle (1pt);
\filldraw[blue] (-0.825,-0.433) circle (1pt);

\end{tikzpicture}} 
\\\vphantom{.}
\hspace{0.3in}

 {\begin{tikzpicture}[baseline=0,scale=1.4]

 \shade[top color=orange!40, bottom color=orange!10]  (0,-1) -- (-1.5,-1) -- (-1.5,1) -- (0,1)-- (0,-1);
\shade[top color=blue!40, bottom color=blue!10]  (0,-1) -- (1.5,-1) -- (1.5,1) -- (0,1)-- (0,-1);
\draw[dgreen,thick,  decoration={markings, mark=at position 0.5 with {\arrow{stealth}}}, postaction={decorate}](0,-1) -- (0,0); 
\draw[dgreen,thick,  decoration={markings, mark=at position 0.5 with {\arrow{stealth}}}, postaction={decorate}](0,0) -- (0,1); 

\draw [red, thick, -<-=0.5] (0,0) to[out=0,in=90 ]  (1,-1);
\node[dgreen, above] at (0,0.9) {\footnotesize{$\Sigma^{[\mathbf{w}],\rho}$}};
\node[dgreen, below] at (0,-1) {\footnotesize{$\Sigma^{[\mathbf{w}],\rho}$}};
\node[red, below] at (1,-1) {\footnotesize{$\cA_{A}$}};

\filldraw[red] (0,0) circle (1.5pt);
\node[red, left] at (0,0) {\footnotesize{$\mu_R^{A^{x-I}-A^x}$}};

   \node[orange,right] at (-1.5,0.7) {\footnotesize{$D_{A^{x-I}}$}};
  \node[blue,left] at (1.5,0.7) {\footnotesize{$D_{A^{x}}$}};
  
\end{tikzpicture}}
\hspace{0.1 in}$\displaystyle =\,\, \frac{1}{M^{r_A/2}} \sum_{\substack{\bw \in [\bw]\\\mathbf{u}\in \mathrm{im}(1-A)}}\sum_{i,j}\, [\cR^{A^{x-I} - A^x}_{ [\bw]}(\bw, \bu)]_{ij} $\,\,
{\begin{tikzpicture}[baseline=0,scale=1.2]

\draw[dgreen,thick,  decoration={markings, mark=at position 0.6 with {\arrow{stealth}}}, postaction={decorate}](0,-1) -- (0,-0.6); 
\draw[blue,thick,  decoration={markings, mark=at position 0.5 with {\arrow{stealth}}}, postaction={decorate}](0,-0.6) -- (0,0); 
\draw[dgreen,thick,  decoration={markings, mark=at position 0.7 with {\arrow{stealth}}}, postaction={decorate}](0,0.6) -- (0,1); 
\draw[blue,thick,  decoration={markings, mark=at position 0.5 with {\arrow{stealth}}}, postaction={decorate}](0,0) -- (0,0.6); 
\draw [blue, thick, -<-=0.3] (0,0) to[out=0,in=90 ]  (1,-1);
\draw [red, thick, -<-=0.5] (0.825,-0.433) to[out=-58,in=90 ]  (1,-1);

\node[dgreen, above] at (0,0.9) {\footnotesize{$\Sigma^{[\mathbf{w}],\rho}$}};
\node[dgreen, below] at (0,-1) {\footnotesize{$\Sigma^{[\mathbf{w}],\rho}$}};
\node[red, below] at (1,-1) {\footnotesize{$\cA_{A}$}};

\node[blue, left] at (0,0.35) {\footnotesize{$L_{\bw + \bu}$}};
\node[blue, left] at (0,-0.3) {\footnotesize{$L_{\bw}$}};
\node[blue, right] at (0.5,0) {\footnotesize{$L_{\bu}$}};

\node[blue, right] at (0,-0.6) {\footnotesize{$i$}};
\node[blue, right] at (0,0.6) {\footnotesize{$j$}};

\filldraw[blue] (0,0) circle (1pt);
\filldraw[blue] (0,-0.6) circle (1pt);
\filldraw[blue] (0,0.6) circle (1pt);
\filldraw[blue] (0.825,-0.433) circle (1pt);

\end{tikzpicture}}
\caption{Left and right junctions $\mu_{L}$, $\mu_{R}$ necessary to specify the interface $\Sigma^{[\bw],\rho}$. On the right-hand side, we do not draw the condensation defects for simplicity.}\label{fig:left_right_junction}
\end{center}
\end{figure}

We begin by expanding the junctions out in terms of simple lines as in Figure \ref{fig:left_right_junction}. Then upon making a change of basis as in  (\ref{eq:Sigmagaugetransf}), we see that 
\bea
\cL^{A^{x-I} - A^x}_{ [\bw]}(\bw, \bu) &\longrightarrow& S^{-1}(\bw)\, \cL^{A^{x-I} - A^x}_{[\bw]}(\bw, \bu) \,S(\bw+\bu)~,
\no\\
\cR^{A^{x-I} - A^x}_{ [\bw]}(\bw, \bu) &\longrightarrow& S^{-1}(\bw)\, \cR^{A^{x-I} - A^x}_{[\bw]}(\bw, \bu) \,S(\bw+\bu)~
\eea
as matrix equations. As a result, neither $\cL^{A^{x-I} - A^x}_{[\bw]}(\bw, \bu) $ nor $\cR^{A^{x-I} - A^x}_{ [\bw]}(\bw, \bu)$ is a gauge invariant piece of data. 
We choose to work in a gauge for which we have 
\bea
\left[ \cL^{A^{x-I} - A^x}_{ [\bw]}(\bw, \bu)\right]_{ij} = \mu_{A^{x-I}}(\bu, \bw) \delta_{ij}~.
\eea
 In this gauge, the consistency condition in Figure \ref{fig:consistency_cond1_left} is trivially satisfied, while the conditions in Figure \ref{fig:consistency_cond1_right} and \ref{fig:consistency_cond1_left_right} give rise to
 \bea
 \left[ \cR^{A^{x-I} - A^x}_{ [\bw]}(\bw, \bu)\right]_{ij} =  \mu_{A^{x-I}}(\bw, \bu) \rho^{A^{x-I} - A^x}_{ij}( \bu)~, 
 \eea
 where the matrix $\rho^{A^{x-I} - A^x}_{ij}( \bu)$ is subject to the following constraint, 
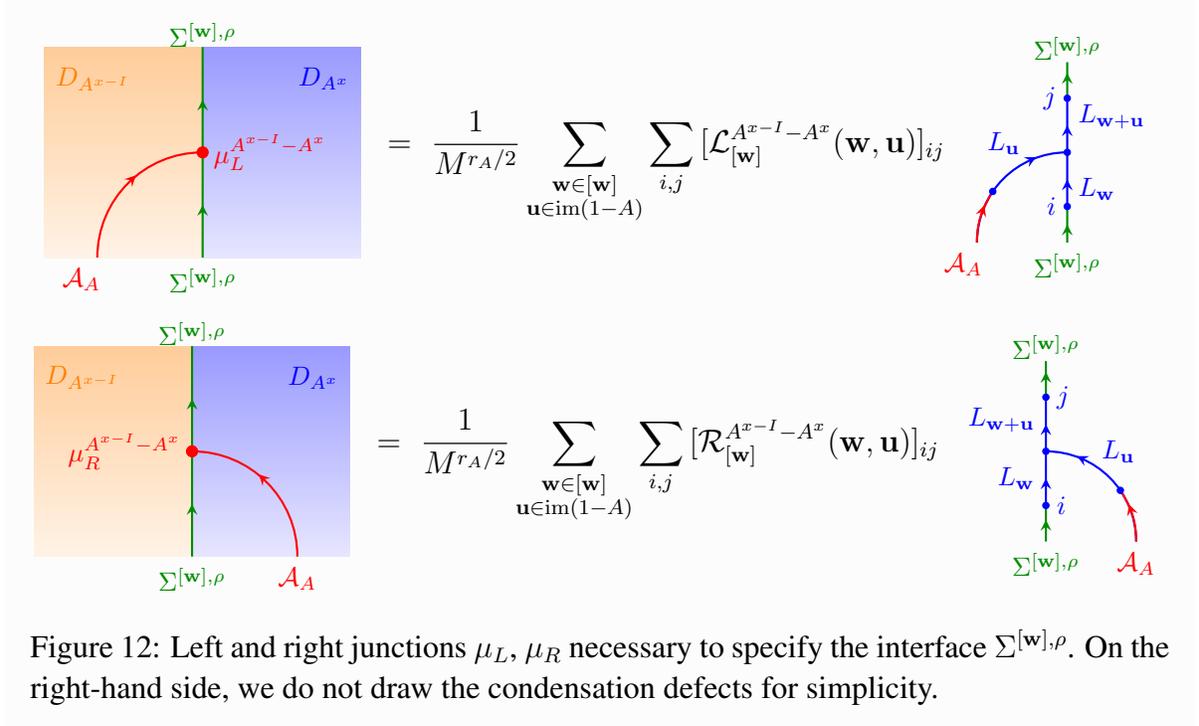
\bea
\label{eq:projrepcond}
 \rho^{A^{x-I} - A^x}(\bu) \cdot  \rho^{A^{x-I} - A^x}( \bu') = {\mu_{A^x}(\bu, \bu') \over \mu_{A^{x-I}}(\bu, \bu')} \rho^{A^{x-I} - A^x}(\bu + \bu')~. 
\eea
In other words, we see that $ \rho^{A^{x-I} - A^x}(\bu)$ is a projective representation of $\mathrm{im}_M(1-A)$, with projective phase given by ${\mu_{A^{x}}(\bu, \bu') \over \mu_{A^{x-I}}(\bu, \bu')}$. 
The superscript $\rho$ that we have been attaching to our interfaces $\Sigma^{[\mathbf{w}],\rho}$ captures this choice of projective representation.

\subsubsection{Little group classification}

It is not obvious that the irreducible projective representation $ \rho^{A^{x-I} - A^x}(\bu)$ (up to similarity transformation) captures \textit{all} of the gauge invariant data of $\mu_L^{A^{x-I}-A^x}$ and $\mu_R^{A^{x-I} - A^x}$, but this is indeed the case. Indeed, it has been shown in the mathematics literature \cite{yamagami2002group}, and described in the physics literature \cite{Choi:2023vgk,Lu:2024lzf} that simple bimodules are in one-to-one correspondence with elements of a certain double coset (in our case, just the label $[\bw]$) and a projective representation of the so-called ``little group'' (in our case, the representation of $\mathrm{im}_M(1-A)$). As such, the labels $[\bw]$ and $\rho$ completely specify the interface $\Sigma_{A^{x-I}-A^x}^{[\mathbf{w}],\rho}$. 

\subsubsection{Simple examples}

Let us give some simple examples. 

\paragraph{$D_{A^{x-I}} = D_{A^x}$:} First, in the case that the defects on the left and right of the interface are identical, the ratio ${\mu_{A^{x}}(\bu, \bu') \over \mu_{A^{x-I}}(\bu, \bu')} = 1$, and hence the label $\rho$ is actually a \textit{linear} (i.e. not projective) representation of $\mathrm{im}_M(1-A)$. For simplicity, let us consider here the case of $M$ being prime, so that $\mathrm{im}_M(1-A) = \ZZ_M^{r_A}$, though similar discussions hold more generally. In this case all representations are one-dimensional, and are specified by $\mathbf{k}= (k_1, \dots, k_{r_A}) \in \ZZ_M^{r_A}$ such that 
\bea
\rho_{\bw,\mathbf{k}}(e_1) = \omega^{k_1}~, \hspace{0.2 in} \rho_{\bw,\mathbf{k}}(e_2) = \omega^{k_2}~,\hspace{0.2 in}  \dots~,\hspace{0.2 in}  \rho_{\bw,\mathbf{k}}(e_{r_A}) = \omega^{k_{r_A}}~, 
\eea
where $e_i$ are the canonical basis of $\ZZ_M^{r_A}$ and $\omega:= e^{2 \pi i \over M}$. On the other hand, $\mathrm{coker}_M(1-A) = \ZZ_M^{n_A}$ with $n_A$ the nullity of $A$, and hence $[\bw]$ takes $M^{n_A}$ possible values. In total then, we have $M^{n_A} \times M^{r_A} = M^{2n}$ interfaces $\Sigma^{{[\bw]}, \rho_\mathbf{k}}$ between any given defect and itself. The physical interpretation of this is as follows. First, the interfaces $\Sigma^{[0], \rho_\mathbf{k}}$ are the generators of the quantum dual $\widehat \ZZ_M^{r_A}$ symmetry on the condensation defect (which arises from the higher-gauging of $\ZZ_M^{r_A}$ used to create the condensation defect). The interfaces $\Sigma^{[\bw], \rho_\mathbf{k}}$ are then obtained from these by fusing $\Sigma^{[0], \rho_\mathbf{k}}$ with bulk lines $L_\bw$ for any $\bw \in [\bw]$ (all $\bw \in [\bw]$ will give rise to the same result, since elements of $\mathrm{im}_M(1-A)$ are simply absorbed by the condensation defect). 

\paragraph{$M=2, r_A = 2$:} We next consider a slightly more involved example. Say that $M=2$ and $A$ is a matrix such that $r_A = 2$, so that $\mathrm{im}_M(1-A) = \ZZ_2^2 = \left\{(0,0), (1,0), (0,1), (1,1) \right\}$. 
If we consider the same condensation defect $D_{A^x}$ on both sides of the interface, then as we have just described the representation $\rho$ will be one-dimensional, and there will be a total of $2^2$ of them, corresponding to the quantum $\widehat\ZZ_2^2$ symmetry generators. We denote them by $\rho_i$ for $i=1,\dots,4$. In addition, there are four possible cosets $[\bw]$, giving a total of 16 interfaces $\Sigma^{[\bw], \rho_i}$. 

Now instead say that we have two \textit{different} condensation defects on the two sides of the interfaces, corresponding to higher gaugings of $\ZZ_2^2$ with different discrete torsions, such that ${\mu_{A^{x}}(\bu, \bu') \over \mu_{A^{x-I}}(\bu, \bu')} = (-1)^{\bu^T \Omega \bu'}$ (the only non-trivial possibility up to gauge transformation). We now see that the condition (\ref{eq:projrepcond}) does not allow $\rho$ to be one-dimensional. Instead, there exists a two-dimensional representation, given explicitly by 
\bea
\rho_{\mathsf{2d}}(0,0) = \mathds{1}~, \hspace{0.2 in} \rho_{\mathsf{2d}}(1,0) = X~, \hspace{0.2 in} \rho_{\mathsf{2d}}(0,1) = Z~, \hspace{0.2 in} \rho_{\mathsf{2d}}(1,1) = i Y~
\eea 
in terms of Pauli matrices $X,Y,Z$.\footnote{We take this to be the representation for $\bw = 0$. For non-zero $\bw$ we could modify the representation by a phase, but this will not be important in the current subsection} This is the only option up to projective equivalence. The physical significance of being in a multi-dimensional representation is that in general, intersections involving the algebra object and the bimodule will, when expanded in terms of simple lines as in Figure \ref{fig:left_right_junction}, necessarily mix different elements of the Hom-space $\mathrm{Hom}(\Sigma^{[\bw], \rho}, L_\bw)$ (which were labelled by the matrix indices $i,j$). 

\subsubsection{Fusion of twist defects and balanced tensor products}
We would next like to understand the in-plane fusion of two twist defects, corresponding to the tensor product of a right $\cA$-module with a left $\cA$-module.  We begin by discussing the global fusions, in particular taking the two twist defects (described by Dirichlet boundary conditions) to be living at the ends of a finite cylinder, as shown below,
\begin{equation}\label{eq:fctt}
    \begin{tikzpicture}[baseline=0,vertex/.style={anchor=base,
    circle,fill=black!25,minimum size=18pt,inner sep=2pt},scale=0.5]
    \shade[top color=orange!60, bottom color=orange!30]  (0,-2) arc(-90:90:1 and 2) -- (4,2) arc(90:-90:1 and 2)--(0,-2);
    \shade[top color=orange!40, bottom color=orange!10] (0,-2) arc(-90:270:1 and 2);
    \draw[thick, black] (0,-2) -- (4,-2);
    \draw[thick, black] (0,+2) -- (4,+2);
    
    \draw[thick, red, ->-=0.25] (0,-2) arc(-90:270:1 and 2);
    \draw[thick, red, ->-=0.5] (4,-2) arc(-90:90:1 and 2);
    \draw[thick, red, dashed] (4,2) arc(90:270:1 and 2);
    
    \node[below, red] at (0,-2) {\footnotesize $\Sigma_A^{[0]}$};
    \node[below, red] at (4,-2) {\footnotesize $\Sigma_A^{[0]}$};
    \node[orange] at (2,0) {\footnotesize $D_A$};
    \end{tikzpicture} \quad.
\end{equation}

One way to understand this configuration is to temporarily forget about the discussion of modules above, and instead simply think of the configuration as describing a condensation defect of $A$ on an interval. We may then evaluate the configuration by introducing a gauge field valued in relative cohomology and explicitly performing a sum. The complex $C_0(X,\partial X)$ is trivial for $X$ being the finite cylinder, and hence $C^0(X,\partial X)$ and consequently $H^0(X,\partial X;\mathcal{A})$ are trivial, so that  ${1 \over |H^0(X,\partial X;\mathcal{A})|} = 1$ and we obtain
\begin{equation}
    \begin{tikzpicture}[baseline=0,vertex/.style={anchor=base,
    circle,fill=black!25,minimum size=18pt,inner sep=2pt},scale=0.5]
    \shade[top color=orange!60, bottom color=orange!30]  (0,-2) arc(-90:90:1 and 2) -- (4,2) arc(90:-90:1 and 2)--(0,-2);
    \shade[top color=orange!40, bottom color=orange!10] (0,-2) arc(-90:270:1 and 2);
    \draw[thick, black] (0,-2) -- (4,-2);
    \draw[thick, black] (0,+2) -- (4,+2);
    
    \draw[thick, red, ->-=0.25] (0,-2) arc(-90:270:1 and 2);
    \draw[thick, red, ->-=0.5] (4,-2) arc(-90:90:1 and 2);
    \draw[thick, red, dashed] (4,2) arc(90:270:1 and 2);
    
    \node[below, red] at (0,-2) {\footnotesize $\Sigma_A^{[0]}$};
    \node[below, red] at (4,-2) {\footnotesize $\Sigma_A^{[0]}$};
    \node[orange] at (2,0) {\footnotesize $D_A$};
    \end{tikzpicture} \quad \quad  \xrightarrow{relative \,\, cohomology} \quad \quad \sum_{\mathbf{u} \in [0]} \quad \begin{tikzpicture}[baseline=0,vertex/.style={anchor=base,
    circle,fill=black!25,minimum size=18pt,inner sep=2pt},scale=0.5]
    \draw[thick, black] (0,-2) arc(-90:270:1 and 2);
    \draw[thick, black] (4,-2) arc(-90:90:1 and 2);
    \draw[thick, black, dashed] (4,2) arc(90:270:1 and 2);
    \draw[thick, black] (0,-2) -- (4,-2);
    \draw[thick, black] (0,+2) -- (4,+2);
    \draw[blue, thick, ->-=0.5] (2,-2) arc(-90:90:0.9 and 2);
    \draw[blue, thick, dashed] (2,2) arc(90:270:0.9 and 2);
    \node[blue, below] at (2,-2) {\footnotesize $L_{\mathbf{u}}$};
    \end{tikzpicture} \quad.
\end{equation}
Thus if we fuse the two twist defects by taking the length of the cylinder to zero, we are left with a sum of invertible lines, leading to the fusion rules
\begin{equation}
    \Sigma_A^{[0]} \times \Sigma_A^{[0]} = \sum_{\mathbf{u} \in [0]} L_{\mathbf{u}} ~.
\end{equation}

On the other hand, let us now try to represent these fusion rules using the language of algebra objects and their modules, discussed above. To do so, we first place the module object along the support of the twist defect. For the configuration \eqref{eq:fctt}, naively we would like to place two algebra objects $\mathcal{A}$ along the two boundaries of the cylinder,
\begin{equation}\label{eq:fcrc}
    \begin{tikzpicture}[baseline={0},vertex/.style={anchor=base,
    circle,fill=black!25,minimum size=18pt,inner sep=2pt},scale=0.5]
    \shade[top color=orange!60, bottom color=orange!30]  (0,-2) arc(-90:90:1 and 2) -- (4,2) arc(90:-90:1 and 2)--(0,-2);
    \shade[top color=orange!40, bottom color=orange!10] (0,-2) arc(-90:270:1 and 2);
    \draw[thick, black] (0,-2) -- (4,-2);
    \draw[thick, black] (0,+2) -- (4,+2);
    
    \draw[thick, red, ->-=0.25] (0,-2) arc(-90:270:1 and 2);
    \draw[thick, red, ->-=0.5] (4,-2) arc(-90:90:1 and 2);
    \draw[thick, red, dashed] (4,2) arc(90:270:1 and 2);
    
    \node[below, red] at (0,-2) {\footnotesize $\Sigma_A^{[0]}$};
    \node[below, red] at (4,-2) {\footnotesize $\Sigma_A^{[0]}$};
    \node[orange] at (2,0) {\footnotesize $D_A$};
    \end{tikzpicture} \quad \quad \xrightarrow{\quad \quad ? \quad \quad} \quad \quad \begin{tikzpicture}[baseline=0,vertex/.style={anchor=base,
    circle,fill=black!25,minimum size=18pt,inner sep=2pt},scale=0.5]
    \draw[thick, black] (0,-2) arc(-90:270:1 and 2);
    \draw[thick, black] (4,-2) arc(-90:90:1 and 2);
    \draw[thick, black, dashed] (4,2) arc(90:270:1 and 2);
    \draw[thick, black] (0,-2) -- (4,-2);
    \draw[thick, black] (0,+2) -- (4,+2);

    \draw[thick, red, ->-=0.25] (0,-2) arc(-90:270:1 and 2);
    \draw[thick, red, ->-=0.5] (4,-2) arc(-90:90:1 and 2);
    \draw[thick, red, dashed] (4,2) arc(90:270:1 and 2);
    
    \node[below, red] at (0,-2) {\footnotesize $\mathcal{A}$};
    \node[below, red] at (4,-2) {\footnotesize $\mathcal{A}$};
    \end{tikzpicture} \quad.
\end{equation}
However, as one can check, this does not lead to the expected fusion rules. This is because, when fusing a right $\mathcal{A}$-module with a left $\mathcal{A}$-module, one must take the \textit{balanced tensor product}, which quotients out the $\mathcal{A}$ action in between. This is realized by inserting the $\mathcal{A}$ object stretching between the two module objects \cite{Lou:2020gfq} as below,
\begin{equation}\label{eq:fcam}
    \begin{tikzpicture}[baseline=0,vertex/.style={anchor=base,
    circle,fill=black!25,minimum size=18pt,inner sep=2pt},scale=0.5]
    \shade[top color=orange!60, bottom color=orange!30]  (0,-2) arc(-90:90:1 and 2) -- (4,2) arc(90:-90:1 and 2)--(0,-2);
    \shade[top color=orange!40, bottom color=orange!10] (0,-2) arc(-90:270:1 and 2);
    \draw[thick, black] (0,-2) -- (4,-2);
    \draw[thick, black] (0,+2) -- (4,+2);
    
    \draw[thick, red, ->-=0.25] (0,-2) arc(-90:270:1 and 2);
    \draw[thick, red, ->-=0.5] (4,-2) arc(-90:90:1 and 2);
    \draw[thick, red, dashed] (4,2) arc(90:270:1 and 2);
    
    \node[below, red] at (0,-2) {\footnotesize $\Sigma_A^{[0]}$};
    \node[below, red] at (4,-2) {\footnotesize $\Sigma_A^{[0]}$};
    \node[orange] at (2,0) {\footnotesize $D_A$};
    \end{tikzpicture} \quad \quad \xrightarrow{\quad \quad algebras \,\, \& \,\, modules \quad \quad} \quad \quad \begin{tikzpicture}[baseline=0,vertex/.style={anchor=base,
    circle,fill=black!25,minimum size=18pt,inner sep=2pt},scale=0.5]
    \draw[thick, black] (0,-2) arc(-90:270:1 and 2);
    \draw[thick, black] (4,-2) arc(-90:90:1 and 2);
    \draw[thick, black, dashed] (4,2) arc(90:270:1 and 2);
    \draw[thick, black] (0,-2) -- (4,-2);
    \draw[thick, black] (0,+2) -- (4,+2);

    \draw[thick, red, ->-=0.15] (0,-2) arc(-90:270:1 and 2);
    \draw[thick, red, ->-=0.4] (4,-2) arc(-90:90:1 and 2);
    \draw[thick, red, dashed] (4,2) arc(90:270:1 and 2);

    \draw[thick, red, -<-=0.4] (1,0) -- (5,0); 
    
    \node[below, red] at (0,-2) {\footnotesize $\mathcal{A}$};
    \node[below, red] at (4,-2) {\footnotesize $\mathcal{A}$};
    \node[below, red] at (2,0) {\footnotesize $\mathcal{A}$};
    \end{tikzpicture} ~.
\end{equation}
To see that this leads to the correct fusion rules, we must show that \eqref{eq:fcam} simplifies to \eqref{eq:fcrc}. To do so, we recall that it is always possible to equip $\mathcal{A}$ with a co-multiplication and a co-unit in a unique way to make it a symmetric, $\Delta$-separable, Frobenious algebra, and consistent gauging requires those additional structures \cite{Choi:2023vgk, Choi:2023xjw}; see e.g. Figure \ref{fig:Aconsistency} above. The diagram of \eqref{eq:fcam} can therefore be simplified as
\begin{equation}
    \begin{tikzpicture}[baseline={([yshift=-1ex]current bounding box.center)},vertex/.style={anchor=base,
    circle,fill=black!25,minimum size=18pt,inner sep=2pt},scale=0.4]
    \draw[thick, black] (0,-2) arc(-90:270:1 and 2);
    \draw[thick, black] (4,-2) arc(-90:90:1 and 2);
    \draw[thick, black, dashed] (4,2) arc(90:270:1 and 2);
    \draw[thick, black] (0,-2) -- (4,-2);
    \draw[thick, black] (0,+2) -- (4,+2);

    \draw[thick, red, ->-=0.15] (0,-2) arc(-90:270:1 and 2);
    \draw[thick, red, ->-=0.4] (4,-2) arc(-90:90:1 and 2);
    \draw[thick, red, dashed] (4,2) arc(90:270:1 and 2);

    \draw[thick, red, -<-=0.4] (1,0) -- (5,0); 
    \end{tikzpicture} \quad = \quad \begin{tikzpicture}[baseline={([yshift=-1ex]current bounding box.center)},vertex/.style={anchor=base,
    circle,fill=black!25,minimum size=18pt,inner sep=2pt},scale=0.4]
    \draw[thick, black] (0,-2) arc(-90:270:1 and 2);
    \draw[thick, black] (4,-2) arc(-90:90:1 and 2);
    \draw[thick, black, dashed] (4,2) arc(90:270:1 and 2);
    \draw[thick, black] (0,-2) -- (4,-2);
    \draw[thick, black] (0,+2) -- (4,+2);

    \draw[thick, red] (0,-2) arc(-90:-315:1 and 2);
    \draw[thick, red] (0,-2) arc(-90:-45:1 and 2);
    \draw[thick, red] (4,-2) arc(-90:-45:1 and 2);
    \draw[thick, red] (4,2) arc(90:45:1 and 2);
    \draw[thick, red, dashed] (4,2) arc(90:270:1 and 2);
    \draw[thick, red, ->-=0.5] (0.707,-1.414) -- (2,-0.586);
    \draw[thick, red, ->-=0.5] (4.707,-1.414) -- (2,-0.586);
    \draw[thick, red, -<-=0.5] (0.707,1.414) -- (2,0.586);
    \draw[thick, red, -<-=0.5] (4.707,1.414) -- (2,0.586);
    \draw[thick, red, ->-=0.5] (2,-0.586) arc(-18:18:1 and 2);
    \end{tikzpicture} \quad = \quad \begin{tikzpicture}[baseline={([yshift=-1ex]current bounding box.center)},vertex/.style={anchor=base,
    circle,fill=black!25,minimum size=18pt,inner sep=2pt},scale=0.4]
    \draw[thick, black] (0,-2) arc(-90:270:1 and 2);
    \draw[thick, black] (4,-2) arc(-90:90:1 and 2);
    \draw[thick, black, dashed] (4,2) arc(90:270:1 and 2);
    \draw[thick, black] (0,-2) -- (4,-2);
    \draw[thick, black] (0,+2) -- (4,+2);
    \draw[red, thick, ->-=0.6] (2,-2) arc(-90:-30:0.8 and 2);
    \draw[red, thick, -<-=0.6] (2,2) arc(90:30:0.8 and 2);
    \draw[red, thick, dashed] (2,2) arc(90:270:0.8 and 2);
    
    \draw[red, thick, ->-=0.6] (2.69,-1) arc(-90:90:0.3 and 1);
    \draw[red, thick, ->-=0.6] (2.69,-1) arc(-90:-270:0.3 and 1);
    
    \end{tikzpicture} \quad = \quad \begin{tikzpicture}[baseline={([yshift=-1ex]current bounding box.center)},vertex/.style={anchor=base,
    circle,fill=black!25,minimum size=18pt,inner sep=2pt},scale=0.4]
    \draw[thick, black] (0,-2) arc(-90:270:1 and 2);
    \draw[thick, black] (4,-2) arc(-90:90:1 and 2);
    \draw[thick, black, dashed] (4,2) arc(90:270:1 and 2);
    \draw[thick, black] (0,-2) -- (4,-2);
    \draw[thick, black] (0,+2) -- (4,+2);
    \draw[red, thick, ->-=0.5] (2,-2) arc(-90:90:0.8 and 2);
    \draw[red, thick, dashed] (2,2) arc(90:270:0.8 and 2);
    \end{tikzpicture} ~,
\end{equation}
where we have omitted the labels of $\mathcal{A}$ for simplicity. The first equality uses the Frobenius property of $\mathcal{A}$, the second equality is simply a topological move of the diagram, and the last equality follows from the $\Delta$-separable condition. Finally, because the algebra object $\displaystyle \mathcal{A} = \bigoplus_{\mathbf{u}\in [0]} L_{\mathbf{u}}$, we conclude that \eqref{eq:fcam} is indeed equivalent to \eqref{eq:fcrc}, thus producing the same fusion rules
\begin{equation}\label{eqn:FusionTwistDefects}
    \Sigma_A^{[0]} \times \Sigma_A^{[0]} = \sum_{\mathbf{u} \in [0]} L_{\mathbf{u}} \quad \Longleftrightarrow \quad \mathcal{A} \boxtimes_{\mathcal{A}} \mathcal{A} = \mathcal{A} ~.
\end{equation}
To summarize, while we can describe the twist defects in terms of the module objects over the condensable algebra, it is important to notice that the tensor product of the twist defects is actually captured by the $\mathcal{A}$-balanced tensor product instead of the usual tensor product in the fusion category.

We next describe local fusions of the twist defects. To do so, recall that the global fusion rules of two simple objects $a,b$ in a fusion category,
\begin{equation}
    a \times b = \sum N_{ab}^c c ~,
\end{equation}
are related to local fusions via the first equation of (\ref{eq:basisconventions}). 
Given that the twist defects are described by a $G$-crossed braided tensor category (which is a fusion category with additional structures), this relation should hold for twist defects as well.

To see how this appears in the language of the algebras and modules, let $\mathcal{M}_L$ be an $\mathcal{A}_L-\mathcal{A}_I$ bimodule and ${\mathcal{M}}_R$ be an $\mathcal{A}_I-\mathcal{A}_R$ bimodule. The balanced tensor product leads to the global fusion rule
\begin{equation}
    \mathcal{M}_L \boxtimes_{\mathcal{A}_I} {\mathcal{M}}_R = \sum_{i} n_i \mathcal{N}_i ~, \quad m_i \in \mathbb{Z}_{>0}
\end{equation}
where $\mathcal{N}_i$ are $\mathcal{A}_L-\mathcal{A}_R$ bimodules. Our previous discussion indicates that the correct global fusion is captured by the following diagram and relates the local fusion junctions in the following way,
\begin{equation}
    \begin{tikzpicture}[baseline={([yshift=-1ex]current bounding box.center)},vertex/.style={anchor=base,
    circle,fill=black!25,minimum size=18pt,inner sep=2pt},scale=0.7]
    \shade[top color=orange!40, bottom color=orange!10]  (0,-1) -- (-2,-1) -- (-2,2) -- (0,2)-- (0,-1);
    \shade[top color=blue!40, bottom color=blue!10]  (0,-1) -- (2,-1) -- (2,2) -- (0,2)-- (0,-1);
    \shade[top color=red!40, bottom color=red!10]  (2,-1) -- (4,-1) -- (4,2) -- (2,2)-- (2,-1);
    \draw[dgreen, line width = 0.4mm, ->-=0.5] (0,-1) -- (0,0.5); 
    \draw[dgreen, line width = 0.4mm, ->-=0.5] (0,0.5) -- (0,2); 

    \draw[dgreen, line width = 0.4mm, ->-=0.5] (2,-1) -- (2,0.5); 
    \draw[dgreen, line width = 0.4mm, ->-=0.5] (2,0.5) -- (2,2);

    \node[below, dgreen] at (0,-1) {\footnotesize $\mathcal{M}_L$};
    \node[below, dgreen] at (2,-1) {\footnotesize ${\mathcal{M}}_R$};
    \node[above, dgreen] at (0,2) {\footnotesize $\mathcal{M}_L$};
    \node[above, dgreen] at (2,2) {\footnotesize ${\mathcal{M}}_R$};

    \draw[red, line width = 0.4mm, ->-=0.5] (2,0.5) -- (0,0.5);
    \node[below, red] at (1,0.5) {\footnotesize $\mathcal{A}_I$};
    \end{tikzpicture} = \sum_{\mathcal{N}_i}\sum_{\alpha = 1}^{n_i} \sqrt{\frac{d_{\mathcal{N}_i}}{d_{\mathcal{M}_L}d_{{\mathcal{M}}_R}}}\,\,\,\begin{tikzpicture}[baseline={([yshift=-1ex]current bounding box.center)},vertex/.style={anchor=base,
    circle,fill=black!25,minimum size=18pt,inner sep=2pt},scale=0.7]
    \shade[top color=orange!40, bottom color=orange!10]  (0,-1) -- (-2,-1) -- (-2,2) -- (0,2) -- (1,1) -- (1,0) -- (0,-1);
    \shade[top color=blue!40, bottom color=blue!10]  (0,-1) -- (1,0) -- (2,-1) -- (0,-1);
    \shade[top color=blue!40, bottom color=blue!10]  (0,2) -- (1,1) -- (2,2) -- (0,2);
    \shade[top color=red!40, bottom color=red!10]  (2,-1) -- (4,-1) -- (4,2) -- (2,2) --  (1,1) -- (1,0) -- (2,-1);

    \draw[dgreen, line width = 0.4mm, ->-=0.5] (0,-1) -- (1,0); 
    \draw[dgreen, line width = 0.4mm, ->-=0.5] (1,1) -- (0,2); 
    \draw[dgreen, line width = 0.4mm, ->-=0.5] (2,-1) -- (1,0); 
    \draw[dgreen, line width = 0.4mm, ->-=0.5] (1,1) -- (2,2);
    \draw[dgreen, line width = 0.4mm, ->-=0.5] (1,0) -- (1,1);

    \node[below, dgreen] at (0,-1) {\footnotesize $\mathcal{M}_L$};
    \node[below, dgreen] at (2,-1) {\footnotesize ${\mathcal{M}}_R$};
    \node[above, dgreen] at (0,2) {\footnotesize $\mathcal{M}_L$};
    \node[above, dgreen] at (2,2) {\footnotesize ${\mathcal{M}}_R$};

    \node[right, dgreen] at (1,0.5) {\footnotesize $\mathcal{N}_i$};

    \filldraw[violet!60] (1,0) circle (2pt);
    \filldraw[violet!60] (1,1) circle (2pt);

    \node[below, violet!60] at (1,0) {\footnotesize $\alpha$};
    \node[above, violet!60] at (1,1) {\footnotesize $\overline\alpha$};
    \end{tikzpicture} \quad.
\end{equation}
Notice that for the left diagram, we have inserted an algebra object $\mathcal{A}_I$ stretching between the two bimodules. This is sometimes referred to as a projector in the literature, since  stretching two $\mathcal{A}_I$ between the bimodules can be simplified into a single $\mathcal{A}_I$ using the bimodule conditions,
\begin{equation}
    \begin{tikzpicture}[baseline={([yshift=-1ex]current bounding box.center)},vertex/.style={anchor=base,
    circle,fill=black!25,minimum size=18pt,inner sep=2pt},scale=0.7]
    \shade[top color=orange!40, bottom color=orange!10]  (0,-1) -- (-2,-1) -- (-2,2) -- (0,2)-- (0,-1);
    \shade[top color=blue!40, bottom color=blue!10]  (0,-1) -- (2,-1) -- (2,2) -- (0,2)-- (0,-1);
    \shade[top color=red!40, bottom color=red!10]  (2,-1) -- (4,-1) -- (4,2) -- (2,2)-- (2,-1);
    \draw[dgreen, line width = 0.4mm, ->-=0.5] (0,-1) -- (0,0); 
    \draw[dgreen, line width = 0.4mm, ->-=0.5] (0,0) -- (0,1); 
    \draw[dgreen, line width = 0.4mm, ->-=0.5] (0,1) -- (0,2); 

    \draw[dgreen, line width = 0.4mm, ->-=0.5] (2,-1) -- (2,0); 
    \draw[dgreen, line width = 0.4mm, ->-=0.5] (2,0) -- (2,1); 
    \draw[dgreen, line width = 0.4mm, ->-=0.5] (2,1) -- (2,2);

    \node[below, dgreen] at (0,-1) {\footnotesize $\mathcal{M}_L$};
    \node[below, dgreen] at (2,-1) {\footnotesize ${\mathcal{M}}_R$};
    \node[left, dgreen] at (0,0.5) {\footnotesize $\mathcal{M}_L$};
    \node[right, dgreen] at (2,0.5) {\footnotesize ${\mathcal{M}}_R$};
    \node[above, dgreen] at (0,2) {\footnotesize $\mathcal{M}_L$};
    \node[above, dgreen] at (2,2) {\footnotesize ${\mathcal{M}}_R$};

    \draw[red, line width = 0.4mm, ->-=0.5] (2,1) -- (0,1);
    \draw[red, line width = 0.4mm, ->-=0.5] (2,0) -- (0,0);
    \node[below, red] at (1,0) {\footnotesize $\mathcal{A}_I$};
    \node[below, red] at (1,1) {\footnotesize $\mathcal{A}_I$};
    \end{tikzpicture} \hspace{0.2 in} = \hspace{0.2 in} \begin{tikzpicture}[baseline={([yshift=-1ex]current bounding box.center)},vertex/.style={anchor=base,
    circle,fill=black!25,minimum size=18pt,inner sep=2pt},scale=0.7]
    \shade[top color=orange!40, bottom color=orange!10]  (0,-1) -- (-2,-1) -- (-2,2) -- (0,2)-- (0,-1);
    \shade[top color=blue!40, bottom color=blue!10]  (0,-1) -- (2,-1) -- (2,2) -- (0,2)-- (0,-1);
    \shade[top color=red!40, bottom color=red!10]  (2,-1) -- (4,-1) -- (4,2) -- (2,2)-- (2,-1);
    \draw[dgreen, line width = 0.4mm, ->-=0.5] (0,-1) -- (0,0.5); 
    \draw[dgreen, line width = 0.4mm, ->-=0.5] (0,0.5) -- (0,2); 

    \draw[dgreen, line width = 0.4mm, ->-=0.5] (2,-1) -- (2,0.5); 
    \draw[dgreen, line width = 0.4mm, ->-=0.5] (2,0.5) -- (2,2);

    \node[below, dgreen] at (0,-1) {\footnotesize $\mathcal{M}_L$};
    \node[below, dgreen] at (2,-1) {\footnotesize ${\mathcal{M}}_R$};
    \node[above, dgreen] at (0,2) {\footnotesize $\mathcal{M}_L$};
    \node[above, dgreen] at (2,2) {\footnotesize ${\mathcal{M}}_R$};

    \draw[red, line width = 0.4mm, ->-=0.5] (2,0.5) -- (0,0.5);
    \node[below, red] at (1,0.5) {\footnotesize $\mathcal{A}_I$};
    \end{tikzpicture} \quad,
\end{equation}
which is reminiscent of the algebra of a projector.

The local fusion junctions are elements in $\mathrm{Hom}(\mathcal{M}_L\boxtimes {\mathcal{M}}_R,\mathcal{N}_i)$ satisfying the following three conditions,
\bea
\label{eq:threemoduleconditions}
\begin{tikzpicture}[baseline={([yshift=-.5ex]current bounding box.center)},vertex/.style={anchor=base,
    circle,fill=black!25,minimum size=18pt,inner sep=2pt},scale=0.8]
    
     \shade[top color=orange!40, bottom color=orange!10]  (-1.5,-1.5)  -- (-1.5,1.5) --(1.5,1.5)--(-0.5,-0.5)--(-0.5,-0.5)-- (0.5,-1.5)--(-1.5,-1.5);
  \shade[top color=red!30, bottom color=red!10]  (2.5,-1.5) -- (0.5,0.5)  --  (1.5,1.5)  -- (2.5,1.5)-- (2.5,-1.5);
    \shade[top color=blue!40, bottom color=blue!10]  (0.5,-1.5) -- (-0.5,-0.5)--(0.5,0.5) -- (2.5,-1.5) -- (0.5,-1.5) ;
    

\draw[red, thick, ->-=.5] (-1.5,-1.5) -- (-0.5,-0.5);
\draw[dgreen, thick, ->-=.5] (-0.5,-0.5) -- (0.5,0.5);
\draw[dgreen, thick, ->-=.5] (0.5,0.5) -- (1.5,1.5);
\draw[dgreen, thick, ->-=.5] (0.5,-1.5) -- (-0.5,-0.5);
\draw[dgreen, thick, ->-=.5] (2.5,-1.5) -- (0.5,0.5);

\node[red, below] at (-1.5,-1.5) {\scriptsize $\cA_L$};
\node[dgreen, below] at (0.5,-1.5) {\scriptsize $\cM_L$};
\node[dgreen, below] at (2.5,-1.5) {\scriptsize $ \cM_R$};
\node[dgreen, above] at (1.5,1.5) {\scriptsize $\cN_i$};
\node[red,left] at (-0.5,-0.5) {\scriptsize$\mu_L^{A_L-A_I}$};
\filldraw[red] (-0.5,-0.5) circle (1.5pt);
\filldraw[violet!60] (0.5,0.5) circle (2pt);
\node[left,violet!60] at (0.5,0.5) {\scriptsize$\alpha$};

\end{tikzpicture} &=& \begin{tikzpicture}[baseline={([yshift=-.5ex]current bounding box.center)},vertex/.style={anchor=base,circle,fill=black!25,minimum size=18pt,inner sep=2pt},scale=0.8]

     \shade[top color=orange!40, bottom color=orange!10]  (-1.5,-1.5)  -- (-1.5,1.5) --(1.5,1.5)--(0.5,0.5)--(1.5,-0.5)--  (0.5,-1.5)--(-1.5,-1.5);
  \shade[top color=red!30, bottom color=red!10]  (2.5,-1.5) -- (0.5,0.5)  --  (1.5,1.5)  -- (2.5,1.5)-- (2.5,-1.5);
    \shade[top color=blue!40, bottom color=blue!10]  (0.5,-1.5) -- (1.5,-0.5)-- (2.5,-1.5)  --  (0.5,-1.5) ;
    

\draw[red, thick, ->-=.5] (-1.5,-1.5) -- (0.5,0.5);
\draw[dgreen, thick, ->-=.5] (0.5,0.5) -- (1.5,1.5);
\draw[dgreen, thick, ->-=.5] (0.5,-1.5) -- (1.5,-0.5);
\draw[dgreen, thick, ->-=.5] (2.5,-1.5) -- (1.5,-0.5);
\draw[dgreen, thick, ->-=.5] (1.5,-0.5) -- (0.5,0.5);

\node[red, below] at (-1.5,-1.5) {\scriptsize $\cA_L$};
\node[dgreen, below] at (0.5,-1.5) {\scriptsize $\cM_L$};
\node[dgreen, below] at (2.5,-1.5) {\scriptsize $ \cM_R$};
\node[dgreen, above] at (1.5,1.5) {\scriptsize $\cN_i$};
\node[red,left] at (0.5,0.5) {\scriptsize$\mu_L^{A_L-A_R}$};
\filldraw[red] (0.5,0.5) circle (1.5pt);
\filldraw[violet!60] (1.5,-0.5) circle (2pt);
\node[right,violet!60] at (1.5,-0.5) {\scriptsize$\alpha$};

\end{tikzpicture} ~, 
\no\\
\begin{tikzpicture}[baseline={([yshift=-.5ex]current bounding box.center)},vertex/.style={anchor=base,
    circle,fill=black!25,minimum size=18pt,inner sep=2pt},scale=0.8]
    
     \shade[top color=orange!40, bottom color=orange!10]  (-1.5,-1.5)  -- (-1.5,1.5) --(1.5,1.5)--(-1.5,-1.5) ;
  \shade[top color=red!30, bottom color=red!10]  (2.5,-1.5) -- (0.5,0.5)  --  (1.5,1.5)  -- (2.5,1.5)-- (2.5,-1.5);
    \shade[top color=blue!40, bottom color=blue!10]  (-1.5,-1.5)  -- (-0.5,-0.5)--(0.5,0.5) -- (2.5,-1.5) -- (-1.5,-1.5)  ;
    

\draw[dgreen, thick, ->-=.5] (-1.5,-1.5) -- (-0.5,-0.5);
\draw[dgreen, thick, ->-=.5] (-0.5,-0.5) -- (0.5,0.5);
\draw[dgreen, thick, ->-=.5] (0.5,0.5) -- (1.5,1.5);
\draw[red, thick, ->-=.5] (0.5,-1.5) -- (-0.5,-0.5);
\draw[dgreen, thick, ->-=.5] (2.5,-1.5) -- (0.5,0.5);

\node[red, below] at (0.5,-1.5) {\scriptsize $\cA_I$};
\node[dgreen, below] at (-1.5,-1.5)  {\scriptsize $\cM_L$};
\node[dgreen, below] at (2.5,-1.5) {\scriptsize $ \cM_R$};
\node[dgreen, above] at (1.5,1.5) {\scriptsize $\cN_i$};
\node[red,left] at (-0.5,-0.5) {\scriptsize$\mu_R^{A_L-A_I}$};
\filldraw[red] (-0.5,-0.5) circle (1.5pt);
\filldraw[violet!60] (0.5,0.5) circle (2pt);
\node[left,violet!60] at (0.5,0.5) {\scriptsize$\alpha$};

\end{tikzpicture} &=& \begin{tikzpicture}[baseline={([yshift=-.5ex]current bounding box.center)},vertex/.style={anchor=base,circle,fill=black!25,minimum size=18pt,inner sep=2pt},scale=0.8]

   \shade[top color=orange!40, bottom color=orange!10]  (-1.5,-1.5)  -- (-1.5,1.5) --(1.5,1.5)--(-1.5,-1.5) ;
  \shade[top color=red!30, bottom color=red!10]  (2.5,-1.5) -- (0.5,0.5)  --  (1.5,1.5)  -- (2.5,1.5)-- (2.5,-1.5);
    \shade[top color=blue!40, bottom color=blue!10]  (-1.5,-1.5)  -- (-0.5,-0.5)--(0.5,0.5) -- (2.5,-1.5) -- (-1.5,-1.5)  ;
    

\draw[dgreen, thick, ->-=.5] (-1.5,-1.5) -- (0.5,0.5);
\draw[dgreen, thick, ->-=.5] (0.5,0.5) -- (1.5,1.5);
\draw[red, thick, ->-=.5] (0.5,-1.5) -- (1.5,-0.5);
\draw[dgreen, thick, ->-=.5] (2.5,-1.5) -- (1.5,-0.5);
\draw[dgreen, thick, ->-=.5] (1.5,-0.5) -- (0.5,0.5);

\node[dgreen, below] at (-1.5,-1.5) {\scriptsize $\cM_L$};
\node[red, below] at (0.5,-1.5)  {\scriptsize $\cA_L$}; 
\node[dgreen, below] at (2.5,-1.5) {\scriptsize $ \cM_R$};
\node[dgreen, above] at (1.5,1.5) {\scriptsize $\cN_i$};
\node[red,right] at (1.5,-0.3) {\scriptsize$\mu_L^{A_I-A_R}$};
\filldraw[red] (1.5,-0.5)  circle (1.5pt);
\filldraw[violet!60] (0.5,0.5)  circle (2pt);
\node[left,violet!60] at (0.5,0.5) {\scriptsize$\alpha$};
\end{tikzpicture} ~, 
\eea
\bea
\begin{tikzpicture}[baseline={([yshift=-.5ex]current bounding box.center)},vertex/.style={anchor=base,
    circle,fill=black!25,minimum size=18pt,inner sep=2pt},scale=0.8]
    
     \shade[top color=orange!40, bottom color=orange!10]  (-1.5,-1.5)  -- (-1.5,1.5) --(1.5,1.5)--(-1.5,-1.5) ;
  \shade[top color=red!30, bottom color=red!10]  (2.5,-1.5) --(0.5,-1.5) --  (-0.5,-0.5)  --  (1.5,1.5)  -- (2.5,1.5)-- (2.5,-1.5);
    \shade[top color=blue!40, bottom color=blue!10]  (-1.5,-1.5)  -- (-0.5,-0.5)--(0.5,-1.5) -- (-1.5,-1.5)  ;
    

\draw[dgreen, thick, ->-=.5] (-1.5,-1.5) -- (-0.5,-0.5);
\draw[dgreen, thick, ->-=.5] (-0.5,-0.5) -- (0.5,0.5);
\draw[dgreen, thick, ->-=.5] (0.5,0.5) -- (1.5,1.5);
\draw[dgreen, thick, ->-=.5] (0.5,-1.5) -- (-0.5,-0.5);
\draw[red, thick, ->-=.5] (2.5,-1.5) -- (0.5,0.5);

\node[red, below] at (2.5,-1.5) {\scriptsize $\cA_I$};
\node[dgreen, below] at (-1.5,-1.5)  {\scriptsize $\cM_L$};
\node[dgreen, below] at (0.5,-1.5) {\scriptsize $ \cM_R$};
\node[dgreen, above] at (1.5,1.5) {\scriptsize $\cN_i$};
\node[red,left] at(0.5,0.5) {\scriptsize$\mu_R^{A_L-A_R}$};
\filldraw[violet!60] (-0.5,-0.5) circle (2pt);
\filldraw[red] (0.5,0.5) circle (1.5pt);
\node[left,violet!60] at (-0.5,-0.5) {\scriptsize$\alpha$};

\end{tikzpicture} &=& \begin{tikzpicture}[baseline={([yshift=-.5ex]current bounding box.center)},vertex/.style={anchor=base,circle,fill=black!25,minimum size=18pt,inner sep=2pt},scale=0.8]

   \shade[top color=orange!40, bottom color=orange!10]  (-1.5,-1.5)  -- (-1.5,1.5) --(1.5,1.5)--(-1.5,-1.5) ;
  \shade[top color=red!30, bottom color=red!10]  (2.5,-1.5) --(-1.5,-1.5) -- (1.5,-0.5)-- (0.5,0.5)  --  (1.5,1.5)  -- (2.5,1.5)-- (2.5,-1.5);
    \shade[top color=blue!40, bottom color=blue!10]  (-1.5,-1.5)  -- (0.5,0.5)--(1.5,-0.5) -- (0.5,-1.5) --  (-1.5,-1.5) ;
    

\draw[dgreen, thick, ->-=.5] (-1.5,-1.5) -- (0.5,0.5);
\draw[dgreen, thick, ->-=.5] (0.5,0.5) -- (1.5,1.5);
\draw[dgreen, thick, ->-=.5] (0.5,-1.5) -- (1.5,-0.5);
\draw[red, thick, ->-=.5] (2.5,-1.5) -- (1.5,-0.5);
\draw[dgreen, thick, ->-=.5] (1.5,-0.5) -- (0.5,0.5);

\node[dgreen, below] at (-1.5,-1.5)  {\scriptsize $\cM_L$};
\node[red, below] at  (2.5,-1.5) {\scriptsize $\cA_L$}; 
\node[dgreen, below] at   (0.5,-1.5){\scriptsize $ \cM_R$};
\node[dgreen, above] at (1.5,1.5) {\scriptsize $\cN_i$};
\node[red,right] at (1.5,-0.3) {\scriptsize$\mu_R^{A_I-A_R}$};
\filldraw[red] (1.5,-0.5)  circle (1.5pt);
\filldraw[violet!60] (0.5,0.5)  circle (2pt);
\node[left,violet!60] at (0.5,0.5) {\scriptsize$\alpha$};
\end{tikzpicture} ~. 
\eea
Notice that the second equality implements the balanced tensor product.

\section{Boundary $F$-symbols from bulk $F$-symbols}
\label{sec:bulktoboundaryFsymb}

Having described twist defects in detail, we may finally return to our original goal of determining the boundary $F$-symbols from the 
bulk $F$-symbols. To do so, we first use the machinery developed in the previous section to compute the relevant bulk $F$-symbols. 

\subsection{Bulk $F$-symbols}\label{app:derivationofbulkFsymbols1}
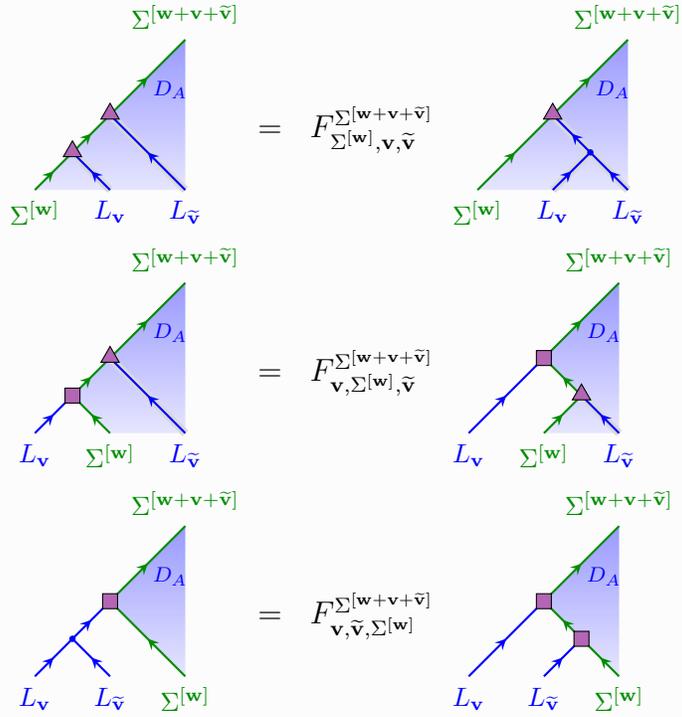
\begin{figure}[tbp]
\begin{center}
    \begin{tikzpicture}[baseline=20,square/.style={regular polygon,regular polygon sides=4},scale=1]
        \shade[top color=blue!40, bottom color=blue!10]  (0,0)--(2,0)--(2,2)--(0,0);
        
        \draw[gray!30, line width=3pt] (1,0) -- (0.5,0.5); 
         \draw[gray!30, line width=3pt] (2,0) -- (1,1); 
         
        \draw[blue,thick, ->-=0.5](1,0) -- (0.5,0.5); 
        \draw[dgreen,thick, ->-=0.5](0,0) -- (0.5,0.5); 
        \draw[dgreen,thick, ->-=0.5] (0.5,0.5)--(1,1); 
        \draw[dgreen,thick, ->-=0.5] (1,1)--(2,2); 
        \draw[blue,thick, ->-=0.5](2,0) -- (1,1); 
        \node[blue,below] at (1.8,1.6) {\scriptsize{$D_{A}$}};
        \node[isosceles triangle,scale=0.4,isosceles triangle apex angle=60,draw,fill=violet!60,rotate=90,minimum size =0.01cm] at (0.5,0.5){};
        \node[isosceles triangle,scale=0.4,isosceles triangle apex angle=60,draw,fill=violet!60,rotate=90,minimum size =0.01cm] at (1,1){};
        \node[dgreen, below] at (0,0) {\footnotesize{$\Sigma^{[\bw]}$}};
        \node[dgreen, above] at (2,2) {\footnotesize{$\Sigma^{[\bw+\bv+\widetilde{\bv}]}$}};
        \node[blue, below] at (1,0) {\footnotesize{$L_\bv$}};
        \node[blue, below] at (2,0) {\footnotesize{$L_{\widetilde{\bv}}$}};
\end{tikzpicture} $=\hspace{0.1 in} F_{\Sigma^{[\bw]}, \bv, \widetilde{\bv}}^{\Sigma^{[\bw+\bv+\widetilde{\bv}]}} $
    \begin{tikzpicture}[baseline=20,square/.style={regular polygon,regular polygon sides=4},scale=1]
        \shade[top color=blue!40, bottom color=blue!10]  (0,0)--(2,0)--(2,2)--(0,0);
        \draw[dgreen,thick, ->-=0.5](0,0) -- (1,1); 
        \draw[dgreen,thick, ->-=0.5](1,1) -- (2,2); 
        
         \draw[gray!30, line width=3pt] (1,0) -- (1.5,0.5); 
          \draw[gray!30, line width=3pt] (2,0) -- (1.5,0.5);
           \draw[gray!30, line width=3pt] (1.5,0.5) -- (1,1); 
           
        \draw[blue,thick, ->-=0.5](1,0) -- (1.5,0.5); 
        \draw[blue,thick, ->-=0.5](2,0) -- (1.5,0.5); 
        \draw[blue,thick, ->-=0.5] (1.5,0.5) -- (1,1); 
        \node[blue,below] at (1.8,1.6) {\scriptsize{$D_{A}$}};
        \node[isosceles triangle,scale=0.4,isosceles triangle apex angle=60,draw,fill=violet!60,rotate=90,minimum size =0.01cm] at (1,1){};
        \filldraw[blue] (1.5,0.5) circle (1pt);
        \node[dgreen, below] at (0,0) {\footnotesize{$\Sigma^{[\bw]}$}};
        \node[dgreen, above] at (2,2) {\footnotesize{$\Sigma^{[\bw+\bv+\widetilde{\bv}]}$}};
        \node[blue, below] at (1,0) {\footnotesize{$L_\bv$}};
        \node[blue, below] at (2,0) {\footnotesize{$L_{\widetilde{\bv}}$}};
\end{tikzpicture} \\
 \begin{tikzpicture}[baseline=20,square/.style={regular polygon,regular polygon sides=4},scale=1]
        \shade[top color=blue!40, bottom color=blue!10]  (1,0)--(2,0)--(2,2)--(0.5,0.5)--(1,0);
        
         \draw[gray!30, line width=3pt] (2,0) -- (1,1); 
         
        \draw[blue,thick, ->-=0.5](0,0) -- (0.5,0.5); 
        \draw[dgreen,thick, ->-=0.5](1,0) -- (0.5,0.5); 
        \draw[dgreen,thick, ->-=0.5] (0.5,0.5)--(1,1); 
        \draw[dgreen,thick, ->-=0.5] (1,1)--(2,2); 
        \draw[blue,thick, ->-=0.5](2,0) -- (1,1); 
        \node[blue,below] at (1.8,1.6) {\scriptsize{$D_{A}$}};

   \node at (0.5,0.5) [square,draw,fill=violet!60,scale=0.5] {}; 

\node[isosceles triangle,scale=0.4,
    isosceles triangle apex angle=60,
    draw,fill=violet!60,
    rotate=90,
    minimum size =0.01cm] at (1,1){};

\node[dgreen, below] at (1,0) {\footnotesize{$\Sigma^{[\bw]}$}};
\node[dgreen, above] at (2,2) {\footnotesize{$\Sigma^{[\bw+\bv+\widetilde{\bv}]}$}};
\node[blue, below] at (0,0) {\footnotesize{$L_\bv$}};
\node[blue, below] at (2,0) {\footnotesize{$L_{\widetilde{\bv}}$}};

\end{tikzpicture} $=\hspace{0.1 in} F^{\Sigma^{[\bw+\bv+\widetilde{\bv}]}}_{\bv,\Sigma^{[\bw]} ,\widetilde{\bv}}$ \begin{tikzpicture}[baseline=20,square/.style={regular polygon,regular polygon sides=4},scale=1]
        \shade[top color=blue!40, bottom color=blue!10]  (1,0)--(2,0)--(2,2)--(1,1)--(1.5,0.5)--(1,0);
           \draw[gray!30, line width=3pt] (2,0) -- (1.5,0.5); 
           
        \draw[blue,thick, ->-=0.5](0,0) -- (1,1); 
        \draw[dgreen,thick, ->-=0.5](1,0) -- (1.5,0.5); 
        \draw[dgreen,thick, ->-=0.5] (1.5,0.5)--(1,1); 
        \draw[dgreen,thick, ->-=0.5] (1,1)--(2,2); 
        \draw[blue,thick, ->-=0.5](2,0) -- (1.5,0.5); 
        \node[blue,below] at (1.8,1.6) {\scriptsize{$D_{A}$}};
        \node at (1,1) [square,draw,fill=violet!60,scale=0.5] {}; 
        \node[isosceles triangle,scale=0.4,isosceles triangle apex angle=60,draw,fill=violet!60,rotate=90,minimum size =0.01cm] at (1.5,0.5){};
        \node[dgreen, below] at (1,0) {\footnotesize{$\Sigma^{[\bw]}$}};
        \node[dgreen, above] at (2,2) {\footnotesize{$\Sigma^{[\bw+\bv+\widetilde{\bv}]}$}};
        \node[blue, below] at (0,0) {\footnotesize{$L_\bv$}};
        \node[blue, below] at (2,0) {\footnotesize{$L_{\widetilde{\bv}}$}};
    \end{tikzpicture} 
    \\
   \begin{tikzpicture}[baseline=20,square/.style={regular polygon,regular polygon sides=4},scale=1]
        \shade[top color=blue!40, bottom color=blue!10]  (2,0)--(1,1)--(2,2)--(2,0);
        \draw[blue,thick, ->-=0.5](0,0) -- (0.5,0.5); 
        \draw[blue,thick, ->-=0.5](1,0) -- (0.5,0.5);
        \draw[blue,thick, ->-=0.5] (0.5,0.5) -- (1,1);
        \draw[dgreen,thick, ->-=0.5] (2,0)--(1,1); 
        \draw[dgreen,thick, ->-=0.5] (1,1)--(2,2); 
        \node[blue,below] at (1.8,1.6) {\scriptsize{$D_{A}$}};
        \node at (1,1) [square,draw,fill=violet!60,scale=0.5] {}; 
        \filldraw[blue] (0.5,0.5) circle (1pt);
        \node[dgreen, below] at (2,0) {\footnotesize{$\Sigma^{[\bw]}$}};
        \node[dgreen, above] at (2,2) {\footnotesize{$\Sigma^{[\bw+\bv+\widetilde{\bv}]}$}};
        \node[blue, below] at (0,0) {\footnotesize{$L_\bv$}};
        \node[blue, below] at (1,0) {\footnotesize{$L_{\widetilde{\bv}}$}};
    \end{tikzpicture} $=\hspace{0.1 in} F^{\Sigma^{[\bw+\bv+\widetilde{\bv}]}}_{\bv ,\widetilde{\bv},\Sigma^{[\bw]}}$ \begin{tikzpicture}[baseline=20,square/.style={regular polygon,regular polygon sides=4},scale=1]
        \shade[top color=blue!40, bottom color=blue!10]  (2,0)--(1,1)--(2,2)--(2,0);
        \draw[blue,thick, ->-=0.5](0,0) -- (1,1); 
        \draw[blue,thick, ->-=0.5](1,0) -- (1.5,0.5);
        \draw[dgreen,thick, ->-=0.5] (2,0)--(1.5,0.5); 
        \draw[dgreen,thick, ->-=0.5] (1.5,0.5)--(1,1); 
        \draw[dgreen,thick, ->-=0.5] (1,1)--(2,2); 
        \node[blue,below] at (1.8,1.6) {\scriptsize{$D_{A}$}};
        \node at (1,1) [square,draw,fill=violet!60,scale=0.5] {}; 
        \node at (1.5,0.5) [square,draw,fill=violet!60,scale=0.5] {}; 
        \node[dgreen, below] at (2,0) {\footnotesize{$\Sigma^{[\bw]}$}};
        \node[dgreen, above] at (2,2) {\footnotesize{$\Sigma^{[\bw+\bv+\widetilde{\bv}]}$}};
        \node[blue, below] at (0,0) {\footnotesize{$L_\bv$}};
        \node[blue, below] at (1,0) {\footnotesize{$L_{\widetilde{\bv}}$}};
    \end{tikzpicture} 

\caption{Bulk $F$-symbols needed to determine boundary $F$-symbols with a single incoming $Q$. }
\label{fig:bulkFsymbols1}
\end{center}
\end{figure}

The bulk $F$-symbols relevant for us here are given in Figure \ref{fig:bulkFsymbols1}, and as we will now derive, they are given (up to gauge transformations) by 
\begin{equation}\label{eq:bulkFsymbols}
\begin{aligned}
    F^{\Sigma^{[\bw+\bv+\widetilde{\bv}]}}_{\Sigma^{[\bw]}, \bv, \widetilde{\bv}} & = \frac{R^{\bv,\Sigma^{[\bw]}}R^{\widetilde{\bv},\Sigma^{[\bw+\bv]}}}{R^{\bv + \widetilde{\bv},\Sigma^{[\bw]}}} \frac{\mu_A(\widetilde{\bv}+(\bv+\bw)_0-(\widetilde{\bv}+\bv+\bw)_0,\bv + \bw_0 - (\bv+\bw)_0)}{R^{\widetilde{\bv},\bv}} ~, \\
    F^{\Sigma^{[\bw+\bv+\widetilde{\bv}]}}_{\bv,\Sigma^{[\bw]} ,\widetilde{\bv}} & = \frac{R^{\widetilde{\bv},\Sigma^{[\bw+\bv]}}}{R^{\widetilde{\bv},\Sigma^{[\bw]}}}\frac{1}{R^{\widetilde{\bv},\bv}} \frac{\mu_A(\widetilde{\bv} + (\bv+\bw)_0 - (\bv+\widetilde{\bv}+\bw)_0,\bv + \bw_0 - (\bv+\bw)_0)}{\mu_A(\bv + (\widetilde{\bv}+\bw)_0 - (\bv+\widetilde{\bv}+\bw)_0,\widetilde{\bv} + \bw_0 - (\widetilde{\bv}+\bw)_0)} ~, \\
    F^{\Sigma^{[\bw+\bv+\widetilde{\bv}]}}_{\bv ,\widetilde{\bv},\Sigma^{[\bw]}} & = \frac{1}{\mu_A(\bv+(\widetilde{\bv}+\bw)_0 - (\bv+\widetilde{\bv}+\bw)_0,\widetilde{\bv}+\bw_0 - (\widetilde{\bv}+\bw)_0)} ~,
\end{aligned}
\end{equation}
where $(\bv + \bw)_0$ is an arbitrary choice of representative in the coset $[\bv + \bw]$. 

Note that in Figure \ref{fig:bulkFsymbols1}, we consider only twist defects, i.e.  boundaries of the symmetry operator $D_A$. Together with the fact that all bulk anyons are invertible, this implies that the little group is trivial, so that the label $\rho$ is trivial as well, and can be dropped. As such, we label the modules only by the right-$\mathcal{A}_A$ coset label $[\bw]$. To present the right-$A$ module structure, we consider fixing a choice of representative in each coset $[\bw]$, and denote it as $\bw_0$. This means that if $[\bw] = [\widetilde{\bw}]$, then $\bw_0 = \widetilde{\bw}_0$. After an explicit gauge choice, the junction expansion for the twist defect $\Sigma^{[\bw]}$ is given by
\begin{equation}
\label{eq:intermedref}
    \begin{tikzpicture}[baseline=0,scale=1.2]
        \shade[top color=blue!40, bottom color=blue!10]  (0,-1) -- (1.5,-1) -- (1.5,1) -- (0,1)-- (0,-1);
        \draw[dgreen,thick, ->-=0.5](0,-1) -- (0,0); 
        \draw[dgreen,thick, ->-=0.5](0,0) -- (0,1); 
        \draw[gray!30, line width=3pt] (0,0) to[out=0,in=90 ]  (1,-1);
        \draw[red, thick, -<-=0.5] (0,0) to[out=0,in=90 ]  (1,-1);
        \node[dgreen, above] at (0,1) {\footnotesize{$\Sigma^{[\bw]}$}};
        \node[dgreen, below] at (0,-1) {\footnotesize{$\Sigma^{[\bw]}$}};
        \node[red, below] at (1,-1) {\footnotesize{$\mathcal{A}_A$}};
        \filldraw[red] (0,0) circle (1.5pt);
        \node[blue,left] at (1.5,0.7) {\footnotesize{$D_{A}$}};
\end{tikzpicture} \hspace{0.1 in} = \hspace{0.1 in} \frac{1}{M^{r_A/2}} \sum_{\bu,\widetilde{\bu}\in\mathcal{A}_A} \mu_{A}(\bu,\widetilde{\bu}) \begin{tikzpicture}[baseline=0,scale=1.2]
      \draw[dgreen,thick,  decoration={markings, mark=at position 0.6 with {\arrow{stealth}}}, postaction={decorate}](0,-1) -- (0,-0.6); 
\draw[blue,thick,  decoration={markings, mark=at position 0.5 with {\arrow{stealth}}}, postaction={decorate}](0,-0.6) -- (0,0); 
\draw[dgreen,thick,  decoration={markings, mark=at position 0.7 with {\arrow{stealth}}}, postaction={decorate}](0,0.6) -- (0,1); 
\draw[blue,thick,  decoration={markings, mark=at position 0.5 with {\arrow{stealth}}}, postaction={decorate}](0,0) -- (0,0.6); 
\draw [blue, thick, -<-=0.3] (0,0) to[out=0,in=90 ]  (1,-1);
\draw [red, thick, -<-=0.5] (0.825,-0.433) to[out=-58,in=90 ]  (1,-1);

\node[dgreen, above] at (0,0.9) {\footnotesize{$\Sigma^{[\mathbf{w}]}$}};
\node[dgreen, below] at (0,-1) {\footnotesize{$\Sigma^{[\mathbf{w}]}$}};
\node[red, below] at (1,-1) {\footnotesize{$\cA_{A}$}};

\node[blue, left] at (0,0.35) {\footnotesize{$L_{\bw_0 + \bu+\widetilde \bu}$}};
\node[blue, left] at (0,-0.3) {\footnotesize{$L_{\bw_0+\bu}$}};
\node[blue, right] at (0.5,0) {\footnotesize{$L_{\widetilde\bu}$}};


\filldraw[blue] (0,0) circle (1pt);
\filldraw[blue] (0,-0.6) circle (1pt);
\filldraw[blue] (0,0.6) circle (1pt);
\filldraw[blue] (0.825,-0.433) circle (1pt);
\end{tikzpicture} ~.
\end{equation}
In order to obtain the $F$-symbols, we must first understand the following purple triangle and square junctions,
\begin{equation}
    \begin{tikzpicture}[baseline=0,scale=1.2,square/.style={regular polygon,regular polygon sides=4}]
        \shade[top color=blue!40, bottom color=blue!10]  (0,-1) -- (1.5,-1) -- (1.5,1) -- (0,1)-- (0,-1);
        \draw[dgreen,thick, ->- = 0.5](0,-1) -- (0,0); 
        \draw[dgreen,thick, ->- = 0.5](0,0) -- (0,1); 
        \draw [blue, thick, -<- = 0.5] (0,0) to[out=180,in=90 ]  (-1,-1);
        \node[dgreen, above] at (0,+1) {\footnotesize{$\Sigma^{[\bw+\bv]}$}};
        \node[dgreen, below] at (0,-1) {\footnotesize{$\Sigma^{[\bw]}$}};
        \node[blue, below] at (-1.15,-1) {\footnotesize{$L_\bv$}};
        \node at (0,0) [square,draw,fill=violet!60,scale=0.5] {}; 
        \node[blue,left] at (1.5,0.7) {\footnotesize{$D_{A}$}};
    \end{tikzpicture} ~~, \hspace{0.5 in} 
    \begin{tikzpicture}[baseline=0,scale=1.2]
        \shade[top color=blue!40, bottom color=blue!10]  (0,-1) -- (1.5,-1) -- (1.5,1) -- (0,1)-- (0,-1);
        \draw[dgreen,thick, ->-=0.5](0,-1) -- (0,0); 
        \draw[dgreen,thick, ->-=0.5](0,0) -- (0,1); 
        \draw [gray!30, line width=3pt] (0,0) to[out=0,in=90 ]  (1,-1);
        \draw [blue, thick, -<-=0.5] (0,0) to[out=0,in=90 ]  (1,-1);
        \node[dgreen, above] at (0,1) {\footnotesize{$\Sigma^{[\bw+\bv]}$}};
        \node[dgreen, below] at (0,-1) {\footnotesize{$\Sigma^{[\bw]}$}};
        \node[blue, below] at (1,-1) {\footnotesize{$L_\bv$}};
        \node[isosceles triangle,scale=0.4, isosceles triangle apex angle=60, draw,fill=violet!60, rotate=90, minimum size =0.01cm] at (0,0){};
        \node[blue,left] at (1.5,0.7) {\footnotesize{$D_{A}$}};
\end{tikzpicture} ~~.
\end{equation}

Notice that braiding relates the square junction to the triangle junction. In the current case, it is easier to compute the square junction first, 
since the anyon $L_{\bv}$ can be viewed as a bimodule of the trivial algebra object, and hence the square junction is nothing but the fusion junction of two bimodules, constrained by the diagram
\begin{equation}
    \begin{tikzpicture}[baseline=0,scale=1.2,square/.style={regular polygon,regular polygon sides=4}]
        \shade[top color=blue!40, bottom color=blue!10]  (0,-1) -- (1.5,-1) -- (1.5,1) -- (0,1)-- (0,-1);
        \draw[gray!30, line width=3pt]  (0,0.5) to[out=0,in=90] (1,-1);
        \draw[dgreen,thick, ->- = 0.5](0,-1) -- (0,0); 
        \draw[dgreen,thick, ->- = 0.75](0,0) -- (0,1); 
        \draw[blue, thick, -<- = 0.5] (0,0) to[out=180,in=90 ]  (-1,-1);
        \draw[red, thick, -<- = 0.5] (0,0.5) to[out=0,in=90] (1,-1);
          \filldraw[red] (0,0.5) circle (1.5pt);
        \node[dgreen, above] at (0,+1) {\footnotesize{$\Sigma^{[\bw+\bv]}$}};
        \node[dgreen, below] at (0,-1) {\footnotesize{$\Sigma^{[\bw]}$}};
        \node[blue, below] at (-1.15,-1) {\footnotesize{$L_\bv$}};
        \node at (0,0) [square,draw,fill=violet!60,scale=0.5] {}; 
        \node[blue, left] at (1.5,0.7) {\footnotesize{$D_{A}$}};
        \node[red, below] at (1,-1) {\footnotesize $\mathcal{A}_A$};
    \end{tikzpicture} \quad  = \quad \begin{tikzpicture}[baseline=0,scale=1.2,square/.style={regular polygon,regular polygon sides=4}]
        \shade[top color=blue!40, bottom color=blue!10]  (0,-1) -- (1.5,-1) -- (1.5,1) -- (0,1)-- (0,-1);
        \draw[gray!30, line width=3pt] (0,-0.5) to[out=0,in=90] (1,-1);
        \draw[dgreen,thick, ->- = 0.25](0,-1) -- (0,0); 
        \draw[dgreen,thick, ->- = 0.5](0,0) -- (0,1); 
        \draw[blue, thick, -<- = 0.5] (0,0) to[out=180,in=90]  (-1,-1);
        \draw[red, thick, -<- = 0.5] (0,-0.5) to[out=0,in=90] (1,-1);
          \filldraw[red] (0,-0.5) circle (1.5pt);
        \node[dgreen, above] at (0,+1) {\footnotesize{$\Sigma^{[\bw+\bv]}$}};
        \node[dgreen, below] at (0,-1) {\footnotesize{$\Sigma^{[\bw]}$}};
        \node[blue, below] at (-1.15,-1) {\footnotesize{$L_\bv$}};
        \node at (0,0) [square,draw,fill=violet!60,scale=0.5] {}; 
        \node[blue, left] at (1.5,0.7) {\footnotesize{$D_{A}$}};
        \node[red, below] at (1,-1) {\footnotesize $\mathcal{A}_A$};
    \end{tikzpicture} ~.
\end{equation}
Expanding the square junction as follows,\footnote{Note that the coefficient $\beta_{\bv, [\bw]}$ depends on the particular representative $\bw_0 \in [\bw]$ that we use to expand on the right-hand side, but for simplicity we will refrain from writing this label explicitly.}
\begin{equation}\label{eq:junc_vSw}
    \begin{tikzpicture}[baseline=0,scale=1.2,square/.style={regular polygon,regular polygon sides=4}]
        \shade[top color=blue!40, bottom color=blue!10]  (0,-1) -- (1.5,-1) -- (1.5,1) -- (0,1)-- (0,-1);
        \draw[dgreen,thick, ->- = 0.5](0,-1) -- (0,0); 
        \draw[dgreen,thick, ->- = 0.5](0,0) -- (0,1); 
        \draw [blue, thick, -<- = 0.5] (0,0) to[out=180,in=90 ]  (-1,-1);
        \node[dgreen, above] at (0,+1) {\footnotesize{$\Sigma^{[\bw+\bv]}$}};
        \node[dgreen, below] at (0,-1) {\footnotesize{$\Sigma^{[\bw]}$}};
        \node[blue, below] at (-1.15,-1) {\footnotesize{$L_\bv$}};
        \node at (0,0) [square,draw,fill=violet!60,scale=0.5] {}; 
        \node[blue,left] at (1.5,0.7) {\footnotesize{$D_{A}$}};
    \end{tikzpicture} \,\,\,=\,\,\,   \sum_{\bu \in \cA_{A}} \beta_{\bv,[\bw]}(\bu) \begin{tikzpicture}[baseline=0,scale=1.2]
        \draw[dgreen,thick, ->-=0.5](0,-1) -- (0,-0.6); 
        \draw[blue,thick, ->-=0.5](0,-0.6) -- (0,0); 
        \draw[blue,thick, ->-=0.5](0,0) -- (0,0.6); 
        \draw[dgreen,thick, ->-=0.5](0,0.6) -- (0,1); 
        \draw [blue, thick, -<-=0.5] (0,0) to[out=180,in=90 ]  (-1,-1);
        \node[blue, right] at (0,0.4) {\footnotesize{$L_{\bw_0 + \bv + \bu}$}};
        \node[blue, right] at (0, -0.4) {\footnotesize{$L_{\bw_0 + \bu}$}};
        \node[dgreen, above] at (0,1.0) {\footnotesize{$\Sigma^{[\bw+\bv]}$}};
        \node[blue, below] at (-1.15,-1) {\footnotesize{$L_\bv$}};
        \node[dgreen, below] at (0,-1) {\footnotesize{$\Sigma^{[\bw]}$}};
        \filldraw[blue] (0,0) circle (1pt);
        \filldraw[blue] (0,-0.6) circle (1pt);
        \filldraw[blue] (0,0.6) circle (1pt);
    \end{tikzpicture} ~,
\end{equation}  
the right-module condition leads to the following constraint,
\begin{equation}
    \beta_{\bv,[\bw]}(\bu+\widetilde{\bu}) \mu_{A}(\bu, \widetilde{\bu}) = \beta_{\bv,[\bw]}(\bu) \mu_{A}(\bv + \bw_0 - (\bv + \bw)_0 + \bu, \widetilde{\bu}) ~,
\end{equation}
which is solved by
\begin{equation}
    \beta_{\bv,[\bw]}(\bu) = \beta_{\bv,[\bw]}(0)\,  \mu_A(\bv + \bw_0 - (\bv+\bw)_0,\bu) ~.
\end{equation}
Notice that $\bv + \bw_0 - (\bv+\bw)_0 \in \mathcal{A}_A= \mathrm{im}_M(1-A)$, and we can view the $\beta_{\bv,[\bw]}(0)$ as the overall normalization of the junction, which we will take to be $1$ for convenience. As a result, we find
\begin{equation}\label{eq:uSs1}
    \begin{tikzpicture}[baseline=0,scale=1.2,square/.style={regular polygon,regular polygon sides=4}]
        \shade[top color=blue!40, bottom color=blue!10]  (0,-1) -- (1.5,-1) -- (1.5,1) -- (0,1)-- (0,-1);
        \draw[dgreen,thick, ->- = 0.5](0,-1) -- (0,0); 
        \draw[dgreen,thick, ->- = 0.5](0,0) -- (0,1); 
        \draw [blue, thick, -<- = 0.5] (0,0) to[out=180,in=90 ]  (-1,-1);
        \node[dgreen, above] at (0,+1) {\footnotesize{$\Sigma^{[\bw+\bv]}$}};
        \node[dgreen, below] at (0,-1) {\footnotesize{$\Sigma^{[\bw]}$}};
        \node[blue, below] at (-1.15,-1) {\footnotesize{$L_\bv$}};
        \node at (0,0) [square,draw,fill=violet!60,scale=0.5] {}; 
        \node[blue,left] at (1.5,0.7) {\footnotesize{$D_{A}$}};
    \end{tikzpicture} \,\,=   \sum_{\bu \in \cA_{A}} \mu_A(\bv + \bw_0 - (\bv+\bw)_0,\bu) \begin{tikzpicture}[baseline=0,scale=1.2]
        \draw[dgreen,thick, ->-=0.5](0,-1) -- (0,-0.6); 
        \draw[blue,thick, ->-=0.5](0,-0.6) -- (0,0); 
        \draw[blue,thick, ->-=0.5](0,0) -- (0,0.6); 
        \draw[dgreen,thick, ->-=0.5](0,0.6) -- (0,1); 
        \draw [blue, thick, -<-=0.5] (0,0) to[out=180,in=90 ]  (-1,-1);
        \node[blue, right] at (0,0.4) {\footnotesize{$L_{\bw_0 + \bv + \bu}$}};
        \node[blue, right] at (0,-0.4) {\footnotesize{$L_{\bw_0 + \bu}$}};
        \node[dgreen, above] at (0,1.0) {\footnotesize{$\Sigma^{[\bw+\bv]}$}};
        \node[blue, below] at (-1.15,-1) {\footnotesize{$L_\bv$}};
        \node[dgreen, below] at (0,-1) {\footnotesize{$\Sigma^{[\bw]}$}};
        \filldraw[blue] (0,0) circle (1pt);
        \filldraw[blue] (0,-0.6) circle (1pt);
        \filldraw[blue] (0,0.6) circle (1pt);
    \end{tikzpicture} ~.
\end{equation}  
To derive the triangle junction, we use the relation\footnote{The coefficients $R^{\bv,\Sigma^{[\bw]}}$ are gauge-dependent, and in particular cannot be fixed by the hexagon identities. }
\begin{equation}
\label{eq:RvSigmadef}
    R^{\bv,\Sigma^{[\bw]}} \begin{tikzpicture}[baseline=0,square/.style={regular polygon,regular polygon sides=4}]
        \draw[dgreen,thick, ->-=0.5](0,-1) -- (0,0); 
        \draw[dgreen,thick, ->-=0.5](0,0) -- (0,1); 
        \draw[blue, thick, -<-=0.5] (0,0) to[out=180,in=90 ]  (-1,-1);
        \node at (0,0) [square,draw,fill=violet!60,scale=0.5] {}; 
        \node[below, blue] at (-1,-1) {\footnotesize $L_{\bv}$};
        \node[dgreen, below] at (0,-1) {\footnotesize $\Sigma^{[\bw]}$};
        \node[dgreen, above] at (0,+1) {\footnotesize $\Sigma^{[\bw+\bv]}$};
    \end{tikzpicture} 
    \quad = \quad \begin{tikzpicture}[baseline=0,square/.style={regular polygon,regular polygon sides=4}]
        \draw[dgreen,thick, ->-=0.75](0,-0.45) -- (0,1); 
        \draw[dgreen,thick](0,-0.55) -- (0,-1); 
        \draw[blue, thick] (0,0) to[out=0,in=0,distance=0.2 in]  (0,-0.5);
        \draw[blue, thick, -<-=0.5] (0,-0.5) to[out=180,in=90,distance=0.2 in]  (-1,-1);
        \node[isosceles triangle,scale=0.4,isosceles triangle apex angle=60,draw,fill=violet!60, rotate=90,minimum size =0.01cm] at (0,0){};
        \node[below, blue] at (-1,-1) {\footnotesize $L_{\bv}$};
        \node[dgreen, below] at (0,-1) {\footnotesize $\Sigma^{[\bw]}$};
        \node[dgreen, above] at (0,+1) {\footnotesize $\Sigma^{[\bw+\bv]}$};
   \end{tikzpicture} ~,
\end{equation}
which implies that
\begin{equation}\label{eq:TriangleJunction}
    \begin{tikzpicture}[baseline=0,scale=1.2]
        \shade[top color=blue!40, bottom color=blue!10]  (0,-1) -- (1.5,-1) -- (1.5,1) -- (0,1)-- (0,-1);
        \draw[dgreen,thick, ->-=0.5](0,-1) -- (0,0); 
        \draw[dgreen,thick, ->-=0.5](0,0) -- (0,1); 
        \draw [gray!30, line width=3pt] (0,0) to[out=0,in=90 ]  (1,-1);
        \draw [blue, thick, -<-=0.5] (0,0) to[out=0,in=90 ]  (1,-1);
        \node[dgreen, above] at (0,1) {\footnotesize{$\Sigma^{[\bw+\bv]}$}};
        \node[dgreen, below] at (0,-1) {\footnotesize{$\Sigma^{[\bw]}$}};
        \node[blue, below] at (1,-1) {\footnotesize{$L_\bv$}};
        \node[isosceles triangle,scale=0.4, isosceles triangle apex angle=60, draw,fill=violet!60, rotate=90, minimum size =0.01cm] at (0,0){};
        \node[blue,left] at (1.5,0.7) {\footnotesize{$D_{A}$}};
\end{tikzpicture} \,\,=   \sum_{\bu \in \mathcal{A}_{A}} \frac{R^{\bv,\Sigma^{[\bw]}}}{R^{\bv,\bw_0 + \bu}} \mu_A(\bv + \bw_0 - (\bv+\bw)_0,\bu) 
    \begin{tikzpicture}[baseline=0,scale=1.2]
        \draw[dgreen,thick, ->-=0.5](0,-1) -- (0,-0.6); 
        \draw[blue,thick, ->-=0.5](0,-0.6) -- (0,0); 
        \draw[blue,thick, ->-=0.5](0,0) -- (0,0.6); 
        \draw[dgreen,thick, ->-=0.5](0,0.6) -- (0,1); 
        \draw[blue, thick, -<-=0.5] (0,0) to[out=0,in=90] (1,-1);
        \node[blue, right] at (0,0.4) {\footnotesize{$L_{\bw_0 + \bv + \bu}$}};
        \node[blue, left] at (0,-0.4) {\footnotesize{$L_{\bw_0 + \bu}$}};
        \node[dgreen, above] at (0,0.9) {\footnotesize{$\Sigma^{[\bw+\bv]}$}};
        \node[blue, below] at (1,-1) {\footnotesize{$L_\bv$}};
        \node[dgreen, below] at (0,-1) {\footnotesize{$\Sigma^{[\bw]}$}};
        \filldraw[blue] (0,0) circle (1pt);
        \filldraw[blue] (0,-0.6) circle (1pt);
        \filldraw[blue] (0,0.6) circle (1pt);
\end{tikzpicture} ~.
\end{equation}

It is now straightforward to compute the desired bulk $F$-symbols. We begin with 
\begin{equation}
    \begin{tikzpicture}[baseline=40,square/.style={regular polygon,regular polygon sides=4},scale=1.2]
        \shade[top color=blue!40, bottom color=blue!10]  (0,0)--(2,0)--(2,2)--(0,0);
         \draw [gray!30, line width=3pt] (1,0) -- (0.5,0.5); 
         \draw [gray!30, line width=3pt] (2,0) -- (1,1); 
        \draw[blue,thick, ->-=0.5](1,0) -- (0.5,0.5); 
        \draw[dgreen,thick, ->-=0.5](0,0) -- (0.5,0.5); 
        \draw[dgreen,thick, ->-=0.5] (0.5,0.5)--(1,1); 
        \draw[dgreen,thick, ->-=0.5] (1,1)--(2,2); 
        \draw[blue,thick, ->-=0.5](2,0) -- (1,1); 
        \node[blue,below] at (1.75,1.6) {\footnotesize{$D_{A}$}};
        \node[isosceles triangle,scale=0.4,isosceles triangle apex angle=60,draw,fill=violet!60,rotate=90,minimum size =0.01cm] at (0.5,0.5){};
        \node[isosceles triangle,scale=0.4,isosceles triangle apex angle=60,draw,fill=violet!60,rotate=90,minimum size =0.01cm] at (1,1){};
        \node[dgreen, below] at (0,0) {\footnotesize{$\Sigma^{[\bw]}$}};
        \node[dgreen, above] at (2,2) {\footnotesize{$\Sigma^{[\bw+\bv+\widetilde{\bv}]}$}};
        \node[blue, below] at (1,0) {\footnotesize{$L_\bv$}};
        \node[blue, below] at (2,0) {\footnotesize{$L_{\widetilde{\bv}}$}};
\end{tikzpicture} = \,\, F_{\Sigma^{[\bw]}, \bv ,\widetilde{\bv}}^{\Sigma^{[\bw+\bv+\widetilde{\bv}]}} 
    \begin{tikzpicture}[baseline=40,square/.style={regular polygon,regular polygon sides=4},scale=1.2]
        \shade[top color=blue!40, bottom color=blue!10]  (0,0)--(2,0)--(2,2)--(0,0);
        \draw[dgreen,thick, ->-=0.5](0,0) -- (1,1); 
        \draw[dgreen,thick, ->-=0.5](1,1) -- (2,2); 
        \draw [gray!30, line width=3pt] (1,0) -- (1.5,0.5); 
        \draw [gray!30, line width=3pt] (2,0) -- (1.5,0.5); 
        \draw [gray!30, line width=3pt] (1.5,0.5) -- (1,1); 
        \draw[blue,thick, ->-=0.5](1,0) -- (1.5,0.5); 
        \draw[blue,thick, ->-=0.5](2,0) -- (1.5,0.5); 
        \draw[blue,thick, ->-=0.5] (1.5,0.5) -- (1,1); 
        \node[blue,below] at (1.75,1.6) {\footnotesize{$D_{A}$}};
        \node[isosceles triangle,scale=0.4,isosceles triangle apex angle=60,draw,fill=violet!60,rotate=90,minimum size =0.01cm] at (1,1){};
        \filldraw[blue] (1.5,0.5) circle (1pt);
        \node[dgreen, below] at (0,0) {\footnotesize{$\Sigma^{[\bw]}$}};
        \node[dgreen, above] at (2,2) {\footnotesize{$\Sigma^{[\bw+\bv+\widetilde{\bv}]}$}};
        \node[blue, below] at (1,0) {\footnotesize{$L_\bv$}};
        \node[blue, below] at (2,0) {\footnotesize{$L_{\widetilde{\bv}}$}};
\end{tikzpicture} ~,
\end{equation}
which upon expansion using the triangle junction leads to
\bea
 &\vphantom{.}&    \frac{R^{\bv,\Sigma^{[\bw]}}}{R^{\bv,\bw_0 + \bu}}  \mu_A(\bv+\bw_0-(\bv+\bw)_0,\bu) 
 \no\\
 &\vphantom{.}& \hspace{0.3 in}\times \frac{R^{\widetilde{\bv},\Sigma^{[\bw+\bv]}}}{R^{\widetilde{\bv},\bw_0 + \bv + \bu}} \mu_A(\widetilde{\bv} + (\bv+\bw)_0 - (\widetilde{\bv} + \bv + \bw)_0,\bv + \bw_0 - (\bv+\bw)_0 + \bu)\no \\
   &\vphantom{.}&  \hspace{0.7 in}= F^{\Sigma^{[\bw+\bv+\widetilde{\bv}]}}_{\Sigma^{[\bw]}, \bv ,\widetilde{\bv}} \frac{R^{\bv + \widetilde{\bv},\Sigma^{[\bw]}}}{R^{\bv + \widetilde{\bv},\bw_0 + \bu}} \mu_A(\bv+\widetilde{\bv} + \bw_0 - (\bv+\widetilde{\bv}+\bw)_0,\bu) ~,
\eea
and hence
\begin{equation}
    F^{\Sigma^{[\bw+\bv+\widetilde{\bv}]}}_{\Sigma^{[\bw]}, \bv, \widetilde{\bv}} = \frac{R^{\bv,\Sigma^{[\bw]}}R^{\widetilde{\bv},\Sigma^{[\bw+\bv]}}}{R^{\bv + \widetilde{\bv},\Sigma^{[\bw]}}} \frac{\mu_A(\widetilde{\bv}+(\bv+\bw)_0-(\widetilde{\bv}+\bv+\bw)_0,\bv + \bw_0 - (\bv+\bw)_0)}{R^{\widetilde{\bv},\bv}} ~.
\end{equation}
Likewise, upon expanding
\begin{equation}
    \begin{tikzpicture}[baseline=30,square/.style={regular polygon,regular polygon sides=4},scale=1.2]
        \shade[top color=blue!40, bottom color=blue!10]  (1,0)--(2,0)--(2,2)--(0.5,0.5)--(1,0);
        \draw[blue,thick, ->-=0.5](0,0) -- (0.5,0.5); 
        \draw[dgreen,thick, ->-=0.5](1,0) -- (0.5,0.5); 
        \draw[dgreen,thick, ->-=0.5] (0.5,0.5)--(1,1); 
        \draw[dgreen,thick, ->-=0.5] (1,1)--(2,2); 
         \draw [gray!30, line width=3pt] (2,0) -- (1,1);
        \draw[blue,thick, ->-=0.5](2,0) -- (1,1); 
        \node[blue,below] at (1.75,1.6) {\footnotesize{$D_{A}$}};

   \node at (0.5,0.5) [square,draw,fill=violet!60,scale=0.5] {}; 

\node[isosceles triangle,scale=0.4,
    isosceles triangle apex angle=60,
    draw,fill=violet!60,
    rotate=90,
    minimum size =0.01cm] at (1,1){};

\node[dgreen, below] at (1,0) {\footnotesize{$\Sigma^{[\bw]}$}};
\node[dgreen, above] at (2,2) {\footnotesize{$\Sigma^{[\bw+\bv+\widetilde{\bv}]}$}};
\node[blue, below] at (0,0) {\footnotesize{$L_\bv$}};
\node[blue, below] at (2,0) {\footnotesize{$L_{\widetilde{\bv}}$}};

\end{tikzpicture} = \,\, F^{\Sigma^{[\bw+\bv+\widetilde{\bv}]}}_{\bv,\Sigma^{[\bw]}, \widetilde{\bv}} \begin{tikzpicture}[baseline=30,square/.style={regular polygon,regular polygon sides=4},scale=1.2]
        \shade[top color=blue!40, bottom color=blue!10]  (1,0)--(2,0)--(2,2)--(1,1)--(1.5,0.5)--(1,0);
        \draw[blue,thick, ->-=0.5](0,0) -- (1,1); 
        \draw[dgreen,thick, ->-=0.5](1,0) -- (1.5,0.5); 
        \draw[dgreen,thick, ->-=0.5] (1.5,0.5)--(1,1); 
        \draw[dgreen,thick, ->-=0.5] (1,1)--(2,2); 
        \draw [gray!30, line width=3pt] (2,0) -- (1.5,0.5); 
        \draw[blue,thick, ->-=0.5](2,0) -- (1.5,0.5); 
        \node[blue,below] at (1.75,1.6) {\footnotesize{$D_{A}$}};
        \node at (1,1) [square,draw,fill=violet!60,scale=0.5] {}; 
        \node[isosceles triangle,scale=0.4,isosceles triangle apex angle=60,draw,fill=violet!60,rotate=90,minimum size =0.01cm] at (1.5,0.5){};
        \node[dgreen, below] at (1,0) {\footnotesize{$\Sigma^{[\bw]}$}};
        \node[dgreen, above] at (2,2) {\footnotesize{$\Sigma^{[\bw+\bv+\widetilde{\bv}]}$}};
        \node[blue, below] at (0,0) {\footnotesize{$L_\bv$}};
        \node[blue, below] at (2,0) {\footnotesize{$L_{\widetilde{\bv}}$}};
    \end{tikzpicture} ~,
\end{equation}  
we obtain the condition
\bea
    &\vphantom{.}& \mu_A(\bv + \bw_0 - (\bv+\bw)_0,\bu)
    \no\\
     &\vphantom{.}&\hspace{0.2 in} \times \frac{R^{\widetilde{\bv},\Sigma^{[\bw+\bv]}}}{R^{\widetilde{\bv},\bu + \bv + \bw_0}} \mu_{A}(\widetilde{\bv} + (\bv+\bw)_0 - (\bv+\widetilde{\bv}+\bw)_0,\bu+\bv + \bw_0 - (\bv+\bw)_0) \no\\
  &\vphantom{.}&\hspace{0.6 in}= F^{\Sigma^{[\bw+\bv+\widetilde{\bv}]}}_{\bv,\Sigma^{[\bw]}, \widetilde{\bv}} \, \frac{R^{\widetilde{\bv},\Sigma^{[\bw]}}}{R^{\widetilde{\bv},\bu+\bw_0}} \mu_A(\widetilde{\bv} + \bw_0 - (\widetilde{\bv}+\bw)_0,\bu)
\\
   &\vphantom{.}&\hspace{0.9 in}\times\, \mu_A(\bv + (\widetilde{\bv}+\bw)_0 - (\bv+\widetilde{\bv}+\bw)_0,\bu + \widetilde{\bv} + \bw_0 - (\widetilde{\bv}+\bw)_0) ~,  \no
\eea
which implies that
\begin{equation}
    F^{\Sigma^{[\bw+\bv+\widetilde{\bv}]}}_{\bv,\Sigma^{[\bw]}, \widetilde{\bv}} = \frac{R^{\widetilde{\bv},\Sigma^{[\bw+\bv]}}}{R^{\widetilde{\bv},\Sigma^{[\bw]}}}\frac{1}{R^{\widetilde{\bv},\bv}} \frac{\mu_A(\widetilde{\bv} + (\bv+\bw)_0 - (\bv+\widetilde{\bv}+\bw)_0,\bv + \bw_0 - (\bv+\bw)_0)}{\mu_A(\bv + (\widetilde{\bv}+\bw)_0 - (\bv+\widetilde{\bv}+\bw)_0,\widetilde{\bv} + \bw_0 - (\widetilde{\bv}+\bw)_0)}  ~.
\end{equation}
Finally, expanding 
\begin{equation}
    \begin{tikzpicture}[baseline=40,square/.style={regular polygon,regular polygon sides=4},scale=1.2]
        \shade[top color=blue!40, bottom color=blue!10]  (2,0)--(1,1)--(2,2)--(2,0);
        \draw[blue,thick, ->-=0.5](0,0) -- (0.5,0.5); 
        \draw[blue,thick, ->-=0.5](1,0) -- (0.5,0.5);
        \draw[blue,thick, ->-=0.5] (0.5,0.5) -- (1,1);
        \draw[dgreen,thick, ->-=0.5] (2,0)--(1,1); 
        \draw[dgreen,thick, ->-=0.5] (1,1)--(2,2); 
        \node[blue,below] at (1.75,1.6) {\footnotesize{$D_{A}$}};
        \node at (1,1) [square,draw,fill=violet!60,scale=0.5] {}; 
        \filldraw[blue] (0.5,0.5) circle (1pt);
        \node[dgreen, below] at (2,0) {\footnotesize{$\Sigma^{[\bw]}$}};
        \node[dgreen, above] at (2,2) {\footnotesize{$\Sigma^{[\bw+\bv+\widetilde{\bv}]}$}};
        \node[blue, below] at (0,0) {\footnotesize{$L_\bv$}};
        \node[blue, below] at (1,0) {\footnotesize{$L_{\widetilde{\bv}}$}};
    \end{tikzpicture} = \,\, F^{\Sigma^{[\bw+\bv+\widetilde{\bv}]}}_{\bv ,\widetilde{\bv},\Sigma^{[\bw]}} \begin{tikzpicture}[baseline=40,square/.style={regular polygon,regular polygon sides=4},scale=1.2]
        \shade[top color=blue!40, bottom color=blue!10]  (2,0)--(1,1)--(2,2)--(2,0);
        \draw[blue,thick, ->-=0.5](0,0) -- (1,1); 
        \draw[blue,thick, ->-=0.5](1,0) -- (1.5,0.5);
        \draw[dgreen,thick, ->-=0.5] (2,0)--(1.5,0.5); 
        \draw[dgreen,thick, ->-=0.5] (1.5,0.5)--(1,1); 
        \draw[dgreen,thick, ->-=0.5] (1,1)--(2,2); 
        \node[blue,below] at (1.75,1.6) {\footnotesize{$D_{A}$}};
        \node at (1,1) [square,draw,fill=violet!60,scale=0.5] {}; 
        \node at (1.5,0.5) [square,draw,fill=violet!60,scale=0.5] {}; 
        \node[dgreen, below] at (2,0) {\footnotesize{$\Sigma^{[\bw]}$}};
        \node[dgreen, above] at (2,2) {\footnotesize{$\Sigma^{[\bw+\bv+\widetilde{\bv}]}$}};
        \node[blue, below] at (0,0) {\footnotesize{$L_\bv$}};
        \node[blue, below] at (1,0) {\footnotesize{$L_{\widetilde{\bv}}$}};
    \end{tikzpicture} ~,
\end{equation}  
using the definition of the square junctions gives
\bea
   &\vphantom{.}&  \mu_A(\bv + \widetilde{\bv} + \bw_0 - (\bv+\widetilde{\bv}+\bw)_0,\bu) 
   \no\\
  &\vphantom{.}&  \hspace{0.3 in}  = F^{\Sigma^{[\bw+\bv+\widetilde{\bv}]}}_{\bv \widetilde{\bv}\Sigma^{[\bw]}}  \mu_A(\widetilde{\bv} + \bw_0 - (\widetilde{\bv} + \bw)_0,\bu) 
  \\
      &\vphantom{.}&  \hspace{0.6 in} \times \mu_A(\bv + (\widetilde{\bv}+\bw)_0 - (\bv + \widetilde{\bv} + \bw)_0,\bu + \widetilde{\bv} + \bw_0 - (\widetilde{\bv}+\bw)_0) ~, \no
\eea
which implies that
\begin{equation}
    F^{\Sigma^{[\bw+\bv+\widetilde{\bv}]}}_{\bv ,\widetilde{\bv},\Sigma^{[\bw]}} = \frac{1}{\mu_A(\bv+(\widetilde{\bv}+\bw)_0 - (\bv+\widetilde{\bv}+\bw)_0,\widetilde{\bv}+\bw_0 - (\widetilde{\bv}+\bw)_0)} ~.
\end{equation}
This concludes the derivation of the results in (\ref{eq:bulkFsymbols}). 

It is important to notice that the fusion between the twist defects $\Sigma^{[0]}$ and the Abelian anyons condensed on the condensation defects are closed. Therefore, they generate a fusion (sub)category of the full SymSET. The  $F$-symbols above simplify greatly if we restrict to only $\Sigma^{[0]}$, giving
\begin{equation}\label{eq:bulkFsymbols1}
    F^{\Sigma^{[0]}}_{\Sigma^{[0]}, \bu, \widetilde{\bu}} = \frac{R^{\bu,\Sigma^{[0]}}R^{\widetilde{\bu},\Sigma^{[0]}}}{R^{\bu + \widetilde{\bu},\Sigma^{[0]}}} \frac{\mu_A(\widetilde{\bu},\bu)}{R^{\widetilde{\bu},\bu}} ~, \quad F^{\Sigma^{[0]}}_{\bu,\Sigma^{[0]}, \widetilde{\bu}} = \frac{1}{R^{\widetilde{\bu},\bu}} \frac{\mu_A(\widetilde{\bu} ,\bu )}{\mu_A(\bu ,\widetilde{\bu})} ~, \quad
    F^{\Sigma^{[0]}}_{\bu ,\widetilde{\bu},\Sigma^{[0]}} = \frac{1}{\mu_A(\bu,\widetilde{\bu})} ~, 
\end{equation}
where $\bu,\widetilde{\bu} \in \mathcal{A}_A$. 

\subsubsection{Concrete examples }
Let us now consider a couple of concrete examples. In both examples, we will take $n=2$ for simplicity.

\paragraph{$\ST_1$ and $\gcd(M,3) = 1$: } 
As a first example, consider the triality symmetry given by $\ST_1$ and $\gcd(M,3) = 1$. As shown in Section \ref{sec:explicitexamples1}, in this case all the bulk anyons are condensed on the condensation defect, and therefore there is a unique twist defect $\Sigma:= \Sigma^{[0]}  $. It is straightforward to work out the multiplication map for the algebra object $\mathcal{A}_{\ST_1}=\mathcal{A}_{\overline{\ST_1}}$ using \eqref{eq:mufinalresult}, from which we find
\begin{equation}
\label{eq:firstinstancemuST1}
    \mu_{\ST_1}(\bu,\widetilde\bu) = \omega^{-x(u_1\widetilde u_2 + u_1 \widetilde u_3 + u_2 \widetilde u_4 + u_3 \widetilde u_4)}~, \quad \mu_{\overline{\ST_1}}(\bu,\widetilde\bu) = \omega^{x(u_1 \widetilde u_2 - 2u_1 \widetilde u_3 - 2 u_2 \widetilde u_4 + u_3 \widetilde u_4)} ~,
\end{equation}
where $x$ is such that $3x = 1 \mod M$. Using \eqref{eq:bulkFsymbols1}, we then find the following bulk $F$-symbols,
\begin{equation}
\begin{array}{ll}
\hspace{-0.2in}    F^{\Sigma}_{\Sigma ,\bv, \widetilde{\bv}} = \frac{R^{\bv,\Sigma} R^{\widetilde{\bv},\Sigma}}{R^{\bv+\widetilde{\bv},\Sigma}} \omega^{-x(\widetilde{v}_1 v_2 - 2\widetilde{v}_1 v_3 - 2\widetilde{v}_2 v_4 + \widetilde{v}_3 v_4)} ~,  & F^{\oSigma}_{\oSigma, \bv, \widetilde{\bv}} = \frac{R^{\bv,\oSigma} R^{\widetilde{\bv},\oSigma}}{R^{\bv+\widetilde{\bv},\oSigma}} \omega^{x(\widetilde{v}_1 v_2 + \widetilde{v}_1 v_3 + \widetilde{v}_2 v_4 +\widetilde{v}_3 v_4)} ~, \\
  \hspace{-0.2in}   F^{\Sigma}_{\bv,\Sigma,\widetilde{\bv}} = \omega^{x(v_1 \widetilde{v}_2 - \widetilde{v}_1 v_2 + v_1 \widetilde{v}_3 + 2 \widetilde{v}_1 v_3 + v_2 \widetilde{v}_4 + 2 \widetilde{v}_2 v_4 + v_3 \widetilde{v}_4 - \widetilde{v}_3 v_4)}~, & F^{\oSigma}_{\bv, \oSigma ,\widetilde{\bv}} = \omega^{x(\widetilde{v}_1 v_2 - v_1 \widetilde{v}_2 + \widetilde{v}_1 v_3 + 2v_1 \widetilde{v}_3 + \widetilde{v}_2 v_4 + 2v_2 \widetilde{v}_4 + \widetilde{v}_3 v_4 - v_3 \widetilde{v}_4)} ~, \\
  \hspace{-0.2in}   F^{\Sigma}_{\bv,\widetilde{\bv},\Sigma} = \omega^{x(v_1 \widetilde{v}_2 + v_1 \widetilde{v}_3 + v_2 \widetilde{v}_4 + v_3 \widetilde{v}_4)} ~, & F^{\oSigma}_{\bv,\widetilde{\bv},\oSigma} = \omega^{-x(v_1 \widetilde{v}_2 - 2 v_1 \widetilde{v}_3 - 2 v_2 \widetilde{v}_4 + v_3 \widetilde{v}_4)} ~.
\end{array}
\end{equation}

\paragraph{$\ST_2$:} 

As another example, consider the order-three symmetry given by $\ST_2$. Let us first describe the spectrum of twist defects in the bulk. Following the procedure in Section \ref{sec:twistdefects2}, we first re-write the condensation defect as
\begin{equation}
\begin{aligned}
    D_{ST_2}(\Sigma) &= \sum_{\gamma_i} L_{(1,1,-1,0)}(\gamma_1)L_{(-1,0,1,-1)}(\gamma_2) \\
    &= \sum_{\gamma_i} e^{-\frac{2\pi i}{M}\langle \gamma_1,\gamma_2\rangle} L_{(-1,-1,1,0)}(\gamma_1) L_{(0,-1,0,1)}(\gamma_2) \\
    &= \sum_{\gamma_i} e^{-\frac{2\pi i}{M}\langle \gamma_1,\gamma_2\rangle} L_{R_1}(\gamma_1) L_{R_2}(\gamma_2)
\end{aligned}
\end{equation}
where $R_1 = (-1,-1,1,0)$ and $R_2 = (0,-1,0,1)$. Similarly, one finds
\begin{equation}
    D_{\overline{ST_2}}(\Sigma) = \sum_{\gamma_i} L_{R_1}(\gamma_1) L_{R_2}(\gamma_2) ~.
\end{equation}
From this, we see that the twist defects are labeled by $\Sigma^{[u]}$ and $\overline{\Sigma}^{[u]}$ where $[u] = [u+\vec{n}\cdot \vec{R}]$. As pointed out before, the bulk contains a fusion subcategory generated by $L_{\vec{n}\cdot \vec{R}}$, $\Sigma^{[0]}$, and $\oSigma^{[0]}$, with fusion rules matching those for triality defects with $\mathbb{Z}_M \times \mathbb{Z}_M$ invertible symmetry. For reasons to be discussed shortly, in this special case, to get the boundary $F$-symbols we only need the $F$-symbols of this subcategory, which can be obtained using the simplified expression \eqref{eq:bulkFsymbols1}.

In particular,  noting that the corresponding algebra objects $\mathcal{A}_{\ST_2}$ and $\mathcal{A}_{\overline{\ST_2}}$ are the same objects
\begin{equation}
    \cA_{\ST_2} = \mathcal{A}_{\overline{\ST_2}} = \sum_{\vec{n}} (e_1^{-1} e_2^{-1} m_1)^{n_1} (e_2^{-1} m_2)^{n_2} = \sum_{\vec{n}} L_{n_1 R_1 + n_2 R_2} ~,
\end{equation}
with distinct multiplication maps derived using \eqref{eq:mufinalresult},\footnote{Strictly speaking, all of the simple lines of the right-hand side should be thought of as being connected to $\cA_A$ via elements of the appropriate hom-spaces, so that the external legs are the same on both sides, as in e.g. Figure~\ref{fig:left_right_junction}. We will employ the current shorthand notation in much of the rest of the paper. Note that in the current case, all hom-spaces are one-dimensional, so there are no additional labels $i,j$ that we must keep track of. When there are non-trivial hom-spaces---for example between $\Sigma$ and $L_\bu$---we will keep track of this by writing the external line as e.g. $(L_\bu)_i$. \label{footnote:shorthand}}
\begin{equation}
    \begin{tikzpicture}[scale=0.50,baseline = {(0,0)}]
    \draw[thick, red, ->-=.5] (-1.7,-1) -- (0,0);
    \draw[thick, red, -<-=.5] (0,0) -- (1.7,-1);
    \draw[thick, red, -<-=.5] (0,2) -- (0,0);
    \node[red, above] at (0,2) {\scriptsize$\cA_{\ST_2}$};
    \node[red, below] at (-1.7,-1) {\scriptsize$\cA_{\ST_2}$};
    \node[red, below] at (1.7,-1) {\scriptsize$\cA_{\ST_2}$};
    \filldraw[red] (0,0) circle (2pt);
    \node[red, below] at (0,0) {\scriptsize $\mu$};
\end{tikzpicture} = \frac{1}{M} \sum_{\vec{n},\vec{n}'} \quad \begin{tikzpicture}[scale=0.50,baseline = {(0,0)}]
    \draw[thick, blue, ->-=.5] (-1.7,-1) -- (0,0);
    \draw[thick, blue, -<-=.5] (0,0) -- (1.7,-1);
    \draw[thick, blue, -<-=.5] (0,2) -- (0,0);
     \filldraw[blue] (0,0) circle (2pt);
    \node[blue, above] at (0,2) {\scriptsize $L_{(\vec{n}+\vec{n}')\cdot \vec{R}}$};
    \node[blue, below] at (-1.7,-1) {\scriptsize $L_{\vec{n} \cdot \vec{R}}$};
    \node[blue, below] at (1.7,-1) {\scriptsize $L_{\vec{n}' \cdot \vec{R}}$};
\end{tikzpicture} ~, \quad \begin{tikzpicture}[scale=0.50,baseline = {(0,0)}]
    \draw[thick, red, ->-=.5] (-1.7,-1) -- (0,0);
    \draw[thick, red, -<-=.5] (0,0) -- (1.7,-1);
    \draw[thick, red, -<-=.5] (0,2) -- (0,0);
    \node[red, above] at (0,2) {\scriptsize$\cA_{\overline{\ST_2}}$};
    \node[red, below] at (-1.7,-1) {\scriptsize$\cA_{\overline{\ST_2}}$};
    \node[red, below] at (1.7,-1) {\scriptsize$\cA_{\overline{\ST_2}}$};
    \filldraw[red] (0,0) circle (2pt);
    \node[red, below] at (0,0) {\scriptsize $\mu$};
\end{tikzpicture} = \frac{1}{M} \sum_{\vec{n},\vec{n}'} \omega^{-n_1' n_2} \quad \begin{tikzpicture}[scale=0.50,baseline = {(0,0)}]
    \draw[thick, blue, ->-=.5] (-1.7,-1) -- (0,0);
    \draw[thick, blue, -<-=.5] (0,0) -- (1.7,-1);
    \draw[thick, blue, -<-=.5] (0,2) -- (0,0);
     \filldraw[blue] (0,0) circle (2pt);
    \node[blue, above] at (0,2) {\scriptsize $L_{(\vec{n}+\vec{n}')\cdot \vec{R}}$};
    \node[blue, below] at (-1.7,-1) {\scriptsize $L_{\vec{n} \cdot \vec{R}}$};
    \node[blue, below] at (1.7,-1) {\scriptsize $L_{\vec{n}' \cdot \vec{R}}$};
\end{tikzpicture} ~,
\end{equation}
we find from \eqref{eq:bulkFsymbols1} the following $F$-symbols,
\begin{equation}\label{eq:ST2_F_Symbol}
\begin{aligned}
   \hspace{-0.2in} & F_{\Sigma^{[0]},\vec{n}\cdot \vec{R},\vec{n}'\cdot \vec{R}}^{\Sigma^{[0]},} = \frac{R^{\vec{n}\cdot\vec{R},\Sigma^{[0]}}R^{\vec{n}'\cdot\vec{R},\Sigma^{[0]}}}{R^{(\vec{n}+\vec{n}')\cdot\vec{R},\Sigma^{[0]}}} \, \omega^{-n_1 n_1' - n_2 n_2' - n_2 n_1'} ~,   
    F^{\Sigma^{[0]}}_{\vec{n}\cdot \vec{R}, \Sigma^{[0]}, \vec{n}'\cdot \vec{R}} = \omega^{-n_1 n_1' - n_2 n_2' - n_2 n_1'} ~,   F^{\Sigma^{[0]}}_{\vec{n}\cdot \vec{R}, \vec{n}'\cdot \vec{R},\Sigma^{[0]}} = 1 ~,
    \\
   \hspace{-0.2in}     & F_{\oSigma^{[0]}, \vec{n}\cdot \vec{R},\vec{n}'\cdot \vec{R}}^{\oSigma^{[0]}} = \frac{R^{\vec{n}\cdot\vec{R},\oSigma^{[0]}}R^{\vec{n}'\cdot\vec{R},\oSigma^{[0]}}}{R^{(\vec{n}+\vec{n}')\cdot\vec{R},\oSigma^{[0]}}} \omega^{-(n_1+n_2)(n_1'+n_2')} ~,  
    F^{\oSigma^{[0]}}_{\vec{n}\cdot\vec{R}, \oSigma^{[0]}, \vec{n}'\cdot\vec{R}} = \omega^{- n_1 n_1' - n_2 n_2' - n_1 n_2'} ~,  
    F^{\oSigma^{[0]}}_{\vec{n}\cdot\vec{R}, \vec{n}'\cdot\vec{R},\oSigma^{[0]}} = \omega^{n_2 n_1'} ~.
\end{aligned}
\end{equation}

\subsection{Taking the boundary limit via anyon condensation}

To go from the bulk results in the previous subsection to the corresponding boundary $F$-symbols, we must now understand how the boundary invertible lines and $N$-ality defects arise from the bulk lines, and in particular how their trivalent junctions may be expressed in terms of the data appearing in the bulk trivalent junctions. This is done by the procedure of anyon condensation, and below we will first give a generic prescription for the computation, after which we consider some concrete examples.

Our starting point is a $G$-crossed braided extension $\mathcal{B}_{G}^\times$ (for more discussions, see e.g. \cite{Etingof:2009yvg,gelaki2009centers,Barkeshli:2014cna}) of a unitary modular tensor category $\mathcal{B}$. Assuming the 3d TQFT $\mathcal{B}$ admits a gapped boundary acquired by condensing some Lagrangian algebra $\mathcal{L}$, the same anyon condensation will also lead to a gapped boundary for the SET corresponding to $\mathcal{B}_{G}^\times$, described by the $G$-graded fusion category $(\mathcal{B}_G^\times)_{\mathcal{L}}$ of (right) $\mathcal{L}$-modules in $\mathcal{B}_G^\times$. As usual, the braiding in the bulk allows us to equip a canonical left $\mathcal{L}$-module structure to any given right $\mathcal{L}$-module, thus turning it into an $\mathcal{L}$-$\mathcal{L}$-bimodule. This leads to a monoidal structure on the category $(\mathcal{B}_G^\times)_{\mathcal{L}}$, where the tensor product is the balanced tensor product over the algebra $\mathcal{L}$. Next, because the Lagrangian algebra $\mathcal{L}$ lies in the trivial grading component $\mathcal{B}$ of $\mathcal{B}_{G}^\times$, any indecomposable $\mathcal{L}$-modules would belong to some particular grading component of $\mathcal{B}_{G}^\times$. This naturally induces a $G$-grading on $(\mathcal{B}_G^\times)_{\mathcal{L}}$ and turns it to a $G$-graded fusion category. The computation of simple objects and their fusion rules, as well as the $F$-symbols of $(\mathcal{B}_G^\times)_{\mathcal{L}}$, is completely identical to what was previously described. 

\

In the case relevant to $N$-ality defects corresponding to gauging an Abelian group, we choose our topological boundary condition to be the Dirichlet boundary condition for the gauge field $\mathbf{a}$, which corresponds to the Lagrangian algebra $\mathcal{L} := \sum_{\mathbf{e}}L_{\mathbf{e}}$ with trivial multiplication map. Let us first consider the trivial-grading component of the boundary fusion category,  which consists of all the invertible symmetries. The lines in this component are labeled by magnetic charges $\mathbf{m}$, and we use $[\mathbf{m}]$ to denote the corresponding $\mathcal{L}$-modules. The trivalent junctions involving the boundary invertible symmetry and the algebra object $\cL$ are given by
\begin{equation}
\begin{aligned}
    \begin{tikzpicture}[baseline=0,scale=1]
        \draw[dgreen,thick, ->-=0.5](0,-1) -- (0,0); 
        \draw[dgreen,thick, ->-=0.5](0,0) -- (0,1); 
        \draw [red, thick, -<-=0.5] (0,0) to[out=0,in=90 ]  (1,-1);
        \node[dgreen, above] at (0,1) {\footnotesize{$[\mathbf{m}]$}};
        \node[dgreen, below] at (0,-1) {\footnotesize{$[\mathbf{m}]$}};
        \node[red, below] at (1,-1) {\footnotesize{$\mathcal{L}$}};
        \filldraw[red] (0,0) circle (1.5pt);
\end{tikzpicture} \quad = \quad  \frac{1}{M^{n/2}} \sum_{\mathbf{e},\widetilde{\mathbf{e}}}\begin{tikzpicture}[baseline=0,scale=1]
        \draw[blue,thick, ->-=0.5](0,-1) -- (0,0); 
        \draw[blue,thick, ->-=0.5](0,0) -- (0,1); 
        \draw[blue, thick, -<-=0.5] (0,0) to[out=0,in=90 ]  (1,-1);
         \filldraw[blue] (0,0) circle (1.5pt);
        \node[blue, above] at (0,1) {\footnotesize{$L_{\mathbf{m} + \mathbf{e} + \widetilde{\mathbf{e}}}$}};
        \node[blue, below] at (0,-1) {\footnotesize{$L_{\mathbf{m} + \mathbf{e}}$}};
        \node[blue, below] at (1,-1) {\footnotesize{$L_{\widetilde{\mathbf{e}}}$}};
\end{tikzpicture} ~, \hspace{0.3 in} \begin{tikzpicture}[baseline=0,scale=1,square/.style={regular polygon,regular polygon sides=4}]
        \draw[dgreen,thick, ->- = 0.5](0,-1) -- (0,0); 
        \draw[dgreen,thick, ->- = 0.5](0,0) -- (0,1); 
        \draw[red, thick, -<- = 0.5] (0,0) to[out=180,in=90 ]  (-1,-1);
        \node[dgreen, above] at (0,+1) {\footnotesize{$[\mathbf{m}]$}};
        \node[dgreen, below] at (0,-1) {\footnotesize{$[\mathbf{m}]$}};
        \node[red, below] at (-1.15,-1) {\footnotesize{$\mathcal{L}$}};
      \filldraw[red] (0,0) circle (1.5pt);
    \end{tikzpicture} \quad = \quad \frac{1}{M^{n/2}} \sum_{\mathbf{e},\widetilde{\mathbf{e}}} R^{\mathbf{e},\mathbf{m}} \begin{tikzpicture}[baseline=0,scale=1,square/.style={regular polygon,regular polygon sides=4}]
        \draw[blue,thick, ->- = 0.5](0,-1) -- (0,0); 
        \draw[blue,thick, ->- = 0.5](0,0) -- (0,1); 
        \draw[blue, thick, -<- = 0.5] (0,0) to[out=180,in=90 ]  (-1,-1);
         \filldraw[blue] (0,0) circle (1.5pt);
        \node[blue, above] at (0,+1) {\footnotesize{$L_{\mathbf{m} + \mathbf{e} + \widetilde{\mathbf{e}}}$}};
        \node[blue, below] at (0,-1) {\footnotesize{$L_{\mathbf{m} + \widetilde{\mathbf{e}}}$}};
        \node[blue, below] at (-1.15,-1) {\footnotesize{$L_{\mathbf{e}}$}};
    \end{tikzpicture} ~,
\end{aligned}
\end{equation}
where the first is a gauge choice and the second follows by relating the two junctions via half-braiding. The fusion junctions between invertible lines $[\mathbf{m}]$ are given by
\begin{equation}
    \begin{tikzpicture}[scale=0.50,baseline = {(0,0)}]
    \draw[thick, dgreen, ->-=.5] (-1.7,-1) -- (0,0);
    \draw[thick, dgreen, -<-=.5] (0,0) -- (1.7,-1);
    \draw[thick, dgreen, -<-=.5] (0,2) -- (0,0);
    \node[dgreen, above] at (0,2) {\scriptsize$[\mathbf{m} + \widetilde{\mathbf{m}}]$};
    \node[dgreen, below] at (-1.7,-1) {\scriptsize$[{\mathbf{m}}]$};
    \node[dgreen, below] at (1.7,-1) {\scriptsize$[\widetilde{\mathbf{m}}]$};
    \filldraw[dgreen] (0,0) circle (3pt);
\end{tikzpicture} = \frac{1}{M^{n/2}} \Omega_{\mathbf{m},\widetilde{\mathbf{m}}}\sum_{\mathbf{e},\widetilde{\mathbf{e}}} R^{\mathbf{e},\widetilde{\mathbf{m}}} \begin{tikzpicture}[scale=0.50,baseline = {(0,0)}]
    \draw[thick, blue, ->-=.5] (-1.7,-1) -- (0,0);
    \draw[thick, blue, -<-=.5] (0,0) -- (1.7,-1);
    \draw[thick, blue, -<-=.5] (0,2) -- (0,0);
     \filldraw[blue] (0,0) circle (3pt);
    \node[blue, above] at (0,2) {\scriptsize$L_{\mathbf{m} + \widetilde{\mathbf{m}} + \mathbf{e} + \widetilde{\mathbf{e}}}$};
    \node[blue, below] at (-1.7,-1) {\scriptsize$L_{\mathbf{m} + \mathbf{e}}$};
    \node[blue, below] at (1.7,-1) {\scriptsize$L_{\widetilde{\mathbf{m}} + \widetilde{\mathbf{e}}}$};
\end{tikzpicture} ~,
\end{equation}
as follows from the same sort of module conditions shown in (\ref{eq:threemoduleconditions}). Here $\Omega_{\mathbf{m},\widetilde{\mathbf{m}}}$ is a 2-cocycle which is not fixed by consistency conditions---it is simply a choice of gauge. It is straightforward to check that with this gauge choice, the boundary $F$-symbols of the invertible symmetries are explicitly $1$, as ensured by the fact that $\Omega_{\mathbf{m},\widetilde{\mathbf{m}}}$ is a 2-cocycle.

The $N$-ality defect $Q$ corresponds to an indecomposable $\mathcal{L}$-module with support on the twist defect $\Sigma$. Note that such an indecomposable $\mathcal{L}$-module must be unique, since the assumption of maximal gauging ensures that there is a unique simple object in each non-trivial grading component of the boundary category. This means that as an object in the bulk, 
\begin{equation}
    Q = K \sum_{[\bu]} \Sigma^{[\bu]} ~, \hspace{0.5 in} K \in \mathbb{Z}_{>0} ~,
\end{equation}
and the junction expansions take the form
\bea
   \begin{tikzpicture}[baseline=0,scale=0.9]
        \draw[dgreen,thick, ->-=0.5](0,-1) -- (0,0); 
        \draw[dgreen,thick, ->-=0.5](0,0) -- (0,1); 
        \draw [red, thick, -<-=0.5] (0,0) to[out=0,in=90 ]  (1,-1);
        \node[dgreen, above] at (0,1) {\footnotesize{$Q$}};
        \node[dgreen, below] at (0,-1) {\footnotesize{$Q$}};
        \node[red, below] at (1,-1) {\footnotesize{$\mathcal{L}$}};
       \filldraw[red] (0,0) circle (1.5pt);
\end{tikzpicture} &=& \frac{1}{M^{n/2}} \sum_{i,j,\mathbf{e},[\bu]} \alpha([\bu],\mathbf{e})_i{}^j \begin{tikzpicture}[baseline=0,scale=0.9]
        \draw[dgreen,thick, ->-=0.5](0,-1) -- (0,0); 
        \draw[dgreen,thick, ->-=0.5](0,0) -- (0,1); 
        \draw[blue, thick, -<-=0.5] (0,0) to[out=0,in=90 ]  (1,-1);
        \node[dgreen, above] at (0,1) {\footnotesize{$(\Sigma^{[\bu+\mathbf{e}]})_j$}};
        \node[dgreen, below] at (0,-1) {\footnotesize{$(\Sigma^{[\bu]})_i$}};
        \node[blue, below] at (1,-1) {\footnotesize{$L_{\mathbf{e}}$}};
        \node[isosceles triangle,scale=0.4, isosceles triangle apex angle=60, draw,fill=violet!60, rotate=90, minimum size =0.01cm] at (0,0){};
\end{tikzpicture} ~, 
\no\\
 \begin{tikzpicture}[baseline=0,scale=0.9,square/.style={regular polygon,regular polygon sides=4}]
        \draw[dgreen,thick, ->- = 0.5](0,-1) -- (0,0); 
        \draw[dgreen,thick, ->- = 0.5](0,0) -- (0,1); 
        \draw[red, thick, -<- = 0.5] (0,0) to[out=180,in=90 ]  (-1,-1);
        \node[dgreen, above] at (0,+1) {\footnotesize{$Q$}};
        \node[dgreen, below] at (0,-1) {\footnotesize{$Q$}};
        \node[red, below] at (-1.15,-1) {\footnotesize{$\mathcal{L}$}};
      \filldraw[red] (0,0) circle (1.5pt);
    \end{tikzpicture} &=& \frac{1}{M^{n/2}} \sum_{i,j,\mathbf{e},[\bu]} R^{\mathbf{e},\Sigma^{[\bu]}} \alpha([\bu],\mathbf{e})_i{}^j \begin{tikzpicture}[baseline=0,scale=0.9,square/.style={regular polygon,regular polygon sides=4}]
        \draw[dgreen,thick, ->- = 0.5](0,-1) -- (0,0); 
        \draw[dgreen,thick, ->- = 0.5](0,0) -- (0,1); 
        \draw[blue, thick, -<- = 0.5] (0,0) to[out=180,in=90 ]  (-1,-1);
        \node[dgreen, above] at (0,+1) {\footnotesize{$(\Sigma^{[\bu+\mathbf{e}]})_j$}};
        \node[dgreen, below] at (0,-1) {\footnotesize{$(\Sigma^{[\bu]})_i$}};
        \node[blue, below] at (-1.15,-1) {\footnotesize{$L_{\mathbf{e}}$}};
        \node at (0,0) [square,draw,fill=violet!60,scale=0.5] {}; 
    \end{tikzpicture} ~,
\eea
where $i,j = 1,\cdots,K$ and $(\Sigma^{[\bu]})_i$ is shorthand notation for the junction between $Q$ and $\Sigma^{[\bu]}$ given by the element of $\mathrm{Hom}(Q, \Sigma^{[\bu]})$ labelled by $i$; see footnote \ref{footnote:shorthand}. From the right module condition,
\begin{equation}
    \begin{tikzpicture}[baseline={([yshift=-.5ex]current bounding box.center)},square/.style={regular polygon,regular polygon sides=4}]
        \draw[red,thick, ->-=0.5](1,0) -- (0.5,0.5); 
        \draw[dgreen,thick, ->-=0.5](0,0) -- (0.5,0.5); 
        \draw[dgreen,thick, ->-=0.5] (0.5,0.5)--(1,1); 
        \draw[dgreen,thick, ->-=0.5] (1,1)--(2,2); 
        \draw[red,thick, ->-=0.5](2,0) -- (1,1); 
          \filldraw[red] (0.5,0.5) circle (1.5pt);
             \filldraw[red] (1,1) circle (1.5pt);
        \node[dgreen, below] at (0,0) {\footnotesize{$Q$}};
        \node[dgreen, above] at (2,2) {\footnotesize{$Q$}};
        \node[red, below] at (1,0) {\footnotesize{$\cL$}};
        \node[red, below] at (2,0) {\footnotesize{$\cL$}};
\end{tikzpicture} \hspace{0.2 in}=\hspace{0.2 in} \begin{tikzpicture}[baseline={([yshift=-.5ex]current bounding box.center)},square/.style={regular polygon,regular polygon sides=4}]
        \draw[dgreen,thick, ->-=0.5](0,0) -- (1,1); 
        \draw[dgreen,thick, ->-=0.5](1,1) --(2,2); 
        \draw[red,thick, ->-=0.5](1,0) -- (1.5,0.5); 
        \draw[red,thick, ->-=0.5](2,0) -- (1.5,0.5); 
        \draw[red,thick, ->-=0.5](1.5,0.5) -- (1,1); 
           \filldraw[red] (1,1) circle (1.5pt);
             \filldraw[red] (1.5,0.5) circle (1pt);
        \node[dgreen, below] at (0,0) {\footnotesize{$Q$}};
        \node[dgreen, above] at (2,2) {\footnotesize{$Q$}};
        \node[red, below] at (1,0) {\footnotesize{$\cL$}};
        \node[red, below] at (2,0) {\footnotesize{$\cL$}};
\end{tikzpicture} ~
\end{equation}
we see that the junction coefficients $\alpha([\bu],\mathbf{e})_i{}^j$ must satisfy
\begin{equation}
    \alpha([\bu],\mathbf{e})_i{}^j \alpha([\bu+\mathbf{e}],\widetilde{\mathbf{e}})_j{}^k = F_{\Sigma^{[\bu]},\mathbf{e},\widetilde{\mathbf{e}}}^{\Sigma^{[\bu+\mathbf{e}+\widetilde{\mathbf{e}}]}} \alpha([\bu],\mathbf{e}+\widetilde{\mathbf{e}})_i{}^k ~.
\end{equation}
The next step is then to derive the fusion junctions $Q\otimes [\mathbf{m}] \rightarrow Q$ and $[\mathbf{m}] \otimes Q \rightarrow Q$, from which the $F$-symbols follow. However, at this point the concrete formulas depend on the details of the symmetry. Below, we will demonstrate this with two concrete examples, highlighting certain simplifications that arise along the way. 

\subsubsection{Example I: $A$ has unique twist defect}
Let us first consider the case in which the twist defect $\Sigma$ for a given bulk symmetry $A$ is unique. This happens when the condensation defect of $A$ is obtained from 1-gauging of all Abelian anyons in the SymTFT. In this case, we drop the $[\mathbf{0}]$ superscript on $\Sigma$. The junction expansions are
\begin{equation}
    \begin{tikzpicture}[baseline=0,scale=1]
        \draw[dgreen,thick, ->-=0.5](0,-1) -- (0,0); 
        \draw[dgreen,thick, ->-=0.5](0,0) -- (0,1); 
        \draw [red, thick, -<-=0.5] (0,0) to[out=0,in=90 ]  (1,-1);
        \node[dgreen, above] at (0,1) {\footnotesize{$Q$}};
        \node[dgreen, below] at (0,-1) {\footnotesize{$Q$}};
        \node[red, below] at (1,-1) {\footnotesize{$\mathcal{L}$}};
          \filldraw[red] (0,0) circle (1.5pt);
\end{tikzpicture} = \frac{1}{M^{n/2}} \sum_{i,j,\mathbf{e}} \alpha(\mathbf{e})_i{}^j \begin{tikzpicture}[baseline=0,scale=1]
        \draw[dgreen,thick, ->-=0.5](0,-1) -- (0,0); 
        \draw[dgreen,thick, ->-=0.5](0,0) -- (0,1); 
        \draw[blue, thick, -<-=0.5] (0,0) to[out=0,in=90 ]  (1,-1);
        \node[dgreen, above] at (0,1) {\footnotesize{$\Sigma_j$}};
        \node[dgreen, below] at (0,-1) {\footnotesize{$\Sigma_i$}};
        \node[blue, below] at (1,-1) {\footnotesize{$L_{\mathbf{e}}$}};
        \node[isosceles triangle,scale=0.4, isosceles triangle apex angle=60, draw,fill=violet!60, rotate=90, minimum size =0.01cm] at (0,0){};
\end{tikzpicture} ~, \hspace{0.3 in} \begin{tikzpicture}[baseline=0,scale=1,square/.style={regular polygon,regular polygon sides=4}]
        \draw[dgreen,thick, ->- = 0.5](0,-1) -- (0,0); 
        \draw[dgreen,thick, ->- = 0.5](0,0) -- (0,1); 
        \draw[red, thick, -<- = 0.5] (0,0) to[out=180,in=90 ]  (-1,-1);
        \node[dgreen, above] at (0,+1) {\footnotesize{$Q$}};
        \node[dgreen, below] at (0,-1) {\footnotesize{$Q$}};
        \node[red, below] at (-1.15,-1) {\footnotesize{$\mathcal{L}$}};
        \filldraw[red] (0,0) circle (1.5pt); 
    \end{tikzpicture} \quad = \quad \frac{1}{M^{n/2}} \sum_{i,j,\mathbf{e}} R^{\mathbf{e},\Sigma} \alpha(\mathbf{e})_i{}^j \begin{tikzpicture}[baseline=0,scale=1,square/.style={regular polygon,regular polygon sides=4}]
        \draw[dgreen,thick, ->- = 0.5](0,-1) -- (0,0); 
        \draw[dgreen,thick, ->- = 0.5](0,0) -- (0,1); 
        \draw[blue, thick, -<- = 0.5] (0,0) to[out=180,in=90 ]  (-1,-1);
        \node[dgreen, above] at (0,+1) {\footnotesize{$\Sigma_j$}};
        \node[dgreen, below] at (0,-1) {\footnotesize{$\Sigma_i$}};
        \node[blue, below] at (-1.15,-1) {\footnotesize{$L_{\mathbf{e}}$}};
        \node at (0,0) [square,draw,fill=violet!60,scale=0.5] {}; 
    \end{tikzpicture}
\end{equation}
where $i,j$ run over the space $\mathrm{Hom}(Q,\Sigma)$ and the coefficients $\alpha(\mathbf{e})_i{}^j$ satisfy
\begin{equation}
\label{eq:constraintonalpha}
    \alpha(\mathbf{e})_i{}^j \alpha(\widetilde{\mathbf{e}})_j{}^k = \frac{R^{\mathbf{e} + \widetilde{\mathbf{e}},\Sigma}}{R^{\mathbf{e},\Sigma} R^{\widetilde{\mathbf{e}},\Sigma}} \frac{1}{\mu_A(\widetilde{\mathbf{e}},\mathbf{e})} \alpha(\mathbf{e} + \widetilde{\mathbf{e}})_i{}^k ~.
\end{equation}
Note importantly that the space $\mathrm{Hom}(Q,\Sigma)$ need not be one-dimensional. That the boundary defect $Q$ is unique in this grading component requires that $\mu_A(\widetilde{\mathbf{e}},\mathbf{e})$ be non-degenerate, ensuring there is a unique solution of $\alpha$ up to similarity transformation (namely, up to gauge transformations of the boundary). 

Next, we construct the fusion junction $Q \boxtimes_{\mathcal{L}} [\mathbf{m}] = Q$. In this case, we can expand the junction as\footnote{Note that $\rho_\mathbf{m}$ appearing heare is not related to the little group representation of $\Sigma$, which is trivial in the current case.}
\begin{equation}
    \begin{tikzpicture}[scale=0.50,baseline = {(0,0)}]
    \draw[thick, dgreen, ->-=.5] (-1.7,-1) -- (0,0);
    \draw[thick, dgreen, -<-=.5] (0,0) -- (1.7,-1);
    \draw[thick, dgreen, -<-=.5] (0,2) -- (0,0);
    \node[dgreen, above] at (0,2) {\scriptsize$Q$};
    \node[dgreen, below] at (-1.7,-1) {\scriptsize$Q$};
    \node[dgreen, below] at (1.7,-1) {\scriptsize$[\mathbf{m}]$};
    \filldraw[dgreen] (0,0) circle (2pt);
\end{tikzpicture} =  \frac{1}{M^{n/2}} \sum_{i,j,\mathbf{e}} \rho_{\mathbf{m}}(\mathbf{e})_i{}^j \quad \begin{tikzpicture}[scale=0.50,baseline = {(0,0)}]
    \draw[thick, dgreen, ->-=.5] (-1.7,-1) -- (0,0);
    \draw[thick, blue, -<-=.5] (0,0) -- (1.7,-1);
    \draw[thick, dgreen, -<-=.5] (0,2) -- (0,0);
    \node[dgreen, above] at (0,2) {\scriptsize $\Sigma_j$};
    \node[dgreen, below] at (-1.7,-1) {\scriptsize $\Sigma_i$};
    \node[isosceles triangle,scale=0.4, isosceles triangle apex angle=60, draw,fill=violet!60, rotate=90, minimum size =0.01cm] at (0,0){};
    \node[blue, below] at (1.7,-1) {\scriptsize $L_{\mathbf{m} + \mathbf{e}}$};
\end{tikzpicture} ~,
\end{equation}
where the junction coefficients can be acquired from solving the following three equations
\begin{equation}
\begin{tikzpicture}[baseline={([yshift=-.5ex]current bounding box.center)},vertex/.style={anchor=base,
    circle,fill=black!25,minimum size=18pt,inner sep=2pt},scale=0.4]
\draw[red, thick, ->-=.5] (-1.5,-1.5) -- (-0.5,-0.5);
\draw[dgreen, thick, ->-=.5] (-0.5,-0.5) -- (0.5,0.5);
\draw[dgreen, thick, ->-=.5] (0.5,0.5) -- (1.5,1.5);
\draw[dgreen, thick, ->-=.5] (0.5,-1.5) -- (-0.5,-0.5);
\draw[dgreen, thick, ->-=.5] (2.5,-1.5) -- (0.5,0.5);

\node[red, below] at (-1.5,-1.5) {\scriptsize $\cL$};
\node[dgreen, below] at (0.5,-1.5) {\scriptsize $Q$};
\node[dgreen, below] at (2.5,-1.5) {\scriptsize $[\mathbf{m}]$};
\node[dgreen, above] at (1.5,1.5) {\scriptsize $Q$};

   \filldraw[red] (-0.5,-0.5) circle (2pt); 
   
\end{tikzpicture} = \begin{tikzpicture}[baseline={([yshift=-.5ex]current bounding box.center)},vertex/.style={anchor=base,circle,fill=black!25,minimum size=18pt,inner sep=2pt},scale=0.4]
\draw[red, thick, ->-=.5] (-1.5,-1.5) -- (0.5,0.5);
\draw[dgreen, thick, ->-=.5] (0.5,0.5) -- (1.5,1.5);
\draw[dgreen, thick, ->-=.5] (0.5,-1.5) -- (1.5,-0.5);
\draw[dgreen, thick, ->-=.5] (2.5,-1.5) -- (1.5,-0.5);
\draw[dgreen, thick, ->-=.5] (1.5,-0.5) -- (0.5,0.5);
  \filldraw[red] (0.5,0.5) circle (2pt); 
  
\node[red, below] at (-1.5,-1.5) {\scriptsize $\cL$};
\node[dgreen, below] at (0.5,-1.5) {\scriptsize $Q$};
\node[dgreen, below] at (2.5,-1.5) {\scriptsize $[\mathbf{m}]$};
\node[dgreen, above] at (1.5,1.5) {\scriptsize $Q$};
\end{tikzpicture} ~,\hspace{0.1 in} \begin{tikzpicture}[baseline={([yshift=-.5ex]current bounding box.center)},vertex/.style={anchor=base,circle,fill=black!25,minimum size=18pt,inner sep=2pt},scale=0.4]
\draw[dgreen, thick, ->-=.5] (-1.5,-1.5) -- (-0.5,-0.5);
\draw[dgreen, thick, ->-=.5] (-0.5,-0.5) -- (0.5,0.5);
\draw[dgreen, thick, ->-=.5] (0.5,0.5) -- (1.5,1.5);
\draw[red, thick, ->-=.5] (0.5,-1.5) -- (-0.5,-0.5);
\draw[dgreen, thick, ->-=.5] (2.5,-1.5) -- (0.5,0.5);

  \filldraw[red] (-0.5,-0.5) circle (2pt); 
  
\node[dgreen, below] at (-1.5,-1.5) {\scriptsize $Q$};
\node[red, below] at (0.5,-1.5) {\scriptsize $\cL$};
\node[dgreen, below] at (2.5,-1.5) {\scriptsize $[\mathbf{m}]$};
\node[dgreen, above] at (1.5,1.5) {\scriptsize $Q$};
\end{tikzpicture} = \begin{tikzpicture}[baseline={([yshift=-.5ex]current bounding box.center)},vertex/.style={anchor=base,
    circle,fill=black!25,minimum size=18pt,inner sep=2pt},scale=0.4]
\draw[dgreen, thick, ->-=.5] (-1.5,-1.5) -- (0.5,0.5);
\draw[dgreen, thick, ->-=.5] (0.5,0.5) -- (1.5,1.5);
\draw[red, thick, ->-=.5] (0.5,-1.5) -- (1.5,-0.5);
\draw[dgreen, thick, ->-=.5] (2.5,-1.5) -- (1.5,-0.5);
\draw[dgreen, thick, ->-=.5] (1.5,-0.5) -- (0.5,0.5);

  \filldraw[red] (1.5,-0.5) circle (2pt); 
  
\node[dgreen, below] at (-1.5,-1.5) {\scriptsize $Q$};
\node[red, below] at (0.5,-1.5) {\scriptsize $\cL$};
\node[dgreen, below] at (2.5,-1.5) {\scriptsize $[\mathbf{m}]$};
\node[dgreen, above] at (1.5,1.5) {\scriptsize $Q$};
\end{tikzpicture}  ~,\hspace{0.1 in} \begin{tikzpicture}[baseline={([yshift=-.5ex]current bounding box.center)},vertex/.style={anchor=base,circle,fill=black!25,minimum size=18pt,inner sep=2pt},scale=0.4]
\draw[dgreen, thick, ->-=.5] (-1.5,-1.5) -- (-0.5,-0.5);
\draw[dgreen, thick, ->-=.5] (-0.5,-0.5) -- (0.5,0.5);
\draw[dgreen, thick, ->-=.5] (0.5,0.5) -- (1.5,1.5);
\draw[dgreen, thick, ->-=.5] (0.5,-1.5) -- (-0.5,-0.5);
\draw[red, thick, ->-=.5] (2.5,-1.5) -- (0.5,0.5);
  \filldraw[red] (0.5,0.5) circle (2pt); 
  
\node[dgreen, below] at (-1.5,-1.5) {\scriptsize $Q$};
\node[dgreen, below] at (0.5,-1.5) {\scriptsize $[\mathbf{m}]$};
\node[red, below] at (2.5,-1.5) {\scriptsize $\cL$};
\node[dgreen, above] at (1.5,1.5) {\scriptsize $Q$};
\end{tikzpicture} = \begin{tikzpicture}[baseline={([yshift=-.5ex]current bounding box.center)},vertex/.style={anchor=base,circle,fill=black!25,minimum size=18pt,inner sep=2pt},scale=0.4]
\draw[dgreen, thick, ->-=0.5] (-1.5,-1.5) -- (0.5,0.5);
\draw[dgreen, thick, ->-=0.5] (0.5,0.5) -- (1.5,1.5);
\draw[dgreen, thick, ->-=0.5] (0.5,-1.5) -- (1.5,-0.5);
\draw[red, thick, ->-=0.5] (2.5,-1.5) -- (1.5,-0.5);
\draw[dgreen, thick, ->-=0.5] (1.5,-0.5) -- (0.5,0.5);
  \filldraw[red] (1.5,-0.5) circle (2pt); 

\node[dgreen, below] at (-1.5,-1.5) {\scriptsize $Q$};
\node[dgreen, below] at (0.5,-1.5) {\scriptsize $[\mathbf{m}]$};
\node[red, below] at (2.5,-1.5) {\scriptsize $\cL$};
\node[dgreen, above] at (1.5,1.5) {\scriptsize $Q$};
\end{tikzpicture} ~,
\end{equation}
giving rise to
\begin{equation}
    \rho_{\mathbf{m}}(\mathbf{e})_i{}^k = \frac{R^{\mathbf{m},\Sigma} R^{\mathbf{e},\Sigma}}{R^{\mathbf{m} + \mathbf{e},\Sigma}} \frac{\mu_A(\mathbf{e},\mathbf{m})}{R^{\mathbf{e},\mathbf{m}}} (\rho_\mathbf{m})_i{}^j \alpha(\mathbf{e})_j{}^k ~,
\end{equation}
with $\rho_{\mathbf{m}}$ satisfying the matrix equation
\begin{equation}
\label{eq:rhomatrixeq}
    \rho_{\mathbf{m}} \alpha(\mathbf{e}) (\rho_{\mathbf{m}})^{-1} = \frac{\mu_A(\mathbf{m},\mathbf{e})}{\mu_A(\mathbf{e},\mathbf{m})} \frac{1}{R^{\mathbf{m},\mathbf{e}}} \alpha(\mathbf{e}) ~. 
\end{equation}
Similarly, the other fusion junction $[\mathbf{m}] \boxtimes_{\mathcal{L}} Q = Q$ is given by
\begin{equation}
    \begin{tikzpicture}[scale=0.50,baseline = {(0,0)}]
    \draw[thick, dgreen, ->-=.5] (-1.7,-1) -- (0,0);
    \draw[thick, dgreen, -<-=.5] (0,0) -- (1.7,-1);
    \draw[thick, dgreen, -<-=.5] (0,2) -- (0,0);
    \node[dgreen, above] at (0,2) {\scriptsize$Q$};
    \node[dgreen, below] at (-1.7,-1) {\scriptsize$[\mathbf{m}]$};
    \node[dgreen, below] at (1.7,-1) {\scriptsize$Q$};
    \filldraw[dgreen] (0,0) circle (2pt);
\end{tikzpicture} = \frac{1}{M^{n/2}} \sum_{i,j,\mathbf{e}} \lambda_{\mathbf{m}}(\mathbf{e})_i{}^j \quad \begin{tikzpicture}[scale=0.50,baseline = {(0,0)},square/.style={regular polygon,regular polygon sides=4}]
    \draw[thick, blue, ->-=.5] (-1.7,-1) -- (0,0);
    \draw[thick, dgreen, -<-=.5] (0,0) -- (1.7,-1);
    \draw[thick, dgreen, -<-=.5] (0,2) -- (0,0);
     \node at (0,0) [square,draw,fill=violet!60,scale=0.5] {}; 
     
    \node[dgreen, above] at (0,2) {\scriptsize $\Sigma_j$};
    \node[blue, below] at (-1.7,-1) {\scriptsize $L_{\mathbf{m} + \mathbf{e}}$};
    \node[dgreen, below] at (1.7,-1) {\scriptsize $\Sigma_i$};
\end{tikzpicture}
\end{equation}
where the junction coefficients are given by
\begin{equation}
    \lambda_m(\mathbf{e})_i{}^k = \frac{R^{\mathbf{e},\Sigma}}{R^{\mathbf{e},\mathbf{m}}} \mu_A(\mathbf{e},\mathbf{m}) (\lambda_{\mathbf{m}})_i{}^j \alpha(\mathbf{e})_{j}{}^k
\end{equation}
and $\lambda_{\mathbf{m}}$ satisfies the matrix equation
\begin{equation}
\label{eq:lambdamatrixeq}
    \lambda_{\mathbf{m}} \alpha(\mathbf{e}) (\lambda_{\mathbf{m}})^{-1} = \frac{\mu_A(\mathbf{m},\mathbf{e})}{\mu_A(\mathbf{e},\mathbf{m})} R^{\mathbf{e},\mathbf{m}} \alpha(\mathbf{e}) ~.
\end{equation}

With the above results, the desired boundary $F$-symbols,
\begin{equation}
\begin{aligned}
    \begin{tikzpicture}[baseline={([yshift=-.5ex]current bounding box.center)},vertex/.style={anchor=base,
    circle,fill=black!25,minimum size=18pt,inner sep=2pt},scale=0.4]
\draw[dgreen, thick, ->-=.5] (-1.5,-1.5) -- (-0.5,-0.5);
\draw[dgreen, thick, ->-=.5] (-0.5,-0.5) -- (0.5,0.5);
\draw[dgreen, thick, ->-=.5] (0.5,0.5) -- (1.5,1.5);
\draw[dgreen, thick, ->-=.5] (0.5,-1.5) -- (-0.5,-0.5);
\draw[dgreen, thick, ->-=.5] (2.5,-1.5) -- (0.5,0.5);

\node[dgreen, below] at (-1.5,-1.5) {\scriptsize $[\mathbf{m}]$};
\node[dgreen, below] at (0.5,-1.5) {\scriptsize $Q$};
\node[dgreen, below] at (2.5,-1.5) {\scriptsize $[\widetilde{\mathbf{m}}]$};
\node[dgreen, above] at (1.5,1.5) {\scriptsize $Q$};
\end{tikzpicture} & = F_{\mathbf{m},Q,\widetilde{\mathbf{m}}}^Q \begin{tikzpicture}[baseline={([yshift=-.5ex]current bounding box.center)},vertex/.style={anchor=base,circle,fill=black!25,minimum size=18pt,inner sep=2pt},scale=0.4]
\draw[dgreen, thick, ->-=.5] (-1.5,-1.5) -- (0.5,0.5);
\draw[dgreen, thick, ->-=.5] (0.5,0.5) -- (1.5,1.5);
\draw[dgreen, thick, ->-=.5] (0.5,-1.5) -- (1.5,-0.5);
\draw[dgreen, thick, ->-=.5] (2.5,-1.5) -- (1.5,-0.5);
\draw[dgreen, thick, ->-=.5] (1.5,-0.5) -- (0.5,0.5);

\node[dgreen, below] at (-1.5,-1.5) {\scriptsize $[\mathbf{m}]$};
\node[dgreen, below] at (0.5,-1.5) {\scriptsize $Q$};
\node[dgreen, below] at (2.5,-1.5) {\scriptsize $[\widetilde{\mathbf{m}}]$};
\node[dgreen, above] at (1.5,1.5) {\scriptsize $Q$};
\end{tikzpicture} ~, \\
\begin{tikzpicture}[baseline={([yshift=-.5ex]current bounding box.center)},vertex/.style={anchor=base,circle,fill=black!25,minimum size=18pt,inner sep=2pt},scale=0.4]
\draw[dgreen, thick, ->-=.5] (-1.5,-1.5) -- (-0.5,-0.5);
\draw[dgreen, thick, ->-=.5] (-0.5,-0.5) -- (0.5,0.5);
\draw[dgreen, thick, ->-=.5] (0.5,0.5) -- (1.5,1.5);
\draw[dgreen, thick, ->-=.5] (0.5,-1.5) -- (-0.5,-0.5);
\draw[dgreen, thick, ->-=.5] (2.5,-1.5) -- (0.5,0.5);

\node[dgreen, below] at (-1.5,-1.5) {\scriptsize $Q$};
\node[dgreen, below] at (0.5,-1.5) {\scriptsize $[\mathbf{m}]$};
\node[dgreen, below] at (2.5,-1.5) {\scriptsize $[\widetilde{\mathbf{m}}]$};
\node[dgreen, above] at (1.5,1.5) {\scriptsize $Q$};
\end{tikzpicture} &= F_{Q,\mathbf{m},\widetilde{\mathbf{m}}}^Q \begin{tikzpicture}[baseline={([yshift=-.5ex]current bounding box.center)},vertex/.style={anchor=base,
    circle,fill=black!25,minimum size=18pt,inner sep=2pt},scale=0.4]
\draw[dgreen, thick, ->-=.5] (-1.5,-1.5) -- (0.5,0.5);
\draw[dgreen, thick, ->-=.5] (0.5,0.5) -- (1.5,1.5);
\draw[dgreen, thick, ->-=.5] (0.5,-1.5) -- (1.5,-0.5);
\draw[dgreen, thick, ->-=.5] (2.5,-1.5) -- (1.5,-0.5);
\draw[dgreen, thick, ->-=.5] (1.5,-0.5) -- (0.5,0.5);

\node[dgreen, below] at (-1.5,-1.5) {\scriptsize $Q$};
\node[dgreen, below] at (0.5,-1.5) {\scriptsize $[\mathbf{m}]$};
\node[dgreen, below] at (2.5,-1.5) {\scriptsize $[\widetilde{\mathbf{m}}]$};
\node[dgreen, above] at (1.5,1.5) {\scriptsize $Q$};
\end{tikzpicture} ~, \\
\begin{tikzpicture}[baseline={([yshift=-.5ex]current bounding box.center)},vertex/.style={anchor=base,circle,fill=black!25,minimum size=18pt,inner sep=2pt},scale=0.4]
\draw[dgreen, thick, ->-=.5] (-1.5,-1.5) -- (-0.5,-0.5);
\draw[dgreen, thick, ->-=.5] (-0.5,-0.5) -- (0.5,0.5);
\draw[dgreen, thick, ->-=.5] (0.5,0.5) -- (1.5,1.5);
\draw[dgreen, thick, ->-=.5] (0.5,-1.5) -- (-0.5,-0.5);
\draw[dgreen, thick, ->-=.5] (2.5,-1.5) -- (0.5,0.5);

\node[dgreen, below] at (-1.5,-1.5) {\scriptsize $[\mathbf{m}]$};
\node[dgreen, below] at (0.5,-1.5) {\scriptsize $[\widetilde{\mathbf{m}}]$};
\node[dgreen, below] at (2.5,-1.5) {\scriptsize $Q$};
\node[dgreen, above] at (1.5,1.5) {\scriptsize $Q$};
\end{tikzpicture} &= F_{\mathbf{m},\widetilde{\mathbf{m}},Q}^Q \begin{tikzpicture}[baseline={([yshift=-.5ex]current bounding box.center)},vertex/.style={anchor=base,circle,fill=black!25,minimum size=18pt,inner sep=2pt},scale=0.4]
\draw[dgreen, thick, ->-=0.5] (-1.5,-1.5) -- (0.5,0.5);
\draw[dgreen, thick, ->-=0.5] (0.5,0.5) -- (1.5,1.5);
\draw[dgreen, thick, ->-=0.5] (0.5,-1.5) -- (1.5,-0.5);
\draw[dgreen, thick, ->-=0.5] (2.5,-1.5) -- (1.5,-0.5);
\draw[dgreen, thick, ->-=0.5] (1.5,-0.5) -- (0.5,0.5);

\node[dgreen, below] at (-1.5,-1.5) {\scriptsize $[\mathbf{m}]$};
\node[dgreen, below] at (0.5,-1.5) {\scriptsize $[\widetilde{\mathbf{m}}]$};
\node[dgreen, below] at (2.5,-1.5) {\scriptsize $Q$};
\node[dgreen, above] at (1.5,1.5) {\scriptsize $Q$};
\end{tikzpicture} ~
\end{aligned}
\end{equation}
can finally be determined. As an example, consider the $F$-symbol $F^{Q}_{Q,\mathbf{m},\widetilde{\mathbf{m}}}$. In this case expansion of the left-hand side contains the term
\begin{equation}
\begin{aligned}
\begin{tikzpicture}[baseline={([yshift=-.5ex]current bounding box.center)},vertex/.style={anchor=base, circle,fill=black!25,minimum size=18pt,inner sep=2pt},scale=0.4]
    \draw[dgreen, thick, ->-=.5] (-1.5,-1.5) -- (-0.5,-0.5);
    \draw[dgreen, thick, ->-=.5] (-0.5,-0.5) -- (0.5,0.5);
    \draw[dgreen, thick, ->-=.5] (0.5,0.5) -- (1.5,1.5);
    \draw[dgreen, thick, ->-=.5] (0.5,-1.5) -- (-0.5,-0.5);
    \draw[dgreen, thick, ->-=.5] (2.5,-1.5) -- (0.5,0.5);

    \node[dgreen, below] at (-1.5,-1.5) {\scriptsize $Q$};
    \node[dgreen, below] at (0.5,-1.5) {\scriptsize $[\mathbf{m}]$};
    \node[dgreen, below] at (2.5,-1.5) {\scriptsize $[\widetilde{\mathbf{m}}]$};
    \node[dgreen, above] at (1.5,1.5) {\scriptsize $Q$};
\end{tikzpicture} & \supset  \frac{1}{M^n}  \sum_{i,j,k} (\rho_{\mathbf{m}})_i{}^j (\rho_{\widetilde{\mathbf{m}}})_{j}{}^k\begin{tikzpicture}[baseline={([yshift=-.5ex]current bounding box.center)},vertex/.style={anchor=base,circle,fill=black!25,minimum size=18pt,inner sep=2pt},scale=0.4]
    \draw[dgreen, thick, ->-=.5] (-1.5,-1.5) -- (-0.5,-0.5);
    \draw[dgreen, thick, ->-=.5] (-0.5,-0.5) -- (0.5,0.5);
    \draw[dgreen, thick, ->-=.5] (0.5,0.5) -- (1.5,1.5);
    \draw[blue, thick, ->-=.5] (0.5,-1.5) -- (-0.5,-0.5);
    \draw[blue, thick, ->-=.5] (2.5,-1.5) -- (0.5,0.5);

  \node[isosceles triangle,scale=0.3, isosceles triangle apex angle=60, draw,fill=violet!60, rotate=90, minimum size =0.01cm] at (-0.5,-0.5){};
    \node[isosceles triangle,scale=0.3, isosceles triangle apex angle=60, draw,fill=violet!60, rotate=90, minimum size =0.01cm] at (0.5,0.5){};
    
    \node[dgreen, below] at (-1.5,-1.5) {\scriptsize $\Sigma_i$};
    \node[blue, below] at (0.5,-1.5) {\scriptsize $L_{\mathbf{m}}$};
    \node[blue, below] at (2.5,-1.5) {\scriptsize $L_{\widetilde{\mathbf{m}}}$};
    \node[dgreen, above] at (1.5,1.5) {\scriptsize $\Sigma_k$};
\end{tikzpicture} \\
&= \frac{1}{M^n} \sum_{i,j,k} (\rho_{\mathbf{m}})_i{}^j (\rho_{\widetilde{\mathbf{m}}})_{j}{}^k F_{\Sigma,\mathbf{m},\widetilde{\mathbf{m}}}^\Sigma \begin{tikzpicture}[baseline={([yshift=-.5ex]current bounding box.center)},vertex/.style={anchor=base,circle,fill=black!25,minimum size=18pt,inner sep=2pt},scale=0.4]
\draw[dgreen, thick, ->-=0.5] (-1.5,-1.5) -- (0.5,0.5);
\draw[dgreen, thick, ->-=0.5] (0.5,0.5) -- (1.5,1.5);
\draw[blue, thick, ->-=0.5] (0.5,-1.5) -- (1.5,-0.5);
\draw[blue, thick, ->-=0.5] (2.5,-1.5) -- (1.5,-0.5);
\draw[blue, thick, ->-=0.5] (1.5,-0.5) -- (0.5,0.5);
  \node[isosceles triangle,scale=0.3, isosceles triangle apex angle=60, draw,fill=violet!60, rotate=90, minimum size =0.01cm] at (0.5,0.5){};

\node[dgreen, below] at (-1.5,-1.5) {\scriptsize $\Sigma_i$};
\node[blue, below] at (0.5,-1.5) {\scriptsize $L_{\mathbf{m}}$};
\node[blue, below] at (2.5,-1.5) {\scriptsize $L_{\widetilde{\mathbf{m}}}$};
\node[dgreen, above] at (1.5,1.5) {\scriptsize $\Sigma_k$};
\end{tikzpicture} \\
&=  \frac{1}{M^n}  \sum_{i,j,k} (\rho_{\mathbf{m}})_i{}^j (\rho_{\widetilde{\mathbf{m}}})_{j}{}^k \frac{R^{\mathbf{m},\Sigma}R^{\widetilde{\mathbf{m}},\Sigma}}{R^{\mathbf{m}+\widetilde{\mathbf{m}},\Sigma}} \mu_A(\widetilde{\mathbf{m}},\mathbf{m}) \begin{tikzpicture}[baseline={([yshift=-.5ex]current bounding box.center)},vertex/.style={anchor=base,circle,fill=black!25,minimum size=18pt,inner sep=2pt},scale=0.4]
\draw[dgreen, thick, ->-=0.5] (-1.5,-1.5) -- (0.5,0.5);
\draw[dgreen, thick, ->-=0.5] (0.5,0.5) -- (1.5,1.5);
\draw[blue, thick, ->-=0.5] (0.5,-1.5) -- (1.5,-0.5);
\draw[blue, thick, ->-=0.5] (2.5,-1.5) -- (1.5,-0.5);
\draw[blue, thick, ->-=0.5] (1.5,-0.5) -- (0.5,0.5);
  \node[isosceles triangle,scale=0.3, isosceles triangle apex angle=60, draw,fill=violet!60, rotate=90, minimum size =0.01cm] at (0.5,0.5){};

\node[dgreen, below] at (-1.5,-1.5) {\scriptsize $\Sigma_i$};
\node[blue, below] at (0.5,-1.5) {\scriptsize $L_{\mathbf{m}}$};
\node[blue, below] at (2.5,-1.5) {\scriptsize $L_{\widetilde{\mathbf{m}}}$};
\node[dgreen, above] at (1.5,1.5) {\scriptsize $\Sigma_k$};
\end{tikzpicture}~,
\end{aligned}
\end{equation}
where we have used (\ref{eq:bulkFsymbols1}) and noted that  $R^{\widetilde{\mathbf{m}}, \mathbf{m}} = 1$ since there is no braiding between purely magnetic lines. 
On the other hand, the right-hand side of the boundary $F$-symbol contains an analogous term 
\begin{equation}
    F_{Q,\mathbf{m},\widetilde{\mathbf{m}}}^Q \begin{tikzpicture}[baseline={([yshift=-.5ex]current bounding box.center)},vertex/.style={anchor=base,circle,fill=black!25,minimum size=18pt,inner sep=2pt},scale=0.4]
    \draw[dgreen, thick, ->-=.5] (-1.5,-1.5) -- (0.5,0.5);
    \draw[dgreen, thick, ->-=.5] (0.5,0.5) -- (1.5,1.5);
    \draw[dgreen, thick, ->-=.5] (0.5,-1.5) -- (1.5,-0.5);
    \draw[dgreen, thick, ->-=.5] (2.5,-1.5) -- (1.5,-0.5);
    \draw[dgreen, thick, ->-=.5] (1.5,-0.5) -- (0.5,0.5);

    \node[dgreen, below] at (-1.5,-1.5) {\scriptsize $Q$};
    \node[dgreen, below] at (0.5,-1.5) {\scriptsize $[\mathbf{m}]$};
    \node[dgreen, below] at (2.5,-1.5) {\scriptsize $[\widetilde{\mathbf{m}}]$};
    \node[dgreen, above] at (1.5,1.5) {\scriptsize $Q$};
\end{tikzpicture} \supset F_{Q,\mathbf{m},\widetilde{\mathbf{m}}}^Q  \frac{1}{M^n}  \Omega_{\mathbf{m},\widetilde{\mathbf{m}}}(\rho_{\mathbf{m} + \widetilde{\mathbf{m}}})_i{}^k \begin{tikzpicture}[baseline={([yshift=-.5ex]current bounding box.center)},vertex/.style={anchor=base,circle,fill=black!25,minimum size=18pt,inner sep=2pt},scale=0.4]
\draw[dgreen, thick, ->-=0.5] (-1.5,-1.5) -- (0.5,0.5);
\draw[dgreen, thick, ->-=0.5] (0.5,0.5) -- (1.5,1.5);
\draw[blue, thick, ->-=0.5] (0.5,-1.5) -- (1.5,-0.5);
\draw[blue, thick, ->-=0.5] (2.5,-1.5) -- (1.5,-0.5);
\draw[blue, thick, ->-=0.5] (1.5,-0.5) -- (0.5,0.5);
  \node[isosceles triangle,scale=0.3, isosceles triangle apex angle=60, draw,fill=violet!60, rotate=90, minimum size =0.01cm] at (0.5,0.5){};

\node[dgreen, below] at (-1.5,-1.5) {\scriptsize $\Sigma_i$};
\node[blue, below] at (0.5,-1.5) {\scriptsize $L_{\mathbf{m}}$};
\node[blue, below] at (2.5,-1.5) {\scriptsize $L_{\widetilde{\mathbf{m}}}$};
\node[dgreen, above] at (1.5,1.5) {\scriptsize $\Sigma_k$};
\end{tikzpicture} ~.
\end{equation}
Comparing the two sides, we find that $F_{Q,\mathbf{m},\widetilde{\mathbf{m}}}^Q$ is determined by
\begin{equation}
\label{eq:boundaryFmethod21}
    F_{Q,\mathbf{m},\widetilde{\mathbf{m}}}^Q \rho_{\mathbf{m} + \widetilde{\mathbf{m}}} = {1\over \Omega_{\mathbf{m},\widetilde{\mathbf{m}}}} \frac{R^{\mathbf{m},\Sigma}R^{\widetilde{\mathbf{m}},\Sigma}}{R^{\mathbf{m}+\widetilde{\mathbf{m}},\Sigma}} \mu_A(\widetilde{\mathbf{m}},\mathbf{m}) \rho_{\mathbf{m}} \rho_{\widetilde{\mathbf{m}}} ~.
\end{equation}
Similarly, the other $F$-symbols are determined by
\bea
\label{eq:boundaryFmethod22}
    \frac{\Omega_{\mathbf{m},\widetilde{\mathbf{m}}}}{\mu_A(\mathbf{m},\widetilde{\mathbf{m}})}\lambda_{\mathbf{m} + \widetilde{\mathbf{m}}} &=& F^{Q}_{\mathbf{m},\widetilde{\mathbf{m}},Q} \lambda_{\widetilde{\mathbf{m}}} \lambda_{\mathbf{m}} ~,
\no\\
    \lambda_{\mathbf{m}}\rho_{\widetilde{\mathbf{m}}} \frac{\mu_A(\widetilde{\mathbf{m}},\mathbf{m})}{\mu_A(\mathbf{m},\widetilde{\mathbf{m}})} &=& F^{Q}_{\mathbf{m},Q,\widetilde{\mathbf{m}}} \rho_{\widetilde{\mathbf{m}}}\lambda_{\mathbf{m}} ~.
\eea
Note that although matrices appear on either side of these equations, the $F$-symbols with a single incoming $Q$ are themselves scalars. Of course, it is not obvious from these expressions that they match with the ones given in (\ref{eq:boundaryFmethod1}) (even up to gauge transformation). We now explicitly check that this is the case in some concrete examples. 

\subsubsection*{$\mathbb{Z}_2 \times \mathbb{Z}_2$ with $\ST_1$}
We first consider the case of $\mathbb{Z}_2 \times \mathbb{Z}_2$ and the triality symmetry being the one constructed from the 
bulk symmetry $A=\ST_1$. On the one hand, in the notation of (\ref{eq:decompofA}) we have 
\bea
\A = - \mathds{1}_{2 \times 2} ~, \hspace{0.2 in} \B = \C = \left( \begin{matrix} 0 & -1 \\ 1 & 0 \end{matrix} \right) ~, \hspace{0.2 in} \D = \mymathbb{0}_{2\times 2}~, 
\eea
from which (\ref{eq:chiphidefs}) gives 
\bea
\label{eq:ST1examplechiphi}
\varphi = \chi =  \left( \begin{matrix} 0 & -1 \\ 1 & 0 \end{matrix} \right)~, \hspace{0.3 in} \widetilde \varphi = \mymathbb{0}_{2\times 2}~. 
\eea
Then from (\ref{eq:boundaryFmethod1}) we compute the following $F$-symbols, 
\bea
\label{eq:ST1examplemethod1}
F_{Q, \bm, \widetilde \bm}^Q = (-1)^{m_1 \widetilde m_2}~, \hspace{0.3in} F_{\bm, Q , \widetilde \bm}^Q = (-1)^{m_1 \widetilde m_2 + \widetilde m_1 m_2 }~, \hspace{0.3 in} F_{\bm, \widetilde \bm,Q}^Q = 1 ~. 
\eea
Repeating the same steps for $A = \overline{\ST_1}$, we likewise find the $F$-symbols for $\oQ$,
\begin{equation}
\label{eq:ST1examplemethod12}
    F_{\oQ, \bm, \widetilde \bm}^{\oQ} = 1 ~, \hspace{0.3in} F_{\bm, \oQ , \widetilde \bm}^{\oQ} = (-1)^{m_1 \widetilde m_2 + \widetilde m_1 m_2 }~, \hspace{0.3 in} F_{\bm, \widetilde \bm,\oQ}^{\oQ} = (-1)^{m_1 \widetilde m_2} ~.
\end{equation}

On the other hand, we may now recompute these results using the second method, in which the $F$-symbols are determined by (\ref{eq:boundaryFmethod21}) and (\ref{eq:boundaryFmethod22}). For $A=\ST_1$, the multiplication coefficients $\mu_A$ are computed using (\ref{eq:mufinalresult}) and give
\begin{equation}
    \mu_{\ST_1}(\bu,\widetilde \bu) = (-1)^{u_1 \widetilde u_2 + u_1 \widetilde u_3 + u_2 \widetilde u_4+ u_3 \widetilde u_4} ~, 
\end{equation}
as was already quoted in (\ref{eq:firstinstancemuST1}). 
We may now use these to evaluate the matrices $\alpha(\mathbf{e})$ satisfying (\ref{eq:constraintonalpha}), i.e.
\begin{equation}
    \alpha(\mathbf{e}) \alpha(\widetilde{\mathbf{e}}) = \frac{R^{\mathbf{e} + \widetilde{\mathbf{e}},\Sigma}}{R^{\mathbf{e},\Sigma} R^{\widetilde{\mathbf{e}},\Sigma}} (-1)^{e_2 \widetilde{e}_1} \alpha(\mathbf{e} + \widetilde{\mathbf{e}}) ~,
\end{equation}
with the solutions being
\begin{equation}
    \alpha(\mathbf{e}) = \frac{1}{R^{\mathbf{e},\Sigma}} X^{e_1} Z^{e_2} ~
\end{equation}
with $X$, $Z$ the corresponding Pauli matrices. %
Using (\ref{eq:rhomatrixeq}) and (\ref{eq:lambdamatrixeq}), namely
\begin{equation}
    \rho_{\mathbf{m}} \alpha(\mathbf{e}) (\rho_{\mathbf{m}})^{-1} = (-1)^{m_1 e_1 + m_2 e_2} \alpha(\mathbf{e}) ~, \quad \lambda_{\mathbf{m}} \alpha_{\mathbf{e}} (\lambda_{\mathbf{m}})^{-1} = \alpha(\mathbf{e}) ~,
\end{equation}
we may now solve for the matrices $\rho_\mathbf{m}$ and $\lambda_{\mathbf{m}}$, with a solution given by
\begin{equation}
    \rho_{\mathbf{m}} = \frac{1}{R^{\mathbf{m},\Sigma}} Z^{m_1} X^{m_2} ~, \quad \lambda_{\mathbf{m}} = \mathds{1} ~.
\end{equation}
Finally, the expressions for $\mu_A$, $\alpha(\mathbf{e})$, $\rho_\bm$, and $\lambda_\bm$ may be inserted into (\ref{eq:boundaryFmethod21}) and (\ref{eq:boundaryFmethod22}) to obtain the $F$-symbols,
\begin{equation}
     F_{Q,\mathbf{m},\widetilde{\mathbf{m}}}^Q = \Omega_{\mathbf{m},\widetilde{\mathbf{m}}}^{-1} ~, \quad F^Q_{\mathbf{m},Q,\widetilde{\mathbf{m}}} = (-1)^{m_1 \widetilde{m}_2 + m_2 \widetilde{m}_1} ~, \quad F^Q_{\mathbf{m},\widetilde{\mathbf{m}},Q} = (-1)^{m_1 \widetilde{m}_2} \Omega_{\mathbf{m},\widetilde{\mathbf{m}}}~.
\end{equation}
Setting $ \Omega_{\mathbf{m},\widetilde{\mathbf{m}}} = (-1)^{m_1 \widetilde m_2}$, we see that this matches with the results of the previous method given in (\ref{eq:ST1examplemethod1}).

Similarly, one can compute the $F$-symbols of $\overline{Q}$ by starting from
\begin{equation}
    \mu_{\overline{\ST_1}}(\bu,\widetilde{\bu}) = (-1)^{u_1 \widetilde{u}_2+u_3 \widetilde{u}_3} ~,
\end{equation}
and choosing
\begin{equation}
    \overline{\alpha}(\mathbf{e}) = \frac{1}{R^{\mathbf{e},\oSigma}} X^{e_2} Z^{e_1} ~, \quad \overline{\rho}_{\bm} = \frac{1}{R^{\bm,\overline{\Sigma}}} \mathds{1} ~, \quad \overline{\lambda}_{\mathbf{m}} = Z^{m_2} X^{m_1} ~,
\end{equation}
from which one obtains
\begin{equation}
    F^{\oQ}_{\oQ,\bm,\widetilde{\bm}} = \Omega_{\bm,\widetilde{\bm}}^{-1} (-1)^{m_1 \widetilde{m}_2} ~, \quad F_{\bm,\oQ,\widetilde{\bm}}^{\oQ} = (-1)^{m_1 \widetilde{m}_2 + m_2 \widetilde{m}_1} ~, \quad F_{\bm,\widetilde{\bm},\oQ}^{\oQ} = \Omega_{\bm,\widetilde{\bm}} ~.
\end{equation}
Once again, choosing the gauge $ \Omega_{\mathbf{m},\widetilde{\mathbf{m}}} = (-1)^{m_1 \widetilde m_2}$ gives a result matching with that in (\ref{eq:ST1examplemethod12}). 

\subsubsection*{$\mathbb{Z}_p \times \mathbb{Z}_p$ with $\ST_1$ where $M$ is an odd prime $>3$}

Let us next consider the case of $\ZZ_p \times \ZZ_p$ symmetry (with $p>3$) and again the triality symmetry corresponding to the 
bulk symmetry $A = \ST_1$. On the one hand, the expressions for $\varphi$, $\chi$, and $\widetilde \varphi$ given in (\ref{eq:ST1examplechiphi}) are unchanged, 
so from (\ref{eq:boundaryFmethod1}) we immediately compute the following $F$-symbols, 
\bea
\label{eq:ST1example2method1}
\begin{array}{lll}
F_{Q, \bm, \widetilde \bm}^Q = \omega^{-m_1 \widetilde m_2} ~, \hspace{0.3in} & F_{\bm, Q , \widetilde \bm}^Q = \omega^{m_1 \widetilde m_2 - \widetilde m_1 m_2 } ~, \hspace{0.3 in} & F_{\bm, \widetilde \bm,Q}^Q = 1 ~, \\
F_{\oQ, \bm, \widetilde \bm}^{\oQ} = 1 ~, \hspace{0.3in} & F_{\bm, \oQ , \widetilde \bm}^{\oQ} = \omega^{m_2 \widetilde m_1 - \widetilde m_2 m_1 }~, \hspace{0.3 in} & F_{\bm, \widetilde \bm,\oQ}^{\oQ} = \omega^{m_1 \widetilde{m}_2} ~,
\end{array}
\eea
with $\omega = e^{2 \pi i \over p}$. 

On the other hand, in order to compute this via the second method, we first compute the multiplication map,
\begin{equation}
\label{eq:muAmuA2}
    \mu_A(\bu,\widetilde \bu) = \omega^{-x (u_1 \widetilde u_2 + u_1 \widetilde u_3 + u_2 \widetilde u_4+ u_3 \widetilde u_4)} ~, 
\end{equation}
where $x$ is the mod $p$ inverse of 3, i.e. $3 x = 1$ mod $p$.\footnote{A unique such $x$ exists since $\mathrm{gcd}(3,p) = 1$. } This is then used to compute $\alpha(\mathbf{e})$ via (\ref{eq:constraintonalpha}),
\begin{equation}
    \alpha(\mathbf{e}) \alpha(\widetilde{\mathbf{e}}) = \frac{R^{\mathbf{e} + \widetilde{\mathbf{e}},\Sigma}}{R^{\mathbf{e},\Sigma} R^{\widetilde{\mathbf{e}},\Sigma}} \omega^{x\widetilde{e}_1 e_2} \alpha(\mathbf{e} + \widetilde{\mathbf{e}}) ~,
\end{equation}
which may be solved in terms of clock and shift matrices $S,V$ such that $S^p = V^p = \mathbf{1}$ and $SV = \omega VS$, so that
\begin{equation}
    \alpha(\mathbf{e}) = \frac{1}{R^{\mathbf{e},\Sigma}} V^{e_1} S^{x e_2} ~.
\end{equation}
Next, the matrices $\rho_{\mathbf{m}}$ and $\lambda_{\mathbf{m}}$ are determined via (\ref{eq:rhomatrixeq}) and (\ref{eq:lambdamatrixeq}), giving
\begin{equation}
    \rho_{\mathbf{m}} \alpha(\mathbf{e}) (\rho_{\mathbf{m}})^{-1} = \omega^{x(m_1 e_1 + m_2 e_2)} \alpha(\mathbf{e}) ~, \quad \lambda_{\mathbf{m}} \alpha(\mathbf{e}) (\lambda_{\mathbf{m}})^{-1} = \omega^{-2x(m_1 e_1 + m_2 e_2)} \alpha(\mathbf{e}) ~
\end{equation}
and we take the solutions to be
\begin{equation}
    \rho_{\mathbf{m}} = \frac{1}{R^{\mathbf{m},\Sigma}} S^{x m_1}V^{-m_2} ~, \quad \lambda_{\mathbf{m}} = S^{-2x m_1} V^{2m_2} ~.
\end{equation}
Finally, plugging the above into (\ref{eq:boundaryFmethod21}) and (\ref{eq:boundaryFmethod22}), we determine the $F$-symbols,
\begin{equation}
    F_{Q,\mathbf{m},\widetilde{\mathbf{m}}}^Q = \Omega_{\mathbf{m},\widetilde{\mathbf{m}}}^{-1}  ~, \quad F_{\mathbf{m},Q,\widetilde{\mathbf{m}}}^Q = \omega^{m_1 \widetilde{m}_2 - m_2 \widetilde{m}_1} ~, \quad F_{\mathbf{m},\widetilde{\mathbf{m}},Q}^Q = \Omega_{\mathbf{m},\widetilde{\mathbf{m}}}\, \omega^{-m_1 \widetilde{m}_2} ~.
\end{equation}
The computation for $\oQ$ is completely identical, where we use
\begin{equation}
    \mu_{\overline{\ST_1}}(\bu,\widetilde{\bu}) = \omega^{x(u_1 \widetilde{u}_2 - 2 u_1 \widetilde{u}_3 - 2 u_2 \widetilde{u}_4 + u_3 \widetilde{u}_4)} ~,
\end{equation}
and choose
\begin{equation}
    \overline{\alpha}(\mathbf{e}) = \frac{1}{R^{\mathbf{e},\overline{\Sigma}}} V^{e_1} S^{-x e_2} ~, \quad \overline{\rho}_{\bm} = \frac{\omega^{-m_1 m_2}}{R^{\mathbf{m},\oSigma}} S^{2x m_1} V^{2m_2} ~, \quad \overline{\lambda}_{\mathbf{m}} = S^{-xm_1} V^{-m_2} ~,
\end{equation}
and find
\begin{equation}
    F^{\oQ}_{\oQ,\mathbf{m},\widetilde{\mathbf{m}}} = \Omega_{\mathbf{m},\widetilde{\mathbf{m}}}^{-1} \omega^{m_1 \widetilde{m}_2} ~, \quad F^{\oQ}_{\mathbf{m},\oQ,\widetilde{\mathbf{m}}} = \omega^{-m_1 \widetilde{m}_2 + m_2 \widetilde{m}_1} ~, \quad F_{\mathbf{m},\widetilde{\mathbf{m}},\oQ}^{\oQ} = \Omega_{\mathbf{m},\widetilde{\mathbf{m}}} ~.
\end{equation}
Upon choosing $\Omega_{\mathbf{m},\widetilde{\mathbf{m}}}=\omega^{m_1 \widetilde m_2}$, we see that these match with the result in (\ref{eq:ST1example2method1}). 
Notice that the $F_{\mathbf{m},Q,\widetilde{\mathbf{m}}}^Q$ only needs to be non-degenerate, and does not have to be symmetric in general.

\subsubsection{Example II: Triality from $\ST_2$}
As another example, we consider the triality defect from $\ST_2$ in \eqref{eq:STdef}. This example has the special feature that the boundary $F$-symbols are identical (up to gauge transformation) to the $F$-symbols of a specific bulk subsector.

We want to show that the $F$-symbols of this fusion subcategory are gauge equivalent to $F$-symbols of the triality fusion category on the $e$-condensed boundary. For the fusion junctions of the invertible lines $[\bm]$, we can always work in a gauge in which the following specific term has junction coefficient $1$,
\begin{equation}
    \begin{tikzpicture}[scale=0.50,baseline = {(0,0)}]
    \draw[thick, dgreen, ->-=.5] (-1.7,-1) -- (0,0);
    \draw[thick, dgreen, -<-=.5] (0,0) -- (1.7,-1);
    \draw[thick, dgreen, -<-=.5] (0,2) -- (0,0);
    \node[dgreen, above] at (0,2) {\scriptsize$[\mathbf{m} + \widetilde{\mathbf{m}}]$};
    \node[dgreen, below] at (-1.7,-1) {\scriptsize$[{\mathbf{m}}]$};
    \node[dgreen, below] at (1.7,-1) {\scriptsize$[\widetilde{\mathbf{m}}]$};
    \filldraw[dgreen] (0,0) circle (2pt);
\end{tikzpicture} \hspace{0.2 in}=\hspace{0.2 in} \frac{1}{M^{n/2}} \begin{tikzpicture}[scale=0.50,baseline = {(0,0)}]
    \draw[thick, blue, ->-=.5] (-1.7,-1) -- (0,0);
    \draw[thick, blue, -<-=.5] (0,0) -- (1.7,-1);
    \draw[thick, blue, -<-=.5] (0,2) -- (0,0);
    \node[blue, above] at (0,2) {\scriptsize$L_{(m_i + \widetilde{m}_i) R_i}$};
    \node[blue, below] at (-1.7,-1) {\scriptsize$L_{m_i R_i}$};
    \node[blue, below] at (1.7,-1) {\scriptsize$L_{\widetilde{m}_i R_i}$};
\end{tikzpicture} + \hspace{0.2 in} \cdots \,\,\,~.
\end{equation}
For the twist defects, we see that the fusion of any pure electric line $L_{\mathbf{e}}$ with $\Sigma^{[0]}$ (or $\oSigma^{[0]}$) leads to a distinct twist defect, and conversely any twist defect $\Sigma^{[\mathbf{e}]}$ (or $\oSigma^{[\mathbf{e}]}$) can be obtained by fusing some $L_{\mathbf{e}}$ with $\Sigma^{[0]}$. Therefore, on the boundary, all of the $\Sigma^{[\mathbf{e}]}$ (resp. $\oSigma^{[\mathbf{e}]}$) together form an $\mathcal{L}_e$-module which we identify as the boundary triality defect $Q$ (resp. $\overline{Q}$),
\begin{equation}
    Q = \bigoplus_{\mathbf{e}} \Sigma^{[\mathbf{e}]} ~, \hspace{0.3 in} \overline{Q} = \bigoplus_{\mathbf{e}} \oSigma^{[\mathbf{e}]} ~.
\end{equation}
When constructing the fusion junctions involving $Q, \overline{Q}$, one can also use a gauge transformation to set a specific term in each junction to have trivial coefficients,
\begin{equation}
\begin{aligned}
    & \begin{tikzpicture}[scale=0.50,baseline = {(0,0)}]
    \draw[thick, dgreen, ->-=.5] (-1.7,-1) -- (0,0);
    \draw[thick, dgreen, -<-=.5] (0,0) -- (1.7,-1);
    \draw[thick, dgreen, -<-=.5] (0,2) -- (0,0);
    \node[dgreen, above] at (0,2) {\scriptsize$Q$};
    \node[dgreen, below] at (-1.7,-1) {\scriptsize$[{\mathbf{m}}]$};
    \node[dgreen, below] at (1.7,-1) {\scriptsize$Q$};
    \filldraw[dgreen] (0,0) circle (2pt);
\end{tikzpicture} \,\,=\,\, \frac{1}{M^{n/2}}  \begin{tikzpicture}[scale=0.50,baseline = {(0,0)},square/.style={regular polygon,regular polygon sides=4}]
    \draw[thick, blue, ->-=.5] (-1.7,-1) -- (0,0);
    \draw[thick, dgreen, -<-=.5] (0,0) -- (1.7,-1);
    \draw[thick, dgreen, -<-=.5] (0,2) -- (0,0);
         \node at (0,0) [square,draw,fill=violet!60,scale=0.5] {}; 
    \node[dgreen, above] at (0,2) {\scriptsize $\Sigma^{[0]}$};
    \node[blue, below] at (-1.7,-1) {\scriptsize $L_{m_i R_i}$};
    \node[dgreen, below] at (1.7,-1) {\scriptsize $\Sigma^{[0]}$};
\end{tikzpicture} + \cdots ~, \quad \begin{tikzpicture}[scale=0.50,baseline = {(0,0)}]
    \draw[thick, dgreen, ->-=.5] (-1.7,-1) -- (0,0);
    \draw[thick, dgreen, -<-=.5] (0,0) -- (1.7,-1);
    \draw[thick, dgreen, -<-=.5] (0,2) -- (0,0);
    \node[dgreen, above] at (0,2) {\scriptsize$Q$};
    \node[dgreen, below] at (-1.7,-1) {\scriptsize$Q$};
    \node[dgreen, below] at (1.7,-1) {\scriptsize$[{\mathbf{m}}]$};
    \filldraw[dgreen] (0,0) circle (2pt);
\end{tikzpicture} \,\,=\,\, \frac{1}{M^{n/2}}  \begin{tikzpicture}[scale=0.50,baseline = {(0,0)}]
    \draw[thick, dgreen, ->-=.5] (-1.7,-1) -- (0,0);
    \draw[thick, blue, -<-=.5] (0,0) -- (1.7,-1);
    \draw[thick, dgreen, -<-=.5] (0,2) -- (0,0);
          \node[isosceles triangle,scale=0.4, isosceles triangle apex angle=60, draw,fill=violet!60, rotate=90, minimum size =0.01cm] at (0,0){};
    \node[dgreen, above] at (0,2) {\scriptsize $\Sigma^{[0]}$};
    \node[dgreen, below] at (-1.7,-1) {\scriptsize $\Sigma^{[0]}$};
    \node[blue, below] at (1.7,-1) {\scriptsize $L_{m_i R_i}$};
\end{tikzpicture} + \cdots ~, \\
    & \begin{tikzpicture}[scale=0.50,baseline = {(0,0)}]
    \draw[thick, dgreen, ->-=.5] (-1.7,-1) -- (0,0);
    \draw[thick, dgreen, -<-=.5] (0,0) -- (1.7,-1);
    \draw[thick, dgreen, -<-=.5] (0,2) -- (0,0);
    \node[dgreen, above] at (0,2) {\scriptsize$\oQ$};
    \node[dgreen, below] at (-1.7,-1) {\scriptsize$[{\mathbf{m}}]$};
    \node[dgreen, below] at (1.7,-1) {\scriptsize$\oQ$};
    \filldraw[dgreen] (0,0) circle (2pt);
\end{tikzpicture} \,\,=\,\, \frac{1}{M^{n/2}}  \begin{tikzpicture}[scale=0.50,baseline = {(0,0)},square/.style={regular polygon,regular polygon sides=4}]
    \draw[thick, blue, ->-=.5] (-1.7,-1) -- (0,0);
    \draw[thick, dgreen, -<-=.5] (0,0) -- (1.7,-1);
    \draw[thick, dgreen, -<-=.5] (0,2) -- (0,0);
         \node at (0,0) [square,draw,fill=violet!60,scale=0.5] {}; 
    \node[dgreen, above] at (0,2) {\scriptsize $\oSigma^{[0]}$};
    \node[blue, below] at (-1.7,-1) {\scriptsize $L_{m_i R_i}$};
    \node[dgreen, below] at (1.7,-1) {\scriptsize $\oSigma^{[0]}$};
\end{tikzpicture} + \cdots ~, \quad \begin{tikzpicture}[scale=0.50,baseline = {(0,0)}]
    \draw[thick, dgreen, ->-=.5] (-1.7,-1) -- (0,0);
    \draw[thick, dgreen, -<-=.5] (0,0) -- (1.7,-1);
    \draw[thick, dgreen, -<-=.5] (0,2) -- (0,0);
    \node[dgreen, above] at (0,2) {\scriptsize$\oQ$};
    \node[dgreen, below] at (-1.7,-1) {\scriptsize$\oQ$};
    \node[dgreen, below] at (1.7,-1) {\scriptsize$[{\mathbf{m}}]$};
    \filldraw[dgreen] (0,0) circle (2pt);
\end{tikzpicture} \,\,=\,\, \frac{1}{M^{n/2}}  \begin{tikzpicture}[scale=0.50,baseline = {(0,0)}]
    \draw[thick, dgreen, ->-=.5] (-1.7,-1) -- (0,0);
    \draw[thick, blue, -<-=.5] (0,0) -- (1.7,-1);
    \draw[thick, dgreen, -<-=.5] (0,2) -- (0,0);
          \node[isosceles triangle,scale=0.4, isosceles triangle apex angle=60, draw,fill=violet!60, rotate=90, minimum size =0.01cm] at (0,0){};
    \node[dgreen, above] at (0,2) {\scriptsize $\oSigma^{[0]}$};
    \node[dgreen, below] at (-1.7,-1) {\scriptsize $\oSigma^{[0]}$};
    \node[blue, below] at (1.7,-1) {\scriptsize $L_{m_i R_i}$};
\end{tikzpicture} + \cdots ~,
\no
\end{aligned}
\end{equation}
\begin{equation}
\begin{aligned}
    & \begin{tikzpicture}[scale=0.50,baseline = {(0,0)}]
    \draw[thick, dgreen, ->-=.5] (-1.7,-1) -- (0,0);
    \draw[thick, dgreen, -<-=.5] (0,0) -- (1.7,-1);
    \draw[thick, dgreen, -<-=.5] (0,2) -- (0,0);
    \node[dgreen, above] at (0,2) {\scriptsize$[\mathbf{m}]$};
    \node[dgreen, below] at (-1.7,-1) {\scriptsize$Q$};
    \node[dgreen, below] at (1.7,-1) {\scriptsize$\oQ$};
    \filldraw[dgreen] (0,0) circle (2pt);
\end{tikzpicture} \,\,=\,\, \frac{1}{M^{n/2}}  \begin{tikzpicture}[scale=0.50,baseline = {(0,0)}]
    \draw[thick, dgreen, ->-=.5] (-1.7,-1) -- (0,0);
    \draw[thick, dgreen, -<-=.5] (0,0) -- (1.7,-1);
    \draw[thick, blue, -<-=.5] (0,2) -- (0,0);
    \filldraw[violet!60] (0,0) circle (4pt);
    \node[blue, above] at (0,2) {\scriptsize $L_{m_i R_i}$};
    \node[dgreen, below] at (-1.7,-1) {\scriptsize $\Sigma^{[0]}$};
    \node[dgreen, below] at (1.7,-1) {\scriptsize $\oSigma^{[0]}$};
\end{tikzpicture} + \cdots ~, \quad \begin{tikzpicture}[scale=0.50,baseline = {(0,0)}]
    \draw[thick, dgreen, ->-=.5] (-1.7,-1) -- (0,0);
    \draw[thick, dgreen, -<-=.5] (0,0) -- (1.7,-1);
    \draw[thick, dgreen, -<-=.5] (0,2) -- (0,0);
    \node[dgreen, above] at (0,2) {\scriptsize$[{\mathbf{m}}]$};
    \node[dgreen, below] at (-1.7,-1) {\scriptsize$\oQ$};
    \node[dgreen, below] at (1.7,-1) {\scriptsize$Q$};
    \filldraw[dgreen] (0,0) circle (2pt);
\end{tikzpicture} \,\,=\,\, \frac{1}{M^{n/2}}  \begin{tikzpicture}[scale=0.50,baseline = {(0,0)}]
    \draw[thick, dgreen, ->-=.5] (-1.7,-1) -- (0,0);
    \draw[thick, dgreen, -<-=.5] (0,0) -- (1.7,-1);
    \draw[thick, blue, -<-=.5] (0,2) -- (0,0);
    \filldraw[violet!60] (0,0) circle (4pt);
    \node[blue, above] at (0,2) {\scriptsize $L_{m_i R_i}$};
    \node[dgreen, below] at (-1.7,-1) {\scriptsize $\oSigma^{[0]}$};
    \node[dgreen, below] at (1.7,-1) {\scriptsize $\Sigma^{[0]}$};
\end{tikzpicture} + \cdots ~, \\
    & \begin{tikzpicture}[scale=0.50,baseline = {(0,0)}]
    \draw[thick, dgreen, ->-=.5] (-1.7,-1) -- (0,0);
    \draw[thick, dgreen, -<-=.5] (0,0) -- (1.7,-1);
    \draw[thick, dgreen, -<-=.5] (0,2) -- (0,0);
    \node[dgreen, above] at (0,2) {\scriptsize$\oQ$};
    \node[dgreen, below] at (-1.7,-1) {\scriptsize$Q$};
    \node[dgreen, below] at (1.7,-1) {\scriptsize$Q$};
    \filldraw[dgreen] (0,0) circle (4pt);
    \node[dgreen, below] at (0,0) {\scriptsize $\alpha$};
\end{tikzpicture} \,\,=\,\, \frac{1}{M^{n/2}}  \begin{tikzpicture}[scale=0.50,baseline = {(0,0)}]
    \draw[thick, dgreen, ->-=.5] (-1.7,-1) -- (0,0);
    \draw[thick, dgreen, -<-=.5] (0,0) -- (1.7,-1);
    \draw[thick, dgreen, -<-=.5] (0,2) -- (0,0);
    \node[dgreen, above] at (0,2) {\scriptsize $\oSigma^{[0]}$};
    \node[dgreen, below] at (-1.7,-1) {\scriptsize $\Sigma^{[0]}$};
    \node[dgreen, below] at (1.7,-1) {\scriptsize $\Sigma^{[0]}$};
 \filldraw[violet!60] (0,0) circle (4pt);
    \node[violet!60, below] at (0,0) {\scriptsize $\alpha$};
\end{tikzpicture} + \cdots ~, \quad \begin{tikzpicture}[scale=0.50,baseline = {(0,0)}]
    \draw[thick, dgreen, ->-=.5] (-1.7,-1) -- (0,0);
    \draw[thick, dgreen, -<-=.5] (0,0) -- (1.7,-1);
    \draw[thick, dgreen, -<-=.5] (0,2) -- (0,0);
    \node[dgreen, above] at (0,2) {\scriptsize$Q$};
    \node[dgreen, below] at (-1.7,-1) {\scriptsize$\oQ$};
    \node[dgreen, below] at (1.7,-1) {\scriptsize$\oQ$};
    \filldraw[dgreen] (0,0) circle (4pt);
    \node[dgreen, below] at (0,0) {\scriptsize $\alpha$};
\end{tikzpicture} \,\,=\,\, \frac{1}{M^{n/2}}  \begin{tikzpicture}[scale=0.50,baseline = {(0,0)}]
    \draw[thick, dgreen, ->-=.5] (-1.7,-1) -- (0,0);
    \draw[thick, dgreen, -<-=.5] (0,0) -- (1.7,-1);
    \draw[thick, dgreen, -<-=.5] (0,2) -- (0,0);
    \node[dgreen, above] at (0,2) {\scriptsize $\Sigma^{[0]}$};
    \node[dgreen, below] at (-1.7,-1) {\scriptsize $\oSigma^{[0]}$};
    \node[dgreen, below] at (1.7,-1) {\scriptsize $\oSigma^{[0]}$};
    \filldraw[violet!60] (0,0) circle (4pt);
    \node[violet!60, below] at (0,0) {\scriptsize $\alpha$};
\end{tikzpicture} + \cdots ~. \\
\end{aligned}
\end{equation}
In other words, we see that in each indecomposable $\mathcal{L}_e$-module on the boundary, each simple object in the expansion appears once, and we have then chosen explicit representatives (namely $L_{m_i R_i}$, $\Sigma^{[0]}$, and $\oSigma^{[0]}$) in each $\mathcal{L}_e$-module and used our gauge freedom to set the junction coefficients of those representatives to be trivial. As mentioned previously, these particular representatives are actually closed under fusion and form a fusion subcategory in the bulk. This implies that the boundary $F$-symbols are given by the $F$-symbols of those representatives. For instance, let us consider the $F$-symbol
\begin{equation}
    \begin{tikzpicture}[baseline={([yshift=-.5ex]current bounding box.center)},vertex/.style={anchor=base,
    circle,fill=black!25,minimum size=18pt,inner sep=2pt},scale=0.6]
    \draw[dgreen, thick, ->-=.5] (-1.5,-1.5) -- (-0.5,-0.5);
    \draw[dgreen, thick, ->-=.5] (-0.5,-0.5) -- (0.5,0.5);
    \draw[dgreen, thick, ->-=.5] (0.5,0.5) -- (1.5,1.5);
    \draw[dgreen, thick, ->-=.5] (0.5,-1.5) -- (-0.5,-0.5);
    \draw[dgreen, thick, ->-=.5] (2.5,-1.5) -- (0.5,0.5);
    
    \node[dgreen, below] at (-1.5,-1.5) {\scriptsize $Q$};
    \node[dgreen, below] at (0.5,-1.5) {\scriptsize $Q$};
    \node[dgreen, below] at (2.5,-1.5) {\scriptsize $Q$};
    \node[dgreen, left] at (0.1,0.3) {\scriptsize $\oQ$};
      \filldraw[dgreen] (-0.5,-0.5) circle (3pt);
    \node[dgreen, below] at (-0.5,-0.5) {\scriptsize $\alpha$};
    \node[dgreen, above] at (1.5,1.5) {\scriptsize $[\mathbf{m}]$};
    \end{tikzpicture} = \sum_\beta \left[F_{Q,Q,Q}^{[\mathbf{m}]}\right]_{\alpha\beta} \begin{tikzpicture}[baseline={([yshift=-.5ex]current bounding box.center)},vertex/.style={anchor=base,circle,fill=black!25,minimum size=18pt,inner sep=2pt},scale=0.6]
    \draw[dgreen, thick, ->-=.5] (-1.5,-1.5) -- (0.5,0.5);
    \draw[dgreen, thick, ->-=.5] (0.5,0.5) -- (1.5,1.5);
    \draw[dgreen, thick, ->-=.5] (0.5,-1.5) -- (1.5,-0.5);
    \draw[dgreen, thick, ->-=.5] (2.5,-1.5) -- (1.5,-0.5);
    \draw[dgreen, thick, ->-=.5] (1.5,-0.5) -- (0.5,0.5);
    
    \node[dgreen, below] at (-1.5,-1.5) {\scriptsize $Q$};
    \node[dgreen, below] at (0.5,-1.5) {\scriptsize $Q$};
    \node[dgreen, below] at (2.5,-1.5) {\scriptsize $Q$};
    \node[dgreen, above] at (1.5,1.5) {\scriptsize $[\mathbf{m}]$};
          \filldraw[dgreen] (1.5,-0.5) circle (3pt);
    \node[dgreen, below] at (1.5,-0.5) {\scriptsize $\beta$};
    \node[dgreen, right] at (0.9,0.3) {\scriptsize $\oQ$};
    \end{tikzpicture} ~.
\end{equation}
After expanding the $\mathcal{L}_e$-modules in terms of simple objects, one can extract the coefficients $\left[F_{Q,Q,Q}^{[\mathbf{m}]}\right]_{\alpha\beta}$ from any diagram with the same incoming and outgoing lines. In particular, this means that 
\begin{equation}
    \begin{tikzpicture}[baseline={([yshift=-.5ex]current bounding box.center)},vertex/.style={anchor=base,
    circle,fill=black!25,minimum size=18pt,inner sep=2pt},scale=0.6]
    \draw[dgreen, thick, ->-=.5] (-1.5,-1.5) -- (-0.5,-0.5);
    \draw[dgreen, thick, ->-=.5] (-0.5,-0.5) -- (0.5,0.5);
    \draw[blue, thick, ->-=.5] (0.5,0.5) -- (1.5,1.5);
    \draw[dgreen, thick, ->-=.5] (0.5,-1.5) -- (-0.5,-0.5);
    \draw[dgreen, thick, ->-=.5] (2.5,-1.5) -- (0.5,0.5);
    
    \node[dgreen, below] at (-1.5,-1.5) {\scriptsize $\Sigma^{[0]}$};
    \node[dgreen, below] at (0.5,-1.5) {\scriptsize $\Sigma^{[0]}$};
    \node[dgreen, below] at (2.5,-1.5) {\scriptsize $\Sigma^{[0]}$};
    \node[dgreen, left] at (0.1,0.3) {\scriptsize $\oSigma^{[0]}$};
      \filldraw[violet!60] (-0.5,-0.5) circle (3pt);
           \filldraw[violet!60] (0.5,0.5) circle (2pt);
    \node[violet!60, below] at (-0.5,-0.5) {\scriptsize $\alpha$};
    \node[blue, above] at (1.5,1.5) {\scriptsize $L_{m_i R_i}$};
    \end{tikzpicture} = \sum_\beta \left[F_{Q,Q,Q}^{[\mathbf{m}]}\right]_{\alpha\beta} \begin{tikzpicture}[baseline={([yshift=-.5ex]current bounding box.center)},vertex/.style={anchor=base,circle,fill=black!25,minimum size=18pt,inner sep=2pt},scale=0.6]
    \draw[dgreen, thick, ->-=.5] (-1.5,-1.5) -- (0.5,0.5);
    \draw[blue, thick, ->-=.5] (0.5,0.5) -- (1.5,1.5);
    \draw[dgreen, thick, ->-=.5] (0.5,-1.5) -- (1.5,-0.5);
    \draw[dgreen, thick, ->-=.5] (2.5,-1.5) -- (1.5,-0.5);
    \draw[dgreen, thick, ->-=.5] (1.5,-0.5) -- (0.5,0.5);
    
    \node[dgreen, below] at (-1.5,-1.5) {\scriptsize $\Sigma^{[0]}$};
    \node[dgreen, below] at (0.5,-1.5) {\scriptsize $\Sigma^{[0]}$};
    \node[dgreen, below] at (2.5,-1.5) {\scriptsize $\Sigma^{[0]}$};
    \node[blue, above] at (1.5,1.5) {\scriptsize $L_{m_i R_i}$};
    \filldraw[violet!60] (1.5,-0.5) circle (3pt);
     \filldraw[violet!60] (0.5,0.5) circle (2pt);
    \node[violet!60, below] at (1.5,-0.5) {\scriptsize $\beta$};
    \node[dgreen, right] at (0.9,0.3) {\scriptsize $\oSigma^{[0]}$};
    \end{tikzpicture} ~,
\end{equation}
where the junction coefficients on both sides are trivial because of our gauge choice. Hence we conclude that
\begin{equation}
    \left[F_{Q,Q,Q}^{[\mathbf{m}]}\right]_{\alpha\beta} = \left[F_{\Sigma^{[0]},\Sigma^{[0]},\Sigma^{[0]}}^{L_{m_i R_i}}\right]_{\alpha\beta} ~.
\end{equation}
It is not hard to see that the same holds for all other $F$-symbols, and this simplifies the computation greatly when compared to the case of $\ST_1$. 
Using \eqref{eq:ST2_F_Symbol}, we identify the boundary $F$-symbols as
\begin{equation}
\begin{aligned}
    & F_{Q,\vec{n},\vec{n}'}^Q = \frac{R^{\vec{n}\cdot\vec{R},\Sigma^{[0]}}R^{\vec{n}'\cdot\vec{R},\Sigma^{[0]}}}{R^{(\vec{n}+\vec{n}')\cdot\vec{R},\Sigma^{[0]}}} \, \omega^{-n_1 n_1' - n_2 n_2' - n_2 n_1'} ~, \quad  
    F^{Q}_{\vec{n}, Q, \vec{n}'} = \omega^{-n_1 n_1' - n_2 n_2' - n_2 n_1'} ~, \quad  F^{Q}_{\vec{n}, \vec{n}',Q} = 1 ~,
    \\
    & F_{\overline{Q}, \vec{n},\vec{n}'}^{\overline{Q}} = \frac{R^{\vec{n}\cdot\vec{R},\oSigma^{[0]}}R^{\vec{n}'\cdot\vec{R},\oSigma^{[0]}}}{R^{(\vec{n}+\vec{n}')\cdot\vec{R},\oSigma^{[0]}}} \omega^{-(n_1+n_2)(n_1'+n_2')} ~, \quad 
    F^{\overline{Q}}_{\vec{n}, \overline{Q}, \vec{n}'} = \omega^{- n_1 n_1' - n_2 n_2' - n_1 n_2'} ~, \quad 
    F^{\overline{Q}}_{\vec{n}, \vec{n}',\overline{Q}} = \omega^{n_2 n_1'} ~.
\end{aligned}
\end{equation}
To see that the above result matches with \eqref{eq:boundaryFmethod2}, we observe that 
\begin{equation}
\begin{aligned}
    & \frac{F_{Q,\vec{n},\vec{n}'}^Q}{F_{Q,\vec{n}',\vec{n}}^Q} = \omega^{n_1 n_2' - n_2 n_1'} ~, \quad F^{Q}_{\vec{n}, Q, \vec{n}'} = \omega^{-n_1 n_1' - n_2 n_2' - n_2 n_1'} ~, \quad \frac{F^{Q}_{\vec{n}, \vec{n}',Q}}{F^{Q}_{\vec{n}', \vec{n},Q}} = 1 ~, \\
    & \frac{F_{\overline{Q}, \vec{n},\vec{n}'}^{\overline{Q}}}{F_{\overline{Q}, \vec{n}',\vec{n}}^{\overline{Q}}} = 1 ~, \quad F^{\overline{Q}}_{\vec{n}, \overline{Q}, \vec{n}'} = \omega^{- n_1 n_1' - n_2 n_2' - n_1 n_2'} ~, \quad \frac{F^{\overline{Q}}_{\vec{n}, \vec{n}',\overline{Q}}}{F^{\overline{Q}}_{\vec{n}', \vec{n},\overline{Q}}} = \omega^{n_2 n_1' - n_1 n_2'} ~,
\end{aligned}   
\end{equation}
and recall that
\begin{equation}
\begin{aligned}
    & \chi_{\ST_2} = \begin{pmatrix} 1 & 0 \\ 1 & 1 \end{pmatrix} ~, \quad \varphi_{\ST_2} = \begin{pmatrix} 0 & 1 \\ -1 & 0 \end{pmatrix} ~, \quad \widetilde{\varphi}_{\ST_2} = 0 ~, \\ 
    & \chi_{\overline{\ST_2}} = \begin{pmatrix} 1 & 1 \\ 0 & 1 \end{pmatrix} ~, \quad \varphi_{\overline{\ST_2}} = 0 ~, \quad \widetilde{\varphi}_{\overline{\ST_2}} = \begin{pmatrix} 0 & -1 \\ 1 & 0 \end{pmatrix} ~.
\end{aligned}
\end{equation}

\section{Obtaining other $F$-symbols from pentagon identities}

\label{sec:pentagons}

In the previous section, we discussed in great detail the $F$-symbols with a single incoming twist defect. In the current section, we show how this largely determines the form of the remaining $F$-symbols by making use of the pentagon identities. Though conceptually straightforward, this is computationally rather intensive, and hence we restrict in this section to the concrete example of triality defects. We find that the $F$-symbols for triality defects are parameterized is a similar manner as for duality defects, with a few distinctions we will highlight later.

\subsection{Classification of triality defects from SymTFT}
Let us first recall how to classify $N$-ality defects using the SymTFT \cite{Etingof:2009yvg,jordan2009classification}. As was already discussed in the introduction, the classification involves specifying a bulk $\mathbb{Z}_N$ symmetry mapping the purely electric Lagrangian algebra to a magnetic Lagrangian algebra, a choice of symmetry fractionalization classes for the chosen $\mathbb{Z}_N$ symmetry, and a discrete torsion for the $\mathbb{Z}_N$ symmetry.

Before proceeding, let us discuss a subtlety in the choice of the bulk $\mathbb{Z}_N$ symmetry. Clearly, to specify the $\mathbb{Z}_N$ symmetry, we only need to specify its generator, given by the matrix $A$ parameterized by the block matrix
\begin{equation}
    A = \begin{pmatrix} \A & \C \\ \B & \D \end{pmatrix}
\end{equation}
as discussed in Section \ref{sec:symmofbulk}. Notice that the component $\B$ must be invertible to ensure that the pure electric Lagrangian algebra is mapped to a magnetic Lagrangian algebra. As pointed out in \cite{jordan2009classification}, conjugating $A$ by a bulk symmetry of the form $U = \sfC_\alpha \sfT_\varphi$ given by \eqref{eq:bsC} and \eqref{eq:bsT} leads to equivalent $N$-ality defects. Indeed, conjugating $A$ by $\sfC_\alpha$ has the effect of relabeling the Abelian group elements by the corresponding group automorphism $\alpha$, and hence does not change the equivalence class of the $N$-ality defects. Likewise, let us consider conjugating a generic $N$-ality symmetry $A$ by some $\sfT_\varphi$, which gives
\begin{equation}
    \sfT_\varphi^{-1} A \,\sfT_\varphi = \begin{pmatrix} \A - \varphi \B & \A - \varphi \B + \C \B - \varphi \D \varphi \\ \B & \B \varphi + \D \end{pmatrix} ~. 
\end{equation}
Since $\B$ is invertible, we can always choose $\varphi = -\B^{-1} \D$ to remove the $\D$ component in $A$. Furthermore, the condition that the matrix $A$ preserves the braiding implies that
\begin{equation}\label{eq:p_ality_A}
    A = \begin{pmatrix} \A & \B^{-1,T} \\ \B & 0 \end{pmatrix} ~, \quad \A^T \B = - \B^T \A ~.
\end{equation}
Comparing this with the decomposition of a generic $A$ and its action on the partition function discussed in \eqref{eq:Adecomposition} and \eqref{eq:act_on_part}, we see that this gauge choice has a clear physical interpretation: it means that we do not stack an SPT before gauging, and instead stack a different SPT after gauging. Notice that the twisted partition function of any theory is ambiguous up to stacking an SPT phase for a background field. Assuming some specific choice, the twisted partition function is invariant under the transformation \eqref{eq:act_on_part}, and one can use such ambiguity to define a new twisted partition function of the same theory that is invariant under a new gauging operation, which does not involve stacking the SPT before gauging. 

\

Having addressed this subtlety, let us now demonstrate the classification procedure using the well-known example of duality defects. First, we must specify a $\mathbb{Z}_2^{em}$-symmetry in the bulk. Requiring that $A$ in \eqref{eq:p_ality_A} squares to the identity matrix implies that
\begin{equation}
    \A = 0 ~, \quad \B = \B^T ~.
\end{equation}
We see that the non-degenerate symmetric matrix $\B$ captures the data of the bilinear characters via $\chi(a,b) = e^{\frac{2\pi i}{M} g^T \B \,h}$. Once a specific $\mathbb{Z}_2^{em}$ symmetry is chosen by specifying $\B$, it can be shown using Shapiro's lemma that the choice of symmetry fractionalization class is unique (see e.g. \cite{Etingof:2009yvg}) and the choice of the discrete torsion $H^3(\mathbb{Z}_2,U(1)) \simeq \mathbb{Z}_2$ corresponds to the well-known choice of the FS indicator. Notice that different choices of the FS indicator can be related to each other by stacking an anomalous $\mathbb{Z}_2$ symmetry generator on the duality defect.

Note that we have yet to use the gauge transformation of conjugating $A$ by the symmetry $\sfC_\alpha$ in \eqref{eq:bsC}. This is related to the fact that in the classification of TY fusion categories, different symmetric non-degenerate bicharacters may lead to equivalent categories. For instance, consider the duality defect of $\mathbb{Z}_2 \times \mathbb{Z}_2$. It is well-known that there are two equivalent choices for the bicharacter, known as the diagonal one and the off-diagonal one. However, as pointed out in the original classification by Tambara and Yamagami \cite{tambara1998tensor}, $\mathbb{Z}_2\times \mathbb{Z}_2$ actually has four symmetric non-degenerate bicharacters, but three of them (including the off-diagonal one) are equivalent to each other under relabeling $\mathbb{Z}_2\times \mathbb{Z}_2$ (in other words, under group automorphism of $\mathbb{Z}_2\times \mathbb{Z}_2$), and thus are considered as equivalent. 

\

The classification of triality defects is similar, but with a few distinctions that we will now comment on. First, in the specification of the bulk $\mathbb{Z}_3$ symmetry, we now require that $A^3 = \mathds{1}$, which implies that
\begin{equation}\label{eq:triAB}
    \A = \B^{-1} \B^T ~, \quad (-\B^{-1}\B^T)^3 = \mathds{1} ~.
\end{equation}
We see that the triality symmetry in the bulk is again parameterized by a non-degenerate matrix $\B$, but now  it satisfies a more exotic condition $(-\B^{-1}\B)^3 = \mathds{1}$. This relation has two important implications. First, in the action of the triality defect on the partition function \eqref{eq:act_on_part}, we see that
\bea
\varphi = \B^{-1} \B^T \B^{-1}~, \hspace{0.3 in} \chi = (\B^{-1})^T ~,  \hspace{0.3 in}  \widetilde \varphi = 0~,
\eea
and the above relation guarantees that $\varphi$ is anti-symmetric as required, since 
\begin{equation}
    (-\B^{-1} \B^T)^3 = \mathds{1} \quad \Longleftrightarrow \quad \B^{-1} \B^T \B^{-1} = -(\B^{-1})^T\B(\B^{-1})^T ~.
\end{equation}
Second, it specifies an order-$3$ automorphisms $-\B^{-1}\B^T$ on the Abelian group, which will play an important role in the $F$-symbols. Note that such a $\B$ again leads to a non-degenerate bicharacter $\Upsilon(g,h) = e^{- \frac{2\pi i}{M}g^T (\B^{-1})^T \, h}$, but unlike the case of duality defects, the condition on $\B$ is not easy to express in terms of $\Upsilon$. Finally, one can ask, given an Abelian group $G$, what are the inequivalent $\B$ after the identification under the group automorphisms of $G$. For the case of $\gcd(|G|,3) = 1$, the classification is carried out in \cite{jordan2009classification}, but for more generic cases the answer is unknown. We will content ourselves with the statement that the bulk $\mathbb{Z}_3$-symmetry is specified by a non-degenerate bicharacter $\Upsilon(g,h) = e^{-\frac{2\pi i}{M}g^T (\B^{-1})^T \, h}$ with $(-\B^{-1}\B^T)^3 = \mathds{1}$.

Once a $\mathbb{Z}_3$ symmetry is chosen, one needs to pick a symmetry fractionalization class. Unlike the case of duality defects, when $\gcd(|G|,3) \neq 1$, the choice will no longer be unique. $F$-symbols with different choices of fractionalization class can also be related to each other using the SymSET as discussed in \cite{Lu:2025gpt}, and we will not dive into the details or give a generic closed-form expression in the current work. Finally, the choices of discrete torsion take values in $H^3(\mathbb{Z}_3,U(1)) = \mathbb{Z}_3$ and are related to each other by stacking an anomalous $\mathbb{Z}_3$ symmetry on the triality defect on the boundary.

\subsection{$F$-symbols from pentagon identities}
We are now ready to spell out the $F$-symbols with a given non-degenerate bicharacter 
\begin{equation}
\label{eq:Upsilondef}
    \Upsilon(g,h) = e^{-\frac{2\pi i}{M}g^T (\B^{-1})^T \, h} ~, \qquad (-\B^{-1} \B^T)^3 = \mathds{1} ~.
\end{equation}
To simplify the expressions, let us introduce a few more auxiliary quantities. First, since $(\B^{-1})^T\B (\B^{-1})^T$ is anti-symmetric, 
\begin{equation}
    \exp\left(\frac{2\pi i}{M} g^T (\B^{-1})^T\B (\B^{-1})^T h \right) ~, \qquad g,h \in G
\end{equation}
specifies a anti-symmetric bilinear character on $G$, and we choose a $2$-cocycle $\overline{\Xi}$ of $G$ constructed from it via
\begin{equation}
\label{eq:Xicondition}
    \frac{\overline{\Xi}(g,h)}{\overline{\Xi}(h,g)} = \exp\left(\frac{2\pi i}{M} g^T(\B^{-1})^T\B (\B^{-1})^T \, h \right) ~.
\end{equation}
Notice that the cohomology class of $\overline{\Xi}$ is determined by this expression. Next, let us choose a projective irreducible representation $\sigma(g)$ of $G$ twisted by the 2-cocycle $\overline{\Xi}$,
\begin{equation}
\label{eq:sigmaprojectivecond}
    \sigma(g) \cdot \sigma(h) = \frac{\overline{\Xi}(g,h)}{\overline{\Xi}(h,g)} \sigma(h) \cdot \sigma(g) ~.
\end{equation}
Notice that $\B$ being non-degenerate ensures the non-degeneracy of $\frac{\overline{\Xi}(g,h)}{\overline{\Xi}(h,g)}$, meaning that there is no central element in the projective representation. This in turn means that the irreducible representation is unique up to similarity transformation.

Let us consider a new projective representation of $G$ given by $\sigma(-\B^T \B^{-1} g)$. It is straightforward to check that 
\begin{equation}
    \sigma(-\B^T \B^{-1} g)\cdot \sigma(-\B^T \B^{-1} h) = \frac{\overline{\Xi}(g,h)}{\overline{\Xi}(h,g)} \sigma(-\B^T \B^{-1} h)\cdot \sigma(-\B^T \B^{-1} g) ~,
\end{equation}
using the above relation of $\overline{\Xi}$ in terms of $\B$. Since the irreducible projective representation is unique up to similarity transformation, there must exist an invertible matrix $S$ such that
\begin{equation}\label{eq:Scond}
    S \cdot \sigma(g) \cdot S^{-1} = \sigma(-\B^T \B^{-1}g) ~.
\end{equation}

The $F$-symbols can finally be written in terms of $\Upsilon,\overline{\Xi},\sigma(g),$ and $S$ as
\begin{equation}
\label{eq:fullFsymbols1}
\begin{aligned}
   & F_{g,h,Q}^Q = F_{\overline{Q},g,h}^{\overline{Q}} = 1 ~, \quad F_{Q,g,h}^Q = \frac{1}{\overline{\Xi}(g,h)} ~, \quad F_{g,h,\overline{Q}}^{\overline{Q}} = \overline{\Xi}(g,h) ~, \quad F_{g,Q, h}^Q = \Upsilon(g,h) ~, \quad F_{g ,\overline{Q}, h}^{\overline{Q}} = \Upsilon(h,g) ~,
   \\
       & F_{g,Q,\overline{Q}}^{h} = 1 ~, \quad F_{Q, g,\overline{Q}}^h = \Upsilon(h,g) ~, \quad F_{Q, \overline{Q}, g}^h = 1 ~, \\
    & F_{g,\overline{Q},Q}^{h} = \overline{\Xi}(h,g) ~, \quad F_{\overline{Q}, g,Q}^h = \Upsilon(g,h) ~, \quad F_{\overline{Q}, Q, g}^h = \frac{1}{\overline{\Xi}(g,h)} ~, \\
    & \left[F_{Q,\overline{Q},Q}^{Q}\right]_{g,h} = \frac{1}{\sqrt{|G|}} \frac{1}{\overline{\Xi}(h,h)\Upsilon(g,h)} ~, \quad \left[F_{\overline{Q},Q,\overline{Q}}^{\overline{Q}}\right]_{g,h} = \frac{1}{\sqrt{|G|}} \frac{\overline{\Xi}(g,g)}{\Upsilon(h,g)} ~,
\\
    & \left[F_{g,Q,Q}^{\overline{Q}}\right]_{\alpha\beta} = \sigma(g)_{\alpha\beta} ~, \quad \left[F_{Q, g,Q}^{\overline{Q}} \right]_{\alpha\beta} = [S \sigma(g)^{-1} S^{-1}]_{\alpha\beta} ~, \quad \left[F^{\overline{Q}}_{Q,Q,g}\right]_{\alpha\beta} = [S^{-1} \sigma(g) S]_{\alpha\beta} \\
    & \left[F_{g,\overline{Q},\overline{Q}}^{Q} \right]_{\alpha\beta} = [\sigma(g)]^{-1,T}_{\alpha\beta} ~, \quad \left[F_{\overline{Q},g,\overline{Q}}^{Q}\right]_{\alpha\beta} = [S^T \sigma(g)^{T} S^{-1,T}]_{\alpha\beta} ~, \quad \left[F_{\overline{Q},\overline{Q},g}^Q\right]_{\alpha\beta} = [S^{-1,T} \sigma(g)^{-1,T} S^T]_{\alpha\beta} ~,
\\
    & \left[F_{Q,Q,Q}^g\right]_{\alpha\beta} = \overline{\Xi}(g,g) [\sigma(g) S]_{\alpha\beta} ~, \quad \left[F_{\overline{Q},\overline{Q},\overline{Q}}^g\right]_{\alpha\beta} = \frac{1}{\overline{\Xi}(g,g)} [\sigma(g)^{-1,T} S^T]_{\alpha\beta} ~, \\
    & \left[F^{Q}_{Q,Q,\overline{Q}}\right]_{\alpha\beta,g} = \frac{1}{\sqrt{|G|}}[S \sigma(g) S^{-1}]_{\alpha\beta} ~,  \quad \left[F_{\overline{Q},\overline{Q},Q}^{\overline{Q}}\right]_{\alpha\beta,g} = \frac{1}{\sqrt{|G|}} \frac{1}{\overline{\Xi}(g,g)} [S^T \sigma(g)^{-1,T} S^{-1,T}]_{\alpha\beta} ~, \\
    & \left[F_{Q,\overline{Q},\overline{Q}}^{\overline{Q}}\right]_{g,\alpha\beta} = [S^{-1} \sigma(g)^{-1}]_{\alpha\beta} ~, \quad \left[F_{\overline{Q},Q,Q}^{Q}\right]_{g,\alpha\beta} = \overline{\Xi}(g,g) [S^{-1,T} \sigma(g)^{T}]_{\alpha\beta} ~.
\end{aligned}
\end{equation}
These results were obtained by directly solving the pentagon identities, though we do not give the details here. Instead, in the following subsection, we will give some simple consistency checks of these results by performing bulk $F$-symbol computations. 

Let us make a few comments on the above solution. First, the above $F$-symbols are only one solution to the pentagon equations---other inequivalent solutions corresponding to the bulk symmetry $\B$ can be acquired by changing the symmetry fractionalization class in the bulk (following the procedure in \cite{Lu:2025gpt}) and stacking anomalous $\mathbb{Z}_3$ symmetries on $Q$.\footnote{To make this precise, note that the choice of auxiliary variables $\overline{\Xi},\sigma(g),S$ is not unique as constrained by \eqref{eq:Xicondition}, \eqref{eq:sigmaprojectivecond}, and \eqref{eq:Scond}. The authors believe that different choices of $\overline{\Xi},\sigma(g),S$ will lead to gauge-equivalent $F$-symbols. This is clear for the case when the choice of fractionalization class is unique, since in that case the only freedom after choosing $\Upsilon$ is stacking with an anomalous $\mathbb{Z}_3$ symmetry, whereas it is straightforward to check that choosing different auxiliary variables does not implement such a stacking in the $F$-symbols. Thus in this case, the two must be gauge equivalent. We leave the explicit confirmation of the same statement in the case where there are multiple choices of fractionalization classes to future work. Nevertheless, even in this case, one can just pick some particular choice of $\overline{\Xi},\sigma(g),S$ and apply the procedure in \cite{Lu:2025gpt} to obtain the full set of gauge-inequivalent solutions.} In the case where the order of the boundary Abelian group is coprime with $3$, it follows from the Zassenhaus theorem that the choice of the symmetry fractionalization class is unique. Hence, any $F$-symbols for a fusion category with the triality fusion rules must be of the above form up to gauge transformation and stacking with an anomalous $\mathbb{Z}_3$ symmetry, even if this fusion category admits additional structure such as a $\mathbb{Z}_3$-crossed braiding structure. In particular, this means the $F$-symbols of the bulk SymSET with  $\ST_1$-symmetry for $\gcd(M,3) =1$, discussed previously, should be of the above form. We will check agreement for some of the bulk $F$-symbols explicitly in the next subsection.

Second, as a consistency check, one can show that the above $F$-symbols corresponding to the bulk symmetry specified by \eqref{eq:p_ality_A} and \eqref{eq:triAB} agree with the relations \eqref{eq:boundaryFmethod1} that we derived previously in Section \ref{sec:globalactionA}. To see this, note that for a bulk $\mathbb{Z}_3$ symmetry of the form \eqref{eq:p_ality_A} and \eqref{eq:triAB}, we have
\begin{equation}
    A = \begin{pmatrix} \B^{-1} \B^T & \B^{-1,T} \\ \B & 0 \end{pmatrix} ~, \hspace{0.3 in} \overline{A} = A^2 = \begin{pmatrix} 0 & \B^{-1} \\ \B^T & \B \B^{-1,T} \end{pmatrix} ~,
\end{equation}
from which we can read off the auxiliary variables \eqref{eq:chiphidefs} for $Q$ and $\oQ$,
\begin{equation}
\begin{array}{lll}
    \varphi_Q = \B^{-1} \B^T \B^{-1} ~, \quad & \chi_Q = \B^{-1,T} ~, \quad & \widetilde{\varphi}_Q = 0 ~, \\
    \varphi_{\oQ} = 0 ~, \quad & \chi_{\oQ} = \B^{-1} ~, \quad & \widetilde{\varphi}_{\oQ} = \B^{-1,T} \B \B^{-1,T} = - \B^{-1} \B^T \B^{-1} ~.
\end{array}
\end{equation}
Then using \eqref{eq:boundaryFmethod1}, we see that the boundary $F$-symbols satisfy
\begin{equation}
\begin{array}{lll}
    \frac{F^{Q}_{Q,g,h}}{F^Q_{Q,h,g}} = \omega^{ g^T \B^{-1} \B^T \B^{-1} h} ~, \quad & F^{Q}_{g,Q,h} = \omega^{-g^T \B^{-1,T} h} ~, \quad & \frac{F^{Q}_{g,h,Q}}{F^{Q}_{h,g,Q}} = 1 ~, \\
    \frac{F^{\oQ}_{\oQ, g,h}}{F^{\oQ}_{\oQ ,h,g}} = 1 ~, \quad & F^{\oQ}_{g, \oQ, h} = \omega^{-g^T\B^{-1}h} ~, \quad & \frac{F^{\oQ}_{g,h,\oQ}}{F^{\oQ}_{h,g,\oQ}} = \omega^{-g^T \B^{-1} \B^T \B^{-1} h} ~,
\end{array}
\end{equation}
which indeed agrees with \eqref{eq:fullFsymbols1} upon using the definitions \eqref{eq:Upsilondef} and \eqref{eq:Xicondition}. Notice that this consistency check also ensures that the $F$-symbols we acquired previously for $\ST_1$ and $\ST_2$ from the bulk in Section \ref{sec:bulktoboundaryFsymb} are consistent with \eqref{eq:boundaryFmethod1}.

Finally, even though the above discussion  focused on $G = \mathbb{Z}_M^{n}$, the $F$-symbols we have obtained hold for a generic Abelian group $G$, as long as $|G|$ is a perfect square. Again, one needs to specify a non-degenerate bicharacter $\Upsilon: G \times G \rightarrow U(1)$. In the more general case, the condition \eqref{eq:Upsilondef} and $\overline{\Xi}$ can be phrased as follows. First, using $\Upsilon$ we define two group homomorphisms $\widehat{\Upsilon}, \widehat{\Upsilon}^T$ from $G$ to $\widehat{G} = \mathrm{Hom}(G,U(1))$ via
\begin{equation}
    \widehat{\Upsilon}(a) := \Upsilon(a,\cdot ):G\rightarrow U(1) ~, \qquad \widehat{\Upsilon}^T(a):= \Upsilon(\cdot,a): G\rightarrow U(1) ~.
\end{equation}
Since $\Upsilon$ is non-degenerate, $\widehat{\Upsilon}^{-1}$ exists, and the condition \eqref{eq:Upsilondef} on $\Upsilon$ becomes that $(\widehat{\Upsilon}^T\circ\widehat{\Upsilon}^{-1})^3$ equals the charge conjugation operation $\widehat{C}$ (which maps any group element to its inverse) on $\widehat{G}$. The group homomorphism $C \circ \widehat{\Upsilon} \circ \widehat{\Upsilon}^{T,-1} \circ \widehat{\Upsilon}:G \rightarrow \widehat{G}$ naturally leads to a bilinear form on $G$ and appears on the RHS of \eqref{eq:Xicondition} as
\begin{equation}
    \frac{\overline{\Xi}(g,h)}{\overline{\Xi}(h,g)} = [C \circ \widehat{\Upsilon} \circ \widehat{\Upsilon}^{T,-1}\circ \widehat{\Upsilon}(g)](h) ~.
\end{equation}

\subsection{Bulk consistency checks}

We now describe how we may use the bulk perspective to check some of the above results. Here we will not be exhaustive---we will content ourselves with checking just a handful of the above $F$-symbols, with the full bulk computation left as an exercise to the reader. We begin by recalling the bulk formulas given in (\ref{eq:bulkFsymbols1}) for $\Sigma = \Sigma^{[0]}$ in the case of $\ST_1$ with $\gcd(M,3) = 1$, which we reproduce here for convenience, 
\bea\label{eq:F1S}
    F^{\Sigma}_{\Sigma, \bu, \widetilde{\bu}} = \frac{R^{\bu,\Sigma}R^{\widetilde{\bu},\Sigma}}{R^{\bu + \widetilde{\bv},\Sigma}} \frac{\mu_A(\widetilde{\bu},\bu)}{R^{\widetilde{\bu},\bu}} ~, \quad F^{\Sigma}_{\bu,\Sigma, \widetilde{\bu}} = \frac{1}{R^{\widetilde{\bu},\bu}} \frac{\mu_A(\widetilde{\bu},\bu)}{\mu_A(\bu,\widetilde{\bu})} ~, \quad F^{\Sigma}_{\bu,\widetilde{\bu},\Sigma} = \mu_A(\bu,\widetilde{\bu})^{-1} ~.
\eea
One also straightforwardly computes their $\overline \Sigma$ counterparts, 
\bea\label{eq:F1oS}
    F^{\overline \Sigma}_{\overline \Sigma, \bu, \widetilde{\bu}} = \frac{R^{\bu,\overline\Sigma}R^{\widetilde{\bu},\overline\Sigma}}{R^{\bu + \widetilde{\bu},\overline\Sigma}} \frac{\mu_{A^2}(\widetilde{\bu},\bu)}{R^{\widetilde{\bu},\bu}} ~, \quad F^{\overline\Sigma}_{\bu,\overline\Sigma, \widetilde{\bu}} = \frac{1}{R^{\widetilde{\bu},\bu}} \frac{\mu_{A^2}(\widetilde{\bu},\bu)}{\mu_{A^2}(\bu,\widetilde{\bu})} ~, \quad F^{\overline\Sigma}_{\bu,\widetilde{\bu},\overline\Sigma} = \mu_{A^2}(\bu,\widetilde{\bu})^{-1} ~.\hspace{0.2 in}
\eea
As pointed out previously, in this case the line operators including the twist defects form a triality fusion category if we neglect the $\mathbb{Z}_3$-crossed braided structure. Therefore, we expect the $F$-symbols computed here to match with \eqref{eq:fullFsymbols1}. At first glance, this does not seem to be the case, but we still have the freedom to perform gauge transformations, and we choose to do so as follows, 
\bea
\label{eq:matchingaugetransf1}
 \begin{tikzpicture}[scale=0.50,baseline = {(0,0)}]
    \draw[thick, blue, ->-=.5] (-1.7,-1) -- (0,0);
    \draw[thick, blue, -<-=.5] (0,0) -- (1.7,-1);
    \draw[thick, blue, -<-=.5] (0,2) -- (0,0);
    \node[blue, above] at (0,2) {\scriptsize $L_{\bu + \widetilde \bu}$};
    \node[blue, below] at (-1.7,-1) {\scriptsize $L_{\bu}$};
    \node[blue, below] at (1.7,-1) {\scriptsize $L_{\widetilde \bu}$};
\end{tikzpicture} \,\,\,\longrightarrow\,\,\, \mu_A(\bu, \widetilde \bu) 
 \begin{tikzpicture}[scale=0.50,baseline = {(0,0)}]
    \draw[thick, blue, ->-=.5] (-1.7,-1) -- (0,0);
    \draw[thick, blue, -<-=.5] (0,0) -- (1.7,-1);
    \draw[thick, blue, -<-=.5] (0,2) -- (0,0);
 \node[blue, above] at (0,2) {\scriptsize $L_{\bu + \widetilde \bu}$};
    \node[blue, below] at (-1.7,-1) {\scriptsize $L_{\bu}$};
    \node[blue, below] at (1.7,-1) {\scriptsize $L_{\widetilde \bu}$};
\end{tikzpicture} ~,
\eea
as well as 
\bea
\label{eq:matchingaugetransf2}
 \begin{tikzpicture}[scale=0.50,baseline = {(0,0)}]
    \draw[thick, dgreen, ->-=.5] (-1.7,-1) -- (0,0);
    \draw[thick, blue, -<-=.5] (0,0) -- (1.7,-1);
    \draw[thick, dgreen, -<-=.5] (0,2) -- (0,0);
     \node[isosceles triangle,scale=0.4, isosceles triangle apex angle=60, draw,fill=violet!60, rotate=90, minimum size =0.01cm] at (0,0){};
    \node[dgreen, above] at (0,2) {\scriptsize $\Sigma$};
    \node[dgreen, below] at (-1.7,-1) {\scriptsize $\Sigma$};
    \node[blue, below] at (1.7,-1) {\scriptsize $L_{ \bu}$};
\end{tikzpicture} \,\,\,\longrightarrow\,\,\, \kappa(\bu) 
 \begin{tikzpicture}[scale=0.50,baseline = {(0,0)}]
    \draw[thick, dgreen, ->-=.5] (-1.7,-1) -- (0,0);
    \draw[thick, blue, -<-=.5] (0,0) -- (1.7,-1);
    \draw[thick, dgreen, -<-=.5] (0,2) -- (0,0);
     \node[isosceles triangle,scale=0.4, isosceles triangle apex angle=60, draw,fill=violet!60, rotate=90, minimum size =0.01cm] at (0,0){};
    \node[dgreen, above] at (0,2) {\scriptsize $\Sigma$};
    \node[dgreen, below] at (-1.7,-1) {\scriptsize $\Sigma$};
    \node[blue, below] at (1.7,-1) {\scriptsize $L_{ \bu}$};
\end{tikzpicture} ~, \hspace{0.5 in}
 \begin{tikzpicture}[scale=0.50,baseline = {(0,0)}]
    \draw[thick, dgreen, ->-=.5] (-1.7,-1) -- (0,0);
    \draw[thick, blue, -<-=.5] (0,0) -- (1.7,-1);
    \draw[thick, dgreen, -<-=.5] (0,2) -- (0,0);
     \node[isosceles triangle,scale=0.4, isosceles triangle apex angle=60, draw,fill=violet!60, rotate=90, minimum size =0.01cm] at (0,0){};
    \node[dgreen, above] at (0,2) {\scriptsize $\overline\Sigma$};
    \node[dgreen, below] at (-1.7,-1) {\scriptsize $\overline\Sigma$};
    \node[blue, below] at (1.7,-1) {\scriptsize $L_{ \bu}$};
\end{tikzpicture} \,\,\,\longrightarrow\,\,\, \overline\kappa(\bu) 
 \begin{tikzpicture}[scale=0.50,baseline = {(0,0)}]
    \draw[thick, dgreen, ->-=.5] (-1.7,-1) -- (0,0);
    \draw[thick, blue, -<-=.5] (0,0) -- (1.7,-1);
    \draw[thick, dgreen, -<-=.5] (0,2) -- (0,0);
     \node[isosceles triangle,scale=0.4, isosceles triangle apex angle=60, draw,fill=violet!60, rotate=90, minimum size =0.01cm] at (0,0){};
    \node[dgreen, above] at (0,2) {\scriptsize $\overline\Sigma$};
    \node[dgreen, below] at (-1.7,-1) {\scriptsize $\overline\Sigma$};
    \node[blue, below] at (1.7,-1) {\scriptsize $L_{ \bu}$};
\end{tikzpicture} ~, \eea
where $\kappa(\bu)$ and $\overline \kappa(\bu)$ are chosen to satisfy 
\bea
\label{eq:kappaconsist}
{\kappa(\bu + \widetilde \bu) \over \kappa(\bu) \kappa(\widetilde \bu)} = {R^{\bu, \Sigma} R^{\widetilde \bu, \Sigma} \over R^{\bu + \widetilde \bu, \Sigma}} {1\over R^{\widetilde \bu, \bu}}{\mu_A(\widetilde \bu, \bu) \over \mu_{A^2}(\bu , \widetilde \bu)}~, \hspace{0.2 in}{\overline\kappa(\bu + \widetilde \bu) \over \overline\kappa(\bu) \overline\kappa(\widetilde \bu)} = {R^{\bu,\overline \Sigma} R^{\widetilde \bu,\overline \Sigma} \over R^{\bu + \widetilde \bu, \overline\Sigma}} {1\over R^{\widetilde \bu, \bu}}{\mu_{A^2}(\widetilde \bu, \bu) \over \mu_{A}(\bu , \widetilde \bu)}~.
\eea
In this gauge, it is straightforward to check that the bulk $F$-symbols take the following form, 
\bea
    F^{\Sigma}_{\Sigma, \bu, \widetilde{\bu}} &=& {\mu_{A^2}(\bu, \widetilde \bu) \over \mu_A(\bu, \widetilde \bu) }~, \quad F^{\Sigma}_{\bu,\Sigma, \widetilde{\bu}} = \frac{1}{R^{\widetilde{\bu},\bu}} \frac{\mu_A(\widetilde{\bu},\bu)}{\mu_A(\bu,\widetilde{\bu})} ~, \quad F^{\Sigma}_{\bu,\widetilde{\bu},\Sigma} = 1 ~,
\no\\
    F^{\overline \Sigma}_{\overline \Sigma, \bu, \widetilde{\bu}} &=& 1 ~, \quad F^{\overline\Sigma}_{\bu,\overline\Sigma, \widetilde{\bu}} = \frac{1}{R^{\widetilde{\bu},\bu}} \frac{\mu_{A^2}(\widetilde{\bu},\bu)}{\mu_{A^2}(\bu,\widetilde{\bu})} ~, \quad F^{\overline\Sigma}_{\bu,\widetilde{\bu},\overline\Sigma} ={ \mu_{A}(\bu,\widetilde{\bu}) \over \mu_{A^2}(\bu,\widetilde{\bu})}~.
\eea
Comparing these to the results in (\ref{eq:fullFsymbols1}), we see that they are of the same general form, with
\bea
\label{eq:explicitXiUpsilon}
\overline \Xi(\bu, \widetilde \bu) = {\mu_A(\bu, \widetilde \bu) \over \mu_{A^2}(\bu, \widetilde \bu)}~, \hspace{0.4 in} \Upsilon(\bu, \widetilde \bu) = {1\over R^{\widetilde \bu, \bu}}{\mu_A(\widetilde \bu, \bu) \over \mu_A(\bu, \widetilde \bu)}~,  
\eea
upon noting the identity 
\bea
{\mu_{A^2}( \bu, \widetilde \bu) \over \mu_{A^2}(\widetilde\bu,  \bu)} = {R^{\bu, \widetilde \bu} \over R^{\widetilde \bu, \bu}} {\mu_A(\widetilde \bu, \bu) \over \mu_A(\bu, \widetilde \bu)}~,
\eea
which can be checked e.g. using the explicit form of $\mu_A$ and $\mu_{A^2}$ given in (\ref{eq:muAmuA2}). 

Having fixed the bulk versions of $\overline \Xi(\bu, \widetilde \bu)$ and $\Upsilon(\bu, \widetilde \bu)$, we may now try to use the bulk perspective to verify other $F$-symbols. Consider, for example, the $F$-symbol $F_{\overline \Sigma, \bu, \Sigma}^{\widetilde \bu}$, corresponding to the following, 
\bea\label{eq:def_FoSuS}
    \begin{tikzpicture}[baseline=20,square/.style={regular polygon,regular polygon sides=4},scale=1.3]
        \shade[top color=orange!40, bottom color=orange!10]  (0,0)--(1,1)--(2,0)--(0,0);
        \draw[dgreen,thick, ->-=0.5] (0,0) --(0.5,0.5); 
        \draw[dgreen,thick, ->-=0.5] (0.5,0.5) --(1,1); 
        \draw[blue,thick, ->-=0.5] (1,1) --(1.5,1.5); 
            \draw [gray!30, line width=3pt] (1,0) -- (0.5,0.5); 
        \draw[blue,thick, ->-=0.5](1,0) -- (0.5,0.5); 
        \draw[dgreen,thick, ->-=0.5](2,0) -- (1,1); 
        \node[orange,below] at (1.5,0.4) {\footnotesize{$D_{A^2}$}};
        \node[isosceles triangle,scale=0.4,isosceles triangle apex angle=60,draw,fill=violet!60,rotate=90,minimum size =0.01cm] at (0.5,0.5){};
        \node[dgreen, below] at (0,0) {\footnotesize{$\overline\Sigma$}};
        \node[blue, above] at (1.5,1.5) {\footnotesize{$L_{\widetilde \bu}$}};
        \node[blue, below] at (1,0) {\footnotesize{$L_\bu$}};
        \node[dgreen, below] at (2,0) {\footnotesize{$\Sigma$}};
\end{tikzpicture} = F_{\overline \Sigma, \bu, \Sigma}^{\widetilde \bu}
    \begin{tikzpicture}[baseline=20,square/.style={regular polygon,regular polygon sides=4},scale=1.3]
               \shade[top color=orange!40, bottom color=orange!10]  (0,0)--(1,1)--(2,0)--(0,0);
        \draw[dgreen,thick, ->-=0.6] (0,0) --(1,1); 
        \draw[blue,thick, ->-=0.5] (1,1) --(1.5,1.5); 
          \draw [gray!30, line width=3pt] (1,0) -- (1.5,0.5); 
            \draw[blue,thick, ->-=0.5](1,0) -- (1.5,0.5); 
        \draw[dgreen,thick, ->-=0.5](2,0) -- (1.5,0.5); 
         \draw[dgreen,thick, ->-=0.5](1.5,0.5) -- (1,1); 
        \node[orange,below] at (0.6,0.4) {\footnotesize{$D_{A^2}$}};
        \node at (1.5,0.5) [square,draw,fill=violet!60,scale=0.5] {}; 
        \node[dgreen, below] at (0,0) {\footnotesize{$\overline\Sigma$}};
        \node[blue, above] at (1.5,1.5) {\footnotesize{$L_{\widetilde \bu}$}};
        \node[blue, below] at (1,0) {\footnotesize{$L_\bu$}};
        \node[dgreen, below] at (2,0) {\footnotesize{$\Sigma$}};
\end{tikzpicture} ~.
\eea
We may now make use of the trivalent vertices (\ref{eq:ex1n1}), (\ref{eq:ex1n2}), and (\ref{eq:ex1n3}) in Appendix \ref{app:trivalentvertices} to expand the diagram in terms of simple lines $L_{\bu}$, the $F$-symbols of which are trivial. This then gives rise to the following result,
\bea
F_{\overline \Sigma, \bu, \Sigma}^{\widetilde \bu} = {R^{\bu, \Sigma} R^{\bu, \overline \Sigma} \over \beta(\bu)} {1\over R^{\bu, \widetilde \bu - \bu}}{\mu_{A^2}(\bu, \widetilde \bu) \over \mu_{A^2}(\widetilde \bu, \bu)}~, \eea
where $\beta(\bu)$ is required to satisfy 
\bea
\label{eq:betaconsist}
{\beta (\bu + \widetilde \bu) \over \beta(\bu) \beta(\widetilde \bu)} = {R^{\bu + \widetilde \bu, \Sigma} \over R^{\bu, \Sigma} R^{\widetilde \bu, \Sigma}} R^{\bu, \widetilde \bu} {\mu_{A^2}(\widetilde \bu, \bu) \over \mu_A(\bu, \widetilde \bu)} ~.
\eea
Let us briefly explain the parameter $\beta(\bu)$ here; for full computational details, the reader can refer to Appendix \ref{app:trivalentvertices}. In order to compute $F_{\oSigma,\bu,\Sigma}^{\widetilde{\bu}}$, in \eqref{eq:def_FoSuS} we use the presentation of $\Sigma$ as a right boundary of the $A^2$-surface operator, in contrast with \eqref{eq:junc_vSw}, where $\Sigma$ appears as a left boundary of the $A$-surface operator. Recall that the module condition can only fix the junction up to an overall phase factor; and from the module conditions, we are left with two phase parameters in the two presentations of the local junction $L_{\bu} \otimes \Sigma \rightarrow \Sigma$. We have made a gauge choice for the presentation in \eqref{eq:junc_vSw}, and the consistency of the $F$-symbols requires the other presentation to be in the same gauge choice. Therefore, the overall phase in the other presentation is no longer a free parameter and we introduce the phase $\beta(\bu)$ to keep track of it.\footnote{This $\beta(\bu)$ is \textit{not} the same as $\beta_{\bv, [\bw]}(\bu)$ in \eqref{eq:junc_vSw}, nor is it related to the coefficients $\beta^i_j$ in (\ref{eq:abdef}). } One way to constrain $\beta(\bu)$ is to require that the $F$-symbols \eqref{eq:F1S} and \eqref{eq:F1oS} computed in the other presentation must remain the same; this leads to the above conditions on $\beta(\bu)$ in \eqref{eq:betaconsist}.

We now shift to the same gauge as before, namely we perform the gauge transformations (\ref{eq:matchingaugetransf1}), (\ref{eq:matchingaugetransf2}), which sends
\bea
F_{\overline \Sigma, \bu, \Sigma}^{\widetilde \bu} \longrightarrow \overline \kappa (\bu) F_{\overline \Sigma, \bu, \Sigma}^{\widetilde \bu}~
\eea
and then choose 
\bea
\overline \kappa(\bu) =  \frac{\beta(\bu)}{R^{\bu, \Sigma} R^{\bu, \overline \Sigma} R^{\bu, \bu}} ~, 
\eea
which is consistent with (\ref{eq:kappaconsist}) and (\ref{eq:betaconsist}). This gives 
\bea
F_{\overline \Sigma, \bu, \Sigma}^{\widetilde \bu} = {1 \over R^{\widetilde \bu, \bu}} {\mu_A(\widetilde \bu, \bu) \over \mu_A(\bu, \widetilde \bu)} = \Upsilon(\bu, \widetilde \bu)~, 
\eea
consistent with the general structure predicted by the pentagon identities. 

Having checked one of the scalar $F$-symbols, we next check one of the matrix $F$-symbols, namely $[F_{\bu, \Sigma, \Sigma}^{\overline \Sigma}]_{\alpha\beta}$, which is captured by the following diagram, 
\bea
    \begin{tikzpicture}[baseline=20,square/.style={regular polygon,regular polygon sides=4},scale=1.3]
        \shade[top color=blue!40, bottom color=blue!10]  (1,0)--(0.5,0.5)--(1,1)--(2,0)--(1,0);
        \shade[top color=orange!40, bottom color=orange!10]  (2,0)--(1,1)--(1.5,1.5)--(2,1.5)--(2,0);
        \draw[blue,thick, ->-=0.5] (0,0) --(0.5,0.5); 
        \draw[dgreen,thick, ->-=0.5] (0.5,0.5) --(1,1); 
        \draw[dgreen,thick, ->-=0.5] (1,1) --(1.5,1.5); 
        \draw[dgreen,thick, ->-=0.5](1,0) -- (0.5,0.5); 
        \draw[dgreen,thick, ->-=0.5](2,0) -- (1,1); 
        \node[blue,below] at (1.5,0.4) {\footnotesize{$D_{A}$}};
         \node[orange,below] at (1.7,1.5) {\footnotesize{$D_{A^2}$}};
        \node at (0.5,0.5) [square,draw,fill=violet!60,scale=0.5] {};
        \node[blue, below] at (0,0) {\footnotesize{$L_{\bu}$}};
        \node[dgreen, above] at (1.5,1.5) {\footnotesize{$\overline \Sigma$}};
        \node[dgreen, below] at (1,0) {\footnotesize{$\Sigma$}};
        \node[dgreen, below] at (2,0) {\footnotesize{$\Sigma$}};
         \filldraw[violet!60] (1,1) circle (1.5pt);
         \node[violet!60,left] at (1,1) {\scriptsize$\alpha$};
\end{tikzpicture} = \sum_\beta \left[F_{ \bu, \Sigma,\Sigma}^{\overline \Sigma}\right]_{\alpha\beta}
    \begin{tikzpicture}[baseline=20,square/.style={regular polygon,regular polygon sides=4},scale=1.3]
         \shade[top color=blue!40, bottom color=blue!10]  (1,0)--(1.5,0.5)--(2,0)--(1,0);
          \shade[top color=orange!40, bottom color=orange!10]  (2,0)--(1,1)--(1.5,1.5)--(2,1.5)--(2,0);
        \draw[blue,thick, ->-=0.6] (0,0) --(1,1); 
        \draw[dgreen,thick, ->-=0.5] (1,1) --(1.5,1.5); 
        \draw[dgreen,thick, ->-=0.5](1,0) -- (1.5,0.5); 
        \draw[dgreen,thick, ->-=0.5](2,0) -- (1.5,0.5); 
         \draw[dgreen,thick, ->-=0.5](1.5,0.5) -- (1,1); 
        \node[blue,below] at (1.5,0.4) {\footnotesize{$D_{A}$}};
          \node[orange,below] at (1.7,1.5) {\footnotesize{$D_{A^2}$}};
        \node at (1,1) [square,draw,fill=violet!60,scale=0.5] {}; 
          \filldraw[violet!60] (1.5,0.5) circle (1.5pt);
          \node[violet!60,right] at (1.5,0.5) {\scriptsize$\beta$};
        \node[blue, below] at (0,0) {\footnotesize{$L_{\bu}$}};
        \node[dgreen, above] at (1.5,1.5) {\footnotesize{$\overline \Sigma $}};
        \node[dgreen, below] at (1,0) {\footnotesize{$\Sigma$}};
        \node[dgreen, below] at (2,0) {\footnotesize{$\Sigma$}};
\end{tikzpicture} ~.
\eea
By using the junctions (\ref{eq:ex2n1}), (\ref{eq:ex2n2}), and (\ref{eq:ex2n3}) given in Appendix \ref{app:trivalentvertices} to expand in terms of simple lines, we obtain the following expression for the $F$-symbols, 
\bea
 \left[F_{ \bu, \Sigma,\Sigma}^{\overline \Sigma}\right]_{\alpha \beta} = {\mu_A(\bu, \bu_1) \mu_A(\bu + \bu_1, \bu_2) \over \mu_A(\bu_1,\bu_2) \mu_{A^2}(\bu, \bu_1+\bu_2)}[\rho(\bu_1+\bu_2) \rho(\bu+\bu_1+\bu_2)^{-1}]_{\beta \alpha} 
\eea
where $\rho(\bu) := \rho^{A-A^2}(\bu)$ is the little group representation satisfying (\ref{eq:projrepcond}), i.e. 
\bea
\label{eq:specialcaserhoproj}
\rho(\bu) \rho(\widetilde \bu) = {\mu_{A^2}(\bu, \widetilde \bu) \over \mu_A(\bu, \widetilde \bu)} \rho(\bu+ \widetilde \bu)~.
\eea
Using this, we may simplify
\bea
\left[F_{ \bu, \Sigma,\Sigma}^{\overline \Sigma}\right]_{\alpha \beta} = [\rho(\bu)]^{-1, T}_{\alpha \beta}
~.
\eea
We would now like to check that this matches with the general solution of the pentagon identity. In that case, we found that $ \left[F_{g,Q,Q}^{\overline{Q}}\right]_{\alpha\beta} = \sigma(g)_{\alpha\beta}$, with $\sigma(g)$ satisfying (\ref{eq:sigmaprojectivecond}). In order for this to match with the current result, we see that we must have 
\bea
\rho(\bu)^{-1, T}\rho(\widetilde\bu)^{-1, T} = {\overline \Xi (\bu, \widetilde \bu) \over \overline \Xi(\widetilde \bu, \bu)} \rho(\widetilde\bu)^{-1, T} \rho(\bu)^{-1, T}~, 
\eea
which, comparing to (\ref{eq:specialcaserhoproj}), is satisfied as long as
\bea
{\overline \Xi (\bu, \widetilde \bu) \over \overline \Xi(\widetilde \bu, \bu)} = {\mu_A(\bu, \widetilde \bu) \over  \mu_{A^2} (\bu, \widetilde \bu)  } { \mu_{A^2}(\widetilde \bu, \bu)  \over \mu_A(\widetilde \bu, \bu) }~. 
\eea
This equality indeed holds for $\overline \Xi (\bu, \widetilde \bu)$ as given in (\ref{eq:explicitXiUpsilon}). 

\section{$(2+1)$d Symmetry TFT for $N$-ality defects }\label{sec:SymTFT}
In this final section, we discuss the gauging of the $\ZZ_N$ symmetry in the bulk $\ZZ_M^n$ gauge theory, thereby giving the SymTFT for $N$-ality defects. We will content ourselves with obtaining the spectrum of line operators of the bulk theory and their fusions for the special case of $N = p$ being a prime number---a more complete discussion of the SymTFT will be given in upcoming work \cite{upcoming}. 

\subsection{Gauging order-$p$ elements of $O(n,n;M)$}
Upon gauging a $\ZZ_{p}$ subgroup of $O(n,n; M)$ generated by an element $A$ of prime order $|A|=p$, we obtain a bulk theory with line operators of the following types:
\begin{itemize}
    \item Invertible lines coming from $L_{\bv}$ for which $\bv$ is left invariant by $A$. In other words, these are lines which are in the mod $M$ kernel of $1-A$. Such lines can be stacked with the generator $K$ of the quantum one-form symmetry $\widehat \ZZ_{p}$. We denote the resulting lines by $\widehat L_{\bv}^k$ for $k=0, \dots, p-1$. There are a total of $p \times |\mathrm{ker}_M(1-A)| = p \times M^{n_A}$ such lines.\footnote{Note that here we restrict to the case where the entries $s_i$ in the Smith normal form of $1-A$ are either multiples of $M$ or coprime with $M$. The more general formula for the size of the kernel is given by \eqref{eqn:KernelSize}.}
    
    \item Lines coming from $L_{\bv}$ for which $\bv  \in \ZZ_M^{2n}$ is not left invariant by $A$. In this case, we must sum over appropriate orbits of lines in the ungauged theory in order to obtain something gauge invariant. Since the order of $A$ is a prime number, the gauge invariant combination must take the form $L_\mathbf{v} + L_{A \mathbf{v}}+ \dots + L_{A^{p-1}\mathbf{v}}$ and we denote it by $\widehat L_{[\![\mathbf{v}]\!]}$. Clearly, $\widehat L_{[\![\mathbf{v}]\!]}$ is the same line as $\widehat L_{[\![\mathbf{v}']\!]}$ if $A^I \bv = \bv'$ for some $I$, and we use the double brackets to indicate this equivalence relation, in order to distinguish it from the notation $[\bw]$ used for elements of $\mathrm{coker}_M(1-A)$. Clearly, it has quantum dimension $p$. In total, there are $\frac{1}{p}\left(M^{2n} - |\mathrm{ker}(1-A)|\right)$ such lines.

    Note that unlike the lines in the first bullet point, $\widehat L_{[\![\mathbf{v}]\!]}$ can absorb factors of $K$---that is, any configuration with an insertion of $\widehat L_{[\![\mathbf{v}]\!]}$ is indistinguishable from the analogous configuration with an insertion of coincident $\widehat L_{[\![\mathbf{v}]\!]}$ and $K$. To see this, note that by definition, when $K$ encircles $n$ units of $A$-flux, it produces a factor of $e^{2 \pi i n \over p}$. On the other hand, when $\widehat L_{[\![\mathbf{v}]\!]}$ encircles such a configuration, it produces a vanishing result, since the individual lines that make up $\widehat L_{[\![\mathbf{v}]\!]}$ (namely $L_{\mathbf{v}}, \dots, L_{A^{p-1}\mathbf{v}}$) are not invariant under any $A^I$ for $I \neq 0 \mod p$, and hence cannot form closed loops. Therefore, the loop of $\widehat L_{[\![\mathbf{v}]\!]}$ vanishes in any correlation function whenever $K$ would produce a non-zero result, and hence the effects of $K$ cannot be detected. In this sense the insertion of $\widehat L_{[\![\mathbf{v}]\!]}$ is indistinguishable from the insertion of coincident $\widehat L_{[\![\mathbf{v}]\!]}$ and $K$; see Figure \ref{fig:LvabsorbK}.

    \begin{figure}[!tbp]
	\centering
\[
 \raisebox{-2em}{\begin{tikzpicture}
 \draw[thick] (0,0) circle (20pt);
  \draw[thick, red] (0,0) circle (18pt);
\node at (-0.8,-0.7) {$\widehat L_{[\![\mathbf{v}]\!]}$};
\node[red] at (-0.05,-0.1) {$K$};
\end{tikzpicture}}
\hspace{0.5 in} = \hspace{0.5 in}
 \raisebox{-2em}{\begin{tikzpicture}
\draw[thick] (0,0) circle (20pt);
\node at (1.1,-0.7) {$\widehat L_{[\![\mathbf{v}]\!]}$};
\end{tikzpicture}}
\]
	\caption{The line $\widehat L_{[\![\mathbf{v}]\!]}$ can absorb the line $K$ since whenever the $K$ loop (red) is non-trivial, the $\widehat L_{[\![\mathbf{v}]\!]}$ loop (black) vanishes.}
	\label{fig:LvabsorbK}
\end{figure}
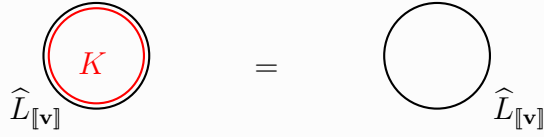

    To unify this case with the case in the first bullet point, we may introduce the spaces
    \bea
    \label{eq:cVdef}
    \cV_{d,M}(A) := \left\{ \mathbf{v} \in \ZZ_M^{2n} \,\Big |\,\, A^{p\over d} \mathbf{v} =   \mathbf{v}   \quad\mathrm{and} \quad A^\ell \mathbf{v} \neq   \mathbf{v}  \,\, \mathrm{for}\,\, \ell <{p\over d}\right\}~,
    \eea
    defined for $d\,\big|\,p$ (i.e. $d = 1$ or $p$). The gauge invariant lines from the Abelian anyons can be collectively denoted as $\widehat L^k_{[\![\mathbf{v}_d]\!]}$ where $\mathbf{v}_d \in \cV_{d,M}(A)$ and $k$ runs from $0,\cdots d-1$. The invariant lines correspond to $d = p$ while the non-invertible lines correspond to $d = 1$.\footnote{Note that it is not hard to generalize the above discussion to show that the same parameterization works for $A$ with generic order as well.}

    \item Finally, there are lines coming from twist defects. Note that the surfaces $D_{A^I}$ become transparent after gauging $D_A$, and hence twist defects become genuine lines. When $|A|$ is prime, any twist defect $\Sigma_{A^I}^{[\mathbf{w}]}$ is invariant under the $A$-action in the pre-gauged theory.\footnote{When $|A|$ is not a prime number, then $A$ can act non-trivially on the twist defects. For instance, consider the case of  $M = 16, n=1$ for which the ungauged  theory is $\ZZ_{16}$ gauge theory, and consider the symmetry $A = \mathrm{diag}(3,11)$. Under this $\mathbb{Z}_4$ symmetry, the twist defect $\Sigma_{A^2}^{[(1,0)]}$ is mapped to $\Sigma_{A^2}^{[(3,0)]}$, and therefore is not invariant under the $A$ action since $[(1,0)] \neq [(3,0)]$ in $\mathrm{coker}(1-A^2)$.} To see this, note that $D_A$ maps $\Sigma_{A^I}^{[\mathbf{w}]}$ to $\Sigma_{A^I}^{[A\mathbf{w}]}$, but because $[A\mathbf{w}] = [\mathbf{w} + (1-A)(-\mathbf{w})] = [\mathbf{w}]$ by the fact that $\mathrm{im}_M(1-A) = \mathrm{im}_M(1-A^I)$ when $|A| = p$, we see that both $[\mathbf{w}]$ and $[A\mathbf{w}]$ parameterize the same class in $\mathrm{coker}_M(1-A^I)$ , and hence $\Sigma_{A^I}^{[\mathbf{w}]} = \Sigma_{A^I}^{[A\mathbf{w}]}$. 
   
    After gauging, the resulting genuine line defects can be stacked with an arbitrary number of $K$ to produce new defects, giving lines $\widehat \Sigma^{[\bw], k}_{A^I}$, where $k= 0, \dots, N-1$ labels the number of factors of $K$ included. In total, there are $p \times \sum_{I=1}^{p-1} |\mathrm{ker}_M(1-{A})| = p(p-1) M^{n_A} $ such lines, each with quantum dimension $\sqrt{\frac{M^{2n}}{|\mathrm{ker}(1-A)|}} = M^{r_{A}\over 2}$. 
\end{itemize}

Let us summarize the discussion so far. The lines can be organized into two types,

\begin{itemize}
\item Line operators $\widehat L^k_{[\![\mathbf{v}_d]\!]}$ with $k = 0, \dots, d-1$, where $\mathbf{v}_d \in \cV_{d,M}(A)$ and $d \,\big|\, p$. 
These have quantum dimension ${p\over d}$.

\item Line operators $\widehat \Sigma^{[\mathbf{w}], k}_{A^I}$ with $k = 0, \dots, N-1$, where $\mathbf{w} \in \mathrm{coker}_M(1-A^I)$. 
These have quantum dimension $M^{r_{A^I}\over 2}$. 
\end{itemize}

As we said above, in the case that $N = p$ is prime, the spectrum of lines of the first type is particularly simple. Indeed, the only divisors of $N = p$ are $d=1$ and $p$ itself, 
and hence there are two types of lines: invertible lines labelled by $\mathbf{v}_p \in \cV_{p,M}(A) = \mathrm{ker}_M(1-A)$, and quantum dimension $p$ lines 
labelled by $\mathbf{v}_1 \in \cV_{1,M}(A) =\ZZ_M^{2n} / \mathrm{ker}_M(1-A)$. The former can be stacked with $K$ lines and come with a label $k=1,\dots, p$, 
whereas the latter absorb all $K$ and hence do not come with an additional label. Including operators $\widehat \Sigma^{[\mathbf{w}], k}_{A^I}$, we find that the 
total quantum dimension is 
\bea
\sum_{a} d_a^2 &=& \overbrace{p \times M^{n_A} \times 1^2}^{\widehat L^k_{[\![\mathbf{v}_p]\!]}} + \overbrace{{1\over p} (M^{2n} - M^{n_A}) \times p^2}^{\widehat L_{[\![\mathbf{v}_1]\!]}} + \overbrace{p(p-1)\times M^{n_A} \times \left(M^{r_{A}\over 2}\right)^2}^{\widehat \Sigma^{[\mathbf{w}], k}_{A^I}}
\no\\
&=& p M^{2n} +  p(p-1) M^{n_A+r_A}  
\no\\
&=& p^2 M^{2n}~,
\eea
where we have made use of the rank-nullity theorem, i.e. $r_A + n_A = 2n$, as well as the fact that for $N = p$, 
the quantities $r_{A^i}$ and $n_{A^i}$ are independent of $i$. 
Thus we obtain precisely the square of the total quantum dimension for the $p$-ality extension of $\ZZ^{n}_M$.

\subsection{Fusion rules }

Having discussed the line operators in the SymTFT, we now give their fusion rules.

\paragraph{Fusion rules involving only $\widehat L^i_{[\![\mathbf{v}]\!]}$:} 

In the ungauged theory, the fusion of $\widehat L^i_{[\![\mathbf{v}_d]\!]} \otimes \widehat L^j_{[\![\mathbf{v}_{d'}]\!]}$ involves $\frac{p}{d}\times \frac{p}{d'}$ distinct junctions between simple lines.  These individual junctions can be labeled according to the subspace in which the outgoing lines reside. After promoting each outgoing line to a gauge-invariant combination and removing redundancies in the construction, we obtain the following fusion rules in the gauged theory,
\bea
\widehat L^i_{[\![\mathbf{v}_p]\!]} \otimes \widehat L^j_{[\![\mathbf{v}'_{p}]\!]}=\widehat L^{i+j}_{[\![\mathbf{v}_{p}+\mathbf{v}'_{p}]\!]}~,
\label{eqn:Fusionpp}
\eea
\bea
\widehat L^i_{[\![\mathbf{v}_p]\!]} \otimes \widehat L_{[\![\mathbf{v}'_{1}]\!]}=\widehat L_{[\![\mathbf{(v+v')}_{1}]\!]}~,
\eea
\bea
\widehat L_{[\![\mathbf{v}_{1}]\!]} \otimes \widehat L_{[\![\mathbf{v}'_{1}]\!]}=\bigoplus_{n=0}^{p-1} \left\{ \begin{matrix}
    \widehat L_{[\![(A^n \mathbf{v}+\mathbf{v}')_{1}]\!]} &&\mathrm{if}~~A^n \mathbf{v}+\mathbf{v}'\notin ~~\mathrm{ker}(1-A)~,
    \\
    \bigoplus_{i=0}^{p-1}\widehat L^i_{[\![(A^n \mathbf{v}+\mathbf{v}')_p]\!]}&&\mathrm{if}~~A^n \mathbf{v}+\mathbf{v}'\in ~~\mathrm{ker}(1-A)~.
\end{matrix}
\right.
\eea
The distribution of $K$ lines is determined by the following observations. First, the junction of three simple lines is invariant under the action of $A$ in the pre-gauged theory, which is detected by wrapping a $D_A$ surface around the junction. Therefore, no Wilson lines, i.e. $K$ lines, can end on the junction after gauging, and thus the number of $K$-lines is conserved in \eqref{eqn:Fusionpp}. Second, note that $\widehat L_{[\![\mathbf{v}_{1}]\!]} \otimes \widehat L_{[\![\mathbf{v}'_{1}]\!]}$ can absorb a $K$ line (since each of the $\widehat L_{[\![\mathbf{v}_{1}]\!]}$ can), and therefore the right-hand side of the fusion must also be able to absorb a $K$-line, either through another $\widehat L_{[\![\mathbf{v}_{1}]\!]}$ or through summing over $q$ labels for $\widehat L^q_{[\![\mathbf{v}_{p}]\!]}$. 

These three fusion channels can be summarized in a single formula,\footnote{Note that this formula together with its derivation also holds for $A$ with generic order $N$.}
\bea
\widehat L^i_{[\![\mathbf{v}_d]\!]} \otimes \widehat L^j_{[\![\mathbf{v}_{d'}]\!]}= \frac{\mathrm{gcd}(d,d')}{p}\bigoplus_{n\in \ZZ_{p/d}, n'\in\ZZ_{{p}/{d'}}} \bigoplus_{k=0}^{{d(n,n')\over \mathrm{gcd}(d,d')}-1}  &\widehat L^{i+j+k\, \mathrm{gcd}(d,d')~\text{mod}~ d(n,n')}_{[\![A^n\mathbf{v}_d +  A^{n'} \mathbf{v}_{d'}]\!]}
\\
\mathrm{where}&\,\,A^n\mathbf{v}_d + A^{n'} \mathbf{v}_{d'} \in \cV_{d(n,n')}  ~,
\no
\eea
where $d(n,n')$ is introduced to keep track of the grading associated with the fusion of two outgoing lines ($d(n,n')=1$ or $p$ in our case). We do not have a closed form expression for the $d(n,n')$, but they can be determined by direct computation in any concrete case. In the pre-gauged theory, there are $p/d(n,n')$ different fusions that all give rise to the same line in the gauged theory. To account for this redundancy, we divide by $N/d(n,n')$. Moreover, when summing over possible stackings with $K$ lines, we further divide by their number $d(n,n')/\mathrm{gcd}(d,d')$.\footnote{Here, $\frac{d(n,n')}{\mathrm{gcd}(d,d')}$ is a positive integer for the following reason. First, $A^{\frac{N}{\mathrm{gcd}(d,d')}}(A^n \bv_d + A^{n'}\bv_{d'}) = A^n \bv_d + A^{n'}\bv_{d'}$ because $\frac{N}{\mathrm{gcd}(d,d')} = \mathrm{lcm}\left(\frac{N}{d},\frac{N}{d'}\right)$. On the other hand, $\frac{N}{d(n,n')}$ is the minimal positive integer such that $A^{\frac{N}{d(n,n')}}(A^n \bv_d + A^{n'}\bv_{d'}) = A^n \bv_d + A^{n'}\bv_{d'}$ by definition. Therefore, $\frac{N}{\mathrm{gcd}(d,d')}$ must be an integer multiple of $\frac{N}{d(n,n')}$, which implies that $\frac{d(n,n')}{\mathrm{gcd}(d,d')}$ is an integer.} Combining these factors yields the final prefactor.
It is easy to check that with this prefactor the quantum dimensions on both sides match---indeed, the quantum dimension on the left is ${p \over d} \times {p \over d'}$, while that on the right-hand side is 
\begin{equation}
    {\mathrm{gcd}(d, d') \over p} \sum_{n=1}^{p/d} \sum_{n' = 1}^{p / d'}  {d (n,n') \over \mathrm{gcd}(d,d')} {p \over d(n, n')}= {p \over d} \times {p \over d'}
\end{equation}
as expected.

\paragraph{Fusion rules involving $\widehat L^i_{[\![\mathbf{v}_d]\!]}$ and $\widehat \Sigma^{[\mathbf{w}],i}_{A^I}$:} 
\label{sec:SigmaLfusion}
The fusion between $\widehat L^i_{[\![\mathbf{v}_d]\!]}$ and $\widehat \Sigma^{[\mathbf{w}],i}_{A^I}$ mostly follows from the fact that, before gauging, 
the lines $L_{\mathbf{v}}$ combine with twist defects to give new twist defects as in (\ref{eq:generaltwistdefectdef}), and takes the form
\bea
\widehat L^i_{[\![\mathbf{v}_d]\!]} \otimes \widehat \Sigma^{[\mathbf{w}],j}_{A^I} = \bigoplus_{k=0}^{{p\over d}-1} \widehat \Sigma^{[\mathbf{v}_d + \mathbf{w}],\,i+j+d k+\delta}_{A^I}~,
\label{eqn:LSigmafusion}
\eea
where $[\mathbf{v}_d]$ and $[\mathbf{w}]$ are elements of $\mathrm{coker}_M(1-A)$, and in particular $[\mathbf{v}_d ]$ is the element of $\mathrm{coker}_M(1-A)$ which has as a representative $\mathbf{v}_d \in \cV_{d,M}(A)$.\footnote{For these fusion rules to be consistent, we must have that $[\bv_d + \bw] \in \mathrm{coker}_M(1-A^I)$ is independent of the choice $\bv_d \in [\![\mathbf{v}_d]\!]$. This is indeed the case for $|A| = p$. For example, for $d = 1$, the coset $[\bv_1 + \bw] \in \mathrm{coker}_M(1-A^I)$ is independent of the choice $\bv_1 \in [\![\mathbf{v}_1]\!]$, since considering a different choice $A^J\bv_1$ ($J\neq 0 \mod p$) gives $[A^J\bv_1 + \bw] = [\bv_1 + \bw + (1-A^J)(-\bv_1)] = [\bv_1+ \bw] $, where we have used the fact that when $|A| = p$ the spaces $\mathrm{im}_M(1-A^x)$ are identical for all $x \neq 0$. Note that we do not have this property for generic order $N$, and hence in that case the fusion rules will become more complicated.} The main novelty of these fusion rules is in the assignment of factors of $K$. First, the sum on the right-hand side is chosen to be 
consistent with the fact that $\widehat L^i_{[\![\mathbf{v}_d]\!]}$ can absorb $K^d$, and hence the right-hand side must also be able to absorb $K^d$. 
The sum on the right-hand side is also consistent with the fact that the quantum dimension of the left-hand side is ${p \over d} \times M^{n_A\over 2}$. Second, the factor of $\delta$ captures a possible shift in naive $K$-line number conservation. In this paper, we will only give a brief discussion on the role of this shift, focusing on situations where $\delta$ is absent, and leave the complete discussion to \cite{upcoming}.

Let us first consider the non-invertible line $\widehat L_{[\![\bv_1]\!]}$, which consists of $p$ lines generated by a $\mathbb{Z}_p$ action of $A$. Its fusion with the twist defects contains fusion channels corresponding to each different representation of the $\mathbb{Z}_p$ quantum symmetry. As a result, $\delta$ does not play any role in these fusion rules, and we always have
\bea
\widehat L_{[\![\mathbf{v}_1]\!]}\otimes \widehat \Sigma^{[\mathbf{w}],i}_{A^I} = \bigoplus_{k=0}^{p-1} \widehat \Sigma^{[\mathbf{v}_1 + \mathbf{w}],\, k}_{A^I}~.
\eea

For $\widehat L_{[\![\bv_p]\!]}^k$ where $\mathbf{v}_p\in \mathrm{ker}_M(1-A)$, first consider the fusion between $\widehat L_{[\![\bv_p]\!]}^k$ and minimal twist defects $\widehat \Sigma^{[ \mathbf{0}],0}_{A^I}$. The outcome of the fusion requires knowledge of the map from $\mathrm{ker}_M(1-A)$ to $\mathrm{coker}_M(1-A)$, 
\bea
\mathbf{v}_p\to [\mathbf{v}_p]~.
\eea
As discussed in \cite{Kaidi:2022cpf}, when this is an injective map, 
every twist defect can be unambiguously defined by this fusion 
\bea
\widehat \Sigma^{ [\mathbf{v}_p],q}_{A^I}:= \widehat L^{q}_{[\![\mathbf{v}_p]\!]}\otimes\widehat \Sigma^{[ \mathbf{0}],0}_{A^I} ~.
\eea
The other fusion rules then follow from this definition, together with the fusion rules between the $\widehat L$, and thus automatically preserve the total number of $K$-lines. 

However, if the map fails to be injective, one encounters an ambiguity: distinct $\mathbf v_p$ may map to the same $[\mathbf v_p]$, leading to multiple ways of defining the same twist defect. In particular, this happens when there exists a non-trivial $\widehat{\mathbf{u}}_p \in \mathrm{ker}_M(1-A)$ such that $[\widehat{\mathbf{u}}_p]=0$. In other words, $\widehat{\mathbf{u}}_p \in \mathrm{ker}_M(1-A) \cap \mathrm{im}_M(1-A)$. This intersection can be non-trivial since we are working modulo $M$. Consider two inequivalent fusions 
\bea
\widehat \Sigma^{[\mathbf{w+v_p}],q'}_{A^I}= \widehat \Sigma^{[\mathbf{w}],q_0}_{A^I}\otimes \widehat L^{q}_{[\![\mathbf{v}_p]\!]} ~~~\mathrm{and}~~ ~\widehat \Sigma^{[\mathbf{w+v_p}],q''}_{A^I}= \widehat \Sigma^{[\mathbf{w}],q_0}_{A^I}\otimes \widehat L^{q}_{[\![\bv_p + \widehat{\mathbf{u}}_p]\!]}~.
\eea 
Only one of them can be fixed by definition (e.g. setting $q'=q+q_0$ for convenience), and the difference $q''-q'$ then becomes a physically meaningful observable. 
To determine the difference, we can first measure the charge $\delta$ of the fusion junction $ \Sigma^{[\mathbf{w}]}= \Sigma^{[\mathbf{w}]}\otimes  L_{\widehat{\mathbf{u}}_p}$ under $D_A$ in the ungauged theory. After gauging, gauge invariance requires attaching $\delta$ of the $K$ lines to the junction.

In what follows, we restrict ourselves to the simpler situation where this ambiguity does not arise, i.e. $\delta$ is trivial, so that $K$-line number conservation always holds. We explicitly determine the fusion rules in the main examples discussed in this paper, namely $\mathsf{ST}_1$ and $\mathsf{ST}_2$ triality defects in $\ZZ_M \times \ZZ_M$ gauge theory. In particular, when $\gcd(M,3)=1$ the intersection contains only the zero vector and thus the ambiguity vanishes.\footnote{More generally, one can show that when $\gcd(M,p)=1$, the intersection between the kernel and image spaces is trivial. To see this, consider $\widehat{\mathbf{u}}_p\in \mathrm{ker}_M(1-A) \cap \mathrm{im}_M(1-A)$, i.e. $(1-A)\widehat{\mathbf{u}}_p=0$ and $\widehat{\mathbf{u}}_p=(1-A)\mathbf{v},\,\mathbf{v}\neq 0$. Using the identity $1-A^k=(1+A+\dots+A^{k-1})(1-A)$ and applying $k=|A|=p$, we find 
\bea
0=(1-A^{|A|})\mathbf{v}=(1+A+\dots+A^{|A|-1})\mathbf{u}=|A|\mathbf{u}~~\Rightarrow~~|A| \mathbf{u}=0~~~~\mathrm{mod}~~M~.
\eea
When $|A|$ and $M$ are coprime, multiplying $|A| \mathbf{u}=0$ by $|A|^{-1}$ shows that $\mathbf{u}$ is the zero vector. Therefore, the intersection is trivial, and hence the ambiguity discussed above cannot occur and the fusion rules can be defined so that the number of $K$-lines is conserved.

On the other hand, when $|A|$ and $M$ have a common divisor, $|A| \mathbf{u}=0~\,\mathrm{mod}~M$ does not force $\mathbf{u}$ to vanish, and we may see a nontrivial intersection. }
The results are then as follows,
\begin{itemize}
\item \textbf{$\mathsf{ST}_1$ for $\gcd(M,3) = 1$}: When $\gcd(M,3) = 1$, the matrix $1-{\mathsf{ST}_1}$ has full rank, and hence the only ${\mathsf{ST}_1}$-invariant line is the trivial one $\widehat L_{[\![\mathbf{0}]\!]}$ and the theory contains only the minimal twist defect. In this case, the fusion rules are given simply by
\bea
\widehat L^i_{[\![\mathbf{0}]\!]}\otimes \widehat\Sigma^{j}_{(\mathsf{ST}_1)^I}=\widehat \Sigma^{i+j}_{(\mathsf{ST}_1)^I}~, \hspace{0.5 in}\widehat L_{[\![\mathbf{v}_1]\!]}\otimes \widehat \Sigma^{i}_{(\mathsf{ST}_1)^I} = \bigoplus_{k=0}^{2} \widehat \Sigma^{k}_{(\mathsf{ST}_1)^I}~.
\eea

\item \textbf{${\mathsf{ST}_2}$ for $\mathrm{gcd}(M,3)=1$:} For $\mathsf{ST}_2$ triality defects, the Smith normal form of $1-{\mathsf{ST}_2}$ is given by $\mathrm{diag}(1,1,0,0)$, and thus $|\mathrm{ker}_M(1-{\mathsf{ST}_2})|=M^2$ (where we choose the basis vectors to be $\mathbf{k}^1 = (1,1,0,1)$ and $\mathbf{k}^2 = (1,0,1,0)$) and $|\mathrm{coker}_M(1-{\mathsf{ST}_2})|=M^2$. The spectrum then contains $3\times M^2$ lines of the form $\widehat L^k_{[\![\mathbf{v}_3]\!]}$, $\tfrac{1}{3}(M^4-M^2)$ lines of the form $\widehat L_{[\![\mathbf{v}_1]\!]}$, and $2\times 3\times M^2$ twist defects $\widehat \Sigma^{[\mathbf{w}],k}_{A^{I}}$. The fusion rules are given by
\bea
\widehat L^i_{[\![\mathbf{v}_3]\!]}\otimes \widehat \Sigma^{[\mathbf{w}],j}_{(\mathsf{ST}_2)^{I}} = \widehat \Sigma^{[\mathbf{w} + \mathbf{v}_3],i+j}_{(\mathsf{ST}_2)^{I}}~,\hspace{0.3 in}
\widehat L_{[\![\mathbf{v}_1]\!]}\otimes \widehat \Sigma^{[\mathbf{w}],i}_{(\mathsf{ST}_2)^{I}} &=& \bigoplus_{k=0}^{2} \widehat \Sigma^{[\mathbf{w}+\mathbf{v}_1],k}_{(\mathsf{ST}_2)^{I}}~.
\eea
\end{itemize}

\paragraph{Fusion rules involving only $\widehat \Sigma^{[\mathbf{v}],i}_{A^I}$:}
We finally turn to the fusions of the twist defects with themselves. The calculation proceeds in three steps. First, we use the fusion rules established in the last section to decompose each twist defect $\widehat \Sigma^{[\mathbf{v}],i}_{A^I}$ into a bare twist defect and an $\widehat L$ line carrying all of the quantum symmetry labels. The fusion of the bare twist defects follows straightforwardly from \eqref{eq:twistdefectfusions} in the ungauged theory, and we may then fuse this back with the $\widehat L$ to get our final result. Here we present some examples,
\begin{itemize}
\item \textbf{$\mathsf{ST}_1$ for $\mathrm{gcd}(M,3)=1$:}
\bea
\widehat \Sigma^{i}_{(\mathsf{ST}_1)^I}\otimes \widehat \Sigma^{j}_{(\mathsf{ST}_1)^J}= \left\{\begin{matrix}\widehat L_{[\![\mathbf{0}]\!]}^{i+j}\oplus\underset{\substack{[\![\mathbf{v}_1]\!],
\\
\,\mathbf{v}_1\in\mathrm{im}_{M}(1-{\mathsf{ST}_1})\backslash\mathrm{ker}_{M}(1-{\mathsf{ST}_1})}}{\bigoplus} \widehat L_{[\![\mathbf{v}_1]\!]} && I+J = 0 \,\,\mathrm{mod}\,\,3~
\\
\\
M^{2}\,\, \widehat\Sigma_{(\mathsf{ST}_1)^{I+J}}^{i+j} && \mathrm{otherwise}~\end{matrix}\right.
\eea
where $\mathrm{im}_{M}(1-{\mathsf{ST}_1})\backslash\mathrm{ker}_{M}(1-{\mathsf{ST}_1})$ denotes the set difference of the image space and kernel space. In particular, we are excluding the intersection $\mathrm{im}_M(1-{\mathsf{ST}_1})\cap\,\mathrm{ker}_M(1-{\mathsf{ST}_1})$ from $\mathrm{im}_M(1-{\mathsf{ST}_1})$ corresponding to the unique $\mathsf{ST}_1$-invariant line, which in this case consists only of the zero vector. All remaining elements group into triples and are identified as lines in $\mathcal{V}_{1,M}(\mathsf{ST}_1)$.
\item \textbf{$\mathsf{ST}_2$ for $\mathrm{gcd}(M,3)=1$:}
\bea
\,&&\widehat \Sigma^{[\mathbf{w}],i}_{(\mathsf{ST}_2)^I}\otimes \widehat \Sigma^{[\mathbf{w}'],j}_{(\mathsf{ST}_2)^J}\nonumber
\\
= &&\left\{\begin{matrix}\widehat L_{[\![\bar{\mathbf{w}} + \bar{\mathbf{w}}']\!]}^{i+j}\oplus\underset{\substack{[\![\mathbf{v}_1]\!],
\\
\,\mathbf{v}_1\in\mathrm{im}_{M}(1-{\mathsf{ST}_2})\backslash\mathrm{ker}_{M}(1-{\mathsf{ST}_2})}}{\bigoplus} \widehat L_{[\![\mathbf{v}_1+\bar{\mathbf{w}}+\bar{\mathbf{w}}']\!]} && I+J = 0 \,\,\mathrm{mod}\,\,3~,
\\
\\
M\,\, \widehat\Sigma_{(\mathsf{ST}_2)^{I+J}}^{[\mathbf{w} + \mathbf{w}'],i+j} && \mathrm{otherwise}~.\end{matrix}\right.
\eea
As discussed above, when the kernel and image of the operator $(1-A)$ intersect trivially, i.e. $\mathrm{ker}_M(1-\mathsf{ST}_2)\cap\mathrm{im}_M(1-\mathsf{ST}_2)=\{\mathbf{0}\}$, a natural isomorphism exists between the kernel and the cokernel. We use $\bar{\mathbf{w}}$ to represent the unique preimage of $[\mathbf{w}]\in\mathrm{coker}_M(1-\mathsf{ST}_2)$ in $\mathrm{ker}_M(1-\mathsf{ST}_2)$ under this isomorphism. Concretely, if we choose a representative of $[\mathbf{w}]\in\mathrm{coker}_M(1-\mathsf{ST}_2)$ to be $\mathbf{w}=(x,y,0,0)$, then we define $\bar{\mathbf{w}}=\frac{2y-x}{3}\mathbf{k}^1+\frac{2x-y}{3}\mathbf{k}^2$, with $\mathbf{k}^1, \mathbf{k}^2$ the basis vectors of $\mathrm{ker}_M(1-A)$ described previously.
Note that in the above formula, we have $[\![\bar{\mathbf{w}}+\bar{\mathbf{w}}']\!]\in \mathcal{V}_{3,M}(\mathsf{ST}_2)$ and $[\![\mathbf{v}_1+\bar{\mathbf{w}}+\bar{\mathbf{w}}']\!]\in \mathcal{V}_{1,M}(\mathsf{ST}_2)$.
\end{itemize}

\section*{Acknowledgements}
JK is supported by the INAMORI Frontier Program at Kyushu University. S.S. is supported by Grant-in-Aid for JSPS Fellows No.~23KJ1533. Z.S. is supported by the Simons Collaboration on Global Categorical Symmetries.

\newpage
\begin{appendix}

\section{Fusion rules of $D_A$}

\label{app:DAfusions}

In this appendix we check that the defects $D_A$ defined in (\ref{eq:defofDA}) and (\ref{eq:QAdefab}) (for which $1-A$ is assumed to be full rank) satisfy the following fusion rules,
\bea
\label{eq:desiredfusions}
D_A(M_2) \times {D_A}^\dagger(M_2) &=& \chi[M_2, \ZZ_M]^{-2n}~,
\no\\
L_{(\mathbf{e}, \mathbf{m})}(\gamma)\times D_A(M_2)   &=& D_A(M_2) \times  L_{A(\mathbf{e}, \mathbf{m})}(\gamma)~.
\eea
In the case in which $1-A$ is not of full rank, the defect takes a different form than the one obtained below, with the generalization given in the main text.
A second derivation of these results in a somewhat more streamlined but less explicit notation is given in Appendix \ref{app:matrixnotation}. 

\subsection{Fusion of $D_A(M_2)$ and $D_A^\dagger(M_2)$ }

 We begin by writing out the defect and its complex conjugate in terms of the basis in (\ref{eq:simplerbasis}),
  \bea
 D_A(M_2) &=& {1\over |H^0(M_2, \ZZ_M)|^{2n}}\sum_{\gamma_1 , \dots, \gamma_{2n} \in H_1(M_2, \ZZ_M)} e^{{2 \pi i \over M}\half  \sum_{i,j=1}^{2n}Q_{ij}^A\langle \g_i, \g_j\rangle }e^{-{2\pi i \over M} \sum_{i<j}^{2n} \b^i \cdot \a^j \langle \g_i, \g_j \rangle}
 \no\\
  &\vphantom{.}& \hspace{1.7 in} \times \prod_{j=1}^n L_j\left(\sum_{i=1}^{2n} \a^i_j \g_i \right) \prod_{j=1}^n \widehat L_j \left(\sum_{i=1}^{2n} \b^i_j \g_i \right) 
 \eea
 as well as,
 \bea
{D_A(M_2)}^\dagger &=& {1\over |H^0(M_2, \ZZ_M)|^{2n}}\sum_{\gamma_1 , \dots, \gamma_{2n} \in H_1(M_2, \ZZ_M)} e^{-{2 \pi i \over M} \half\sum_{i,j=1}^{2n}Q_{ij}^A\langle \g_i, \g_j\rangle }e^{{2\pi i \over M} \sum_{i<j}^{2n} \b^i \cdot \a^j \langle \g_i, \g_j \rangle}
 \no\\
  &\vphantom{.}& \hspace{1.7 in} \times \prod_{j=1}^n \widehat L_j\left(- \sum_{i=1}^{2n} \b^i_j \g_i  \right) \prod_{j=1}^n  L_j \left(-\sum_{i=1}^{2n} \a^i_j \g_i\right)~, \qquad\qquad
 \eea
 where the matrix elements $\a$ and $\beta$ were defined in (\ref{eq:abdef}) and $ \b^i \cdot \a^j := \sum_{k=1}^n \b^i_k \a^j_k$. 
 
  We now compute the product 
 \bea
  D_A(M_2)\times {D_A(M_2)}^\dagger  &=&  {1\over |H^0(M_2, \ZZ_M)|^{4n}}\sum_{\gamma_i, \widetilde \gamma_i}e^{{2 \pi i \over M}\half \sum_{i,j=1}^{2n}Q_{ij}^A\langle \g_i, \g_j\rangle }e^{-{2\pi i \over M} \sum_{i<j}^{2n} \b^i \cdot \a^j \langle \g_i, \g_j \rangle}
  \no\\
 &\vphantom{.}& \hspace{0.5 in}\times  e^{-{2 \pi i \over M}\half \sum_{i,j=1}^{2n}Q_{ij}^A\langle \widetilde \g_i, \widetilde \g_j\rangle }e^{{2\pi i \over M} \sum_{i<j}^{2n} \b^i \cdot \a^j \langle \widetilde \g_i, \widetilde \g_j \rangle}
 \no\\
  &\vphantom{.}& \hspace{0.5 in}\times\prod_{j=1}^n L_j\left(\sum_{i=1}^{2n} \a^i_j \g_i \right) \prod_{j=1}^n \widehat L_j \left(\sum_{i=1}^{2n} \b^i_j \g_i \right) 
  \no\\
  &\vphantom{.}& \hspace{0.5 in}\times  \prod_{j=1}^n \widehat L_j\left(- \sum_{i=1}^{2n} \b^i_j \widetilde\g_i  \right) \prod_{j=1}^n  L_j \left(-\sum_{i=1}^{2n} \a^i_j \widetilde \g_i\right)~. 
 \eea
  We begin by combining the factors of $\widehat L_j$, commuting the $L_j$ to the left, combining the factors of $L_i$, and shifting $\g_i \rightarrow \g_i + \widetilde \g_i$. This gives 
  \bea
    D_A(M_2)\times {D_A(M_2)}^\dagger  &=&  {1\over |H^0(M_2, \ZZ_M)|^{4n}}\sum_{\gamma_i, \widetilde \gamma_i}e^{{2 \pi i \over M}\half \sum_{i,j=1}^{2n}Q_{ij}^A\langle \g_i + \widetilde \g_i, \g_j + \widetilde \g_j \rangle }e^{-{2\pi i \over M} \sum_{i<j}^{2n} \b^i \cdot \a^j \langle \g_i + \widetilde \g_i, \g_j + \widetilde \g_j \rangle}
  \no\\
 &\vphantom{.}& \hspace{0.5 in}\times  e^{-{2 \pi i \over M}\half \sum_{i,j=1}^{2n}Q_{ij}^A\langle \widetilde \g_i, \widetilde \g_j\rangle }e^{{2\pi i \over M} \sum_{i<j}^{2n} \b^i \cdot \a^j \langle \widetilde \g_i, \widetilde \g_j \rangle}e^{ {2 \pi i \over M} \sum_{i,j=1}^{2n} \b^i \cdot \a^j \langle \g_i, \widetilde \g_j\rangle }
 \no\\
  &\vphantom{.}& \hspace{0.5 in}\times \prod_{j=1}^n L_j\left(\sum_{i=1}^{2n} \a^i_j \g_i \right) \prod_{j=1}^n \widehat L_j \left(\sum_{i=1}^{2n} \b^i_j \g_i \right)~.
  \eea
  Some rearrangement then puts this in the following form, 
  \bea
  \label{eq:DADAdagg1}
   D_A(M_2)\times {D_A(M_2)}^\dagger  &=&  {1\over |H^0(M_2, \ZZ_M)|^{4n}}\sum_{\gamma_i, \widetilde \gamma_i}e^{{2 \pi i \over M} \half\sum_{i,j=1}^{2n}Q_{ij}^A\langle \g_i , \g_j  \rangle }e^{-{2\pi i \over M} \sum_{i<j}^{2n} \b^i \cdot \a^j \langle \g_i , \g_j  \rangle}
  \no\\
 &\vphantom{.}& \hspace{0.25 in}\times  e^{{2 \pi i \over M}\half \sum_{i,j=1}^{2n}Q_{ij}^A \left[ \langle  \g_i, \widetilde \g_j\rangle+ \langle\widetilde  \g_i,  \g_j\rangle \right]}e^{{2\pi i \over M} \sum_{i<j}^{2n} (\b^i \cdot \a^j+\b^j \cdot \a^i) \langle  \g_j, \widetilde \g_i \rangle} e^{{2\pi i \over M} \sum_{i=1}^{2n} \beta^i \cdot \a^i \langle \g_i, \widetilde \g_i \rangle}
  \no\\
  &\vphantom{.}& \hspace{0.5 in}\times \prod_{j=1}^n L_j\left(\sum_{i=1}^{2n} \a^i_j \g_i \right) \prod_{j=1}^n \widehat L_j \left(\sum_{i=1}^{2n} \b^i_j \g_i \right)~.
  \eea
Now note that for an $O(n,n;M)$ matrix $A$, we have (modulo $M$), 
\bea
(1-A)^{T} \mathfrak{I} (1-A) = \mathfrak{I}(1-A) + (1-A)^{T}\mathfrak{I}~. 
\eea
  In terms of the components $\a^i_j$, $\b^i_j$, this gives the following equations for $i , j \leq n$, 
  \bea
  \label{eq:abconstraints}
  \a^i \cdot \b^j + \b^i \cdot \a^j &=& \b_i^j + \b^i_j
  \no\\
  \a^i \cdot \b^{j+n} + \b^i \cdot \a^{j+n} &=& \b^{j+n}_i + \a^i_j
  \no\\
  \a^{i+n} \cdot \b^{j+n} + \b^{i+n} \cdot \a^{j+n} & = & \a_j^{i+n} + \a_i^{j+n}~.
  \eea
  From this, we note that 
  \bea
  \sum_{i<j}^{2n} (\b^i \cdot \a^j+\b^j \cdot \a^i) \langle \widetilde \g_i , \g_j \rangle &=& \sum_{j=1}^n \sum_{i=1}^{j-1} \beta_j^i \langle \widetilde \g_i , \g_j\rangle + \sum_{j=1}^n \sum_{i=1}^{j-1} \a_i^{j+n} \langle \widetilde \g_{i+n}, \g_{j+n}\rangle
  \\
  &\vphantom{,}& \hspace{0.5 in}+ \sum_{j=1}^n \sum_{i=1}^{j-1} \beta^j_i \langle \widetilde \g_i, \g_j \rangle + \sum_{j=1}^n \sum_{i=1}^n \b_i^{j+n} \langle \widetilde \g_i, \g_{j+n}\rangle 
  \no\\
   &\vphantom{,}& \hspace{0.5 in}+ \sum_{j=1}^n \sum_{i=1}^n \a_j^i \langle \widetilde \g_i, \g_{j+n}\rangle + \sum_{j=1}^n \sum_{i=1}^{j-1} \a_j^{i+n}\langle \widetilde \g_{i+n}, \g_{j+n}\rangle
   \no
   \no\\
   &=&  \sum_{j=1}^n \sum_{i=1}^{j-1} \beta_j^i \langle \widetilde \g_i , \g_j\rangle + \sum_{j=1}^n \sum_{i=1}^{j-1} \a_i^{j+n} \langle \widetilde \g_{i+n}, \g_{j+n}\rangle
 \no \\
  &\vphantom{,}& \hspace{0.5 in}- \sum_{j=1}^n \sum_{i=j}^n \b_i^j \langle \widetilde \g_i, \g_j \rangle + \sum_{i=1}^n \left \langle \widetilde \g_i, \sum_{j=1}^{2n} \b_i^j \g_j\right \rangle
\no\\
 &\vphantom{,}& \hspace{0.5 in}- \sum_{j=1}^n \sum_{i=j}^n \a_j^{i+n} \langle \widetilde \g_{i+n} , \g_{i+n} \rangle - \sum_{j=1}^n \left \langle \g_{j+n}, \sum_{i=1}^{2n} \a^i_j \widetilde \g_i \right\rangle
 \no\\
 &=& - \sum_{i=1}^n \left \langle \sum_{j=1}^{2n} \b_i^j \g_j, \widetilde \g_i\right \rangle - \sum_{i=1}^n \left \langle \sum_{j=1}^{2n} \a_i^j \g_j, \widetilde \g_{i+n}\right \rangle
 \no\\
   &\vphantom{,}& \hspace{0.5 in} + \sum_{i<j}^n \b^i_j \langle \widetilde \g_i, \g_j\rangle - \sum_{i \geq j}^n \b_i^j \langle \widetilde \g_i, \g_j\rangle 
   \no\\
    &\vphantom{,}& \hspace{0.5 in} - \sum_{i \geq j}^n \a_i^{j+n} \langle \widetilde \g_{i+n}, \g_{j+n}\rangle + \sum_{i<j}^n \a_j^{i+n} \langle \widetilde \g_{i+n}, \g_{j+n}\rangle 
    \no\\
        &\vphantom{,}& \hspace{0.5 in} -\sum_{i,j=1}^n \a_i^j \langle \widetilde \g_{i+n}, \g_j \rangle + \sum_{i,j=1}^n  \a_j^i \langle \widetilde \g_i, \g_{j+n}\rangle  ~.
  \eea
Finally, making use of the definition of $Q_{ij}^A$ in (\ref{eq:QAdefab}), we see that (\ref{eq:DADAdagg1}) simplifies to 
    \bea
   D_A(M_2)\times {D_A(M_2)}^\dagger  &=&  {1\over |H^0(M_2, \ZZ_M)|^{4n}}\sum_{\gamma_i, \widetilde \gamma_i}e^{{2 \pi i \over M}\half \sum_{i,j=1}^{2n}Q_{ij}^A\langle \g_i , \g_j  \rangle }e^{-{2\pi i \over M} \sum_{i<j}^{2n} \b^i \cdot \a^j \langle \g_i , \g_j  \rangle}
  \no\\
 &\vphantom{.}& \hspace{0.5 in}\times  e^{{2 \pi i \over M} \sum_{i=1}^n \left \langle \sum_{j=1}^{2n} \b_i^j \g_j, \widetilde \g_i\right \rangle} e^{{2 \pi i \over M}\sum_{i=1}^n \left \langle \sum_{j=1}^{2n} \a_i^j \g_j, \widetilde \g_{i+n}\right \rangle}
 \no\\
  &\vphantom{.}& \hspace{0.5 in}\times \prod_{j=1}^n L_j\left(\sum_{i=1}^{2n} \a^i_j \g_i \right) \prod_{j=1}^n \widehat L_j \left(\sum_{i=1}^{2n} \b^i_j \g_i \right)~.
  \eea
  Performing the sums over $\widetilde \g_i$ for $i \leq n$ and $n<i\leq 2n$ then gives a series of delta functions 
  \bea
  |H_1(M_2, \ZZ_M)|^{2n} \prod_{i=1}^n \delta\left(\sum_{j=1}^{2n} \b_i^j \g_j\right)\prod_{i=1}^n \delta\left(\sum_{j=1}^{2n} \a_i^j \g_j\right)~
  \eea
  which trivializes the arguments of $L_j$ and $\widehat L_j$, and since $1-A$ was assumed to be full rank we end up with 
  \bea
   D_A(M_2)\times {D_A(M_2)}^\dagger = {  |H^1(M_2, \ZZ_M)|^{2n}  \over |H^0(M_2, \ZZ_M)|^{4n}} = \chi[M_2,  \ZZ_M]^{-2n}~,
  \eea
  as desired.  
  
  Note that when $1-A$ is not of full rank, some of the delta functions are redundant. 
  In this case, the delta functions will still fix the arguments of all line operators to zero, but some of the sums over $\gamma_i$ 
  will remain, and will lead to extra factors of $|H^1(M_2, \ZZ_M)|$. In particular, when $1-A$ is of rank $r_A$, 
  one obtains an extra factor of $|H^1(M_2, \ZZ_M)|^{2n-r_A}$, giving
    \bea
    \label{eq:subtletyinnormalization}
   D_A(M_2)\times {D_A(M_2)}^\dagger =|H^1(M_2, \ZZ_M)|^{2n-r_A}\, \chi[M_2,  \ZZ_M]^{-2n}~, \,\,
  \eea
  where $D_A(M_2)$ is the defect given in (\ref{eq:defofDA}). 
  This tells us that the form of the defect given in (\ref{eq:defofDA}) is no longer the correct one when $1-A$ is not of full rank. 
The appropriate modification is given in (\ref{eq:generaldefofDA}). 
  
\subsection{Commutator of $D_A(M_2)$ and $L_{(\mathbf{e},\mathbf{m})}$}

We next confirm the fusion rules between $D_A(M_2)$ and $L_{(\mathbf{e},\mathbf{m})}$. Note that we may write 
\bea
L_{(\mathbf{e},\mathbf{m})}(\gamma) = \prod_{j=1}^n L_j^{e_j}(\gamma) \prod_{j=1}^n \widehat L_j^{m_j}(\gamma)
\eea
from which we also obtain
\bea
L_{A(\mathbf{e}, \mathbf{m})}(\gamma) = \prod_{j=1}^n L_j\left( \left(e_j - \sum_{i=1}^n (\a_j^i e_i + \a_j^{i+n} m_i)\right)\gamma\right) 
 \prod_{j=1}^n \widehat L_j\left( \left(m_j - \sum_{i=1}^n (\b_j^i e_i + \b_j^{i+n} m_i)\right)\gamma\right)~. \no\\
\eea
A straightforward computation then shows that 
\bea
 L_{(\mathbf{e},\mathbf{m})}(\gamma')\times D_A(M_2)  &=& {1\over |H^0(M_2, \ZZ_M)|^{2n}}\sum_{\g_1, \dots, \g_{2n}} e^{{2\pi i \over M} \half\sum_{i,j=1}^{2n} Q_{ij}^A \langle \g_i, \g_j\rangle} e^{- {2\pi i \over M} \sum_{i< j}^{2n} \beta^i \cdot \a^j \langle \g_i, \g_j\rangle} 
\no\\
&\vphantom{.}& \hspace{0.5 in} \times e^{-{2 \pi i \over M} \sum_{j=1}^n \langle m_j \g', \sum_{i=1}^{2n} \a_j^i \g_i \rangle}e^{{2\pi i \over M} \sum_{j=1}^n \left \langle  \left(- e_j + \sum_{i=1}^n (\a^i_j e_i + \a_j^{i+n} m_i) \right)\g',\, \sum_{i=1}^{2n} \b_j^i \g_i \right\rangle}
\no\\
&\vphantom{.}& \hspace{0.5 in} \times \prod_{j=1}^n L_j \left( \sum_{i=1}^n (\a_j^i e_i + \a_j^{i+n} m_i) \g' + \sum_{i=1}^{2n} \a_j^i \g_i \right)
\no\\
&\vphantom{.}& \hspace{0.5 in} \times \prod_{j=1}^n \widehat L_j \left( \sum_{i=1}^n (\b_j^i e_i + \b_j^{i+n} m_i) \g' + \sum_{i=1}^{2n} \b_j^i \g_i \right)  \times L_{A(\mathbf{e}, \mathbf{m})}(\gamma')  ~.
\eea
Making the change of variables 
\bea
\g_i \rightarrow \g_i - \left\{\begin{matrix} e_i \g' && i \leq n \\ m_{i-n} \g' && i>n \end{matrix}  \right.
\eea
the line operators appearing on the left-hand side reduce to the usual ones $L_j (\sum_{i=1}^{2n} \a_j^i \g_i)$ and 
$\widehat L_j (\sum_{i=1}^{2n} \b_j^i \g_i)$ expected for $D_A(M_2)$. It then remains to show that the exponential factors reduce to those for 
$D_A(M_2)$. In particular, we must have all exponential terms depending on $\langle \g', \gamma_i \rangle$ cancel. 
For $i \leq n$, it can be checked that this is ensured as long as 
\bea
Q_{i j}^A - Q_{j i}^A &=&  2\a^i_{j-n} ~, \hspace{0.5in}  j>n~,
\no\\
Q_{ij}^A - Q_{ji}^A &=& 2\beta^i_j~,\hspace{0.64in} i < j \leq n~,
\no\\
Q_{ij}^A - Q_{ji}^A &=& -2\beta_i^j~,\hspace{0.5in}  j<i~,
\eea
which are indeed satisfied by $Q_{ij}^A$ given in (\ref{eq:QAdefab}). On the other hand, the terms involving $\langle \g', \gamma_i \rangle$ for $i > n$ cancel as long as 
\bea
Q_{ij}^A - Q_{ji}^A &=& - 2\a_{i-n}^j~, \hspace{0.5 in} j \leq n~,
\no\\
Q_{i j}^A - Q_{ji}^A &=& -2 \a_{i-n}^{j}\hspace{0.6 in} n < j < i~,
\no\\
Q_{ij}^A - Q_{ji}^A &=& 2 \a_{j-n}^{i}\hspace{0.75 in} i < j ~,
\eea
which is again satisfied by $Q_{ij}^A$ given in (\ref{eq:QAdefab}). This confirms the second line of (\ref{eq:desiredfusions}).

 \section{Alternative derivation of the fusion rules of $D_A$}
 \label{app:matrixnotation}
In this appendix, we repeat the computations in Appendix \ref{app:DAfusions} in a more streamlined (but less explicit) fashion by introducing a new notation for the defects. 
This enables us to compute the additional fusion rule 
\bea
\label{eq:appfinalfus}
D_{A^{I}}(M_2)\times D_{A^{J}}(M_2)=\chi[M_{2}, \ZZ_{M}]^{-n}D_{A^{I+J}}(M_2)~
\eea
for $N=p$ being prime.

\subsection{Determining $Q^{A}$}
We begin by redetermining the matrix $Q^A$ appearing in the definition of $D_A(M_2)$. Our starting point is the product
\bea\label{eq:middle_Q}
 L_{(\mathbf{e},\mathbf{m})}(\gamma')\times D_A(M_2)  &=& {1\over |H^0(M_2, \ZZ_M)|^{2n}}\sum_{\g_1, \dots, \g_{2n}} e^{{2\pi i \over M} \half\sum_{i,j=1}^{2n} Q_{ij}^A \langle \g_i, \g_j\rangle} e^{- {2\pi i \over M} \sum_{i< j}^{2n} \beta^i \cdot \a^j \langle \g_i, \g_j\rangle} 
\no\\
&\vphantom{.}& \hspace{0.5 in} \times e^{-{2 \pi i \over M} \sum_{j=1}^n \langle m_j \g', \sum_{i=1}^{2n} \a_j^i \g_i \rangle}e^{{2\pi i \over M} \sum_{j=1}^n \left \langle  \left(- e_j + \sum_{i=1}^n (\a^i_j e_i + \a_j^{i+n} m_i) \right)\g',\, \sum_{i=1}^{2n} \b_j^i \g_i \right\rangle}
\no\\
&\vphantom{.}& \hspace{0.5 in} \times \prod_{j=1}^n L_j \left( \sum_{i=1}^n (\a_j^i e_i + \a_j^{i+n} m_i) \g' + \sum_{i=1}^{2n} \a_j^i \g_i \right)
\no\\
&\vphantom{.}& \hspace{0.5 in} \times \prod_{j=1}^n \widehat L_j \left( \sum_{i=1}^n (\b_j^i e_i + \b_j^{i+n} m_i) \g' + \sum_{i=1}^{2n} \b_j^i \g_i \right)  \times L_{A(\mathbf{e}, \mathbf{m})}(\gamma')  ~.
\eea
We may simplify the phase factor by defining
 \bea
 \Gamma := \left(
 \begin{matrix}  \g_1 \\ 
                \g_2 \\
                \vdots\\
                \g_{2n}
 \end{matrix}  
 \right) ~,\hspace{0.5 in} 
\bv := \left(
 \begin{matrix}  e_1 \\ 
                \vdots\\
                e_{n} \\
                 m_1 \\ 
                \vdots\\
                m_{n}
 \end{matrix}  
 \right) ~,\hspace{0.5 in}  
 \Omega := \left( \begin{matrix} 0 & \mathds{1}_{n \times n} \\ 0 & 0 \end{matrix} \right) ~ , 
 \eea
 and decomposing the matrix $(1-A)^{T}\Omega^{T}(1-A)$ as follows,
 \bea\label{eq:decomposition}
    (1-A)^{T}\Omega^{T}(1-A)=X_{R}+X_{L}+X_{D}\ ,
 \eea
 where $X_{R}$ and $X_{L}$ are the upper-right and lower-left submatrices of $(1-A)^{T}\Omega^{T}(1-A)$, respectively, and $X_{D}$ is a diagonal matrix 
 composed of the diagonal entries of $(1-A)^{T}\Omega^{T}(1-A)$.
 We also introduce the following shorthand notation
 \bea
    \langle \Gamma^{\prime T} M\,\Gamma\rangle:=\sum_{i, j=1}^{2n}M_{ij}\langle \g^{\prime}_i, \g_j\rangle\ , 
 \eea
 where $M$ is an arbitrary $2n \times 2n$ matrix. We note that this inner product satisfies the following anti-symmetry property
 \bea
 \langle \Gamma^{\prime T} M\,\Gamma\rangle=-\langle \Gamma^{T}\, M^{T}\Gamma'\rangle\ .
 \eea
 We can then put the phase factor appearing in \eqref{eq:middle_Q} into the following index-free form,\footnote{We drop the factor $2\pi i/M$ for simplicity.}
\bea
     \text{(phase factor)}&=& \frac{1}{2}\langle \Gamma^{T} Q^{A}\,\Gamma\rangle
     -\langle \Gamma^{T} X_{R}\,\Gamma\rangle-\langle \gamma^{\prime} V^{T} \Omega^{T}(1-A)\Gamma\rangle
 \no\\
     &\vphantom{.}& 
     -\langle \gamma^{\prime} V^{T} \Omega (1-A)\Gamma\rangle
     +\langle \gamma^{\prime} V^{T} (1-A)^{T} \Omega (1-A)\Gamma\rangle\ .
 \eea
 In order to be compatible with the desired fusion rules, the above phase factor after the shift $\Gamma \to \Gamma-V\gamma'$ must be equal to $\langle \Gamma^{T} Q^{A}\,\Gamma\rangle/2
     -\langle \Gamma^{T} X_{R}\,\Gamma\rangle$. This condition is enough to determine $Q^{A}$ as follows,
  \bea
 \begin{aligned}
     & \frac{1}{2}\langle \gamma^{\prime} V^{T} \left(Q^{A}-(Q^{A})^{T}\right)\,\Gamma\rangle\\
     &\quad =
     \langle \gamma^{\prime} V^{T} (X_{R}-X_{R}^{T})\,\Gamma\rangle
     -\langle \gamma^{\prime} V^{T} ( \Omega+\Omega^{T})(1-A)\Gamma\rangle
     +\langle \gamma^{\prime} V^{T} (1-A)^{T} \Omega (1-A)\Gamma\rangle \\
     &\quad = \langle \gamma^{\prime} V^{T} (X_{R}-X_{R}^{T})\,\Gamma\rangle
     -\langle \gamma^{\prime} V^{T} \mathfrak{I}(1-A)\Gamma\rangle\\
     &\qquad\qquad\qquad
     +\langle \gamma^{\prime} V^{T} (1-A)^{T} \mathfrak{I} (1-A)\Gamma\rangle
     -\langle \gamma^{\prime} V^{T} (1-A)^{T} \Omega^{T} (1-A)\Gamma\rangle\\
     &\quad = \langle \gamma^{\prime} V^{T} (X_{R}-X_{R}^{T})\,\Gamma\rangle
     +\langle \gamma^{\prime} V^{T} (1-A)^{T} \mathfrak{I} \Gamma\rangle
     -\langle \gamma^{\prime} V^{T} (X_{R}+X_{L}+X_{D}) \Gamma\rangle \\
      &\quad = -\langle \gamma^{\prime} V^{T} (X_{L}+X_{R}^{T}+X_{D}-(1-A)^{T}\mathfrak{I})\,\Gamma\rangle\ ,
 \end{aligned}
 \eea  
 where in the second equality we have used $\Omega+\Omega^{T}=\mathfrak{I}$, and in the third equality we have made use of the following matrix identity
 \bea\label{eq:matrix_identity}
     (1-A)^{T} \mathfrak{I} (1-A)=\mathfrak{I} (1-A)+(1-A)^{T} \mathfrak{I}\ . 
 \eea
 From this, we can deduce the exact form of $Q^{A}$,
 \bea\label{eq:middle_Q1}
 \begin{aligned}
      Q^{A}&=-X_{L}-X_{R}^{T}-X_{D}+(1-A)^{T}\mathfrak{I} \\
      &=X_{R}+X_{L}^{T}+X_{D}-\mathfrak{I}(1-A)\ .
 \end{aligned}
 \eea
 Furthermore, the matrix identity \eqref{eq:matrix_identity} ensures that $X_{R}+X_{L}^{T}+X_{D}$ can be written as
 \bea
 X_{R}+X_{L}^{T}+X_{D}=\left[\mathfrak{I}(1-A)\right]_{R}+\left[(1-A)^{T}\mathfrak{I}\right]_{R}+\left[\mathfrak{I}(1-A)\right]_{D}\ ,
 \eea
 where $[M]_{R, L, D}$ represent the upper-right, lower-left, and diagonal part of the matrix $M$ as defined in a similar way to \eqref{eq:decomposition}. By plugging this into \eqref{eq:middle_Q1}, we arrive at 
 \bea\label{eq:Q}
    Q^{A}=\left[(1-A)^{T}\mathfrak{I}\right]_{R}-\left[\mathfrak{I}(1-A)\right]_{L}\ ,
 \eea
 which is the same result as was given in the main text in (\ref{eq:QAdefab}). 
 \subsection{Fusion of $D_{A^{I}}(M_{2})$ and $D_{A^{J}}(M_{2})$}
We next derive the fusion rule between two condensation defects $D_{A^{I}}(M_{2})$ and $D_{A^{J}}(M_{2})$ where $I, J=1, 2, \cdots, |A|-1$. Here we  assume that the order of $|A|$ is a prime number $p$, and that $A$ is of full rank. In this case, there are no redundancies in the line operators, and the condensation defect $D_{A^{I}}(M_{2})$ is defined by\footnote{For non-full rank cases, the condensation defects should be modified as \eqref{eq:condens_non_full}. The derivations in this appendix, however, can be straightforwardly extended to that case using the relation \eqref{eq:widehatQexpression}.}
 \bea\label{eq:condens_I}
 D_{A^{I}}(M_2) &=& {1\over |H^0(M_2, \ZZ_M)|^{2n}}\sum_{\gamma_1 , \dots, \gamma_{2n} \in H_1(M_2, \ZZ_M)} e^{{2 \pi i \over M}\half  \sum_{i,j=1}^{2n}Q_{ij}^{A^{I}}\langle \g_i, \g_j\rangle }e^{-{2\pi i \over M} \sum_{i<j}^{2n} \b_{I}^i \cdot \a_{I}^j \langle \g_i, \g_j \rangle}
 \no\\
  &\vphantom{.}& \hspace{1.7 in} \times \prod_{j=1}^n L_j\left(\sum_{i=1}^{2n} \a^i_{I\, j} \g_i \right) \prod_{j=1}^n \widehat L_j \left(\sum_{i=1}^{2n} \b^i_{I\, j} \g_i \right) \,,
 \eea
 where $\a^i_{I\, j}$ and $\b^i_{I\, j}$ are the elements of the matrix $1-A^{I}$,
\bea
 \label{eq:abdef_app1}
 1- A^{I} = \left(\begin{matrix}  \a_{I\, 1}^1 & \dots & \a_{I\, 1}^{2n} \\ \vdots & & \vdots \\ \a_{I\, n}^1 &\dots & \a_{I\, n}^{2n} \\ \b^1_{I\, 1} & \dots & \b_{I\, 1}^{2n} \\ \vdots & & \vdots \\ \b_{I\, n}^1 & \dots & \b_{I\, n}^{2n}\end{matrix}  \right) \, . 
 \eea

 We begin by focusing on the case of $I+J\not=0 \mod p$. By using the expression \eqref{eq:condens_I}, we immediately arrive at the following,
 \bea\label{eq:condens_I1}
 \begin{aligned}
 &|H^0(M_2, \ZZ_M)|^{4n}\cdot D_{A^{I}}(M_2)\times D_{A^{J}}(M_2)\\
 &=\quad \sum_{\substack{\gamma_1 , \dots, \gamma_{2n} \in H_1(M_2, \ZZ_M)\\ \gamma'_1 , \dots, \gamma'_{2n} \in H_1(M_2, \ZZ_M)}} \omega^{\half  \sum_{i,j=1}^{2n}\left(Q_{ij}^{A^{I}}\langle \g_i, \g_j\rangle+Q_{ij}^{A^{J}}\langle \g'_i, \g'_j\rangle\right)}\\
 &\qquad\qquad \qquad \cdot \omega^{-\sum_{i<j}^{2n} \b_{I}^i \cdot \a_{I}^j \langle \g_i, \g_j \rangle-\sum_{i<j}^{2n} \b_{J}^i \cdot \a_{J}^j \langle \g'_i, \g'_j \rangle-\sum_{i, j=1}^{2n} \b_{I}^i \cdot \a_{J}^j \langle \g_i, \g'_j \rangle}  \\
  &\qquad\qquad \times \prod_{j=1}^n L_j\left(\sum_{i=1}^{2n} \a^i_{I\, j} \g_i+\sum_{i=1}^{2n} \a^i_{J\, j} \g'_i \right) \prod_{j=1}^n \widehat L_j \left(\sum_{i=1}^{2n} \b^i_{I\, j} \g_i +\sum_{i=1}^{2n} \b^i_{J\, j} \g'_i\right) ~.
  \end{aligned}
 \eea
 As in the previous section, it is convenient to transition to matrix notation, where we have
\bea\label{eq:phase_factor}
    \begin{aligned}
        \text{(phase factor)}&=\half  \sum_{i,j=1}^{2n}\left(Q_{ij}^{A^{I}}\langle \g_i, \g_j\rangle+Q_{ij}^{A^{J}}\langle \g'_i, \g'_j\rangle\right)\\
        &\qquad -\sum_{i<j}^{2n} \b_{I}^i \cdot \a_{I}^j \langle \g_i, \g_j \rangle-\sum_{i<j}^{2n} \b_{J}^i \cdot \a_{J}^j \langle \g'_i, \g'_j \rangle-\sum_{i, j=1}^{2n} \b_{I}^i \cdot \a_{J}^j \langle \g_i, \g'_j \rangle \\
        &=\half \langle \Gamma^{T} Q^{I}\,\Gamma\rangle+\half \langle \Gamma^{\prime T} Q^{J}\,\Gamma'\rangle-\langle \Gamma^{T} X^{I}_{R}\,\Gamma\rangle-\langle \Gamma^{\prime T} X^{J}_{R}\,\Gamma^{\prime}\rangle-\langle \Gamma^{T} (1-A^{I})^{T} \Omega^{T} (1-A^{J})\Gamma^{\prime}\rangle\ ,
    \end{aligned}
\eea
where $Q^{I}:=Q^{A^{I}}$ and $X_{R}^{I}$ is the upper-right part of the matrix $(1-A^{I})^{T} \Omega^{T} (1-A^{I})$ as defined around \eqref{eq:decomposition}. 
To proceed with the computation of the fusion rules, we change the summation variable from $\Gamma'$ to $\Gamma^{\prime\prime}$ defined by 
\bea
\Gamma'=(1-A^{J})^{-1}(1-A^{I+J})\Gamma^{\prime\prime}- (1-A^{J})^{-1}(1-A^{I})\Gamma\ .
\eea
Under this change of variables, the line operators in \eqref{eq:condens_I} become 
 \bea
 \begin{aligned}
     L_j\left(\sum_{i=1}^{2n} \a^i_{I\, j} \g_i+\sum_{i=1}^{2n} \a^i_{J\, j} \g'_i \right)&\longrightarrow& L_j\left(\sum_{i=1}^{2n} \a^i_{I+J\, j} \g_i^{\prime\prime}\right)\ , \\
     \widehat{L}_j\left(\sum_{i=1}^{2n} \b^i_{I\, j} \g_i+\sum_{i=1}^{2n} \b^i_{J\, j} \g'_i \right)&\longrightarrow& \widehat{L}_j\left(\sum_{i=1}^{2n} \b^i_{I+J\, j} \g_i^{\prime\prime}\right)\ ,
 \end{aligned}
 \eea
 while the phase factor in \eqref{eq:phase_factor} is written as\footnote{Throughout these computations, we frequently use the fact that the matrices $A^{I}$ and $(1-A^{J})^{-1}$ commute with each other. This can easily be shown using the following identity,
 \bea
    (1-A^{J})^{-1}=1+A^{K}(1-A^{J})^{-1}A^{J-K}\ , \qquad \forall\, K=0, 1, \cdots, p-1\ , 
 \eea
 which is a special case of the Sherman-Morrison-Woodbury formula.
 }

 \bea
    \begin{aligned}
        \text{(phase factor)}&=\half\langle \Gamma^{T} X\,\Gamma\rangle+\langle \Gamma^{\prime\prime T} Y\,\Gamma\rangle +\langle \Gamma^{\prime\prime T}Z\,\Gamma^{\prime\prime}\rangle\ ,
    \end{aligned}
 \eea
 where $X$, $Y$ and $Z$ are matrices defined by
 \bea
 \begin{aligned}
     X&:=-\left(\mathfrak{I}-(1-A^{I})^{T}\mathfrak{I}(1-A^{J})^{-1}\right)(1-A^{I})\ , \\
     Y&:= \left(1-A^{I+J}\right)^{T}\mathfrak{I}\left(1-(1-A^{J})^{-1}\right)(1-A^{I})\ , \\
     Z&:= \half\left[(1-A^{I+J})^{T}\Omega (1-A^{I+J})-(1-A^{J})^{-T}\mathfrak{I}(1-A^{I+J})-(1-A^{J})^{-T}(1-A^{I+J})^{T}\mathfrak{I}\right]\ .
 \end{aligned}
 \eea
 It is possible to show that $Y=-X$ thanks to the following identity,
 \bea\label{eq:matrix_id_Y}
 \mathfrak{I}-(1-A^{I})^{T}\mathfrak{I}(1-A^{J})^{-1}=\left(1-A^{I+J}\right)^{T}\mathfrak{I}\left(1-(1-A^{J})^{-1}\right)~,
 \eea
 which can in turn be derived by making use of \eqref{eq:matrix_identity} and the following formula
 \bea
 1-A^{I+J}=1-A^{I}+1-A^{J}-(1-A^{I})(1-A^{J})\ .
 \eea
 Together with the fact that $X$ is invertible and skew-symmetric, we arrive at
  \bea\label{eq:condens_I2}
 &\vphantom{.}&\hspace{-0.5 in}|H^0(M_2, \ZZ_M)|^{4n}\cdot D_{A^{I}}(M_2)\times D_{A^{J}}(M_2)\\
 &\vphantom{.}& = \sum_{\Gamma, \Gamma' \in H_1(M_2, \ZZ_M^{2n})} \omega^{\half\langle \Gamma^{T} X\,\Gamma\rangle+\langle \Gamma^{\prime\prime T} Y\,\Gamma\rangle +\langle \Gamma^{\prime\prime T}Z\,\Gamma^{\prime\prime}\rangle}
 \no\\
 &\vphantom{.}&\hspace{0.5 in}\times \prod_{j=1}^n L_j\left[\left((1-A^{I+J})\Gamma^{\prime\prime}\right)_{j}\right] \prod_{j=1}^n \widehat L_j \left[\left((1-A^{I+J})\Gamma^{\prime\prime}\right)_{j+n}\right] \, ,
 \eea
 and can perform the summation over $\Gamma$. To evaluate this summation, we note the following identity,
  \bea
    \begin{aligned}
        \sum_{\Gamma \in H_1(M_2, \ZZ_M^{2n})} \omega^{\half\langle \Gamma^{T} X\,\Gamma\rangle+\langle \Gamma^{\prime\prime T} Y\,\Gamma\rangle}
        &= |H_1(M_2, \ZZ_M)|^{n}\, \omega^{-\half \langle \Gamma^{\prime\prime T} YX^{-T}Y^{T}\,\Gamma^{\prime\prime}\rangle}\ . 
    \end{aligned}
 \eea
 To prove this identity, we first rewrite the left-hand side as 
   \bea
    \begin{aligned}
        \sum_{\Gamma \in H_1(M_2, \ZZ_M^{2n})} \omega^{\half\langle \Gamma^{T} X\,\Gamma\rangle+\langle \Gamma^{\prime\prime T} Y\,\Gamma\rangle}&=\sum_{\Gamma \in H_1(M_2, \ZZ_M^{2n})} \omega^{\half\langle (\Gamma^{T}+\Gamma^{\prime\prime T}Y X^{-1})X(\Gamma+X^{-T}Y^{T}\Gamma^{\prime\prime})\rangle-\half \langle \Gamma^{\prime\prime T} YX^{-T}Y^{T}\,\Gamma^{\prime\prime}\rangle}\\
        &=\sum_{\Gamma \in H_1(M_2, \ZZ_M^{2n})} \omega^{\half\langle \Gamma^{T} X \Gamma \rangle-\half \langle \Gamma^{\prime\prime T} YX^{-T}Y^{T}\,\Gamma^{\prime\prime}\rangle}\ ,
    \end{aligned}
 \eea
 where in the first line we used the skew-symmetric property of the matrix $X$, and in the second line we changed the sum variables as $\Gamma\to\Gamma-X^{-T}Y^{T}\Gamma^{\prime\prime}$. Since the matrix $X$ is both invertible and skew-symmetric, there exists an orthogonal matrix $U\in O(2n, \ZZ_{M})$ such that
   \bea
 \label{eq:abdef_app}
 U^{T}XU = \bigoplus_{i=1}^{n}
 \begin{pmatrix}
0          & \lambda_i \\
-\lambda_i & 0         
\end{pmatrix}\ , \qquad \forall\lambda_{i}\not=0 \ \mod M\ .
\eea
We now use this matrix $U$ to  change the summation variables via $\Gamma\to U\Gamma$, which leads to the desired result,
   \bea
     \begin{aligned}
         \sum_{\Gamma \in H_1(M_2, \ZZ_M^{2n})} \omega^{\half\langle \Gamma^{T} X\,\Gamma\rangle+\langle \Gamma^{\prime\prime T} Y\,\Gamma\rangle}&=\sum_{\Gamma \in H_1(M_2, \ZZ_M^{2n})} \omega^{\half\langle \Gamma^{T} X \Gamma \rangle-\half \langle \Gamma^{\prime\prime T} YX^{-T}Y^{T}\,\Gamma^{\prime\prime}\rangle}\\
         &=\sum_{\gamma_{1}, \cdots, \gamma_{2n}}\omega^{\sum_{i=1}^{n}\lambda_{i}\langle\g_{2i-1},\g_{2i} \rangle}\omega^{-\half \langle \Gamma^{\prime\prime T} YX^{-T}Y^{T}\,\Gamma^{\prime\prime}\rangle}\\
         &=\prod_{i=1}^{n}\left[\sum_{\gamma_{2i-1}, \gamma_{2i}}\omega^{\lambda_{i}\langle\g_{2i-1},\g_{2i} \rangle}\right]\omega^{-\half \langle \Gamma^{\prime\prime T} YX^{-T}Y^{T}\,\Gamma^{\prime\prime}\rangle}\\
         &= |H_1(M_2, \ZZ_M)|^{n}\, \omega^{-\half \langle \Gamma^{\prime\prime T} YX^{-T}Y^{T}\,\Gamma^{\prime\prime}\rangle}\ .
   \end{aligned}
 \eea
 By using this, the fusion in question is reduced to
 \bea\label{eq:fusion_inter}
    \begin{aligned}
        &D_{A^{I}}(M_2)\times D_{A^{J}}(M_2)\\
        &\qquad =\frac{\chi[M_{2}, \ZZ_{M}]^{-n}}{|H^0(M_2, \ZZ_M)|^{2n}}\sum_{\Gamma' \in H_1(M_2, \ZZ_M^{2n})}\omega^{\langle \Gamma^{\prime\prime T}Z\,\Gamma^{\prime\prime}\rangle-\half \langle \Gamma^{\prime\prime T} YX^{-T}Y^{T}\,\Gamma^{\prime\prime}\rangle}\\
&\qquad\qquad\qquad\qquad\qquad\qquad\qquad\qquad \cdot\prod_{j=1}^n L_j\left[\left((1-A^{I+J})\Gamma^{\prime\prime}\right)_{j}\right] \prod_{j=1}^n \widehat L_j \left[\left((1-A^{I+J})\Gamma^{\prime\prime}\right)_{j+n}\right] \ .
    \end{aligned}
 \eea
We finally notice the following identity,
\bea\label{eq:identity_phase}
\langle \Gamma^{\prime\prime T}Z\,\Gamma^{\prime\prime}\rangle-\half \langle \Gamma^{\prime\prime T} YX^{-T}Y^{T}\,\Gamma^{\prime\prime}\rangle=\frac{1}{2}\langle \Gamma^{\prime\prime T} Q^{I+J}\,\Gamma^{\prime\prime}\rangle
     -\langle \Gamma^{\prime\prime T} X^{I+J}_{R}\,\Gamma^{\prime\prime}\rangle\ ,
\eea
which can be proven as follows,
 \bea
    \begin{aligned}
    &\langle \Gamma^{\prime\prime T}Z\,\Gamma^{\prime\prime}\rangle-\half \langle \Gamma^{\prime\prime T} YX^{-T}Y^{T}\,\Gamma^{\prime\prime}\rangle-\left[\frac{1}{2}\langle \Gamma^{\prime\prime T} Q^{I+J}\,\Gamma^{\prime\prime}\rangle
     -\langle \Gamma^{\prime\prime T} X^{I+J}_{R}\,\Gamma^{\prime\prime}\rangle\right]\\
     &\qquad =-\frac{1}{2}\langle \Gamma^{\prime\prime T}(1-A^{I+J})^{T}\mathfrak{I}\left(1-(1-A^{J})^{-1}\right)(1+A^{I})\,\Gamma^{\prime\prime}\rangle\\
     &\qquad =-\frac{1}{2}\langle \Gamma^{\prime\prime T}\left(\mathfrak{I}-(1-A^{I})^{T}\mathfrak{I}(1-A^{J})^{-1}\right)(1+A^{I})\,\Gamma^{\prime\prime}\rangle\\
     &\qquad =-\frac{1}{2}\langle \Gamma^{\prime\prime T}\left(2\mathfrak{I}-(1-A^{I})^{T}\mathfrak{I}(1-A^{J})^{-1}-(1-A^{J})^{-T}\mathfrak{I}(1-A^{I})\right)\,\Gamma^{\prime\prime}\rangle\\
     &\qquad =0\ .
     \end{aligned}
 \eea
In the second equality, we used \eqref{eq:matrix_id_Y}, while in the final equality, we utilized the symmetricity of the middle matrix. By plugging \eqref{eq:identity_phase} into \eqref{eq:fusion_inter}, we finally obtain the expected fusion rule (\ref{eq:appfinalfus}).

As a final note, consider the remaining case $I+J=0 \mod p$~. Since we did not use the invertible property of $1-A^{I+J}$ until equation \eqref{eq:condens_I2}, we may use this equation as our starting point. Fortunately, when $I+J=0 \mod p$~, we find that all line operators in \eqref{eq:condens_I2} are trivial and $X=Y=Z=0$, and hence the summand is trivial, which immediately leads to the following fusion rule,
\bea
D_{A^{I}}(M_2)\times D_{A^{p-I}}(M_2)={ |H^1(M_2, \ZZ_M)|^2 \over |H^0(M_2, \ZZ_M)|^4}=\chi[M_{2}, \ZZ_{M}]^{-2n} ~.
\eea

\section{Trivalent junctions for $\ST_1$ when $\gcd(M,3) = 1$}
\label{app:trivalentvertices}

In this appendix we give the simple line expansions for a number of bulk trivalent junctions relevant for triality defects coming from $\ST_1$ when $\gcd(M,3) = 1$. 
All of these results are obtained by simply imposing the bimodule conditions (\ref{eq:threemoduleconditions}). 
Note that the results given in this appendix are those obtained \textit{before} performing the gauge transformations in (\ref{eq:matchingaugetransf1}), 
(\ref{eq:matchingaugetransf2}). Performing such gauge transformations will in general shift the results below, as discussed in the main text. 

\subsection{Fusion junction between twist defects and anyons}

We begin by considering trivalent junctions involving one incoming simple line $L_{\bv}$ and one twist defect $\Sigma$ or $\overline \Sigma$. 
First, the results for $\Sigma$ are found to be as follows,
\bea
\label{eq:ex2n1}
        \begin{tikzpicture}[baseline=0,scale=1,square/.style={regular polygon,regular polygon sides=4}]
            \shade[top color=blue!40, bottom color=blue!10]  (0,-1) -- (1.5,-1) -- (1.5,1) -- (0,1)-- (0,-1);
            \draw[dgreen,thick, ->- = 0.5](0,-1) -- (0,0); 
            \draw[dgreen,thick, ->- = 0.5](0,0) -- (0,1); 
            \draw [blue, thick, -<- = 0.5] (0,0) to[out=180,in=90 ]  (-1,-1);
            \node[dgreen, above] at (0,+1) {\footnotesize{$\Sigma$}};
            \node[dgreen, below] at (0,-1) {\footnotesize{$\Sigma$}};
            \node[blue, below] at (-1.15,-1) {\footnotesize{$L_\bu$}};
            \node at (0,0) [square,draw,fill=violet!60,scale=0.5] {}; 
            \node[blue,left] at (1.5,0.7) {\footnotesize{$D_{A}$}};
        \end{tikzpicture} &=& \sum_{\bv} \mu_A(\bu,\bv) \begin{tikzpicture}[baseline=0,scale=1]
            \draw[dgreen,thick, ->-=0.5](0,-1) -- (0,-0.6); 
            \draw[blue,thick, ->-=0.5](0,-0.6) -- (0,0); 
            \draw[blue,thick, ->-=0.5](0,0) -- (0,0.6); 
            \draw[dgreen,thick, ->-=0.5](0,0.6) -- (0,1); 
            \draw [blue, thick, -<-=0.5] (0,0) to[out=180,in=90 ]  (-1,-1);
            \node[blue, right] at (0,0.4) {\footnotesize{$L_{\bu + \bv}$}};
            \node[blue, right] at (0,-0.4) {\footnotesize{$L_{\bv}$}};
            \node[dgreen, above] at (0,1.0) {\footnotesize{$\Sigma$}};
            \node[blue, below] at (-1.15,-1) {\footnotesize{$L_\bu$}};
            \node[dgreen, below] at (0,-1) {\footnotesize{$\Sigma$}};
            \filldraw[blue] (0,0) circle (1pt);
            \filldraw[blue] (0,-0.6) circle (1pt);
            \filldraw[blue] (0,0.6) circle (1pt);
        \end{tikzpicture} ~, 
           \\
          \begin{tikzpicture}[baseline=0,scale=1]
            \shade[top color=blue!40, bottom color=blue!10]  (0,-1) -- (1.5,-1) -- (1.5,1) -- (0,1)-- (0,-1);
            \draw[dgreen,thick, ->-=0.5](0,-1) -- (0,0); 
            \draw[dgreen,thick, ->-=0.5](0,0) -- (0,1); 
            \draw [gray!30, line width=3pt] (0,0) to[out=0,in=90 ]  (1,-1);
            \draw [blue, thick, -<-=0.5] (0,0) to[out=0,in=90 ]  (1,-1);
            \node[dgreen, above] at (0,1) {\footnotesize{$\Sigma$}};
            \node[dgreen, below] at (0,-1) {\footnotesize{$\Sigma$}};
            \node[blue, below] at (1,-1) {\footnotesize{$L_\bu$}};
            \node[isosceles triangle,scale=0.4, isosceles triangle apex angle=60, draw,fill=violet!60, rotate=90, minimum size =0.01cm] at (0,0){};
            \node[blue,left] at (1.5,0.7) {\footnotesize{$D_{A}$}};
    \end{tikzpicture} &=& \sum_{\bv} \frac{R^{\bu,\Sigma}}{R^{\bu,\bv}} \mu_A(\bu,\bv) 
        \begin{tikzpicture}[baseline=0,scale=1]
            \draw[dgreen,thick, ->-=0.5](0,-1) -- (0,-0.6); 
            \draw[blue,thick, ->-=0.5](0,-0.6) -- (0,0); 
            \draw[blue,thick, ->-=0.5](0,0) -- (0,0.6); 
            \draw[dgreen,thick, ->-=0.5](0,0.6) -- (0,1); 
            \draw[blue, thick, -<-=0.5] (0,0) to[out=0,in=90] (1,-1);
            \node[blue, right] at (0,0.4) {\footnotesize{$L_{\bu + \mathbf{v}}$}};
            \node[blue, right] at (0,-0.4) {\footnotesize{$L_{\bv}$}};
            \node[dgreen, above] at (0,0.9) {\footnotesize{$\Sigma$}};
            \node[blue, below] at (1,-1) {\footnotesize{$L_\bu$}};
            \node[dgreen, below] at (0,-1) {\footnotesize{$\Sigma$}};
            \filldraw[blue] (0,0) circle (1pt);
            \filldraw[blue] (0,-0.6) circle (1pt);
            \filldraw[blue] (0,0.6) circle (1pt);
    \end{tikzpicture} ~,
   \\
   \label{eq:ex1n1}
        \begin{tikzpicture}[baseline=0,scale=1,square/.style={regular polygon,regular polygon sides=4}]
            \shade[top color=orange!40, bottom color=orange!10]  (0,-1) -- (-1.5,-1) -- (-1.5,1) -- (0,1)-- (0,-1);
            \draw[dgreen,thick, ->- = 0.5](0,-1) -- (0,0); 
            \draw[dgreen,thick, ->- = 0.5](0,0) -- (0,1); 
            \draw[gray!30, line width=3pt] (0,0) to[out=180,in=90 ]  (-1,-1);
            \draw[blue, thick, -<- = 0.5] (0,0) to[out=180,in=90 ]  (-1,-1);
            \node[dgreen, above] at (0,+1) {\footnotesize{$\Sigma$}};
            \node[dgreen, below] at (0,-1) {\footnotesize{$\Sigma$}};
            \node[blue, below] at (-1.15,-1) {\footnotesize{$L_\bu$}};
            \node at (0,0) [square,draw,fill=violet!60,scale=0.5] {}; 
            \node[orange,right] at (-1.5,0.7) {\footnotesize{$D_{A^2}$}};
        \end{tikzpicture} &=& \beta(\bu) \sum_{\bv} \frac{R^{\bu,\bv}}{R^{\bu,\Sigma}} \mu_{A^2}(\bv,\bu) \begin{tikzpicture}[baseline=0,scale=1]
            \draw[dgreen,thick, ->-=0.5](0,-1) -- (0,-0.6); 
            \draw[blue,thick, ->-=0.5](0,-0.6) -- (0,0); 
            \draw[blue,thick, ->-=0.5](0,0) -- (0,0.6); 
            \draw[dgreen,thick, ->-=0.5](0,0.6) -- (0,1); 
            \draw [blue, thick, -<-=0.5] (0,0) to[out=180,in=90 ]  (-1,-1);
            \node[blue, right] at (0,0.4) {\footnotesize{$L_{\bu + \bv}$}};
            \node[blue, right] at (0,-0.4) {\footnotesize{$L_{\bv}$}};
            \node[dgreen, above] at (0,1.0) {\footnotesize{$\Sigma$}};
            \node[blue, below] at (-1.15,-1) {\footnotesize{$L_\bu$}};
            \node[dgreen, below] at (0,-1) {\footnotesize{$\Sigma$}};
            \filldraw[blue] (0,0) circle (1pt);
            \filldraw[blue] (0,-0.6) circle (1pt);
            \filldraw[blue] (0,0.6) circle (1pt);
        \end{tikzpicture} ~, 
    \eea
   \bea
         \begin{tikzpicture}[baseline=0,scale=1]
            \shade[top color=orange!40, bottom color=orange!10]  (0,-1) -- (-1.5,-1) -- (-1.5,1) -- (0,1)-- (0,-1);
            \draw[dgreen,thick, ->-=0.5](0,-1) -- (0,0); 
            \draw[dgreen,thick, ->-=0.5](0,0) -- (0,1); 
            \draw [blue, thick, -<-=0.5] (0,0) to[out=0,in=90 ]  (1,-1);
            \node[dgreen, above] at (0,1) {\footnotesize{$\Sigma$}};
            \node[dgreen, below] at (0,-1) {\footnotesize{$\Sigma$}};
            \node[blue, below] at (1,-1) {\footnotesize{$L_\bu$}};
            \node[isosceles triangle,scale=0.4, isosceles triangle apex angle=60, draw,fill=violet!60, rotate=90, minimum size =0.01cm] at (0,0){};
            \node[orange,right] at (-1.5,0.7) {\footnotesize{$D_{A^2}$}};
    \end{tikzpicture} &=& \beta(\bu) \sum_{\bv} \mu_{A^2}(\bv,\bu) 
        \begin{tikzpicture}[baseline=0,scale=1]
            \draw[dgreen,thick, ->-=0.5](0,-1) -- (0,-0.6); 
            \draw[blue,thick, ->-=0.5](0,-0.6) -- (0,0); 
            \draw[blue,thick, ->-=0.5](0,0) -- (0,0.6); 
            \draw[dgreen,thick, ->-=0.5](0,0.6) -- (0,1); 
            \draw[blue, thick, -<-=0.5] (0,0) to[out=0,in=90] (1,-1);
            \node[blue, right] at (0,0.4) {\footnotesize{$L_{\bu + \bv}$}};
            \node[blue, right] at (0,-0.4) {\footnotesize{$L_{\bv}$}};
            \node[dgreen, above] at (0,0.9) {\footnotesize{$\Sigma$}};
            \node[blue, below] at (1,-1) {\footnotesize{$L_\bu$}};
            \node[dgreen, below] at (0,-1) {\footnotesize{$\Sigma$}};
            \filldraw[blue] (0,0) circle (1pt);
            \filldraw[blue] (0,-0.6) circle (1pt);
            \filldraw[blue] (0,0.6) circle (1pt);
    \end{tikzpicture} ~,
\\
        \begin{tikzpicture}[baseline=0,scale=1.0,square/.style={regular polygon,regular polygon sides=4}]
            \shade[top color=orange!40, bottom color=orange!10]  (0,-1) -- (1.5,-1) -- (1.5,1) -- (0,1)-- (0,-1);
            \shade[top color=blue!40, bottom color=blue!10]  (-1.5,-1) -- (0,-1) -- (0,1) -- (-1.5,1)-- (-1.5,-1);
            \draw[dgreen,thick, ->- = 0.5](0,-1) -- (0,0); 
            \draw[dgreen,thick, ->- = 0.5](0,0) -- (0,1); 
            \draw [gray!30, line width=3pt] (0,0) to[out=180,in=90 ]  (-1,-1);
            \draw [blue, thick, -<- = 0.5] (0,0) to[out=180,in=90 ]  (-1,-1);
            \node[dgreen, above] at (0,+1) {\footnotesize{$\Sigma$}};
            \node[dgreen, below] at (0,-1) {\footnotesize{$\Sigma$}};
            \node[blue, below] at (-1.15,-1) {\footnotesize{$L_\bu$}};
            \node at (0,0) [square,draw,fill=violet!60,scale=0.5] {}; 
            \node[orange,left] at (1.5,0.7) {\footnotesize{$D_{A^2}$}};
            \node[blue,right] at (-1.5,0.7) {\footnotesize{$D_{A}$}};
        \end{tikzpicture} &=& \sum_{\bv} \mu_{A}(\bv,\bu) R^{\bu,\bv} (\alpha_\bu)_i{}^j \begin{tikzpicture}[baseline=0,scale=1.0]
            \draw[dgreen,thick, ->-=0.5](0,-1) -- (0,-0.6); 
            \draw[blue,thick, ->-=0.5](0,-0.6) -- (0,0); 
            \draw[blue,thick, ->-=0.5](0,0) -- (0,0.6); 
            \draw[dgreen,thick, ->-=0.5](0,0.6) -- (0,1); 
            \draw [blue, thick, -<-=0.5] (0,0) to[out=180,in=90 ]  (-1,-1);
            \node[blue, right] at (0,0.4) {\footnotesize{$L_{\bu + \bv}$}};
            \node[blue, right] at (0,-0.4) {\footnotesize{$L_{\bv}$}};
            \node[dgreen, above] at (0,1.0) {\footnotesize{$\Sigma$}};
            \node[blue, below] at (-1.15,-1) {\footnotesize{$L_\bu$}};
            \node[dgreen, below] at (0,-1) {\footnotesize{$\Sigma$}};
            \node[blue, left] at (0,-0.6) {\footnotesize{$i$}};
            \node[blue, left] at (0,0.6) {\footnotesize{$j$}};
            \filldraw[blue] (0,0) circle (1pt);
            \filldraw[blue] (0,-0.6) circle (1pt);
            \filldraw[blue] (0,0.6) circle (1pt);
        \end{tikzpicture} ~,
        \\
        \begin{tikzpicture}[baseline=0,scale=1.0]
            \shade[top color=orange!40, bottom color=orange!10]  (0,-1) -- (1.5,-1) -- (1.5,1) -- (0,1)-- (0,-1);
            \shade[top color=blue!40, bottom color=blue!10]  (-1.5,-1) -- (0,-1) -- (0,1) -- (-1.5,1)-- (-1.5,-1);
            \draw[dgreen,thick, ->-=0.5](0,-1) -- (0,0); 
            \draw[dgreen,thick, ->-=0.5](0,0) -- (0,1); 
            \draw [gray!30, line width=3pt] (0,0) to[out=0,in=90 ]  (1,-1);
            \draw [blue, thick, -<-=0.5] (0,0) to[out=0,in=90 ]  (1,-1);
            \node[dgreen, above] at (0,1) {\footnotesize{$\Sigma$}};
            \node[dgreen, below] at (0,-1) {\footnotesize{$\Sigma$}};
            \node[blue, below] at (1,-1) {\footnotesize{$L_\bu$}};
            \node[isosceles triangle,scale=0.4, isosceles triangle apex angle=60, draw,fill=violet!60, rotate=90, minimum size =0.01cm] at (0,0){};
            \node[orange,left] at (1.5,0.7) {\footnotesize{$D_{A^2}$}};
            \node[blue,right] at (-1.5,0.7) {\footnotesize{$D_{A}$}};
    \end{tikzpicture} &=& \sum_{\bv} R^{\bu,\Sigma} \mu_A(\bv,\bu) (\alpha_{\bu})_i{}^j
        \begin{tikzpicture}[baseline=0,scale=1.0]
            \draw[dgreen,thick, ->-=0.5](0,-1) -- (0,-0.6); 
            \draw[blue,thick, ->-=0.5](0,-0.6) -- (0,0); 
            \draw[blue,thick, ->-=0.5](0,0) -- (0,0.6); 
            \draw[dgreen,thick, ->-=0.5](0,0.6) -- (0,1); 
            \draw[blue, thick, -<-=0.5] (0,0) to[out=0,in=90] (1,-1);
            \node[blue, right] at (0,0.4) {\footnotesize{$L_{\bu + \mathbf{v}}$}};
            \node[blue, right] at (0,-0.4) {\footnotesize{$L_{\bv}$}};
            \node[dgreen, above] at (0,0.9) {\footnotesize{$\Sigma$}};
            \node[blue, below] at (1,-1) {\footnotesize{$L_\bu$}};
            \node[dgreen, below] at (0,-1) {\footnotesize{$\Sigma$}};
            \node[blue, left] at (0,-0.6) {\footnotesize{$i$}};
            \node[blue, left] at (0,0.6) {\footnotesize{$j$}};
            \filldraw[blue] (0,0) circle (1pt);
            \filldraw[blue] (0,-0.6) circle (1pt);
            \filldraw[blue] (0,0.6) circle (1pt);
    \end{tikzpicture} ~,
\eea
where the functions $\beta(\bv)$ and $\alpha(\bv)$ are constrained to satisfy 
\bea\label{eq:Sauxcon}
    \alpha_{\bu} \cdot \rho(\bv)\cdot (\alpha_\bu)^{-1} &=& R^{\bu,\bv} \frac{\mu_A(\bv,\bu)}{\mu_A(\bu,\bv)} \rho(\bv) ~,
    \no\\
    \frac{\beta(\bu+\widetilde{\bu})}{\beta(\bu)\beta(\widetilde{\bu})} &=& \frac{R^{\bu+\widetilde{\bu},\Sigma}}{R^{\bu,\Sigma} R^{\widetilde{\bu},\Sigma}} R^{\bu,\widetilde{\bu}} \frac{\mu_{A^2}(\widetilde{\bu},\bu)}{\mu_{A}(\bu,\widetilde{\bu})} ~,
    \no\\
    \alpha^{-1}_{\bu} \cdot \alpha_{\widetilde{\bu}}^{-1}  &=& \frac{\mu_A(\widetilde{\bu},\bu)}{\mu_A(\bu,\widetilde{\bu})} R^{\bu,\widetilde{\bu}} \alpha^{-1}_{\bu+\widetilde{\bu}} ~,
\eea
and $\rho(\bu)$ is the representation of the little group labeling $\Sigma$, which satisfies $\rho(\bu)\rho(\widetilde \bu) = \frac{\mu_{A^2}(\bu,\widetilde \bu)}{\mu_A(\bu,\widetilde \bu)} \rho(\bu + \widetilde \bu)$. 

Here, the phase $\beta(\bu)$ is introduced to keep track of the gauge choices discussed below \eqref{eq:betaconsist}. In the presentation where $\Sigma$ appears as the interface between $A$ and $A^2$ surface operators, in principle, we want to introduce phases to keep track of the gauge choices as well. However, the junction expansion now contains the matrix parameters $(\alpha_\bu)_i{}^j$, and the bimodule condition given in the first line of \eqref{eq:Sauxcon} can only determine the matrices $\alpha_\bu$ up to overall phases; thus, the explicit choice of $\alpha_\bu$ is sufficient to keep track of this phase. The second and third lines in \eqref{eq:Sauxcon} are obtained by requiring that the $F$-symbols \eqref{eq:F1S} computed in the two other presentations be the same. For example, consider $F_{\Sigma,\bu,\widetilde{\bu}}^\Sigma$ presented as
\begin{equation}
    \begin{tikzpicture}[baseline={([yshift=-.5ex]current bounding box.center)},vertex/.style={anchor=base,
    circle,fill=black!25,minimum size=18pt,inner sep=2pt},scale=0.8]
        \shade[top color=orange!40, bottom color=orange!10]  (-1.5,-1.5)  -- (-1.5,1.5) --(1.5,1.5)--(-1.5,-1.5) ;
        
        \draw[dgreen, thick, ->-=.5] (-1.5,-1.5) -- (-0.5,-0.5);
        \draw[dgreen, thick, ->-=.5] (-0.5,-0.5) -- (0.5,0.5);
        \draw[dgreen, thick, ->-=.5] (0.5,0.5) -- (1.5,1.5);
        \draw[blue, thick, ->-=.5] (0.5,-1.5) -- (-0.5,-0.5);
        \draw[blue, thick, ->-=.5] (2.5,-1.5) -- (0.5,0.5);

        \node[blue, below] at (2.5,-1.5) {\footnotesize $L_{\widetilde{\bu}}$};
        \node[blue, below] at (0.5,-1.5) {\footnotesize $L_{\bu}$};
        \node[dgreen, below] at (-1.5,-1.5)  {\footnotesize $\Sigma$};
        \node[dgreen, above] at (1.5,1.5) {\footnotesize $\Sigma$};
        \node[orange, below] at (-1,1.5) {\footnotesize $D_{A^2}$};

\end{tikzpicture} = F_{\Sigma, \bu, \widetilde{\bu}}^{\Sigma} \begin{tikzpicture}[baseline={([yshift=-.5ex]current bounding box.center)},vertex/.style={anchor=base,circle,fill=black!25,minimum size=18pt,inner sep=2pt},scale=0.8]
    \shade[top color=orange!40, bottom color=orange!10]  (-1.5,-1.5)  -- (-1.5,1.5) --(1.5,1.5)--(-1.5,-1.5) ;

    \draw[dgreen, thick, ->-=.5] (-1.5,-1.5) -- (0.5,0.5);
    \draw[blue, thick, ->-=.5] (0.5,0.5) -- (1.5,1.5);
    \draw[dgreen, thick, ->-=.5] (0.5,-1.5) -- (1.5,-0.5);
    \draw[blue, thick, ->-=.5] (2.5,-1.5) -- (1.5,-0.5);
    \draw[blue, thick, ->-=.5] (1.5,-0.5) -- (0.5,0.5);

    \node[dgreen, below] at (-1.5,-1.5)  {\footnotesize $\Sigma$};
    \node[blue, below] at  (2.5,-1.5) {\footnotesize $L_{\widetilde{\bu}}$}; 
    \node[blue, below] at (0.5,-1.5) {\footnotesize $L_{\bu}$};
    \node[dgreen, above] at (1.5,1.5) {\footnotesize $\Sigma$};
    \node[orange, below] at (-1,1.5) {\footnotesize $D_{A^2}$};
    
\end{tikzpicture} ~.
\end{equation}
Expanding both sides leads to 
\bea
    &\vphantom{,}&  \beta(\bu) \mu_{A^2}(\bv,\bu) \beta(\widetilde{\bu}) \mu_{A^2}(\bv+\bu,\widetilde{\bu}) \,\,\, = \,\,\, F_{\Sigma, \bu, \widetilde{\bu}}^{\Sigma} \beta(\bu+\widetilde{\bu}) \mu_{A^2}(\bv,\bu+\widetilde{\bu}) \\
    &\vphantom{,}& \hspace{0.8 in} \Rightarrow \,\,\,\, F_{\Sigma, \bu, \widetilde{\bu}}^{\Sigma} = \frac{\beta(\bu)\beta(\widetilde{\bu})}{\beta(\bu+\widetilde{\bu})} \frac{1}{\mu_{A^2}(\bu,\widetilde{\bu})} ~.
\eea
Matching the above with \eqref{eq:F1S} leads to the second condition in \eqref{eq:Sauxcon}, and the third condition in \eqref{eq:Sauxcon} on $\alpha_\bu$ can be determined similarly.

The results for trivalent junctions involving $\overline \Sigma$ instead of $\Sigma$ are obtained by making replacements $A \leftrightarrow A^2$, $\Sigma \leftrightarrow \overline \Sigma$, with the final results being, 
\bea
\label{eq:ex2n2}
        \begin{tikzpicture}[baseline=0,scale=1,square/.style={regular polygon,regular polygon sides=4}]
            \shade[top color=orange!40, bottom color=orange!10]  (0,-1) -- (1.5,-1) -- (1.5,1) -- (0,1)-- (0,-1);
            \draw[dgreen,thick, ->- = 0.5](0,-1) -- (0,0); 
            \draw[dgreen,thick, ->- = 0.5](0,0) -- (0,1); 
            \draw [blue, thick, -<- = 0.5] (0,0) to[out=180,in=90 ]  (-1,-1);
            \node[dgreen, above] at (0,+1) {\footnotesize{$\overline\Sigma$}};
            \node[dgreen, below] at (0,-1) {\footnotesize{$\overline\Sigma$}};
            \node[blue, below] at (-1.15,-1) {\footnotesize{$L_\bu$}};
            \node at (0,0) [square,draw,fill=violet!60,scale=0.5] {}; 
            \node[orange,left] at (1.5,0.7) {\footnotesize{$D_{A^2}$}};
        \end{tikzpicture} &=& \sum_{\bv} \mu_{A^2}(\bu,\bv) \begin{tikzpicture}[baseline=0,scale=1]
            \draw[dgreen,thick, ->-=0.5](0,-1) -- (0,-0.6); 
            \draw[blue,thick, ->-=0.5](0,-0.6) -- (0,0); 
            \draw[blue,thick, ->-=0.5](0,0) -- (0,0.6); 
            \draw[dgreen,thick, ->-=0.5](0,0.6) -- (0,1); 
            \draw [blue, thick, -<-=0.5] (0,0) to[out=180,in=90 ]  (-1,-1);
            \node[blue, right] at (0,0.4) {\footnotesize{$L_{\bu + \bv}$}};
            \node[blue, right] at (0,-0.4) {\footnotesize{$L_{\bv}$}};
            \node[dgreen, above] at (0,1.0) {\footnotesize{$\overline\Sigma$}};
            \node[blue, below] at (-1.15,-1) {\footnotesize{$L_\bu$}};
            \node[dgreen, below] at (0,-1) {\footnotesize{$\overline\Sigma$}};
            \filldraw[blue] (0,0) circle (1pt);
            \filldraw[blue] (0,-0.6) circle (1pt);
            \filldraw[blue] (0,0.6) circle (1pt);
        \end{tikzpicture} ~, 
           \\
           \label{eq:ex1n2}
          \begin{tikzpicture}[baseline=0,scale=1]
            \shade[top color=orange!40, bottom color=orange!10]  (0,-1) -- (1.5,-1) -- (1.5,1) -- (0,1)-- (0,-1);
            \draw[dgreen,thick, ->-=0.5](0,-1) -- (0,0); 
            \draw[dgreen,thick, ->-=0.5](0,0) -- (0,1); 
            \draw [gray!30, line width=3pt] (0,0) to[out=0,in=90 ]  (1,-1);
            \draw [blue, thick, -<-=0.5] (0,0) to[out=0,in=90 ]  (1,-1);
            \node[dgreen, above] at (0,1) {\footnotesize{$\overline\Sigma$}};
            \node[dgreen, below] at (0,-1) {\footnotesize{$\overline\Sigma$}};
            \node[blue, below] at (1,-1) {\footnotesize{$L_\bu$}};
            \node[isosceles triangle,scale=0.4, isosceles triangle apex angle=60, draw,fill=violet!60, rotate=90, minimum size =0.01cm] at (0,0){};
            \node[orange,left] at (1.5,0.7) {\footnotesize{$D_{A^2}$}};
    \end{tikzpicture} &=& \sum_{\bv} \frac{R^{\bu,\overline\Sigma}}{R^{\bu,\bv}} \mu_{A^2}(\bu,\bv) 
        \begin{tikzpicture}[baseline=0,scale=1]
            \draw[dgreen,thick, ->-=0.5](0,-1) -- (0,-0.6); 
            \draw[blue,thick, ->-=0.5](0,-0.6) -- (0,0); 
            \draw[blue,thick, ->-=0.5](0,0) -- (0,0.6); 
            \draw[dgreen,thick, ->-=0.5](0,0.6) -- (0,1); 
            \draw[blue, thick, -<-=0.5] (0,0) to[out=0,in=90] (1,-1);
            \node[blue, right] at (0,0.4) {\footnotesize{$L_{\bu + \mathbf{v}}$}};
            \node[blue, right] at (0,-0.4) {\footnotesize{$L_{\bv}$}};
            \node[dgreen, above] at (0,0.9) {\footnotesize{$\overline\Sigma$}};
            \node[blue, below] at (1,-1) {\footnotesize{$L_\bu$}};
            \node[dgreen, below] at (0,-1) {\footnotesize{$\overline\Sigma$}};
            \filldraw[blue] (0,0) circle (1pt);
            \filldraw[blue] (0,-0.6) circle (1pt);
            \filldraw[blue] (0,0.6) circle (1pt);
    \end{tikzpicture} ~.
\\
        \begin{tikzpicture}[baseline=0,scale=1,square/.style={regular polygon,regular polygon sides=4}]
            \shade[top color=blue!40, bottom color=blue!10]  (0,-1) -- (-1.5,-1) -- (-1.5,1) -- (0,1)-- (0,-1);
            \draw[dgreen,thick, ->- = 0.5](0,-1) -- (0,0); 
            \draw[dgreen,thick, ->- = 0.5](0,0) -- (0,1); 
            \draw[gray!30, line width=3pt] (0,0) to[out=180,in=90 ]  (-1,-1);
            \draw[blue, thick, -<- = 0.5] (0,0) to[out=180,in=90 ]  (-1,-1);
            \node[dgreen, above] at (0,+1) {\footnotesize{$\overline\Sigma$}};
            \node[dgreen, below] at (0,-1) {\footnotesize{$\overline\Sigma$}};
            \node[blue, below] at (-1.15,-1) {\footnotesize{$L_\bu$}};
            \node at (0,0) [square,draw,fill=violet!60,scale=0.5] {}; 
            \node[blue,right] at (-1.5,0.7) {\footnotesize{$D_{A}$}};
        \end{tikzpicture} &=& {\overline\beta}(\bu) \sum_{\bv} \frac{R^{\bu,\bv}}{R^{\bu,\overline\Sigma}} \mu_{A}(\bv,\bu) \begin{tikzpicture}[baseline=0,scale=1]
            \draw[dgreen,thick, ->-=0.5](0,-1) -- (0,-0.6); 
            \draw[blue,thick, ->-=0.5](0,-0.6) -- (0,0); 
            \draw[blue,thick, ->-=0.5](0,0) -- (0,0.6); 
            \draw[dgreen,thick, ->-=0.5](0,0.6) -- (0,1); 
            \draw [blue, thick, -<-=0.5] (0,0) to[out=180,in=90 ]  (-1,-1);
            \node[blue, right] at (0,0.4) {\footnotesize{$L_{\bu + \bv}$}};
            \node[blue, right] at (0,-0.4) {\footnotesize{$L_{\bv}$}};
            \node[dgreen, above] at (0,1.0) {\footnotesize{$\overline\Sigma$}};
            \node[blue, below] at (-1.15,-1) {\footnotesize{$L_\bu$}};
            \node[dgreen, below] at (0,-1) {\footnotesize{$\overline\Sigma$}};
            \filldraw[blue] (0,0) circle (1pt);
            \filldraw[blue] (0,-0.6) circle (1pt);
            \filldraw[blue] (0,0.6) circle (1pt);
        \end{tikzpicture} ~, 
   \eea
   \bea
         \begin{tikzpicture}[baseline=0,scale=1]
            \shade[top color=blue!40, bottom color=blue!10]  (0,-1) -- (-1.5,-1) -- (-1.5,1) -- (0,1)-- (0,-1);
            \draw[dgreen,thick, ->-=0.5](0,-1) -- (0,0); 
            \draw[dgreen,thick, ->-=0.5](0,0) -- (0,1); 
            \draw [blue, thick, -<-=0.5] (0,0) to[out=0,in=90 ]  (1,-1);
            \node[dgreen, above] at (0,1) {\footnotesize{$\overline\Sigma$}};
            \node[dgreen, below] at (0,-1) {\footnotesize{$\overline\Sigma$}};
            \node[blue, below] at (1,-1) {\footnotesize{$L_\bu$}};
            \node[isosceles triangle,scale=0.4, isosceles triangle apex angle=60, draw,fill=violet!60, rotate=90, minimum size =0.01cm] at (0,0){};
            \node[blue,right] at (-1.5,0.7) {\footnotesize{$D_{A}$}};
    \end{tikzpicture} &=&  \overline\beta(\bu) \sum_{\bv} \mu_{A}(\bv,\bu) 
        \begin{tikzpicture}[baseline=0,scale=1]
            \draw[dgreen,thick, ->-=0.5](0,-1) -- (0,-0.6); 
            \draw[blue,thick, ->-=0.5](0,-0.6) -- (0,0); 
            \draw[blue,thick, ->-=0.5](0,0) -- (0,0.6); 
            \draw[dgreen,thick, ->-=0.5](0,0.6) -- (0,1); 
            \draw[blue, thick, -<-=0.5] (0,0) to[out=0,in=90] (1,-1);
            \node[blue, right] at (0,0.4) {\footnotesize{$L_{\bu + \bv}$}};
            \node[blue, right] at (0,-0.4) {\footnotesize{$L_{\bv}$}};
            \node[dgreen, above] at (0,0.9) {\footnotesize{$\overline\Sigma$}};
            \node[blue, below] at (1,-1) {\footnotesize{$L_\bu$}};
            \node[dgreen, below] at (0,-1) {\footnotesize{$\overline\Sigma$}};
            \filldraw[blue] (0,0) circle (1pt);
            \filldraw[blue] (0,-0.6) circle (1pt);
            \filldraw[blue] (0,0.6) circle (1pt);
    \end{tikzpicture} ~,
\\
        \begin{tikzpicture}[baseline=0,scale=1.0,square/.style={regular polygon,regular polygon sides=4}]
            \shade[top color=blue!40, bottom color=blue!10]  (0,-1) -- (1.5,-1) -- (1.5,1) -- (0,1)-- (0,-1);
            \shade[top color=orange!40, bottom color=orange!10]  (-1.5,-1) -- (0,-1) -- (0,1) -- (-1.5,1)-- (-1.5,-1);
            \draw[dgreen,thick, ->- = 0.5](0,-1) -- (0,0); 
            \draw[dgreen,thick, ->- = 0.5](0,0) -- (0,1); 
            \draw [gray!30, line width=3pt] (0,0) to[out=180,in=90 ]  (-1,-1);
            \draw [blue, thick, -<- = 0.5] (0,0) to[out=180,in=90 ]  (-1,-1);
            \node[dgreen, above] at (0,+1) {\footnotesize{$\overline\Sigma$}};
            \node[dgreen, below] at (0,-1) {\footnotesize{$\overline\Sigma$}};
            \node[blue, below] at (-1.15,-1) {\footnotesize{$L_\bu$}};
            \node at (0,0) [square,draw,fill=violet!60,scale=0.5] {}; 
            \node[blue,left] at (1.5,0.7) {\footnotesize{$D_{A}$}};
            \node[orange,right] at (-1.5,0.7) {\footnotesize{$D_{A^2}$}};
        \end{tikzpicture} &=&  \sum_{\bv} \mu_{A^2}(\bv,\bu) R^{\bu,\bv} (\overline\alpha_\bu)_i{}^j \begin{tikzpicture}[baseline=0,scale=1.0]
            \draw[dgreen,thick, ->-=0.5](0,-1) -- (0,-0.6); 
            \draw[blue,thick, ->-=0.5](0,-0.6) -- (0,0); 
            \draw[blue,thick, ->-=0.5](0,0) -- (0,0.6); 
            \draw[dgreen,thick, ->-=0.5](0,0.6) -- (0,1); 
            \draw [blue, thick, -<-=0.5] (0,0) to[out=180,in=90 ]  (-1,-1);
            \node[blue, right] at (0,0.4) {\footnotesize{$L_{\bu + \bv}$}};
            \node[blue, right] at (0,-0.4) {\footnotesize{$L_{\bv}$}};
            \node[dgreen, above] at (0,1.0) {\footnotesize{$\overline\Sigma$}};
            \node[blue, below] at (-1.15,-1) {\footnotesize{$L_\bu$}};
            \node[dgreen, below] at (0,-1) {\footnotesize{$\overline\Sigma$}};
            \node[blue, left] at (0,-0.6) {\footnotesize{$i$}};
            \node[blue, left] at (0,0.6) {\footnotesize{$j$}};
            \filldraw[blue] (0,0) circle (1pt);
            \filldraw[blue] (0,-0.6) circle (1pt);
            \filldraw[blue] (0,0.6) circle (1pt);
        \end{tikzpicture} ~,
        \\
        \begin{tikzpicture}[baseline=0,scale=1.0]
            \shade[top color=blue!40, bottom color=blue!10]  (0,-1) -- (1.5,-1) -- (1.5,1) -- (0,1)-- (0,-1);
            \shade[top color=orange!40, bottom color=orange!10]  (-1.5,-1) -- (0,-1) -- (0,1) -- (-1.5,1)-- (-1.5,-1);
            \draw[dgreen,thick, ->-=0.5](0,-1) -- (0,0); 
            \draw[dgreen,thick, ->-=0.5](0,0) -- (0,1); 
            \draw [gray!30, line width=3pt] (0,0) to[out=0,in=90 ]  (1,-1);
            \draw [blue, thick, -<-=0.5] (0,0) to[out=0,in=90 ]  (1,-1);
            \node[dgreen, above] at (0,1) {\footnotesize{$\overline\Sigma$}};
            \node[dgreen, below] at (0,-1) {\footnotesize{$\overline\Sigma$}};
            \node[blue, below] at (1,-1) {\footnotesize{$L_\bu$}};
            \node[isosceles triangle,scale=0.4, isosceles triangle apex angle=60, draw,fill=violet!60, rotate=90, minimum size =0.01cm] at (0,0){};
            \node[blue,left] at (1.5,0.7) {\footnotesize{$D_{A}$}};
            \node[orange,right] at (-1.5,0.7) {\footnotesize{$D_{A^2}$}};
    \end{tikzpicture} &=&  \sum_{\bv} R^{\bu,\overline\Sigma} \mu_{A^2}(\bv,\bu) (\overline\alpha_{\bu})_i{}^j
        \begin{tikzpicture}[baseline=0,scale=1.0]
            \draw[dgreen,thick, ->-=0.5](0,-1) -- (0,-0.6); 
            \draw[blue,thick, ->-=0.5](0,-0.6) -- (0,0); 
            \draw[blue,thick, ->-=0.5](0,0) -- (0,0.6); 
            \draw[dgreen,thick, ->-=0.5](0,0.6) -- (0,1); 
            \draw[blue, thick, -<-=0.5] (0,0) to[out=0,in=90] (1,-1);
            \node[blue, right] at (0,0.4) {\footnotesize{$L_{\bu + \mathbf{v}}$}};
            \node[blue, right] at (0,-0.4) {\footnotesize{$L_{\bv}$}};
            \node[dgreen, above] at (0,0.9) {\footnotesize{$\overline\Sigma$}};
            \node[blue, below] at (1,-1) {\footnotesize{$L_\bu$}};
            \node[dgreen, below] at (0,-1) {\footnotesize{$\overline\Sigma$}};
            \node[blue, left] at (0,-0.6) {\footnotesize{$i$}};
            \node[blue, left] at (0,0.6) {\footnotesize{$j$}};
            \filldraw[blue] (0,0) circle (1pt);
            \filldraw[blue] (0,-0.6) circle (1pt);
            \filldraw[blue] (0,0.6) circle (1pt);
    \end{tikzpicture} ~,
\eea
and where the functions $\overline \beta(\bv)$ and $\overline\alpha(\bv)$are constrained to satisfy 
\bea
        \frac{\overline\beta(\bu+\widetilde{\bu})}{\overline\beta(\bu)\overline\beta(\widetilde{\bu})} &=& \frac{R^{\bu+\widetilde{\bu},\overline\Sigma}}{R^{\bu,\overline\Sigma} R^{\widetilde{\bu},\overline\Sigma}} R^{\bu,\widetilde{\bu}} \frac{\mu_{A}(\widetilde{\bu},\bu)}{\mu_{A^2}(\bu,\widetilde{\bu})} ~,
    \no\\
       \overline \alpha_{\bu} \cdot \overline \rho(\bv)\cdot (\overline \alpha_\bu)^{-1} &=& R^{\bu,\bv} \frac{\mu_{A^2}(\bv,\bu)}{\mu_{A^2}(\bu,\bv)} \overline\rho(\bv) ~,
        \no\\
      \overline \alpha^{-1}_{\bu} \cdot \overline\alpha_{\widetilde{\bu}}^{-1}  &=& \frac{\mu_{A^2}(\widetilde{\bu},\bu)}{\mu_{A^2}(\bu,\widetilde{\bu})} R^{\bu,\widetilde{\bu}} \overline \alpha^{-1}_{\bu+\widetilde{\bu}}~,
 \eea
  and $\overline \rho(\bu)$ is the representation of the little group labeling $\Sigma$, which satisfies $\overline\rho(\bu)\overline\rho(\widetilde \bu) = \frac{\mu_{A}(\bu,\widetilde \bu)}{\mu_{A^2}(\bu,\widetilde \bu)} \overline\rho(\bu + \widetilde \bu)$. 
 
 \subsection{Fusion junctions between two twist defects}
 
 We next give the expressions for fusions involving two incoming twist defects. We only give four of the junctions, which will be used in the main text,
\bea
    \begin{tikzpicture}[baseline=20,square/.style={regular polygon,regular polygon sides=4},scale=1.0]
        \shade[top color=blue!40, bottom color=blue!10]  (0,0)--(2,0)--(1,1)--(0,0);
        \draw[dgreen,thick, ->-=0.5](0,0) --(1,1); 
        \draw[blue,thick, ->-=0.5] (1,1)--(1,2); 
        \draw[dgreen,thick, ->-=0.5] (2,0) -- (1,1); 
        \node[blue] at (1,0.3) {\footnotesize{$D_{A}$}};
        \node[dgreen, below] at (0,0) {\footnotesize{$\Sigma$}};
        \node[dgreen, below] at (2,0) {\footnotesize{$\overline{\Sigma}$}};
        \node[blue, above] at (1,2) {\footnotesize{$L_{\bv}$}};
         \filldraw[violet!60] (1,1) circle (1.5pt);
    \end{tikzpicture} &=& \frac{1}{M} \sum_{\bu} \mu_A(\bu,\bv-\bu) \begin{tikzpicture}[scale=0.50,baseline = {(0,0)}]
    \draw[thick, blue, ->-=.5] (-1.7,-1) -- (0,0);
    \draw[thick, blue, -<-=.5] (0,0) -- (1.7,-1);
    \draw[thick, blue, -<-=.5] (0,2) -- (0,0);
    \node[blue, above] at (0,2) {\scriptsize $L_{\bv}$};
    \node[blue, below] at (-1.7,-1) {\scriptsize $L_{\bu}$};
    \node[blue, below] at (1.7,-1) {\scriptsize $L_{\bv - \bu}$};
\end{tikzpicture} ~,
\\
         \label{eq:ex1n3}
\begin{tikzpicture}[baseline=20,square/.style={regular polygon,regular polygon sides=4},scale=1.0]
        \shade[top color=orange!40, bottom color=orange!10]  (0,0)--(2,0)--(1,1)--(0,0);
        \draw[dgreen,thick, ->-=0.5](0,0) --(1,1); 
        \draw[blue,thick, ->-=0.5] (1,1)--(1,2); 
        \draw[dgreen,thick, ->-=0.5] (2,0) -- (1,1); 
        \node[orange] at (1,0.3) {\footnotesize{$D_{A^2}$}};
        \node[dgreen, below] at (0,0) {\footnotesize{$\overline{\Sigma}$}};
        \node[dgreen, below] at (2,0) {\footnotesize{$\Sigma$}};
        \node[blue, above] at (1,2) {\footnotesize{$L_{\bv}$}};
         \filldraw[violet!60] (1,1) circle (1.5pt);
    \end{tikzpicture} &=& \frac{1}{M} \sum_{\bu} \mu_{A^2}(\bu,\bv-\bu) \begin{tikzpicture}[scale=0.50,baseline = {(0,0)}]
    \draw[thick, blue, ->-=.5] (-1.7,-1) -- (0,0);
    \draw[thick, blue, -<-=.5] (0,0) -- (1.7,-1);
    \draw[thick, blue, -<-=.5] (0,2) -- (0,0);
      \node[blue, above] at (0,2) {\scriptsize $L_{\bv}$};
    \node[blue, below] at (-1.7,-1) {\scriptsize $L_{\bu}$};
    \node[blue, below] at (1.7,-1) {\scriptsize $L_{\bv - \bu}$};
\end{tikzpicture} ~,
\\
\label{eq:ex2n3}
 \begin{tikzpicture}[baseline=40,square/.style={regular polygon,regular polygon sides=4},scale=1.2,baseline = {(0,1)}]
        \shade[top color=blue!40, bottom color=blue!10]  (0,0)--(2,0)--(1,1)--(0,0);
        \shade[top color=orange!40, bottom color=orange!10]  (2,0)--(2,2)--(1,2) -- (1,1) -- (2,0);
        \draw[dgreen,thick, ->-=0.5](0,0) -- (1,1); 
        \draw[dgreen, thick, ->-=0.5](1,1) -- (1,2); 
        \draw[dgreen,thick, ->-=0.5](2,0) -- (1,1); 
        \filldraw[violet!60] (1,1) circle (1.5pt);
        \node[orange, below] at (1.7,1.8) {\footnotesize {$D_{A^2}$}};
        \node[blue] at (1,0.3) {\footnotesize {$D_{A}$}};
        \node[dgreen, below] at (0,0) {\footnotesize{$\Sigma$}};
        \node[dgreen, above] at (1,2) {\footnotesize{$\overline{\Sigma}$}};
        \node[dgreen, below] at (2,0) {\footnotesize{$\Sigma$}};
        \node[violet!60, left] at (1,1) {\scriptsize $\alpha$};
\end{tikzpicture} &=& \frac{1}{M^{3/2}} \sum_{\bu,\widetilde\bu,i} \mu_A(\bu,\widetilde\bu) [\rho(\bu+\widetilde\bu)^{-1}]_{i\alpha} \begin{tikzpicture}[scale=0.50,baseline = {(0,0)}]
    \draw[thick, blue, ->-=.5] (-1.7,-1) -- (0,0);
    \draw[thick, blue, -<-=.5] (0,0) -- (1.7,-1);
    \draw[thick, blue, -<-=.5] (0,2) -- (0,0);
    \node[blue, above] at (0,2) {\scriptsize $L_{\bu + \widetilde\bu}$};
    \node[blue, below] at (-1.7,-1) {\scriptsize $L_{\bu}$};
    \node[blue, below] at (1.7,-1) {\scriptsize $(L_{\widetilde\bu})_i$};
\end{tikzpicture} ~,
\\
    \begin{tikzpicture}[baseline=40,square/.style={regular polygon,regular polygon sides=4},scale=1.2,baseline = {(0,1)}]
           \shade[top color=orange!40, bottom color=orange!10]  (0,0)--(2,0)--(1,1)--(0,0);
        \shade[top color=blue!40, bottom color=blue!10]  (2,0)--(2,2)--(1,2) -- (1,1) -- (2,0);
        \draw[dgreen,thick, ->-=0.5](0,0) -- (1,1); 
        \draw[dgreen, thick, ->-=0.5](1,1) -- (1,2); 
        \draw[dgreen,thick, ->-=0.5](2,0) -- (1,1); 
           \filldraw[violet!60] (1,1) circle (1.5pt);
        \node[blue, below] at (1.7,1.8) {\footnotesize {$D_{A}$}};
        \node[orange] at (1,0.3) {\footnotesize {$D_{A^2}$}};
        \node[dgreen, above] at (1,2) {\footnotesize{$\Sigma$}};
        \node[dgreen, below] at (0,0) {\footnotesize{$\overline{\Sigma}$}};
        \node[dgreen, below] at (2,0) {\footnotesize{$\overline{\Sigma}$}};
          \node[violet!60, left] at (1,1) {\scriptsize $\alpha$};
\end{tikzpicture} &=& \frac{1}{M^{3/2}} \sum_{\bu,\widetilde\bu,i} \mu_{A^2}(\bu,\widetilde\bu) [\overline{\rho}(\bu+\widetilde\bu)^{-1}]_{i\alpha} \begin{tikzpicture}[scale=0.50,baseline = {(0,0)}]
    \draw[thick, blue, ->-=.5] (-1.7,-1) -- (0,0);
    \draw[thick, blue, -<-=.5] (0,0) -- (1.7,-1);
    \draw[thick, blue, -<-=.5] (0,2) -- (0,0);
    \node[blue, above] at (0,2) {\scriptsize $L_{\bu + \widetilde\bu}$};
    \node[blue, below] at (-1.7,-1) {\scriptsize $L_\bu$};
    \node[blue, below] at (1.7,-1) {\scriptsize $(L_{\widetilde\bu})_i$};
\end{tikzpicture} ~,
\eea
where $(L_\bu)_i$ is shorthand notation for the junction between $\Sigma$ (resp. $\overline \Sigma$) and $L_\bu$ given by the element of $\mathrm{Hom}(\Sigma, L_\bu)$ (resp. $\mathrm{Hom}(\overline\Sigma, L_\bu)$) labeled by $i$; see footnote \ref{footnote:shorthand}.

\end{appendix}

\bibliographystyle{ytamsalpha}
\def\arxivfont{\rm}
\bibliography{ref}

\newcommand{\etalchar}[1]{$^{#1}$}
\providecommand{\bysame}{\leavevmode\hbox to3em{\hrulefill}\thinspace}
\providecommand{\MR}{\relax\ifhmode\unskip\space\fi MR }
\providecommand{\MRhref}[2]{%
  \href{http://www.ams.org/mathscinet-getitem?mr=#1}{#2}
}
\providecommand{\href}[2]{#2}
\providecommand{\doihref}[2]{\href{#1}{#2}}
\providecommand{\arxivfont}{\tt}
\begin{thebibliography}{BSNTW25}

\bibitem[ABBSN22]{Apruzzi:2022rei}
F.~Apruzzi, I.~Bah, F.~Bonetti, and S.~Schafer-Nameki, \emph{{Non-Invertible
  Symmetries from Holography and Branes}},
  \href{http://arxiv.org/abs/2208.07373}{{\arxivfont arXiv:2208.07373
  [hep-th]}}.

\bibitem[ABC{\etalchar{+}}22]{Antinucci:2022vyk}
A.~Antinucci, F.~Benini, C.~Copetti, G.~Galati, and G.~Rizi, \emph{{The
  holography of non-invertible self-duality symmetries}},
  \doihref{http://dx.doi.org/10.1007/JHEP03(2025)052}{JHEP \textbf{03} (2025)
  052}, \href{http://arxiv.org/abs/2210.09146}{{\arxivfont arXiv:2210.09146
  [hep-th]}}.

\bibitem[ABE{\etalchar{+}}21]{Apruzzi:2021nmk}
F.~Apruzzi, F.~Bonetti, I.~n.~G. Etxebarria, S.~S. Hosseini, and
  S.~Schafer-Nameki, \emph{{Symmetry TFTs from String Theory}},
  \href{http://arxiv.org/abs/2112.02092}{{\arxivfont arXiv:2112.02092
  [hep-th]}}.

\bibitem[ACGR22]{Antinucci:2022cdi}
A.~Antinucci, C.~Copetti, G.~Galati, and G.~Rizi,
  \emph{{{\textquotedblleft}Zoology{\textquotedblright} of non-invertible
  duality defects: the view from class $ \mathcal{S} $}},
  \doihref{http://dx.doi.org/10.1007/JHEP04(2024)036}{JHEP \textbf{04} (2024)
  036}, \href{http://arxiv.org/abs/2212.09549}{{\arxivfont arXiv:2212.09549
  [hep-th]}}.

\bibitem[AGR22]{Antinucci:2022eat}
A.~Antinucci, G.~Galati, and G.~Rizi, \emph{{On Continuous 2-Category
  Symmetries and Yang-Mills Theory}},
  \href{http://arxiv.org/abs/2206.05646}{{\arxivfont arXiv:2206.05646
  [hep-th]}}.

\bibitem[AHKZ25]{Albert:2025umy}
J.~Albert, Y.~Honda, J.~Kaidi, and Y.~Zheng, \emph{{Haagerup Symmetry in
  $(E_8)_1$?}}, \href{http://arxiv.org/abs/2512.08225}{{\arxivfont
  arXiv:2512.08225 [hep-th]}}.

\bibitem[And24]{Ando:2024hun}
T.~Ando, \emph{{A journey on self-$G$-ality}},
  \href{http://arxiv.org/abs/2405.15648}{{\arxivfont arXiv:2405.15648
  [cond-mat.str-el]}}.

\bibitem[Apr22]{Apruzzi:2022dlm}
F.~Apruzzi, \emph{{Higher Form Symmetries TFT in 6d}},
  \href{http://arxiv.org/abs/2203.10063}{{\arxivfont arXiv:2203.10063
  [hep-th]}}.

\bibitem[ASNW24]{Apruzzi:2024cty}
F.~Apruzzi, S.~Schafer-Nameki, and A.~Warman, \emph{{Non-Invertible Symmetries
  in 6d from Green-Schwarz Automorphisms}},
  \href{http://arxiv.org/abs/2411.09674}{{\arxivfont arXiv:2411.09674
  [hep-th]}}.

\bibitem[ATRG22]{Arias-Tamargo:2022nlf}
G.~Arias-Tamargo and D.~Rodriguez-Gomez, \emph{{Non-Invertible Symmetries from
  Discrete Gauging and Completeness of the Spectrum}},
  \href{http://arxiv.org/abs/2204.07523}{{\arxivfont arXiv:2204.07523
  [hep-th]}}.

\bibitem[BBCW14]{Barkeshli:2014cna}
M.~Barkeshli, P.~Bonderson, M.~Cheng, and Z.~Wang, \emph{{Symmetry
  Fractionalization, Defects, and Gauging of Topological Phases}},
  \doihref{http://dx.doi.org/10.1103/PhysRevB.100.115147}{Phys. Rev. B
  \textbf{100} (2019) 115147},
  \href{http://arxiv.org/abs/1410.4540}{{\arxivfont arXiv:1410.4540
  [cond-mat.str-el]}}.

\bibitem[BBFP22a]{Bartsch:2022mpm}
T.~Bartsch, M.~Bullimore, A.~E.~V. Ferrari, and J.~Pearson,
  \emph{{Non-invertible Symmetries and Higher Representation Theory I}},
  \href{http://arxiv.org/abs/2208.05993}{{\arxivfont arXiv:2208.05993
  [hep-th]}}.

\bibitem[BBFP22b]{Bartsch:2022ytj}
\bysame, \emph{{Non-invertible symmetries and higher representation theory
  II}}, \doihref{http://dx.doi.org/10.21468/SciPostPhys.17.2.067}{SciPost Phys.
  \textbf{17} (2024) 067}, \href{http://arxiv.org/abs/2212.07393}{{\arxivfont
  arXiv:2212.07393 [hep-th]}}.

\bibitem[BBG23]{Bartsch:2023wvv}
T.~Bartsch, M.~Bullimore, and A.~Grigoletto, \emph{{Representation theory for
  categorical symmetries}}, \href{http://arxiv.org/abs/2305.17165}{{\arxivfont
  arXiv:2305.17165 [hep-th]}}.

\bibitem[BBPSN23]{Bhardwaj:2023idu}
L.~Bhardwaj, L.~E. Bottini, D.~Pajer, and S.~Sch{\"a}fer-Nameki, \emph{{Gapped
  phases with non-invertible symmetries: (1+1)d}},
  \doihref{http://dx.doi.org/10.21468/SciPostPhys.18.1.032}{SciPost Phys.
  \textbf{18} (2025) 032}, \href{http://arxiv.org/abs/2310.03784}{{\arxivfont
  arXiv:2310.03784 [hep-th]}}.

\bibitem[BBSNT22]{Bhardwaj:2022yxj}
L.~Bhardwaj, L.~Bottini, S.~Schafer-Nameki, and A.~Tiwari,
  \emph{{Non-Invertible Higher-Categorical Symmetries}},
  \href{http://arxiv.org/abs/2204.06564}{{\arxivfont arXiv:2204.06564
  [hep-th]}}.

\bibitem[BCDP22]{Benini:2022hzx}
F.~Benini, C.~Copetti, and L.~Di~Pietro, \emph{{Factorization and global
  symmetries in holography}},
  \href{http://arxiv.org/abs/2203.09537}{{\arxivfont arXiv:2203.09537
  [hep-th]}}.

\bibitem[BCP13]{Brunner:2013xna}
I.~Brunner, N.~Carqueville, and D.~Plencner, \emph{{A quick guide to defect
  orbifolds}}, \doihref{http://dx.doi.org/10.1090/pspum/088/01456}{Proc. Symp.
  Pure Math. \textbf{88} (2014) 231--242},
  \href{http://arxiv.org/abs/1310.0062}{{\arxivfont arXiv:1310.0062 [hep-th]}}.

\bibitem[BDSNY24]{Bhardwaj:2024xcx}
L.~Bhardwaj, T.~D{\'e}coppet, S.~Schafer-Nameki, and M.~Yu, \emph{{Fusion
  3-Categories for Duality Defects}},
  \doihref{http://dx.doi.org/10.1007/s00220-025-05388-1}{Commun. Math. Phys.
  \textbf{406} (2025) 208}, \href{http://arxiv.org/abs/2408.13302}{{\arxivfont
  arXiv:2408.13302 [math.CT]}}.

\bibitem[BDZH22]{Bashmakov:2022jtl}
V.~Bashmakov, M.~Del~Zotto, and A.~Hasan, \emph{{On the 6d Origin of
  Non-invertible Symmetries in 4d}},
  \href{http://arxiv.org/abs/2206.07073}{{\arxivfont arXiv:2206.07073
  [hep-th]}}.

\bibitem[BDZHK22]{Bashmakov:2022uek}
V.~Bashmakov, M.~Del~Zotto, A.~Hasan, and J.~Kaidi, \emph{{Non-invertible
  symmetries of class S theories}},
  \doihref{http://dx.doi.org/10.1007/JHEP05(2023)225}{JHEP \textbf{05} (2023)
  225}, \href{http://arxiv.org/abs/2211.05138}{{\arxivfont arXiv:2211.05138
  [hep-th]}}.

\bibitem[BDZM24]{Bonetti:2024etn}
F.~Bonetti, M.~Del~Zotto, and R.~Minasian, \emph{{SymTFTs and non-invertible
  symmetries of 6d (2,0) SCFTs of type D from M-theory}},
  \doihref{http://dx.doi.org/10.1007/JHEP02(2025)156}{JHEP \textbf{02} (2025)
  156}, \href{http://arxiv.org/abs/2412.07842}{{\arxivfont arXiv:2412.07842
  [hep-th]}}.

\bibitem[BGH{\etalchar{+}}25]{Bhardwaj:2025jtf}
L.~Bhardwaj, Y.~Gai, S.-J. Huang, K.~Inamura, S.~Schafer-Nameki, A.~Tiwari, and
  A.~Warman, \emph{{Gapless Phases in (2+1)d with Non-Invertible Symmetries}},
  \href{http://arxiv.org/abs/2503.12699}{{\arxivfont arXiv:2503.12699
  [cond-mat.str-el]}}.

\bibitem[BKN21]{Burbano:2021loy}
I.~M. Burbano, J.~Kulp, and J.~Neuser, \emph{{Duality Defects in $E_8$}},
  \href{http://arxiv.org/abs/2112.14323}{{\arxivfont arXiv:2112.14323
  [hep-th]}}.

\bibitem[Bro92]{Brown1992-ds}
W.~C. Brown, \emph{Matrices over commutative rings}, Chapman \& Hall Pure and
  Applied Mathematics, CRC Press, Boca Raton, FL, November 1992.

\bibitem[BSN24]{Bottini:2025hri}
L.~E. Bottini and S.~Schafer-Nameki, \emph{{Construction of a Gapless Phase
  with Haagerup Symmetry}},
  \doihref{http://dx.doi.org/10.1103/PhysRevLett.134.191602}{Phys. Rev. Lett.
  \textbf{134} (2025) 191602},
  \href{http://arxiv.org/abs/2410.19040}{{\arxivfont arXiv:2410.19040
  [hep-th]}}.

\bibitem[BSNTW25]{Bhardwaj:2025piv}
L.~Bhardwaj, S.~Schafer-Nameki, A.~Tiwari, and A.~Warman, \emph{{Gapped Phases
  in (2+1)d with Non-Invertible Symmetries: Part II}},
  \href{http://arxiv.org/abs/2502.20440}{{\arxivfont arXiv:2502.20440
  [hep-th]}}.

\bibitem[BSNW22]{Bhardwaj:2022lsg}
L.~Bhardwaj, S.~Schafer-Nameki, and J.~Wu, \emph{{Universal Non-Invertible
  Symmetries}}, \href{http://arxiv.org/abs/2208.05973}{{\arxivfont
  arXiv:2208.05973 [hep-th]}}.

\bibitem[BT17]{Bhardwaj:2017xup}
L.~Bhardwaj and Y.~Tachikawa, \emph{{On Finite Symmetries and Their Gauging in
  Two Dimensions}}, \doihref{http://dx.doi.org/10.1007/JHEP03(2018)189}{JHEP
  \textbf{03} (2018) 189},
\href{http://arxiv.org/abs/1704.02330}{{\arxivfont arXiv:1704.02330 [hep-th]}}.

\bibitem[CCH{\etalchar{+}}21]{Choi:2021kmx}
Y.~Choi, C.~Cordova, P.-S. Hsin, H.~T. Lam, and S.-H. Shao,
  \emph{{Non-Invertible Duality Defects in 3+1 Dimensions}},
  \href{http://arxiv.org/abs/2111.01139}{{\arxivfont arXiv:2111.01139
  [hep-th]}}.

\bibitem[CCH{\etalchar{+}}22]{Choi:2022zal}
\bysame, \emph{{Non-invertible Condensation, Duality, and Triality Defects in
  3+1 Dimensions}}, \href{http://arxiv.org/abs/2204.09025}{{\arxivfont
  arXiv:2204.09025 [hep-th]}}.

\bibitem[CDH{\etalchar{+}}21]{Chen:2021xuc}
X.~Chen, A.~Dua, P.-S. Hsin, C.-M. Jian, W.~Shirley, and C.~Xu, \emph{{Loops in
  4+1d Topological Phases}}, \href{http://arxiv.org/abs/2112.02137}{{\arxivfont
  arXiv:2112.02137 [cond-mat.str-el]}}.

\bibitem[CLS{\etalchar{+}}18]{Chang:2018iay}
C.-M. Chang, Y.-H. Lin, S.-H. Shao, Y.~Wang, and X.~Yin, \emph{{Topological
  Defect Lines and Renormalization Group Flows in Two Dimensions}},
  \doihref{http://dx.doi.org/10.1007/JHEP01(2019)026}{JHEP \textbf{01} (2019)
  026}, \href{http://arxiv.org/abs/1802.04445}{{\arxivfont arXiv:1802.04445
  [hep-th]}}.

\bibitem[CLS22a]{Choi:2022jqy}
Y.~Choi, H.~T. Lam, and S.-H. Shao, \emph{{Non-invertible Global Symmetries in
  the Standard Model}}, \href{http://arxiv.org/abs/2205.05086}{{\arxivfont
  arXiv:2205.05086 [hep-th]}}.

\bibitem[CLS22b]{Choi:2022rfe}
\bysame, \emph{{Non-invertible Time-reversal Symmetry}},
  \href{http://arxiv.org/abs/2208.04331}{{\arxivfont arXiv:2208.04331
  [hep-th]}}.

\bibitem[CLS23]{Choi:2023vgk}
Y.~Choi, D.-C. Lu, and Z.~Sun, \emph{{Self-duality under gauging a
  non-invertible symmetry}},
  \doihref{http://dx.doi.org/10.1007/JHEP01(2024)142}{JHEP \textbf{01} (2024)
  142}, \href{http://arxiv.org/abs/2310.19867}{{\arxivfont arXiv:2310.19867
  [hep-th]}}.

\bibitem[CO22]{Cordova:2022ieu}
C.~Cordova and K.~Ohmori, \emph{{Non-Invertible Chiral Symmetry and Exponential
  Hierarchies}}, \href{http://arxiv.org/abs/2205.06243}{{\arxivfont
  arXiv:2205.06243 [hep-th]}}.

\bibitem[CR12]{Carqueville:2012dk}
N.~Carqueville and I.~Runkel, \emph{{Orbifold completion of defect
  bicategories}}, \doihref{http://dx.doi.org/10.4171/qt/76}{Quantum Topol.
  \textbf{7} (2016) 203--279},
  \href{http://arxiv.org/abs/1210.6363}{{\arxivfont arXiv:1210.6363
  [math.QA]}}.

\bibitem[CRSS23]{Choi:2023xjw}
Y.~Choi, B.~C. Rayhaun, Y.~Sanghavi, and S.-H. Shao, \emph{{Remarks on
  boundaries, anomalies, and noninvertible symmetries}},
  \doihref{http://dx.doi.org/10.1103/PhysRevD.108.125005}{Phys. Rev. D
  \textbf{108} (2023) 125005},
  \href{http://arxiv.org/abs/2305.09713}{{\arxivfont arXiv:2305.09713
  [hep-th]}}.

\bibitem[CW22a]{Chatterjee:2022kxb}
A.~Chatterjee and X.-G. Wen, \emph{{Algebra of local symmetric operators and
  braided fusion $n$-category -- symmetry is a shadow of topological order}},
  \href{http://arxiv.org/abs/2203.03596}{{\arxivfont arXiv:2203.03596
  [cond-mat.str-el]}}.

\bibitem[CW22b]{Chatterjee:2022tyg}
\bysame, \emph{{Holographic theory for the emergence and the symmetry
  protection of gaplessness and for continuous phase transitions}},
  \href{http://arxiv.org/abs/2205.06244}{{\arxivfont arXiv:2205.06244
  [cond-mat.str-el]}}.

\bibitem[DAGV22]{Damia:2022bcd}
J.~A. Damia, R.~Argurio, and E.~Garcia-Valdecasas, \emph{{Non-Invertible
  Defects in 5d, Boundaries and Holography}},
  \href{http://arxiv.org/abs/2207.02831}{{\arxivfont arXiv:2207.02831
  [hep-th]}}.

\bibitem[DAT22]{Damia:2022rxw}
J.~A. Damia, R.~Argurio, and L.~Tizzano, \emph{{Continuous Generalized
  Symmetries in Three Dimensions}},
  \href{http://arxiv.org/abs/2206.14093}{{\arxivfont arXiv:2206.14093
  [hep-th]}}.

\bibitem[DZGE22]{DelZotto:2022ras}
M.~Del~Zotto and I.~n. Garc\'\i{}a~Etxebarria, \emph{{Global Structures from
  the Infrared}}, \href{http://arxiv.org/abs/2204.06495}{{\arxivfont
  arXiv:2204.06495 [hep-th]}}.

\bibitem[DZHRG25]{DelZotto:2025yoy}
M.~Del~Zotto, A.~Hasan, and E.~Riedel~G{\r{a}}rding, \emph{{SymTFT, Protected
  Gaplessness, and Spontaneous Breaking of Non-invertible Symmetries}},
  \href{http://arxiv.org/abs/2504.18501}{{\arxivfont arXiv:2504.18501
  [hep-th]}}.

\bibitem[DZMM24]{DelZotto:2024tae}
M.~Del~Zotto, S.~N. Meynet, and R.~Moscrop, \emph{{Remarks on geometric
  engineering, symmetry TFTs and anomalies}},
  \doihref{http://dx.doi.org/10.1007/JHEP07(2024)220}{JHEP \textbf{07} (2024)
  220}, \href{http://arxiv.org/abs/2402.18646}{{\arxivfont arXiv:2402.18646
  [hep-th]}}.

\bibitem[ENOM09]{Etingof:2009yvg}
P.~Etingof, D.~Nikshych, V.~Ostrik, and w.~a. a. b.~E. Meir, \emph{{Fusion
  categories and homotopy theory}},
  \href{http://arxiv.org/abs/0909.3140}{{\arxivfont arXiv:0909.3140
  [math.QA]}}.

\bibitem[FFRS04]{Frohlich:2004ef}
J.~Frohlich, J.~Fuchs, I.~Runkel, and C.~Schweigert, \emph{{Kramers-Wannier
  duality from conformal defects}},
  \doihref{http://dx.doi.org/10.1103/PhysRevLett.93.070601}{Phys. Rev. Lett.
  \textbf{93} (2004) 070601},
  \href{http://arxiv.org/abs/cond-mat/0404051}{{\arxivfont
  arXiv:cond-mat/0404051}}.

\bibitem[FFRS06]{Frohlich:2006ch}
\bysame, \emph{{Duality and defects in rational conformal field theory}},
  \doihref{http://dx.doi.org/10.1016/j.nuclphysb.2006.11.017}{Nucl. Phys. B
  \textbf{763} (2007) 354--430},
  \href{http://arxiv.org/abs/hep-th/0607247}{{\arxivfont
  arXiv:hep-th/0607247}}.

\bibitem[FFRS09]{Frohlich:2009gb}
\bysame, \doihref{http://dx.doi.org/10.1142/9789814304634_0056}{\emph{{Defect
  lines, dualities, and generalised orbifolds}}}, {16th International Congress
  on Mathematical Physics}, 9 2009.
  \href{http://arxiv.org/abs/0909.5013}{{\arxivfont arXiv:0909.5013
  [math-ph]}}.

\bibitem[FMT22]{Freed:2022qnc}
D.~S. Freed, G.~W. Moore, and C.~Teleman, \emph{{Topological symmetry in
  quantum field theory}}, \href{http://arxiv.org/abs/2209.07471}{{\arxivfont
  arXiv:2209.07471 [hep-th]}}.

\bibitem[FPSV15]{fuchs2015brauer}
J.~Fuchs, J.~Priel, C.~Schweigert, and A.~Valentino, \emph{On the brauer groups
  of symmetries of abelian dijkgraaf--witten theories}, Communications in
  Mathematical Physics \textbf{339} (2015) 385--405.

\bibitem[FRS02]{Fuchs:2002cm}
J.~Fuchs, I.~Runkel, and C.~Schweigert, \emph{{TFT construction of RCFT
  correlators 1. Partition functions}},
  \doihref{http://dx.doi.org/10.1016/S0550-3213(02)00744-7}{Nucl. Phys. B
  \textbf{646} (2002) 353--497},
  \href{http://arxiv.org/abs/hep-th/0204148}{{\arxivfont
  arXiv:hep-th/0204148}}.

\bibitem[FT12]{Freed:2012bs}
D.~S. Freed and C.~Teleman, \emph{{Relative quantum field theory}},
  \doihref{http://dx.doi.org/10.1007/s00220-013-1880-1}{Commun. Math. Phys.
  \textbf{326} (2014) 459--476},
  \href{http://arxiv.org/abs/1212.1692}{{\arxivfont arXiv:1212.1692 [hep-th]}}.

\bibitem[FT18]{Freed:2018cec}
\bysame, \emph{{Topological dualities in the Ising model}},
  \href{http://arxiv.org/abs/1806.00008}{{\arxivfont arXiv:1806.00008
  [math.AT]}}.

\bibitem[GE22]{GarciaEtxebarria:2022vzq}
I.~n. Garc\'\i{}a~Etxebarria, \emph{{Branes and Non-Invertible Symmetries}},
  \href{http://arxiv.org/abs/2208.07508}{{\arxivfont arXiv:2208.07508
  [hep-th]}}.

\bibitem[GK20]{Gaiotto:2020iye}
D.~Gaiotto and J.~Kulp, \emph{{Orbifold groupoids}},
  \doihref{http://dx.doi.org/10.1007/JHEP02(2021)132}{JHEP \textbf{02} (2021)
  132}, \href{http://arxiv.org/abs/2008.05960}{{\arxivfont arXiv:2008.05960
  [hep-th]}}.

\bibitem[GNN09]{gelaki2009centers}
S.~Gelaki, D.~Naidu, and D.~Nikshych, \emph{Centers of graded fusion
  categories}, \doihref{http://dx.doi.org/10.2140/ant.2009.3.959}{Algebra
  Number Theory \textbf{3} (2009) 959--990},
  \href{http://arxiv.org/abs/0905.3117}{{\arxivfont arXiv:0905.3117
  [math.QA]}}.

\bibitem[HHTZ22]{Heckman:2022muc}
J.~J. Heckman, M.~H\"ubner, E.~Torres, and H.~Y. Zhang, \emph{{The Branes
  Behind Generalized Symmetry Operators}},
  \href{http://arxiv.org/abs/2209.03343}{{\arxivfont arXiv:2209.03343
  [hep-th]}}.

\bibitem[HLO{\etalchar{+}}21]{Huang:2021nvb}
T.-C. Huang, Y.-H. Lin, K.~Ohmori, Y.~Tachikawa, and M.~Tezuka,
  \emph{{Numerical Evidence for a Haagerup Conformal Field Theory}},
  \doihref{http://dx.doi.org/10.1103/PhysRevLett.128.231603}{Phys. Rev. Lett.
  \textbf{128} (2022) 231603},
  \href{http://arxiv.org/abs/2110.03008}{{\arxivfont arXiv:2110.03008
  [cond-mat.stat-mech]}}.

\bibitem[HLS21]{Huang:2021zvu}
T.-C. Huang, Y.-H. Lin, and S.~Seifnashri, \emph{{Construction of
  two-dimensional topological field theories with non-invertible symmetries}},
  \doihref{http://dx.doi.org/10.1007/JHEP12(2021)028}{JHEP \textbf{12} (2021)
  028}, \href{http://arxiv.org/abs/2110.02958}{{\arxivfont arXiv:2110.02958
  [hep-th]}}.

\bibitem[HT22]{Hayashi:2022fkw}
Y.~Hayashi and Y.~Tanizaki, \emph{{Non-invertible self-duality defects of
  Cardy-Rabinovici model and mixed gravitational anomaly}},
  \href{http://arxiv.org/abs/2204.07440}{{\arxivfont arXiv:2204.07440
  [hep-th]}}.

\bibitem[Ina22]{Inamura:2022lun}
K.~Inamura, \emph{{Fermionization of fusion category symmetries in 1+1
  dimensions}}, \href{http://arxiv.org/abs/2206.13159}{{\arxivfont
  arXiv:2206.13159 [cond-mat.str-el]}}.

\bibitem[Izu01]{izumi2001structure}
M.~Izumi, \emph{The structure of sectors associated with {L}ongo--{R}ehren
  inclusions {II}: examples},
  \doihref{http://dx.doi.org/10.1142/S0129055X01000818}{Reviews in Mathematical
  Physics \textbf{13} (2001) 603--674}.

\bibitem[JL08]{jordan2009classification}
D.~{Jordan} and E.~{Larson}, \emph{{On the classification of certain fusion
  categories}}, \doihref{http://dx.doi.org/10.48550/arXiv.0812.1603}{arXiv
  e-prints (2008) arXiv:0812.1603},
  \href{http://arxiv.org/abs/0812.1603}{{\arxivfont arXiv:0812.1603
  [math.QA]}}.

\bibitem[JT25]{Jia:2025bui}
Q.~Jia and J.~Tian, \emph{{Symmetry, Symmetry Topological Field Theory and von
  Neumann Algebra}}, \href{http://arxiv.org/abs/2507.17103}{{\arxivfont
  arXiv:2507.17103 [hep-th]}}.

\bibitem[KLW{\etalchar{+}}20]{Kong:2020cie}
L.~Kong, T.~Lan, X.-G. Wen, Z.-H. Zhang, and H.~Zheng, \emph{{Algebraic higher
  symmetry and categorical symmetry -- a holographic and entanglement view of
  symmetry}},
  \doihref{http://dx.doi.org/10.1103/PhysRevResearch.2.043086}{Phys. Rev. Res.
  \textbf{2} (2020) 043086}, \href{http://arxiv.org/abs/2005.14178}{{\arxivfont
  arXiv:2005.14178 [cond-mat.str-el]}}.

\bibitem[KNY21]{Koide:2021zxj}
M.~Koide, Y.~Nagoya, and S.~Yamaguchi, \emph{{Non-invertible topological
  defects in 4-dimensional $\mathbb{Z}_2$ pure lattice gauge theory}},
  \href{http://arxiv.org/abs/2109.05992}{{\arxivfont arXiv:2109.05992
  [hep-th]}}.

\bibitem[KNZZ23]{Kaidi:2023maf}
J.~Kaidi, E.~Nardoni, G.~Zafrir, and Y.~Zheng, \emph{{Symmetry TFTs and
  anomalies of non-invertible symmetries}},
  \doihref{http://dx.doi.org/10.1007/JHEP10(2023)053}{JHEP \textbf{10} (2023)
  053}, \href{http://arxiv.org/abs/2301.07112}{{\arxivfont arXiv:2301.07112
  [hep-th]}}.

\bibitem[KORS20]{Komargodski:2020mxz}
Z.~Komargodski, K.~Ohmori, K.~Roumpedakis, and S.~Seifnashri, \emph{{Symmetries
  and strings of adjoint QCD$_{2}$}},
  \doihref{http://dx.doi.org/10.1007/JHEP03(2021)103}{JHEP \textbf{03} (2021)
  103}, \href{http://arxiv.org/abs/2008.07567}{{\arxivfont arXiv:2008.07567
  [hep-th]}}.

\bibitem[KOZ21]{Kaidi:2021xfk}
J.~Kaidi, K.~Ohmori, and Y.~Zheng, \emph{{Kramers-Wannier-like duality defects
  in (3+1)d gauge theories}},
  \href{http://arxiv.org/abs/2111.01141}{{\arxivfont arXiv:2111.01141
  [hep-th]}}.

\bibitem[KOZ22]{Kaidi:2022cpf}
\bysame, \emph{{Symmetry TFTs for Non-invertible Defects}},
  \doihref{http://dx.doi.org/10.1007/s00220-023-04859-7}{Commun. Math. Phys.
  \textbf{404} (2023) 1021--1124},
  \href{http://arxiv.org/abs/2209.11062}{{\arxivfont arXiv:2209.11062
  [hep-th]}}.

\bibitem[KSSS]{upcoming}
J.~Kaidi, X.~Shi, S.~Shimamori, and Z.~Sun, \emph{{The SymTFT for N-ality
  defects: Part II}}, to appear .

\bibitem[KZZ22]{Kaidi:2022uux}
J.~Kaidi, G.~Zafrir, and Y.~Zheng, \emph{{Non-Invertible Symmetries of
  $\mathcal{N}=4$ SYM and Twisted Compactification}},
  \href{http://arxiv.org/abs/2205.01104}{{\arxivfont arXiv:2205.01104
  [hep-th]}}.

\bibitem[LDOV21]{Lootens:2021tet}
L.~Lootens, C.~Delcamp, G.~Ortiz, and F.~Verstraete, \emph{{Dualities in
  one-dimensional quantum lattice models: symmetric Hamiltonians and matrix
  product operator intertwiners}},
  \href{http://arxiv.org/abs/2112.09091}{{\arxivfont arXiv:2112.09091
  [quant-ph]}}.

\bibitem[LOST22]{Lin:2022dhv}
Y.-H. Lin, M.~Okada, S.~Seifnashri, and Y.~Tachikawa, \emph{{Asymptotic Density
  of States in 2D CFTs with Non-Invertible Symmetries}},
  \doihref{http://dx.doi.org/10.1007/JHEP03(2023)094}{JHEP \textbf{03} (2023)
  094}, \href{http://arxiv.org/abs/2208.05495}{{\arxivfont arXiv:2208.05495
  [hep-th]}}.

\bibitem[LRS22]{Lin:2022xod}
L.~Lin, D.~G. Robbins, and E.~Sharpe, \emph{{Decomposition, condensation
  defects, and fusion}}, \href{http://arxiv.org/abs/2208.05982}{{\arxivfont
  arXiv:2208.05982 [hep-th]}}.

\bibitem[LS22]{Lu:2022ver}
D.-C. Lu and Z.~Sun, \emph{{On Triality Defects in 2d CFT}},
  \href{http://arxiv.org/abs/2208.06077}{{\arxivfont arXiv:2208.06077
  [hep-th]}}.

\bibitem[LSCH20]{Lou:2020gfq}
J.~Lou, C.~Shen, C.~Chen, and L.-Y. Hung, \emph{{A (dummy\textquoteright{}s)
  guide to working with gapped boundaries via (fermion) condensation}},
  \doihref{http://dx.doi.org/10.1007/JHEP02(2021)171}{JHEP \textbf{02} (2021)
  171}, \href{http://arxiv.org/abs/2007.10562}{{\arxivfont arXiv:2007.10562
  [hep-th]}}.

\bibitem[LSY24]{Lu:2024ytl}
D.-C. Lu, Z.~Sun, and Y.-Z. You, \emph{{Realizing triality and $p$-ality by
  lattice twisted gauging in (1+1)d quantum spin systems}},
  \doihref{http://dx.doi.org/10.21468/SciPostPhys.17.5.136}{SciPost Phys.
  \textbf{17} (2024) 136}, \href{http://arxiv.org/abs/2405.14939}{{\arxivfont
  arXiv:2405.14939 [cond-mat.str-el]}}.

\bibitem[LSZ24]{Lu:2024lzf}
D.-C. Lu, Z.~Sun, and Z.~Zhang, \emph{{Exploring $G$-ality defects in 2-dim
  QFTs}}, \href{http://arxiv.org/abs/2406.12151}{{\arxivfont arXiv:2406.12151
  [hep-th]}}.

\bibitem[LSZ25]{Lu:2025gpt}
\bysame, \emph{{SymSETs and self-dualities under gauging non-invertible
  symmetries}}, \href{http://arxiv.org/abs/2501.07787}{{\arxivfont
  arXiv:2501.07787 [hep-th]}}.

\bibitem[LYZ23]{Lawrie:2023tdz}
C.~Lawrie, X.~Yu, and H.~Y. Zhang, \emph{{Intermediate defect groups,
  polarization pairs, and noninvertible duality defects}},
  \doihref{http://dx.doi.org/10.1103/PhysRevD.109.026005}{Phys. Rev. D
  \textbf{109} (2024) 026005},
  \href{http://arxiv.org/abs/2306.11783}{{\arxivfont arXiv:2306.11783
  [hep-th]}}.

\bibitem[MMT22]{Moradi:2022lqp}
H.~Moradi, S.~F. Moosavian, and A.~Tiwari, \emph{{Topological holography:
  Towards a unification of Landau and beyond-Landau physics}},
  \doihref{http://dx.doi.org/10.21468/SciPostPhysCore.6.4.066}{SciPost Phys.
  Core \textbf{6} (2023) 066},
  \href{http://arxiv.org/abs/2207.10712}{{\arxivfont arXiv:2207.10712
  [cond-mat.str-el]}}.

\bibitem[MO25]{Maeda:2025rxc}
J.~Maeda and T.~Oishi, \emph{{$N$-ality symmetry and SPT phases in (1+1)d}},
  \href{http://arxiv.org/abs/2504.20151}{{\arxivfont arXiv:2504.20151
  [hep-th]}}.

\bibitem[NR14]{nikshych2014categorical}
D.~Nikshych and B.~Riepel, \emph{Categorical lagrangian grassmannians and
  brauer--picard groups of pointed fusion categories}, Journal of Algebra
  \textbf{411} (2014) 191--214.

\bibitem[PR25]{Putrov:2025xmw}
P.~Putrov and R.~Radhakrishnan, \emph{{Braidings on topological operators,
  anomaly of higher-form symmetries and the SymTFT}},
  \href{http://arxiv.org/abs/2503.13633}{{\arxivfont arXiv:2503.13633
  [hep-th]}}.

\bibitem[PZ00]{Petkova:2000ip}
V.~B. Petkova and J.~B. Zuber, \emph{{Generalized twisted partition
  functions}}, \doihref{http://dx.doi.org/10.1016/S0370-2693(01)00276-3}{Phys.
  Lett. B \textbf{504} (2001) 157--164},
  \href{http://arxiv.org/abs/hep-th/0011021}{{\arxivfont
  arXiv:hep-th/0011021}}.

\bibitem[QSC{\etalchar{+}}25]{Qi:2025tal}
M.~Qi, R.~Sohal, X.~Chen, D.~T. Stephen, and A.~Prem, \emph{{The Symmetry Taco:
  Equivalences between Gapped, Gapless, and Mixed-State SPTs}},
  \href{http://arxiv.org/abs/2507.05335}{{\arxivfont arXiv:2507.05335
  [cond-mat.str-el]}}.

\bibitem[RSS22]{Roumpedakis:2022aik}
K.~Roumpedakis, S.~Seifnashri, and S.-H. Shao, \emph{{Higher Gauging and
  Non-invertible Condensation Defects}},
  \href{http://arxiv.org/abs/2204.02407}{{\arxivfont arXiv:2204.02407
  [hep-th]}}.

\bibitem[SNTWZ25]{Schafer-Nameki:2025fiy}
S.~Schafer-Nameki, A.~Tiwari, A.~Warman, and C.~Zhang, \emph{{SymTFT Approach
  for Mixed States with Non-Invertible Symmetries}},
  \href{http://arxiv.org/abs/2507.05350}{{\arxivfont arXiv:2507.05350
  [quant-ph]}}.

\bibitem[Tac17]{Tachikawa:2017gyf}
Y.~Tachikawa, \emph{{On gauging finite subgroups}},
  \doihref{http://dx.doi.org/10.21468/SciPostPhys.8.1.015}{SciPost Phys.
  \textbf{8} (2020) 015}, \href{http://arxiv.org/abs/1712.09542}{{\arxivfont
  arXiv:1712.09542 [hep-th]}}.

\bibitem[THF15]{2015arXiv150306812T}
J.~C.~Y. {Teo}, T.~L. {Hughes}, and E.~{Fradkin}, \emph{{Theory of Twist
  Liquids: Gauging an Anyonic Symmetry}}, arXiv e-prints (2015)
  arXiv:1503.06812, \href{http://arxiv.org/abs/1503.06812}{{\arxivfont
  arXiv:1503.06812 [cond-mat.str-el]}}.

\bibitem[TW19]{Thorngren:2019iar}
R.~Thorngren and Y.~Wang, \emph{{Fusion Category Symmetry I: Anomaly In-Flow
  and Gapped Phases}}, \href{http://arxiv.org/abs/1912.02817}{{\arxivfont
  arXiv:1912.02817 [hep-th]}}.

\bibitem[TW21]{Thorngren:2021yso}
\bysame, \emph{{Fusion Category Symmetry II: Categoriosities at $c$ = 1 and
  Beyond}}, \href{http://arxiv.org/abs/2106.12577}{{\arxivfont arXiv:2106.12577
  [hep-th]}}.

\bibitem[TY98]{tambara1998tensor}
D.~Tambara and S.~Yamagami, \emph{Tensor categories with fusion rules of
  self-duality for finite abelian groups}, Journal of Algebra \textbf{209}
  (1998) 692--707.

\bibitem[Ver88]{verlinde1988fusion}
E.~Verlinde, \emph{Fusion rules and modular transformations in 2d conformal
  field theory}, Nuclear Physics B \textbf{300} (1988) 360--376.

\bibitem[Wit03]{Witten:2003ya}
E.~Witten, \emph{{SL(2,Z) action on three-dimensional conformal field theories
  with Abelian symmetry}}, {From Fields to Strings: Circumnavigating
  Theoretical Physics: A Conference in Tribute to Ian Kogan}, 7 2003,
  pp.~1173--1200. \href{http://arxiv.org/abs/hep-th/0307041}{{\arxivfont
  arXiv:hep-th/0307041}}.

\bibitem[WY21]{Wang:2021vki}
J.~Wang and Y.-Z. You, \emph{{Gauge Enhanced Quantum Criticality Between Grand
  Unifications: Categorical Higher Symmetry Retraction}},
  \href{http://arxiv.org/abs/2111.10369}{{\arxivfont arXiv:2111.10369
  [hep-th]}}.

\bibitem[Yam02]{yamagami2002group}
S.~Yamagami, \emph{Group symmetry in tensor categories and duality for
  orbifolds}, Journal of Pure and Applied Algebra \textbf{167} (2002) 83--128.

\end{thebibliography}

\end{document}